# Terrestrial Analogs to Titan for Geophysical Research

Conor A. Nixon[1], Samuel Birch[2], Audrey Chatain[3], Charles Cockell[4], Kendra K. Farnsworth[1, 5], Peter M. Higgins[6], Stéphane Le Mouélic[7], Rosaly M.C. Lopes[8], Michael J. Malaska[9], Mohit Melwani Daswani[8], Kelly E. Miller[10], Catherine D. Neish[11], Olaf G. Podlaha[12], Jani Radebaugh[13], Lauren R. Schurmeier[14], Ashley Schoenfeld[8], Krista M. Soderlund[15], Anezina Solomonidou[16], Christophe Sotin[7], Nicholas A. Teanby[17], Tetsuya Tokano[18], Steven D. Vance[8]

[1] Solar System Exploration Division, NASA Goddard Space Flight Center, Greenbelt, MD 20771, USA
[2] Department of Earth, Environmental, and Planetary Sciences, Brown University, Providence, RI 02912, USA
[3] LATMOS/IPSL, UVSQ Université Paris-Saclay, Sorbonne Université, CNRS, Guyancourt, France
[4] School of Physics and Astronomy, University of Edinburgh, Scotland, UK
[5] University of Maryland, Baltimore County, MD, 21250 USA
[6] Department of Earth and Planetary Sciences, Harvard University, Cambridge, MA, USA
[7] Nantes Université, Univ. Angers, CNRS, UMR 6112, Laboratoire de Planétologie et Géosciences, F-44000, Nantes, France
[8] Jet Propulsion Laboratory, California Institute of Technology, CA 91109, USA
[9] Blue Marble Space Institute of Science, 600 1st Avenue, 1st Floor, Seattle, Washington 98104, USA
[10] Southwest Research Institute, San Antonio, TX 78238, USA,
[11] Department of Earth Sciences, University of Western Ontario, London, Canada
[12] Institute for Earth Sciences, Ruhr-University, Bochum, Germany
[13] Department of Geological Sciences, Brigham Young University, Provo, UT, USA
[14] Hawaiʻi Institute of Geophysics and Planetology, University of Hawaiʻi at Mānoa, Honolulu, HI 96822, USA
[15] Institute for Geophysics, Jackson School of Geosciences, The University of Texas at Austin, Austin, TX, USA
[16] Hellenic Space Center, 178 Kifisias Avenue, Chalandri 15231, Athens, Greece
[17] School of Earth Sciences, University of Bristol, Bristol, BS8 1RJ, UK
[18] Institut für Geophysik und Meteorologie, Universität zu Köln, 50923 Köln, Germany

Corresponding author: Conor A. Nixon (conor.a.nixon@nasa.gov)

## Key Points

- Titan and Earth are the only two worlds in the solar system today to exhibit processes such as rainfall, rivers and surface seas.
- The many parallels between Earth and Titan are yet to be fully exploited scientifically to learn how geophysical processes operate.
- With few missions to Titan, we can gain insights about Titan and Earth by greater use of field analog research alongside other techniques.

## Abstract

Saturn's moon Titan exhibits remarkable parallels to the Earth in many geophysical and geological processes not found elsewhere in the solar system at the present day. These include a nitrogen atmosphere with a condensible gas - methane - replacing the Earth's water, leading to an active meteorology with rainfall and surface manifestations including rivers, lakes and seas, and the dissolution of karstic terrain. Other phenomena such as craters, dunes, and tectonic features are found elsewhere - e.g. on Mars and Venus - but their continuing alteration by pluvial, fluvial and lacustrine processes can be studied only on Earth and Titan. Meanwhile Titan also hosts an interior liquid water ocean with similarities to the Earth as well as to ocean worlds such as Europa and Enceladus. Our focus in this review paper is twofold: to describe the geophysical and geological parallels between Earth and Titan, and to evaluate the yet-underexploited possibilities for field analog research to gain new knowledge about these processes. To date, Titan's much colder temperature and different atmospheric and crustal materials have led to a skepticism that useful analogs can be found on Earth. Our conclusion, however, is that a much larger range of useful analog field work is possible and this work will substantially enhance our knowledge of both worlds. Such investigation will supplement the existing sparse data for Titan returned by space missions, will greatly enhance our understanding of such datasets, and will help to provide science impetus and goals for future missions.

## Plain Language Summary

Saturn's largest moon, Titan, is not like the other moons of the solar system, and in some ways more closely resembles the Earth. Like Earth, Titan has a dense nitrogen atmosphere, with the condensible gas methane acting similarly to water on the Earth, leading to meteorological processes. In this review paper we describe the many similar landforms between Earth and Titan, such as rivers, lakes, craters, dunes and mountains. We focus our attention specifically on the potential for field work on Earth to learn more about both worlds through investigation of analog sites - or locations of specific interest due to exhibiting close similarities in geophysical properties. Our conclusion is that much greater scope for such work exists than has currently been realized, leading to many new opportunities for scientific investigation. Field analog work is particularly valuable in the case of Titan, which is distant and visited by few space missions directly, and will be particularly important for planning measurements and interpreting results from NASA's forthcoming Dragonfly mission expected to reach Titan in 2034.

## 1.0 Introduction

The term ‘terrestrial planets’ is commonly used to refer to the four inner ‘rocky’ planets of the solar system - Mercury, Venus, Earth and Mars - to distinguish them from the outer ‘gas’ planets - Jupiter, Saturn, Uranus and Neptune that lack a well defined solid surface. These general categories mask some notable differences between their constituent members - e.g. that Mercury doesn’t have a dense atmosphere or weather (e.g. clouds), unlike Venus, Earth and Mars; or that Uranus and Neptune are different enough from Jupiter and Saturn to arguably form a separate sub-class of gas planets, and are therefore sometimes referred to as the ‘ice giants’. These canonical categories also ignore some strong similarities with smaller bodies that perhaps *should* be included in these classes - e.g. that Earth’s Moon has substantial enough similarities in appearance to Mercury that the two bodies can easily be mistaken for each other by a non-specialist. Or, that Saturn’s moon Titan is in some respects more similar to Earth, Venus and Mars than to the other moons of the solar system, by virtue of having a dense atmosphere with weather, including clouds and rain, leading to surface features such as rivers and lakes.

In this review paper we explore the geophysical parallels between Earth and Titan in more detail, including both the general processes occurring, and also specific locations on both planets where examples of these occurrences can be found - known as ‘field analogs’ (Hodges and Schmitt 2011). This leads us to our second topic: the potential for analog field measurements on Earth, to learn more about similar processes on Titan, and in the wider sense how these processes operate across terrestrial planets in general ('comparative planetology').

The ordering of the paper is as follows: in the remainder of Section 1 we present some important background and context information, on Titan's global characteristics (1.1) and on previous use of field investigations for comparative planetology (1.2). Then in Section 2 we proceed to examine the principal atmospheric, surface, and subsurface terrains in greater detail, to evaluate potential Earth analog sites for these locations, and to discuss field investigations and measurements that could be carried out. In Section 3 we discuss limitations to terrestrial analog work for Titan, while in Section 4 we offer our summary, conclusions, and recommendations for further work. A Glossary of some less commonly encountered geophysical terms is included at the end that may be referred to as needed.

## 1.1 Global Characteristics of Titan

Titan is the largest moon of Saturn, with an orbital period of 383 hours (~16 Earth days) (Sinclair 1977, Jacobson 2022). Since Titan is tidally locked to Saturn, it has leading and trailing hemispheres, as well as Saturn-facing and anti-Saturn hemispheres, and one orbit is one Titan 'day'. Titan is a 'regular' moon, orbiting counterclockwise as seen from the north celestial pole, with an orbit that is close to Saturn's ring plane, and also no axial tilt (obliquity) relative to its orbit. However, since Saturn itself has a 26.73° obliquity and orbits the Sun in 29.5 Earth years, Titan also inherits this annual cycle, with seasons ~7.4 terrestrial years in length. The Cassini-Huygens mission (Matson et al.

2002; Lebreton et al. 2002) arrived at the Saturn system in June 2004 during southern summer, with Saturn and Titan reaching northern spring equinox in August 2009 and northern summer solstice in June 2017, just before the end of the mission in September that year.

Titan has a diameter of 5150 km, and little oblateness, with a difference between the polar and equatorial radius of 410 m. Titan's bulk density is 1.881 g/cm$^3$ (Iess et al. 2010), indicating an interior composition of approximately equal parts of rock, liquid water ("interior ocean") and an ice crust. Modeling of Titan's gravity field indicates that it has largely differentiated, with an outer hydrosphere overlying a rocky core (Iess et al. 2012), although the amount of volatiles that may still be trapped in the core is currently a matter of debate (e.g., Glein et al. 2015; Miller et al. 2019). Heating from the mantle leads to a liquid water ocean below an outer icy crust and, potentially, below the ocean is a layer of high pressure ices that separates the ocean from the rocky core (Kalousova and Sotin 2018; Journaux et al. 2020).

The composition of Titan's crust has been proposed to consist (from top to bottom or from exterior to interior) of a porous icy crust partially covered with fluvially derived sediments of a variety of sizes, organic in composition (Barnes et al., 2015). This outermost region may consist of layered and fractured deposits (Barnes et al. 2015; Radebaugh et al. 2011) and is overlying a liquid hydrocarbon saturated, porous icy crust. Below this layer a potentially vertically expanding hydrocarbon clathrate porous icy crust layer has been proposed, which itself again is overlying a non-porous icy crust (Tobie et al. 2006).

The most distinctive characteristic of Titan as a moon is that it possesses a dense atmosphere, approximately 95-98% nitrogen and 5-2% methane (Niemann et al. 2005, 2010). The atmosphere above 600 km is ionized by solar Extreme UV on the dayside ($5\times10^{10}$ eV cm$^{-2}$ s$^{-1}$), superimposed on both day and night sides by the deposition of electrons, protons and ions precipitating from the magnetosphere of Saturn (electrons: ~$2.3\times10^{9}$ eV cm$^{-2}$ s$^{-1}$, others: 1 to 23 $\times10^{8}$ eV cm$^{-2}$ s$^{-1}$ depending on solar activity; Gronoff et al., 2009). Galactic Cosmic Rays are more penetrating and deposit at ~65 km, creating a secondary lower ionosphere (~$1.4\times10^{9}$ eV cm$^{-2}$ s$^{-1}$ ; Gronoff et al., 2009). This deposition of energy in the nitrogen-methane atmosphere leads to complex photochemical and ion reactions producing a plethora of organic molecules and haze particles (e.g Yung et al. 1984), which then sediment to the surface (with a deposition rate of ~$2\times10^{-13}$ g cm$^{-2}$ s$^{-1}$ for liquid hydrocarbons, ~$5\times10^{-14}$ g cm$^{-2}$ s$^{-1}$ for solid simple hydrocarbons, and ~$3\times10^{-14}$ g cm$^{-2}$ s$^{-1}$ for aerosols; Vuitton et al., 2019).

The presence of the atmosphere has a number of repercussions and manifestations on the cold (~93 K) surface (Jennings et al. 2009; 2011, 2016; 2019), including seas/lakes of condensed methane/ethane/nitrogen (Stofan et al. 2006; Lopes et al. 2007; Hayes 2016) in the polar regions. There are also extensive surface coverings, including: dune fields with solid sands that may originate either (i) from atmospherically derived organic molecules (Barnes et al. 2015); (ii) from erosion of Titan's crust (Barnes et al. 2015) or

(iii) are externally derived (Bottke et al. 2024); vast plains that appear fairly uniform (Lopes et al., 2016); and fluvially incised terrains known as labyrinth terrains (Malaska et al. 2020). The major geological units and their distribution are discussed in Lopes et al. (2020).

## 1.2 Field Analog Research

Missions that land on planets, moons, and small bodies throughout the Solar System provide the ultimate 'ground-truth' information with which to test our hypotheses about the formation and evolution of planetary bodies and to search for life outside of the Earth's ecosystems. However, such missions typically come with enormous costs, complexity and time delay to receive data. While waiting for new missions to be developed, planetary research proceeds through techniques such as modeling, remote observation, and laboratory experiments. A fourth and increasingly used approach in recent decades is *field analog* research: the sampling and analysis of natural sites on the Earth that in some way mimic the environments or processes that occur on other Solar System bodies (Hodges and Schmitt 2011). Field analog sites used by planetary scientists to date have included deserts, ocean depths, icy environs, caves, volcanoes, and more (Preston and Dartnell 2014). While remaining imperfect replicas of actual planetary environments, such field environs nevertheless provide important means to test experimental techniques and prove instruments, and to gather data about biotic and abiotic processes that occur in extreme environments, expanding the parameter space of our knowledge towards the greater and more extreme limits that occur on other worlds.

Field research has existed since at least the time of the Scientific Revolution (1500s), when scientifically motivated explorers began to investigate the most extreme climates on the Earth: to better understand the origins and limits of life, and the intricacies of geophysical processes. However, analog fieldwork and research as a means of learning about other planetary bodies became a distinct field with the beginning of planetary exploration during the early years of the space race.

*Lunar analogs*: In the 1960s, Apollo astronauts were being trained in geological techniques relevant for lunar exploration by field work on Earth, including sites such as Marathon Basin of West Texas (Dasch 2016), Sudbury Crater in Canada (Lofgren et al. 2011), Ries Crater in Germany (Sauro et al. 2022), the Pinacates volcanic field of northern Mexico (Darack 2018), and Iceland (Skjöld et al. 2024). In northern Arizona astronauts trained at multiple sites, including Meteor Crater and Cinder Lakes crater field, a field made up of debris from Sunset Crater where artificial craters were made to mimic an area within Mare Tranquillitatis (Vaughan et al. 2019, see **Fig. 1**). However, beyond its pedagogical purpose, this work began to generate new geological investigations into the parallels and differences between terrestrial and lunar rocks, leading to new ideas about the origin and development of the Earth-Moon system (Haliday 2008; Paniello et al. 2012; Borg and Carlson 2023). In particular, the work at Meteor Crater shifted the paradigm away from volcanoes to impacts as the mechanism for forming all the circular craters we see on the Moon (Shoemaker 1962).

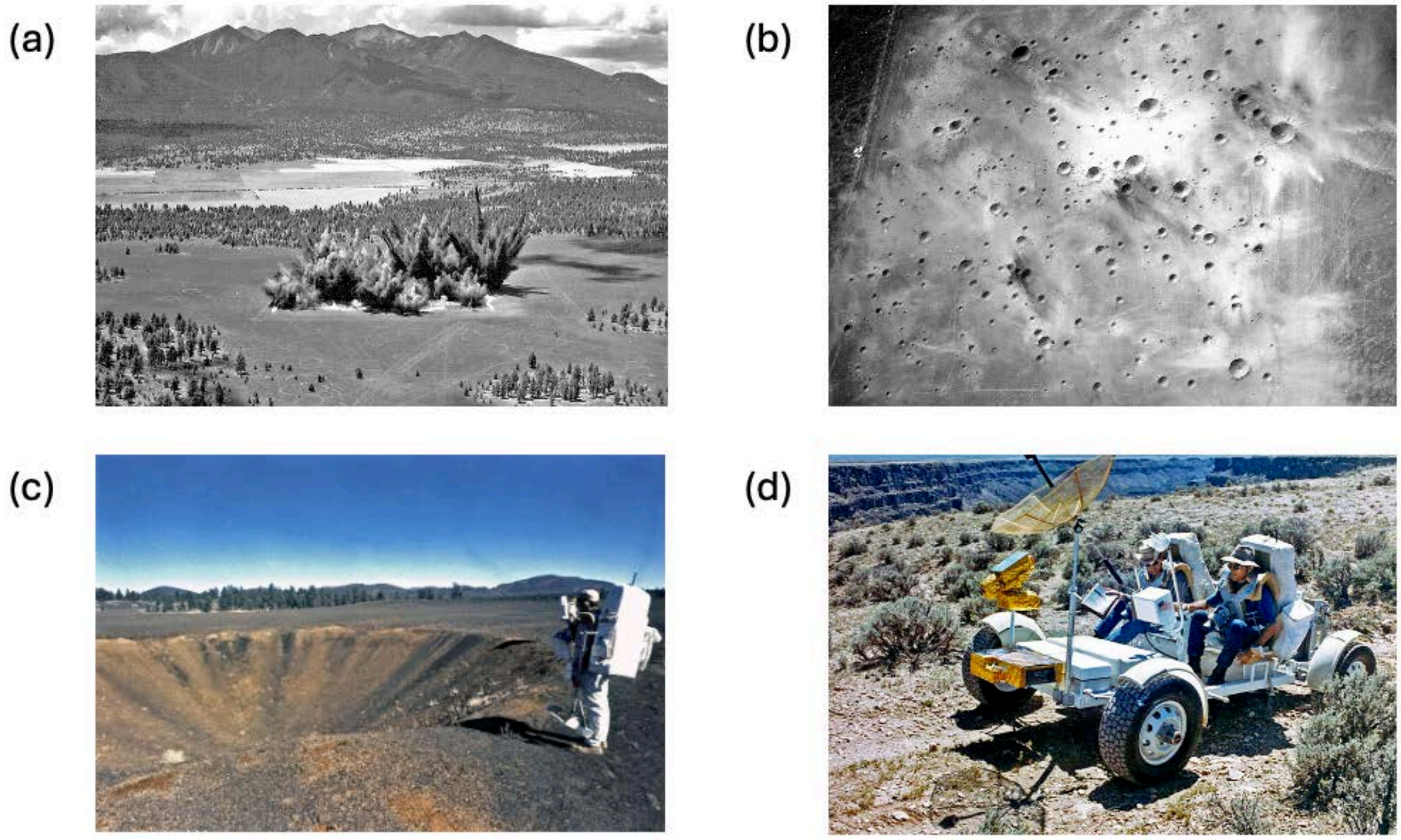

**Figure 1**: (a) Creation of Crater Field #2 at Crater Lake, AZ on July 27, 1968. A total of 354 craters were generated by explosives. (b) Aerial view of the completed Cinder Lake Crater Field #2, showing very light ejecta caused by excavation of clay beds immediately below black basaltic cinders. (c) Apollo 17 astronauts Harrison Schmidt and Gene Cernan view a crater at the Cinder Lake Training Field. (d) Schmidt and Cernan driving the 'Grover' lunar rover prototype in AZ. All images: USGS. See Schaber (2005) and Data Availability for details.

*Mars and Venus*: On the heels of this lunar research, as space agencies turned their eyes to Mars in the 1970s, field work followed as it became clear that the surfaces of Mars and Venus showed aeolian processes common to Earth in addition to volcanism and impact cratering observed on the Moon. Active dust storms, along with dunes and other landforms related to wind, were observed on Mars (Masursky et al. 1972). Wind measurements on the surface of Venus suggested aeolian landforms might exist there, even before the surface was imaged by Magellan's radar and dunes were found there (Avduevskii et al. 1977). Even in the 1970s there were suggestions that since Titan has an atmosphere, it might have aeolian processes as well (Danielson et al. 1973; Hunten 1977; Atreya et al. 1978). The Planetary Geology Field Conference on Aeolian Processes brought scientists together to the Mojave desert to exchange ideas on aeolian and other processes. This meeting resulted in a Comparative Planetary Geology Guidebook, one of the first and more comprehensive guidebooks on this topic (Greeley et al. 1978).

After Mariner 9 and Viking showed large channels on Mars (McCauley et al. 1972, Blasius et al. 1977) and debate started about their formation, field work provided evidence that the channels were formed by large quantities of running water. A field conference on the Channeled Scablands occurred in 1978 to discuss the hydrodynamics of high-velocity flood erosion and morphological comparisons with Martian features (Baker and Nummedal, 1978). Mariner 9 and Viking images also revealed enormous volcanoes on Mars, which appeared to be basaltic, leading to comparative field studies in Hawaii (e.g. Carr and Greeley, 1980) and other volcanic terrains such as Santorini, Greece (Pantazidis et al. 2019). Mars analog field work has continued to be a thriving field (Hipkin et al. 2013; Hays et al. 2017; Preston and Dartnell 2014), and includes locations such as the Dry Valleys of Antarctica (Horowitz et al. 1972; Marchant and Head 2007;  Heldmann et al. 2007; Dartnell et al. 2010), the Atacama Desert of northern Chile (Navarro-González et al. 2003; Ruff et al. 2016), the Altiplano-Puna plateau of Argentina (Bridges et al. 2015), Svalbard island in the high Arctic (Steele et al. 2007; Hausrath et al. 2008; Ulrich et al. 2011), Sudbury Crater in Canada (Lofgren et al. 2011), and many other cold or arid environments. See **Fig. 2**.

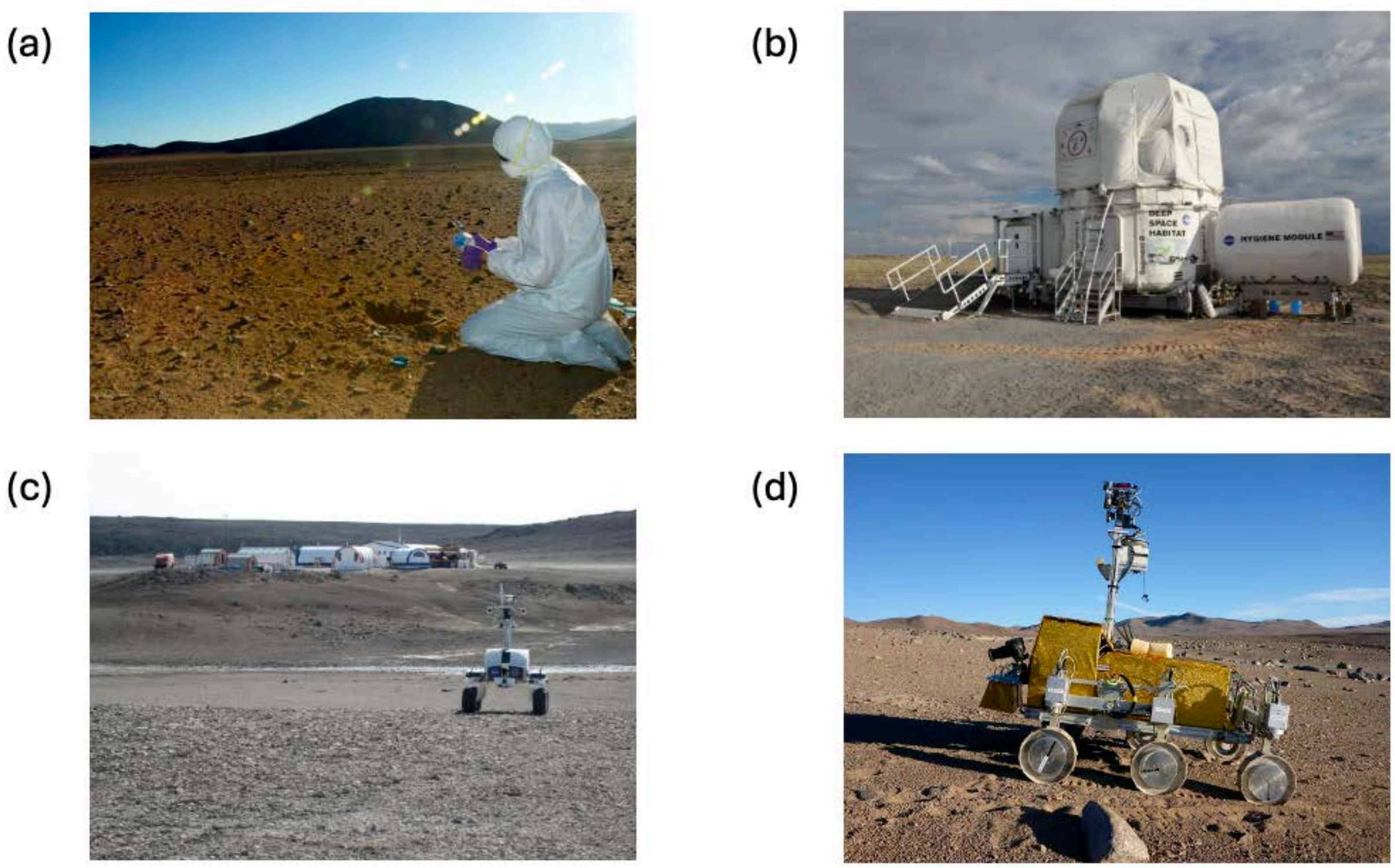

**Figure 2:** (a) NASA scientist Mary Beth Wilhelm, wearing a cleanroom suit, collects soil samples containing microorganisms from Chile’s Atacama Desert. (b) The 2011 version of the deep space habitat at the Desert Research and Technology Studies (Desert RATS) analog field test, AZ. (c) The K-10 ‘Red’ rover in Haughton Crater, Devon Island, Nunavut, in the Canadian High Arctic. (d) The Bridget rover taking part in the five-day SAFER field trial in the Atacama Desert, fitted with prototype ExoMars instruments. All NASA images except (d) ESA.

*Giant Planet Satellites*: Exploration of the moons of Jupiter and Saturn by Voyager, Galileo, and Cassini opened up the outer Solar System to comparative field studies. The vivid colors of volcanic Io shown by Voyager led to the suggestion that sulfur volcanism was taking place there (Sagan 1979). Although sulfur was later found to be a minor compositional component for Io effusive volcanism in general, it led to studies of rare field sites where secondary sulfur volcanism occurred such as near the summit of Mauna Loa in Hawaii (Greeley et al., 1984). Towards the end of the Galileo mission, high resolution images of Io revealed that several of its volcanic paterae (calderas) house lava lakes (Lopes et al., 2004), a conclusion based on observations of lava lakes on Earth, characterized in the infrared by hot edges where the lava lake crust breaks against the caldera wall. Field studies have been conducted at active lava lakes and lava flows in the interest of understanding fine-scale lava thermal properties expressed in spacecraft imagery of Io (Radebaugh et al. 2016; Lopes et al. 2018).

The discovery of thriving ecosystems in Earth's deep ocean (including at crustal faults such as mid-ocean ridges (Kelley et al. 2005)) has revealed chemically driven habitats with biological cycling in oceans underneath ice shelves (Martínez-Pérez et al. 2022). The existence of these unique environments, combined with discoveries about oceans inside the moons of the outer solar system (e.g., Nimmo and Pappalardo 2016) has led to a new line of analog research aimed at the search for life in ocean worlds (Lorenz et al. 2011, Schmidt 2018, 2020; Paton 2024).

*Titan*: One of the most exotic and singular worlds in the solar system is Saturn's moon Titan, with its dense organic-rich atmosphere (Hanel et al. 1981, Coustenis et al. 1989, Coustenis et al. 2007), surface hydrocarbon seas and dunes (Hayes 2016, Stofan et al. 2007, Lorenz et al. 2006), icy crust (Lopes et al. 2020), complex geological surface processes, and deep subsurface water ocean (Iess et al. 2012, Durante et al. 2019). Finding appropriate analog sites for such a cold, anoxic, but chemically rich environment is much more difficult than for a rocky inner planet such as Mars, with its more direct similarities to the polar regions of the Earth. For this reason, Titan field analog research has received generally much less attention than that for Moon and Mars. However, following the geophysical revelations of the Cassini-Huygens mission, Titan analog field work is starting to receive interest, with recent research into dunes (Lorenz et al. 2006; Radebaugh et al. 2010, **Fig. 3)**, lakes (e.g. Cornet et al. 2012; Lorenz et al. 2010; Harrison 2012; Sharma and Byrne 2011), rivers (Burr et al. 2013), alluvial fans (Radebaugh et al. 2018) and impact craters (Neish et al. 2018, Perkins et al. 2023). Excellent work in these areas notwithstanding, it is clear that many more Titan analog investigations are possible than currently take place, and the subject of this article is to systematically consider what characteristics of Titan may be replicated on the Earth and where these may be found.

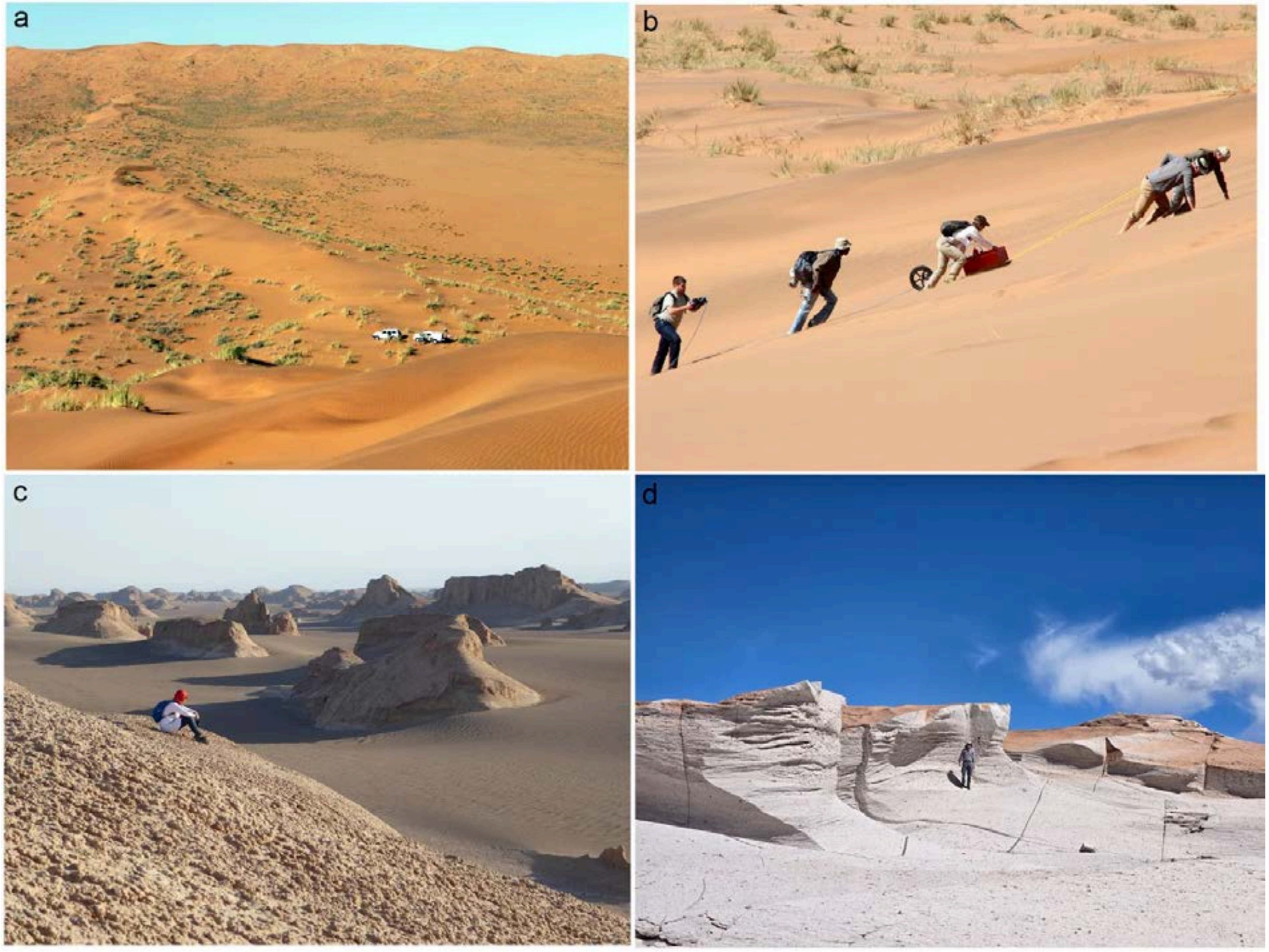

**Figure 3**: a. Dune of the Namib Sand Sea and crossing interdune. Note the trucks at bottom for scale. These are similar in size to the dunes of Titan. b. Field research of a dune crest using a Ground Penetrating Radar (GPR). c. Yardangs of the Lut desert, Iran. These are disjointed and eroded from mass wasting near the field margins. The surface is rough clay. d. Yardangs of the Argentine Puna, made of volcanic ash. Features are smaller here, and wind sculpting is visible. Images by authors; see also Chandler et al. (2015).

## 2.0 Titan Environments and Terrestrial Analogs

As described above in Section 1.1, Titan is a complex, diverse world with many unique properties on the global scale. We now proceed to describe how the global properties are manifested on a local scale, including lakes and seas, mountains and craters, rivers, labyrinth terrains, and more. **Fig. 4** shows a schematic diagram of the principal Titan environmental types for which we will later identify terrestrial analogs, including: six surface manifestations, two sub-surface manifestations, and for completeness, the atmosphere. **Fig. 5** shows example locations on Titan manifesting the principal surface terrain types shown in **Fig. 4**, superimposed on a recent VIMS-ISS mosaic of Titan (Le Mouélic et al., 2019; Seignovert et al. 2019).

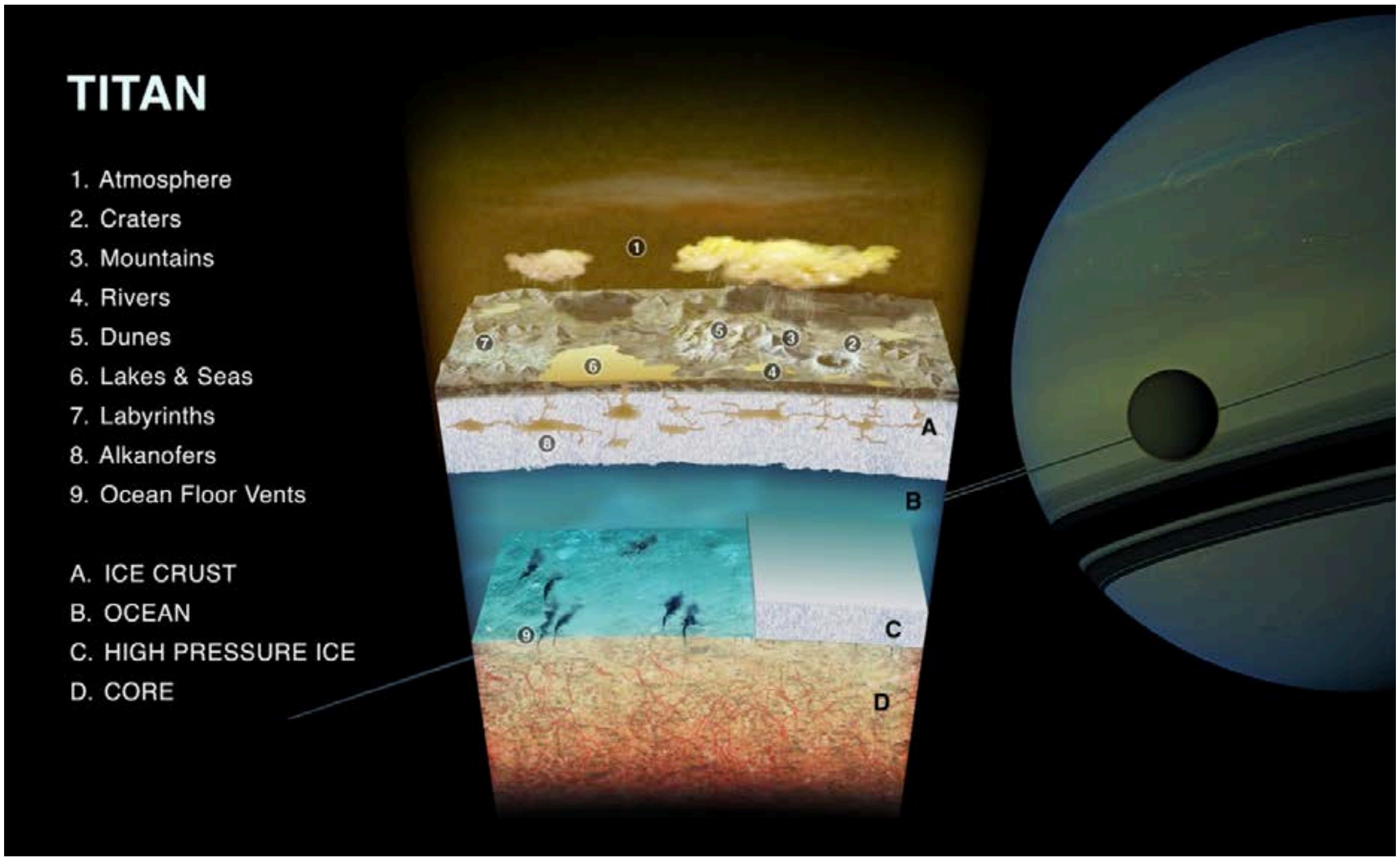

**Figure 4**: Cross section of Titan atmosphere and interior, showing diverse environments on Titan for which terrestrial analogs exist (NASA/Theophilus B. Griswold).

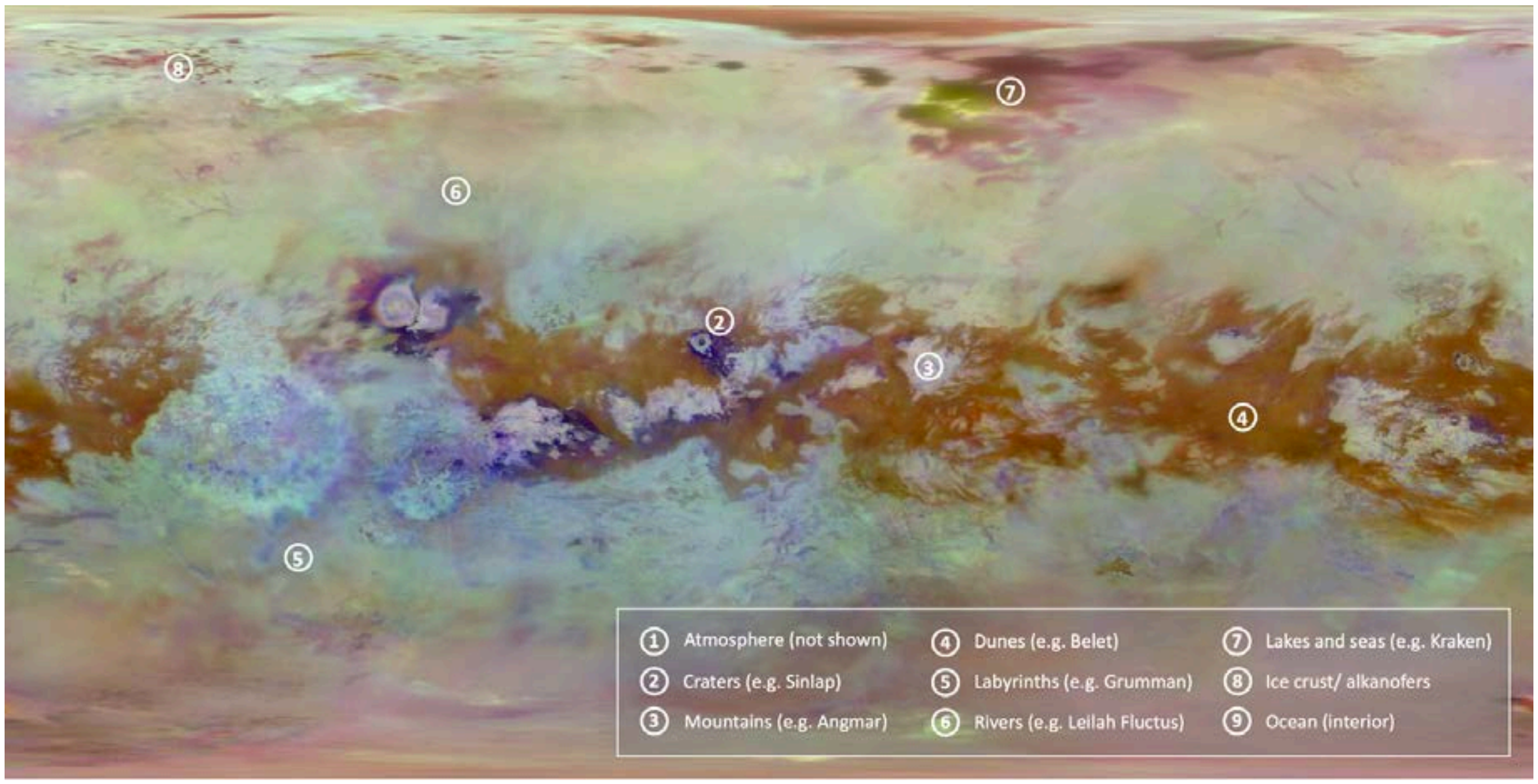

**Figure 5:** Example locations of various environmental types shown in Fig. 4. Titan VIMS-ISS mosaic from Seignovert et al. (2019) using the VIMS color ratio map described in Le Mouélic et al. (2019).

**Table 1** summarizes each of the major environments in detail, and for each case, describes potential terrestrial analogs.

| **Table 1**: Summary of Titan-Earth Environmental Analogs Considered | | |
|---|---|---|
| | **Earth feature** | **Titan feature** |

| **Table 1**: Summary of Titan-Earth Environmental Analogs Considered | | |
|---|---|---|
| **Atmosphere** | Nitrogen-dominated oxygen-rich atmosphere, water as condensable species. Weather, climate, radiative processes and chemistry. | Nitrogen-dominated reducing atmosphere, methane as condensable species. Weather, climate, radiative processes and chemistry. |
| **Craters** | Craters formed by hypervelocity impact into silicate substrate, and modified by erosion and tectonic processes. | Craters formed by hypervelocity impact into icy substrate, and modified by erosion and viscous relaxation. |
| **Surface icy materials (mountains, craters)** | Tectonically lifted silicate crust bedrock exposures. | Water ice crust bedrock exposures. |
| **Surface dunes and solid hydrocarbons** | Loose granular materials piled into dunes by eolian activity. Granular materials are relatively resistant to physical erosion depending on transport distance. Composition includes quartz, gypsum, and ice. | Loose granular materials piled into dunes by eolian activity. Granular materials are relatively resistant to physical erosion depending on transport distance. Composition is likely organic materials. |
| **Labyrinths** | Dissected plateaux from fluvial, fluviokarstic, or karstic processes, driven by liquid water dissolving or eroding uplifted plateau materials. | Dissected plateaux from fluvial, fluviokarstic, or karstic processes driven by liquid hydrocarbons dissolving or eroding uplifted organic plateau materials. |

| **Table 1**: Summary of Titan-Earth Environmental Analogs Considered | | |
|---|---|---|
| **Rivers** | Liquid water, flowing over (and transporting) silicate rocks. Also supraglacial rivers over ice sheets. | Hydrocarbon rivers flowing over (and transporting) organic materials and/or water ice. |
| **Surface lakes and seas** | Liquid water, filling depressions or inset into a mantled karstic plain. | Hydrocarbon liquids, filling depressions or inset into a mantled karstic plain. |
| **Alkanofers** | Porous & permeable (several orders of magnitude) siliceous or carbonate sedimentary strata deposited in various terrestrial or marine environments, containing thermally or biogenically generated complex hydrocarbon mixtures. Laterally and vertically confined by geological structures or stratal facies changes of variable permeability. Involved in marine (seabottom) and terrestrial lacustrine (liquid table) processes. | Porous & permeable organic or water ice strata, containing pore-filling liquid hydrocarbon components such as nitrogen/methane, ethane, propane, and 1-butene. Nitrogen/methane is a cycling volatile analogous to terrestrial water. Likely laterally and vertically constrained by impermeable layers (equivalent to terrestrial aquitards). Involved in karstic and liquid table (lacustrine) processes. |
| **Interior ocean** | Briny water ocean covering ~70% of planetary surface; in contact with atmosphere above and silicate crust below. | Interior, global water ocean of uncertain composition; in contact with water ice/clathrate layer above and |

| Table 1: Summary of Titan-Earth Environmental Analogs Considered | | |
|---|---|---|
| | | either high pressure ice or rocky material at lower boundary. |

These environments are now considered in turn.

## 2.1 Atmosphere

Although not a typical 'site' of field/analog study, the Earth's atmosphere is much more accessible than Titan's, and is host to many similar processes. Means of study include direct measurements of the lower atmosphere by meteorological stations, aircraft or balloon, of the middle atmosphere by balloon, and the upper atmosphere by rockets; as well as direct and indirect measurements of these layers by remote sensing (weather satellites), radar experiments and others. For this reason, we include the Earth's atmosphere as a 'terrestrial analog site' for Titan in this paper.

*Titan*: The atmosphere of Titan has several physical, compositional and meteorological similarities to the Earth including surface pressure (1.46 bar vs 1.01 bar on the Earth), composition (both nitrogen dominated) and a hydrological cycle (phase change of methane on Titan, water on Earth) (Hörst 2017). Akin to the atmospheric structure of Earth, Titan possesses distinct atmospheric layers (Nixon 2024). In the troposphere, convection takes place due to heating of the atmosphere from the surface and the temperature decreases near-adiabatically with altitude. Meanwhile, in the stratosphere, the temperature increases with altitude due to strong absorption of solar radiation by the organic haze, which plays a similar role to the ozone layer and oxygen in Earth's stratosphere for the radiative balance (McKay et al. 1991). In the mesosphere, the temperature decreases with altitude due to radiative cooling. The thermosphere (non-collisional regime, high temperatures), ionosphere (plasma dominated), and exosphere (escaping regime) are observed at higher altitudes (**Fig. 6**).

However, significant differences must not be discounted. The surface temperature on Titan is much lower than the Earth (~93 K vs ~293 K, Fulchignoni et al. 2005). Titan's atmosphere lacks large amounts of molecular oxygen, which dramatically changes the chemistry (Vuitton et al. 2014). Titan's annual cycle is 30 times slower and its diurnal cycle is 16 times slower than Earth's. The slow planetary rotation leads to significantly different atmospheric dynamics as the Coriolis force is substantially weaker than on Earth (Tokano 2011; Mitchell and Lora 2016). This places Titan's atmosphere in the cyclostrophic instead of geostrophic dynamical regime. Precipitation patterns are also different on Titan. For instance, there are no substantial mid-latitude weather systems but the tropical rainbelt approaches the polar region in summer.

The low temperatures cause extremely long radiative time constants in Titan's lower stratosphere and troposphere (Achterberg et al 2011; Bézard et al 2018), which results in an equator-to-pole temperature variation of a few K at most (Cottini et al. 2012; Jennings et al. 2009, 2011, 2016, 2019). In addition, Titan has multiple condensable gases in the stratosphere including the hydrocarbons ethane, acetylene, propane, and other materials that may lead to either cloud layers of different compositions (Sagan and Thompson 1984; Coustenis et al. 1999), and/or ices and droplets that are co-condensed from multiple substances (Anderson et al. 2018).

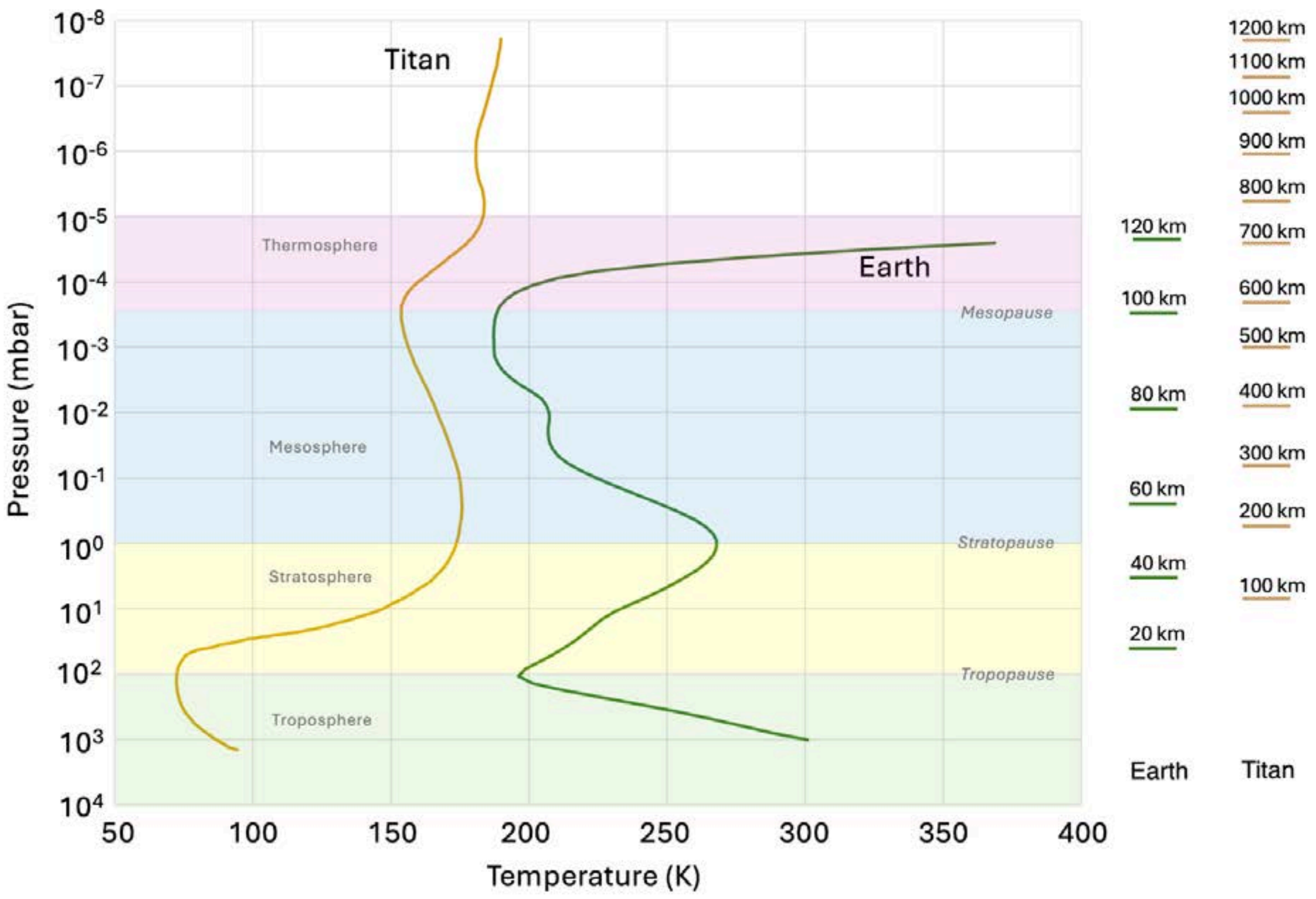

**Figure 6**: Comparison of typical Earth and Titan atmosphere temperature profiles. Although Titan's atmosphere is significantly colder, both have similar layers and transitions (inflection points) at similar pressure levels, due to the physics of atmospheric radiation and thermodynamics. Titan's atmosphere however is much more extended than the Earth's atmosphere due to the lower gravity (1.35 $m/s^2$ vs 9.81 $m/s^2$).

Interestingly, the temperatures at ~$10^{-4}$-$10^{-3}$ mbar are similar on both Earth and Titan (Müller-Wodarg et al 2008 - Fig. 6), which corresponds to the lower thermosphere and mesopause region. Some processes such as interaction with solar UV in this critical region may be more directly comparable than elsewhere in the atmosphere.

*Earth analogs:* An experimental comparison of layers in the terrestrial atmosphere to those on Titan (**Table 2**) may yield new insights into both worlds.

**Table 2**: Terrestrial analogs to Titan Atmosphere

| **Titan location(s)** | **Terrestrial Analog Site(s)** | **Analog Process Type** | **Measurement** | **Notes/Limitations** |
|---|---|---|---|---|
| Exosphere | Exosphere | Escape | Molecular and atomic fluxes | Titan has no magnetic field. |
| Ionosphere | Ionosphere | Ion chemistry | Ions, energetics | In addition to Solar UV Saturn's magnetospheric electrons are also important on Titan. |
| Thermosphere | Thermosphere | Non-LTE processes | Molecules, temperature | Constraints are currently sparse on these processes on both planets. |
| Mesosphere | Mesosphere | Cooling processes | Molecules, temperature | Different molecules involved in radiative cooling processes on Titan compared to the Earth. |
| Stratosphere | Stratosphere | Condensation process, ozone layer | Ice clouds, organic haze | The major heating source on Earth is solar absorption of UV by $O_2$ and $O_3$, whereas on Titan haze takes this role. |
| Troposphere | Troposphere | Meteorology | Humidity, clouds, rain | Different condensible species and different dynamical regime effects morphology of weather features. |

In the exosphere, escape processes on Titan may benefit from knowledge of these processes on the Earth. Non-thermal escape processes, such as sputtering by solar wind particles and ion pickup, are important on both Earth (Johnson 1994) and especially on Titan with its lower gravity (Lammer and Bauer 1991, 1993, Lammer et al. 1998, Shematovich et al. 2003, Johnson 2009, Gu et al. 2019, Hsu and Ip 2019, Snowden and Higgins 2021). Thermal escape, especially Jeans escape where atoms and molecules are lost from the high-energy tail of the Maxwellian, is an important loss source for light atoms such as H and He (Cui 2008, Tucker and Johnson 2009, Volkov et al. 2011) and magnetic pressure may add to this (Edberg et al.  2011). Hydrodynamic escape, when loss occurs to the slower molecules as well as they are pushed through the exobase by expansion of the atmosphere, may be occurring on Titan today (Yelle et al. 2008; 2012; Strobel 2008, 2009, Schaufelberger et al. 2012, Salem et al. 2017) along with impacts (Korycansky and Zahnle 2011, Schlichting and Mukhopadhyay 2018).

Hydrodynamic escape was likely to have been important on Earth in the past (Kasting 1993, Johnson et al. 2016, Wang et al. 2022) .

Although Titan and the Earth have nitrogen-dominated atmospheres , the trace gases in Titan's atmosphere (Coustenis et al. 1989, 2007) are quite different from those in Earth's atmosphere (Porcelli and Ballentine 2002, Goldstein and Galbally 2007). However, ion-phase processes initiated by sunlight and solar wind impacts are common to both (Torr and Torr 1979, Waite et al. 2005; Vuitton et al. 2007; Cui et al. 2009; Vuitton et al. 2009, Danilov 2012, Shuman et al. 2015). In the past, our understanding of the Earth's ionospheric chemistry has aided our predictions and data interpretations of Titan's atmosphere (Strobel 1979; Yung et al. 1984), while in future the reciprocal (Titan studies informing knowledge of Earth) may become important.

Lower down, in the thermosphere and mesosphere, radiative processes dominate on both the Earth and Titan (Fuller-Rowell and Rees 1980, Friedson and Yung 1984; Lellouch et al. 1990; Roble 1995; Müller-Wodarg et al. 2000; Koskinen et al. 2011, Fuller-Rowell 2014). While in the mesosphere it is rapid radiative cooling of molecules in LTE (local thermodynamic equilibrium) that predominates, in the thermosphere it is the radiatively pumped energy combined with relatively long radiative timescales in the absence of LTE that leads to elevated temperatures (Roble 1995; Müller-Wodarg et al. 2000). Understanding radiative processes for molecules in both atmospheres is therefore of mutual benefit.

The stratospheres of both the Earth and Titan show stable stratification and rising temperatures with increasing altitude (Flasar and Achterberg 2009; Butchart 2022) and are regions where cirrus/ice clouds are seen, especially over the winter pole (McCormick et al. 1982; Steele et al. 1983; Tabazadeh et al. 1994; Peter 1997; Griffith et al. 2006; Rodriguez et al., 2011; Turtle et al., 2018; Le Mouélic et al., 2018, Tritscher et al. 2021). However, the prevailing zonal winds are qualitatively different in Earth's and Titan's stratospheres. Earth's stratosphere is characterized by easterly (westward) winds in summer and westerly (eastward) winds in winter, since the Coriolis force acting on the equator-straddling Brewer-Dobson circulation determines the wind direction (e.g. Holton 2004). On the other hand, Titan's stratosphere is strongly super-rotating, i.e. the wind is westerly in both winter and summer hemispheres and presumably also in the equatorial region (Bird et al. 2005; Achterberg et al. 2008). The superrotation is likely related to Titan's slow rotation and the resulting cyclostrophic rather than geostrophic wind balance, yet the mechanisms giving rise to the atmospheric super-rotation on Titan are still an area of active research (Read and Lebonnois 2018; Imamura et al. 2020).

Cross-equatorial meridional over-turning also occurs, leading to polar concentrations of gases on Titan which are then confined by circumpolar jets (Teanby et al. 2008; Mitchell et al. 2021; Guendelman et al. 2022; Waugh 2023). Much of our understanding of Titan's polar dynamics and chemistry has been informed and aided by pre-existing knowledge of these processes on the Earth (Thompson 2008), and continued study of the stratosphere and polar atmosphere of the Earth is likely to lead to further insights.

Polar Stratospheric Clouds (PSCs) on Earth (Schoeberl and Hartmann 1991; McCormick et al. 1982; Tabazadeh et al. 1994, **Fig. 7**(d)) bear some similarities to hydrocarbon and nitrile ice clouds observed in Titan's polar winter (**Fig. 7**(c); de Kok et al 2014; Vinatier et al. 2018; Le Mouélic et al. 2012, 2018; Vinatier et al. 2018). These types of clouds have the effect of removing key species from the atmosphere - ethane, HCN and benzene ices have been proposed on Titan - but could also form sites for exotic heterogenous surface chemistry, allowing more complex chemical product synthesis than in the gas phase alone. Such processes are known to be critical in Earth's ozone destruction cycle (Wallace and Hobbs 2006) and there may be analogues for different chemistries occurring on Titan, such as dicyanoacetylene ($C_4N_2$) ice production (Anderson et al. 2016).

The tropospheres of both Titan and Earth are characterized by dynamic weather phenomena. Titan and Earth form a unique pair in the sense of having clouds leading to rainfall onto a solid surface (Rodriguez et al. 2011; Turtle et al. 2011a, 2018), albeit different molecules are involved ($H_2O$ on Earth and $CH_4$ on Titan) (Tokano et al. 2001; Lunine and Atreya 2008). On Earth $H_2O$ rainfall occurs in warm areas/seasons, while snowfall occurs in cold areas/seasons (Gu et al. 2015). On Titan, volatile cycling has been inferred from surface darkening, presumably caused by rainfall (Turtle et al. 2011a), but hail from convective clouds is also considered possible (Graves et al. 2008) and temporary brightening of the surface (Barnes et al. 2013; Dhingra et al. 2019). Titan's cloud images first obtained from ground-based telescopes (e.g. Brown et al. 2002) and later by the Cassini imaging instruments (Turtle et al. 2011a, Rodriguez et al.

2011, Le Mouélic et al. 2018) were intuitively recognized as methane clouds rather than volcanic plumes or dust storms, by analogy with terrestrial water clouds. Cloud morphology tells us much about the local humidity, thermal structure and atmospheric dynamics (Udelhofen et al. 1995; Rusch et al. 2009; Houze 2014), so that studying the morphology of various Earth clouds may be extremely helpful in understanding Titan's meteorology.

However, Titan also occasionally exhibits unusual cloud patterns not typically seen on Earth. For example, a large, chevron-shaped cloud near the equator may point to planetary-scale waves (Mitchell et al. 2011, **Fig. 7**(a)), or may be analogous to a 'derecho' type storm system seen on Earth (Šepić et al. 2012; Corfidi et al. 2016; **Fig. 7**(b)).

An intercomparison of tropospheric winds on Earth and Titan is difficult at present because tropospheric wind data from Titan are extremely scarce (Bird et al. 2005), but such information will be collected by the *Dragonfly* mission (Barnes et al. 2021).

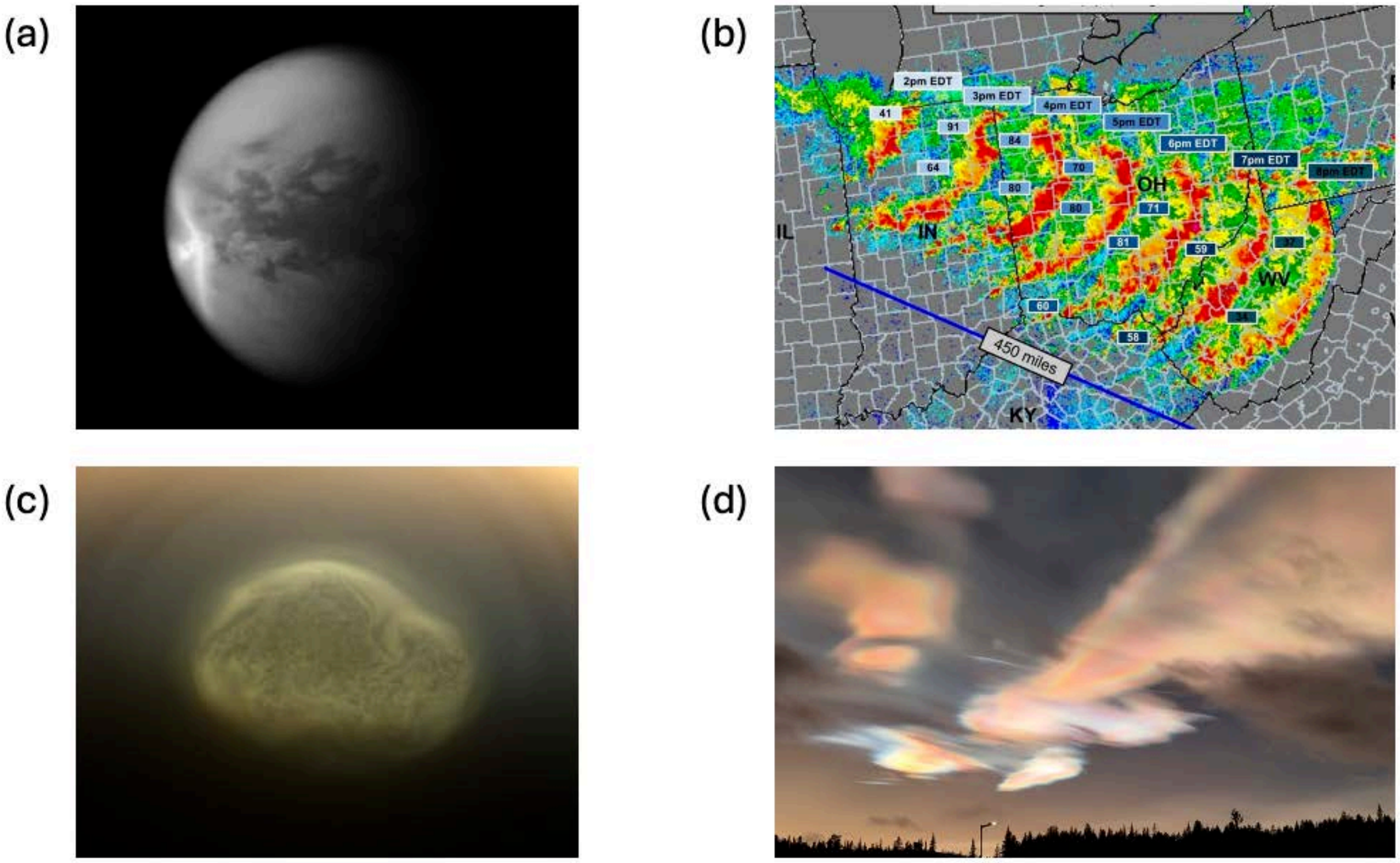

**Figure 7**: Titan and Earth Atmospheric Features. (a) a very large, chevron-shaped storm system seen on Titan 09/27/2010, measuring 1200 km east-west and with north and south-reaching extensions each 1500 km in length (NASA/JPL/SSI/CICLOPS). (b) Time-evolution of a Derecho-type storm system that moved across the NE United States June 29th 2012 (NOAA NWS). (c) Titan south polar stratospheric cloud seen June 27$^{th}$ 2012 (NASA/JPL/SSI/CICLOPS). (d) Polar stratospheric clouds seen over Kiruna, Sweden, January 19th 2025 (ESA).

*Field measurements*: In the space age, global and many regional atmospheric measurements can most conveniently and comprehensively be made from satellites. However, direct sampling by aircraft, rocket and balloon still retains importance for measurements requiring either small-scale precision, or of properties difficult to obtain from remote sensing. Aircraft measurements can proceed up to altitudes of ~21 km, e.g. with NASA's ER-2 aircraft (modified U2, Scott et al, 1990), although more typically at lower altitudes up to the tropopause (12 km). With a high payload capacity, aircraft are ideally suited to make in situ measurements, especially inside clouds and storms which are difficult to measure with remote sensing (Rosenfeld et al. 2006). Weather balloons can access higher altitudes up to 37 km or more in some cases, and have the advantage of being able to sensitively track wind motions (unlike rockets and aircraft), with the disadvantage of less control over trajectory. The time spent at altitude by a balloon is equal, or greater to that obtainable with a plane, and significantly greater than a rocket on a parabolic trajectory. See Kräuchi et al. (2016) for an overview. For higher altitudes (>37 km) up the Karman line at 100 km or beyond, sub-orbital sounding rockets allow direct measurements of the upper atmosphere (Kuminov et al. 2021).

## 2.2 Impact Craters

*Titan:* Since early in the Cassini mission, scientists have sought to create size-number histograms of impact craters on Titan, with the major goal of assessing the surface age. As the mission progressed, the crater counts have continuously been updated (Lorenz et al. 2007; Wood et al. 2010; Neish and Lorenz 2012; Hedgepeth et al. 2020), although all remain consistent with the initial age estimated from just three craters. Thus far, 90 possible impact craters have been identified on Titan, although only 37 of these are considered “certain” or “nearly certain” (Hedgepeth et al., 2020). The relatively low crater counts found in all works indicate a surprisingly young age of less than 1 Gyr (Neish and Lorenz 2012), possibly as young as ~300 Myr (Wakita et al., 2024). However, these age estimates are based on craters observed through remote sensing observations; if additional craters are buried or eroded beyond recognition, the age of its surface could be substantially older (Shah et al., 2025).

Almost all of the impact craters on Titan are considered “complex”. Complex craters are characterized by terraced crater rims, flat crater floors, and the presence of a central uplift (Melosh, 1989). They are expected to form on Titan at diameters between ~3 to 150 km, based on observations of Ganymede and Callisto (Schenk, 2002). The more simple, bowl-shaped craters are not formed on Titan because smaller impactors will break apart in Titan’s thick atmosphere before reaching the ground (Artemieva and Lunine, 2003; Korycansky and Zahnle, 2005). The largest crater on Titan - Menrva - may be the one example of a multiring basin seen there (Williams et al. 2011, Crosta et al. 2021). If other large craters are present, they are too heavily modified to observe from orbit.

Much of the work done to study craters on Titan has focused on understanding the inter-related topics of the spatial distribution of craters (Neish and Lorenz 2014) and the modification processes/erosion that operate on Titan (Le Mouelic et al. 2008; Neish et al. 2013, 2015, 2016; Solomonidou et al. 2020b; 2024). Most impact craters observed on Titan are found in the equatorial highlands, often in or near the sand seas, with a few in the mid-latitude plains, and almost none observed in the polar regions (Neish and Lorenz, 2014). The topography and morphology of the impact craters suggest they have primarily been modified by aeolian infill (Neish et al., 2013), with fluvial modification playing a secondary role near the equator (Hedgepeth et al., 2020) and possibly a more important role near the poles (Neish et al., 2016).

Increased fluvial erosion with increasing latitude may be consistent with the spectra of Titan's impact craters (Solomonidou et al., 2020b). Impact craters found in the mid-latitudes show more spectral evidence for water ice in the infrared, which may be caused by methane rain storms "washing clean" their surfaces of sand and other atmospheric deposits (Werynski et al., 2019; Solomonidou et al., 2020b). Alternatively, the relative lack of impact craters near Titan's poles may be due to impacts into a former polar ocean or wet, unconsolidated sediments (Neish and Lorenz, 2014; Wakita et al., 2022). These so-called "marine impacts" cause slumping of material back into the crater cavity, limiting their topographic expression and making them difficult to observe from orbit (e.g., Collins and Wünnemann, 2005; Bray et al., 2022). Viscous relaxation also plays a major role in crater modification on Titan if clathrates are present in the subsurface (Schurmeier et al., 2024). Impacts into a methane clathrate layer produce craters that are several kilometers shallower than similarly sized craters formed in a pure water-ice substrate (Schurmeier et al. 2025). Intriguingly, although craters on Titan should possess central uplifts, few craters show any topographic or morphologic evidence of them (with Ksa being a rare exception - Hedgepeth et al., 2020). This does not appear to be a result of viscous relaxation (Brouwer et al., 2025), but it may be related to fluvial erosion (Neish et al., 2016).

In summary, the morphologic appearance, topography, and distribution of impact craters on Titan suggest moderate to significant modification by exogenic processes, similar to those experienced by impact craters on Earth. Titan is also the only other world in the solar system with liquids presently on its surface, making it the only other place where marine impacts may be possible.

*Earth analogs:* Earth, like Titan, has a notable lack of impact craters observed on its surface (Osinski et al., 2022). Only ~200 impact craters have been identified on Earth through a combination of remote sensing techniques, geophysical exploration, field work, and drilling. Unlike Titan, there are a large number of simple craters on Earth, with diameters less than ~3 km depending on the substrate composition (Dence, 1972; Osinski et al., 2022). Due to its higher gravity, the transition from complex craters to peak-ring craters and multi-ring basins occurs at smaller diameters on Earth than on Titan; their morphologies also differ somewhat (Morgan et al., 2016). As a result, the best terrestrial analogues are complex craters that range in size from ~3 to 30 km in diameter. Of this subset, the best analogues are those craters that are unvegetated and only moderately eroded, such as the Haughton Impact Structure (Fig. 8; Osinski et al., 2005), Goat Paddock (Fig. 9; Harms et al., 1980), and Jebel Waqf as Suwwan (Salameh et al., 2008). However, some smaller craters may also serve as useful morphologic analogues, like the Roter Kamm simple crater in Namibia, which is infilled by sand like many Titan craters (Fig. 10; Grant et al., 1997). Lawn Hill (Lees and O'Donohue, 2024) serves as a useful analogue for any possible marine impacts on Titan (Neish and Lorenz, 2014).

All of Earth's impact craters have been modified in some way through exogenic and endogenic processes. Examples of terrestrial impact structures modified under contrasting climatic and erosional regimes, such as Bosumtwi in Ghana (Jones et al., 1981), Lonar in India (Fudali and Pani, 2013), and Mistastin in Canada (e.g., Currie, 1971; Osinski et al., 2011), provide useful analogs for assessing the role of post-impact erosion, sediment infill, and climate in shaping crater morphology. Indeed, fluvial erosion and aeolian infilling are two processes that have modified many craters on Earth (see **Figs. 8, 9, 10**). Fluvial erosion tends to erode the crater rim and fill in the crater floor, while aeolian modification mostly fills in the crater interior, leaving the rim intact (Forsberg-Taylor et al., 2004). Many marine impacts have also been identified through geophysical exploration and drilling campaigns (e.g., Poag et al., 1994). It is likely that the same crater modification processes also occur on Titan, and may obscure their presence in remote sensing data (Shah et al., 2025). For example, if marine craters are present on Titan, it would not be possible to identify them from Cassini images due to their lack of surface expression. Endogenic processes such as tectonism also play a large role in shaping terrestrial impact craters (e.g., Spray et al., 2004). Observations suggest this plays a relatively limited role on Titan, although the shape of some craters like Selk seem to have been influenced by pre-existing faults (Soderblom et al., 2010), similar to the square morphology of Meteor Crater (e.g., Poelchau et al., 2009).

On Titan, the bulk of the craters are found in relatively sediment-poor areas such as dunes, low latitude plains, and areas with extensive exposed icy crust. Very few impact craters have been identified in high latitude areas or areas where thick organic sediments have been invoked. Where impact craters on Titan are observed, one notable difference between impact craters on Earth versus Titan is the presence of ejecta blankets. Many impact craters on Titan show obvious ejecta blankets in remote sensing data (e.g., Wood et al., 2010), but this feature is rarely seen around craters on Earth (e.g., Osinski et al., 2011; Shah et al., 2025). This observation suggests that the amount of erosion on Titan in these areas is likely less than that observed on Earth.

(a)

(b)

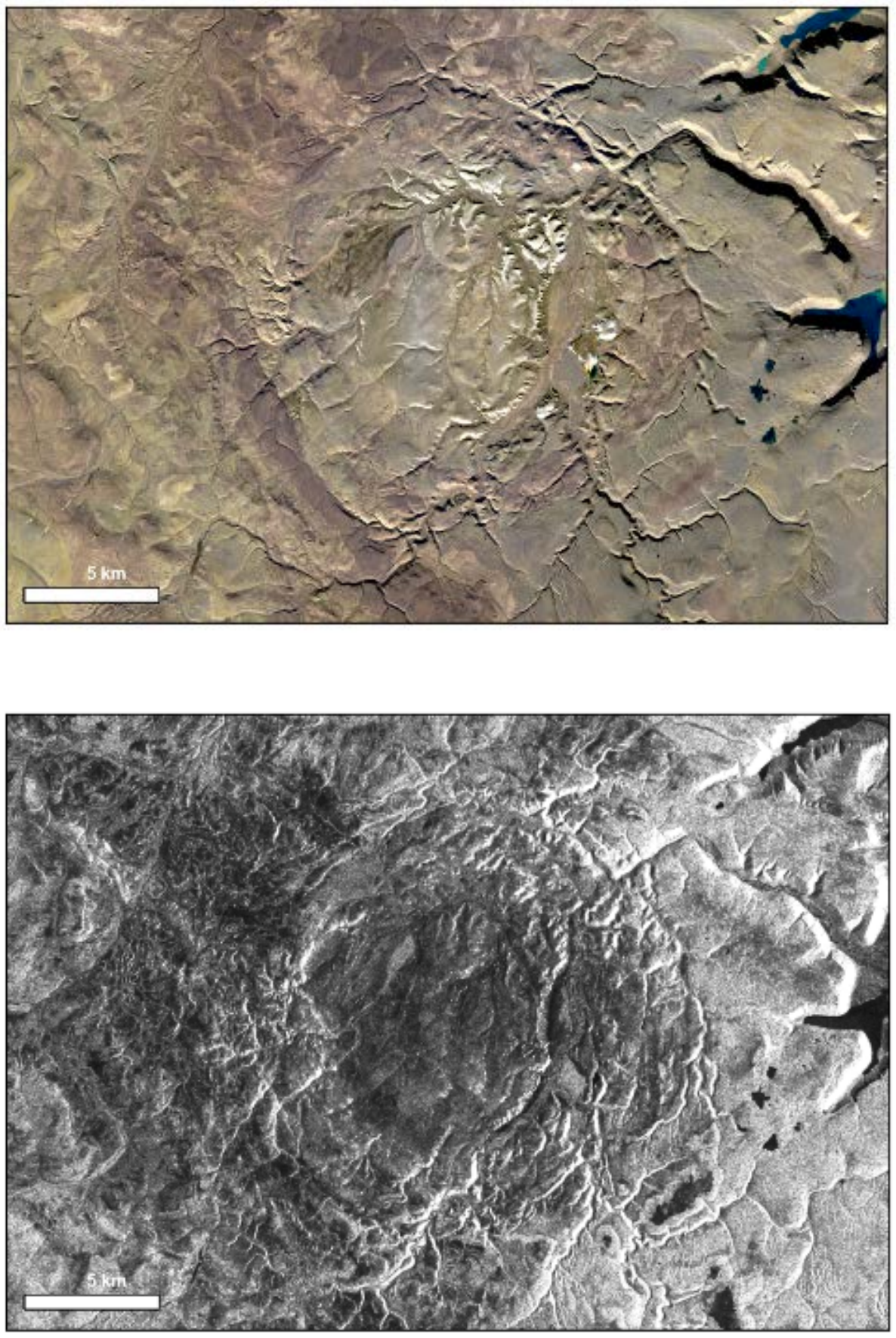

(c)

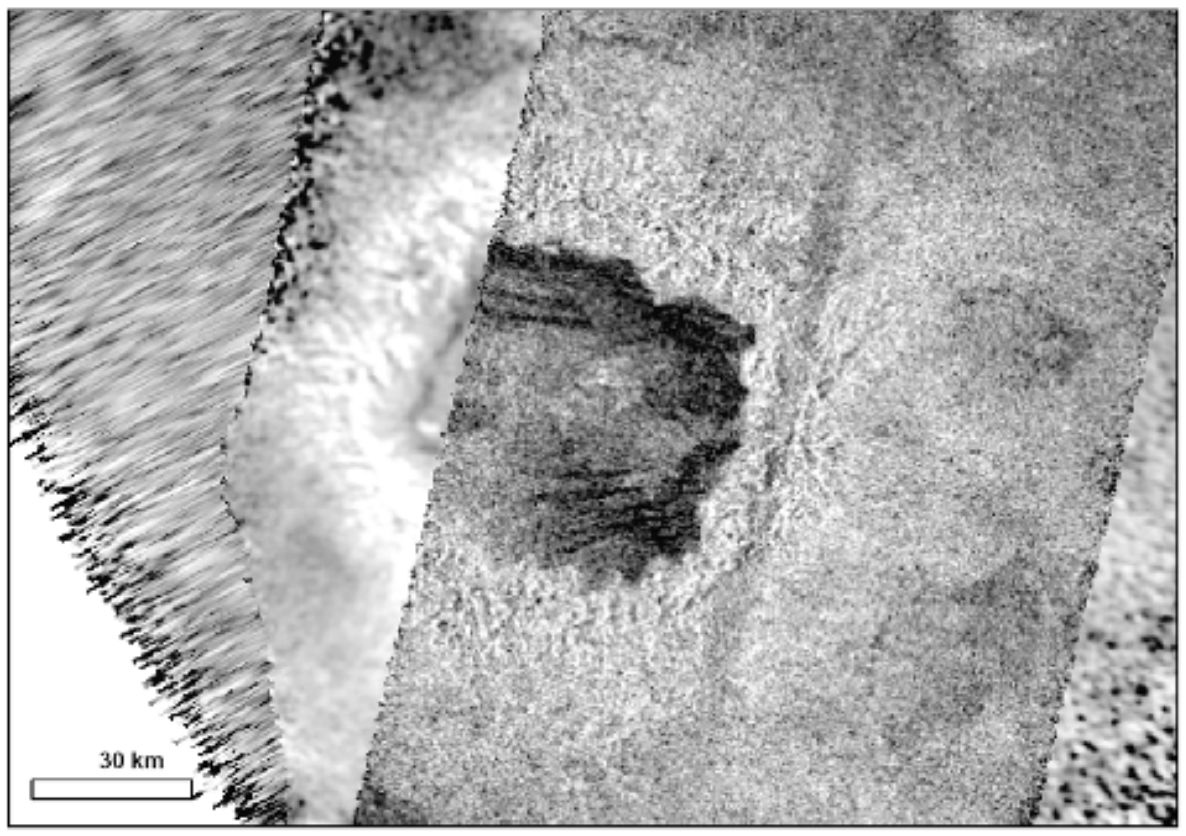

**Figure 8**: (a) Landsat-8 OLI true color image of the Haughton Impact Structure on Devon Island, NU, Canada. (b) Sentinel-1 C-Band (5.6 cm) HH radar image of the same impact structure. (c) Cassini Ku-Band (2.2 cm) HH RADAR mosaic of Selk crater on Titan. Both craters show moderate amounts of fluvial erosion and a radar dark crater interior, although the rim of Selk is more defined and it has aeolian deposits in its interior.

(a)

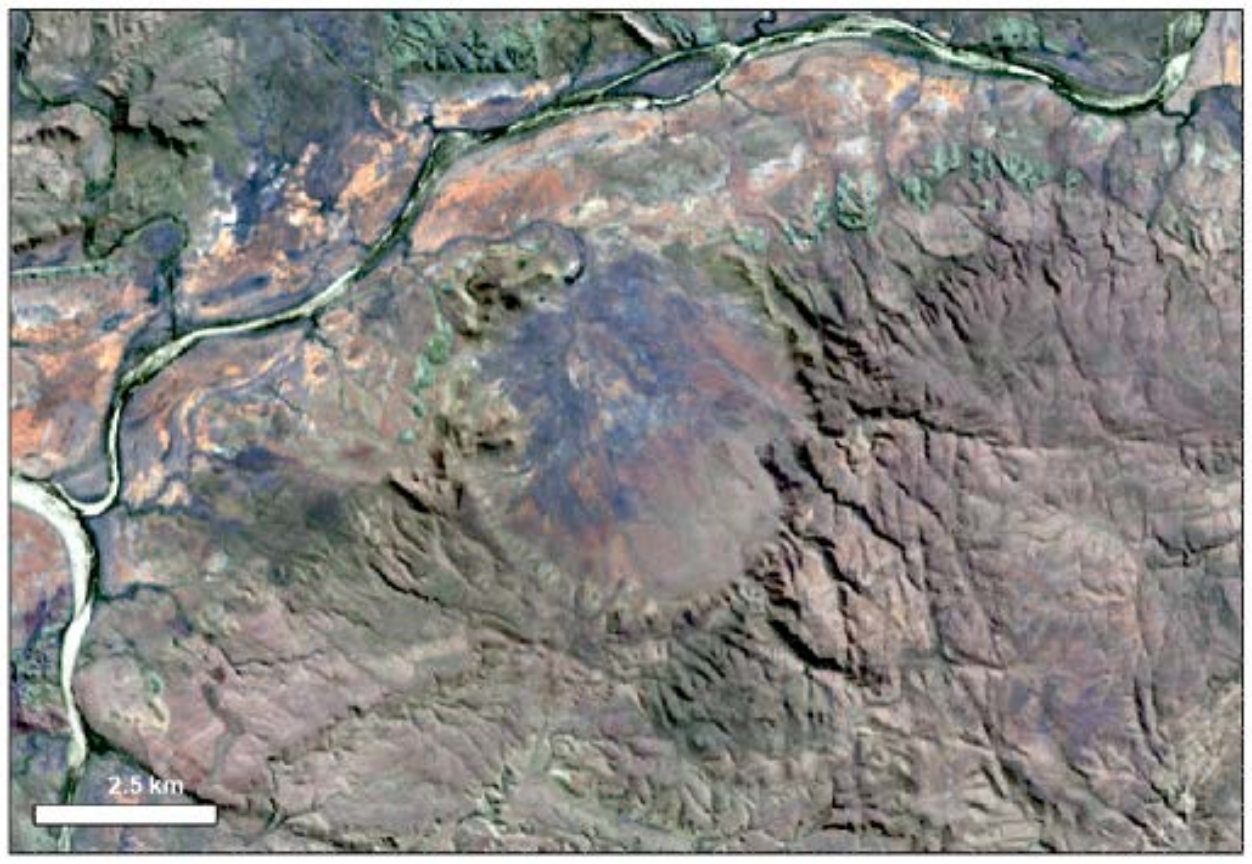

(b)

(c)

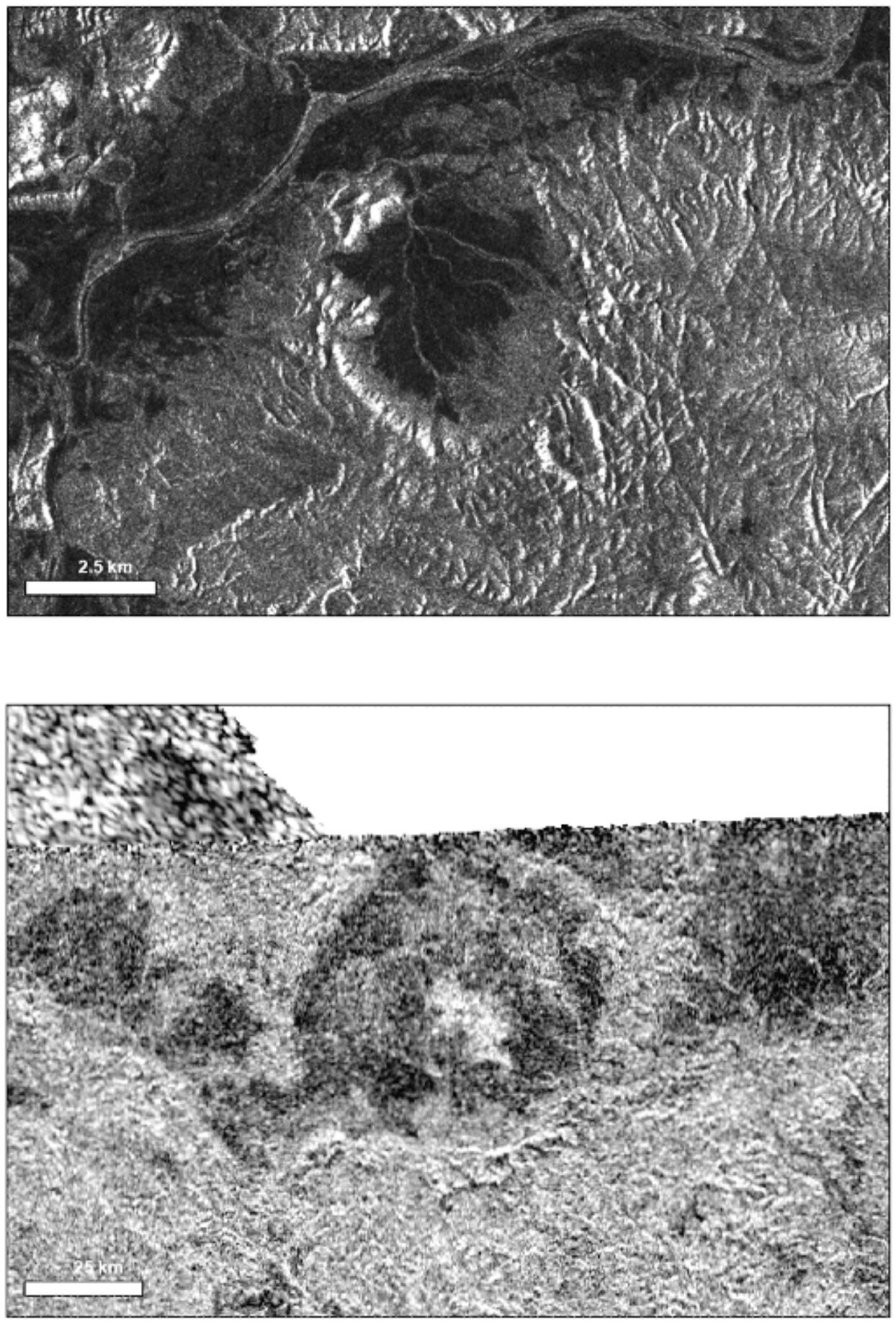

**Figure 9**: (a) Landsat-8 OLI true color image of the Goat Paddock impact crater in Western Australia. (b) Sentinel-1 C-Band (5.6 cm) VV radar image of the same impact structure. The crater is drained through a gap cut through the northern rim, fed by radar-bright streams. (c) Cassini Ku-Band (2.2 cm) HH RADAR mosaic of “nearly certain” Crater 26 on Titan from Wood et al. (2010). This crater shows evidence for fluvial erosion, with a bright stream channel traversing the eastern half of the dark crater floor.

(a)

(b)

(c)

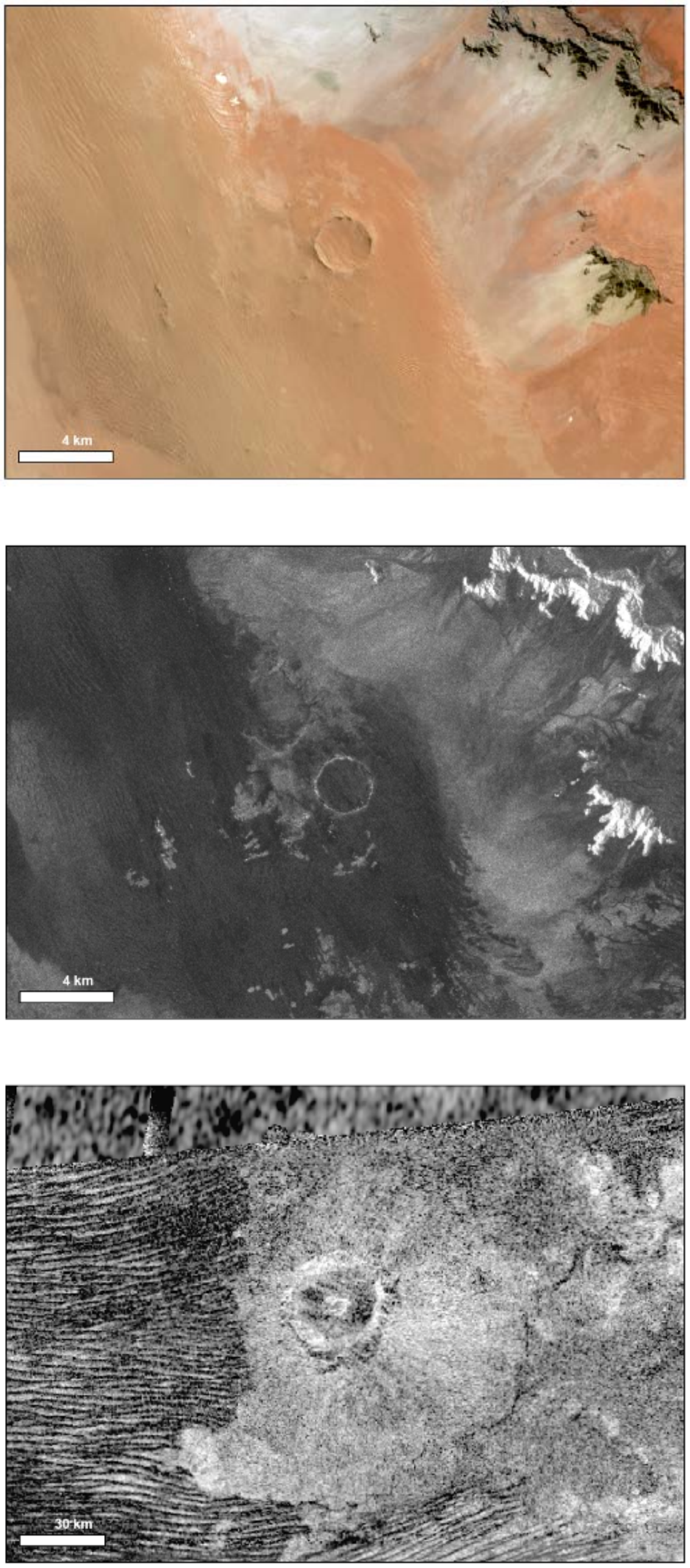

**Figure 10**: (a) Landsat-8 OLI true color image of the Roter Kamm impact crater in the Namib Desert. (b) Sentinel-1 C-Band (5.6 cm) VV radar image of the same impact structure. (c) Cassini Ku-Band (2.2 cm) HH RADAR mosaic of Ksa crater on Titan surrounded by the Fensal sand sea. Note the dark, sand-filled interiors in both craters, and the bright, circular crater rim standing up above the dunes.

Some impact craters on Earth have also been shaped by glacial erosion (e.g., Grieve, 1988, Hergarten and Kenkmann 2019), but this is not a process that can occur in Titan's current nitrogen-rich atmospheric conditions. On Earth, some craters are filled to form lakes (e.g., El'gygytgyn, Gurov et al. 2007), but no obvious crater lakes have been detected on Titan (Hayes 2016). In addition, although viscous relaxation modifies craters on Titan, it is unlikely to have any noticeable impact on craters formed in silicate substrates (Johnson and McGetchin, 1973). Nonetheless, given the similarities in erosional processes occurring on both worlds, Earth remains the best analogue environment for understanding the modification of craters on Titan.

*Field measurements*: In situ measurements at craters are useful to determine their origin, mechanism of formation, and subsequent history. On Titan, the *Dragonfly* mission will explore Selk crater to characterize the composition of any impact melt deposits there (Neish et al., 2018; Barnes et al. 2021). Prebiotic compounds may have developed in these deposits when Titan's surface organics mixed with the liquid water present shortly after impact (Neish et al., 2010b). Thus, the focus of the mission will be to identify and characterize melt deposits at Selk. In the past, drilling/coring at terrestrial craters has helped to determine the properties of the impactor and details of the impact event (e.g., Urrutia-Fucugauch et al. 2004). This will not be possible with the suite of instruments on *Dragonfly*. Instead, they will use colour cameras to distinguish different material types, stereophotogrammetry to determine surface topography, a gamma-ray and neutron spectrometer to determine the bulk composition of the regolith, and a mass spectrometer to determine detailed chemical composition (Turtle and Lorenz, 2024). Analog field measurements on Earth (**Table 3**) should thus focus on differentiating rock type based on colour (e.g., Tornabene et al., 2005), measurements of geochemistry and geochemical alteration (e.g., Koeberl 2002), determining local stratigraphy (e.g., O'Keefe et al. 1999), and acquiring high-resolution topography (e.g., Perkins et al., 2023). The aim of these measurements would be to determine the best approach for identifying and characterizing impact melt deposits using *Dragonfly*-like instruments.

**Table 3**: Terrestrial analogs to Titan Impact Craters

| Titan location(s) | Terrestrial Analog Site(s) | Analog Type | Measurement | Notes/limitations |
| --- | --- | --- | --- | --- |
| Sinlap, Selk | Haughton Impact Structure (Canada), Jebel Waqf as Suwwan (Jordan) | Morphological | Geochemistry; stratigraphy; topography; etc | Moderately eroded complex craters into a sedimentary target. Substrate silicate rock, not organic-rich ice. Smaller, since most complex craters on Earth are smaller than those on Titan. |
| Ksa | Roter Kamm (Namibia) | Morphological | Geochemistry; stratigraphy; topography; etc | Crater infilled with aeolian sediments. Simple crater on Earth vs. complex crater on Titan. |
| Crater 26 from Wood et al. (2010) | Goat Paddock (Australia) | Morphological | Geochemistry; stratigraphy; topography; etc | Complex crater with extensive fluvial erosion, including streams that cut through the crater rim and flow onto the crater floor. Smaller, since most complex craters on Earth are smaller than those on Titan. |
| Soi | Lawn Hill (Australia) | Morphological | Geochemistry; stratigraphy; topography; etc | Complex crater with minimal topographic expression but clear radar backscatter differences between the rim and the floor. |

## 2.3 Surface crustal ice landforms: ridges/mountains

*Titan:* Interpreting tectonic structures on Titan is complicated by two factors. First, poor Synthetic Aperture Radar (SAR) resolution combined with the lack of topographic data coverage (i.e., SARTopo and altimetry) means that tectonic indicators on Titan are largely identified via indirect approaches. This includes inferring tectonic fabric through analysis of drainage network patterns (Burr et al., 2009, 2013), and inferring uplift processes from planform geometries (e.g., Black et al., 2012; Liu et al., 2016a). Secondly, Titan's surface is subjected to an unknown amount of erosion and sediment deposition, obfuscating structural expression that might otherwise be diagnostic. Faults are impossible to identify on Titan at present and must instead be inferred.

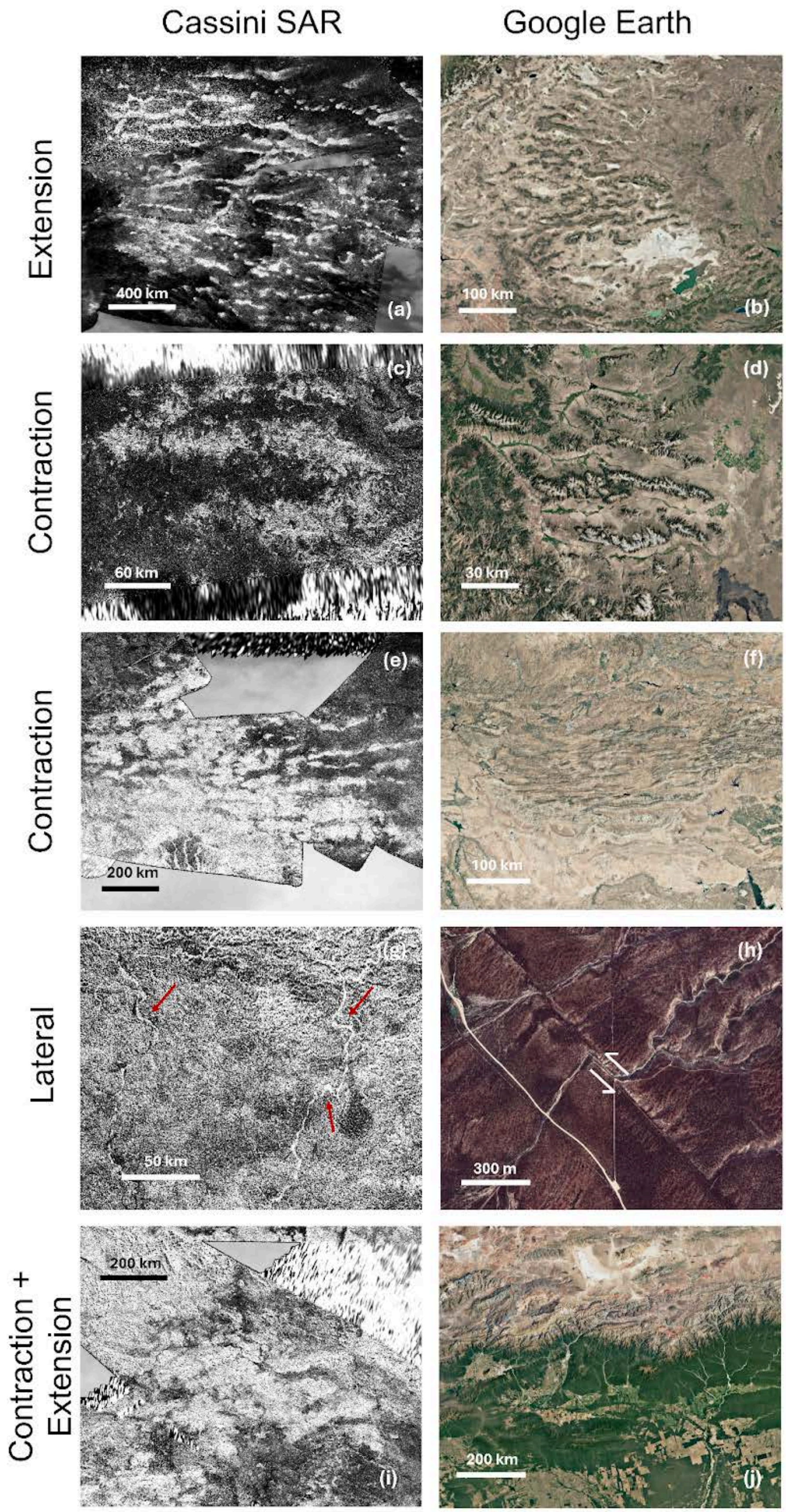

**Figure 11**: Titan and associated terrestrial analogues from **Table 4**. a) Adiri Mountain Region in Cassini SAR (centered 144°E, 8°S). b) Basin and Range Province, USA, in visible imagery (Google Earth). c) North polar mountains in SAR (centered on 50°E, 50°N). d) The Lost River, Lehmi, and Beaverhead Ranges, Idaho, USA, in visible imagery (Google Earth). e) Xanadu Annex in SAR (centered on 62°W, 30°S). f) Zagros Mountains, Iran, in visible imagery (Google Earth). g) SW Xanadu in SAR (centered on 140°W, 10°S). h) Wallace Creek, California, USA, in visible imagery (Google Earth). (i) Eastern Xanadu in SAR (centered on 97°W, 17°S). (j) Salta Rift Basin, Argentina, in visible imagery (Google Earth).

Despite these complications, the tectonics of Titan has been studied and appears rich in diversity. Global morphological analysis reveals the presence of ridges, mountainous regions, and elevated linear features, many of which resemble terrestrial fold mountains and contractional structures (Solomonidou et al. 2013; **Fig. 11a, 11c, 11e**). Recent large-scale geomorphological mapping by Schoenfeld et al. (2021; 2023) further detailed the spatial distribution, morphology, and stratigraphic relationships of these features, emphasizing the regional tectonic and sedimentary evolution across Titan's equatorial regions.

The preponderance of equatorial mountain belts is perhaps the strongest argument for Titan's tectonic verisimilitude. They appear long and narrow in planform, are oriented predominantly E–W, and present bright-dark pairing in the SAR that suggests a ridge line (Radebaugh et al., 2007, Cook-Hallett et al., 2015, Liu et al., 2016a; **Fig. 11a, 11c, 11e**). Comparison with a global topographic map (interpolated from altimetry and SARTopo) indicates that the equatorial ridges are preferentially associated with higher-than-average elevations (Liu et al., 2016a). Their arcuate morphology (Radebaugh et al., 2011, Solomonidou et al., 2013, Liu et al., 2016a), and low slope and relief (Radebaugh et al., 2007, Liu et al., 2016a, Mitri et al., 2010), have led to these mountains being interpreted as thrust belts having formed by global contraction.

A model by Mitri et al. (2010) demonstrated that volume change from internal cooling and growth of the high pressure ice layer at the base of Titan's ocean would result in contractional folds on Titan. However, this model does not account for the predominantly E-W orientation and would instead result in ridge formation with no preferred orientation. Cook-Hallett et al. (2015) suggests that either global contraction coupled with spin-up or global expansion coupled with spin-down could explain the pattern if Titan's polar lithosphere were thin. Cook-Hallet et al. (2015) map mountains with N–S orientation between 60° latitude and the poles. Liu et al. (2016a) contests this, asserting that, while ridges are concentrated at the equator, ridges trend E–W globally. They instead propose an alternate model where global contraction, in addition to an initially thin equatorial lithosphere (Beuthe, 2010), is posited to explain the global tectonic pattern.

Global contraction of the moon would generate stresses of magnitude <0.1 MPa (Cook-Hallett et al., 2015), lower than needed to induce failure under compression and thrust fault formation. However, Liu et al (2016b) demonstrate that liquid hydrocarbons in near subsurface pore spaces can lead to fluid overpressures that would lower the threshold for failure, facilitating compressional deformation at smaller stresses (<1 MPa). This is also true for the Earth, where pore fluids weaken the lithosphere and asthenosphere. A thinner lithosphere could also form if Titan's crust contains thick deposits of highly insulating surface materials: i.e., clathrates and organic-rich plains or sands. These along with pore fluids could lower the threshold for failure in both compression and extension to under 1 MPa near the surface, potentially enabling failure in more directions (Schurmeier and Fagents, 2025).

Another notable tectonic province on Titan is the Xanadu region. Located on Titan's leading hemisphere, Xanadu is characterized by a vast, rugged landscape of linear features, extensive fluvial networks, and craters (**Fig. 11e, 11g**). Despite the wide distribution of mountains and high local lineations, topographic data suggests that overall, the region has lower than average elevation compared to its surroundings (Radebaugh et al., 2011; Corlies et al., 2017; Kirk et al., 2012; Lorenz et al., 2013; Mitri et al., 2014; Stiles et al., 2009; Zebker et al., 2009). A model by Radebaugh et al. (2011) proposed that Xanadu experienced two phases of deformation: initial compression and then subsequent extension. In the first phase, north-south compression would have induced crustal thickening and E-W mountain building. In the second phase, the region underwent extension, resulting in regional lowering and horst and graben formation. The region was then reworked by fluvial processes, attributing to its rugged morphology and leveled topography (Langhans et al., 2013). Matteoni et al. (2020) further posits that SW Xanadu is a transtensional basin, bordered to the west and east by large-scale normal faulting and to the north and south by thrust faulting, with strike-slip faulting accommodating shear within.

*Earth*: Terrestrial deformation is described by the kinematic model known as plate tectonics (McKenzie and Parker, 1967; Morgan, 1968; Le Pichon, 1968). The origin of tectonic stress for the Earth is density and gravity: internally generated heat within the Earth’s interior manifests as mantle convection, plate subduction, plate divergence, all of which are processes that work to resolve density instabilities (Tackley, 2000; Conrad and Lithgow-Bertelloni, 2002; Bercovici, 2003; Bercovici and Ricard, 2014). Relative motions driven by convection result in three types of plate boundaries: convergent, divergent, and transform. Each boundary is associated with characteristic deformation and crustal fractures, known as faults. Convergent boundaries are associated with thrust (reverse) faults resulting from compression (**Fig. 11d, 11f**), divergent boundaries are associated with normal faulting resulting from tension (**Fig. 11b)**, and transform faults are associated with strike-slip faults resulting from lateral motion (**Fig. 11h)**. These fault types are classified according to kinematic indicators, such as direction of slip and angle of the fault.

In addition to faults, major tectonic features on Earth include mountains, ridges, escarpments, rifts, grabens and other linear terrains that are subjected to erosion (e.g., Scheidegger, 2004). Some of these features may be more associated with one stress regime over another (i.e., rifts and grabens are often formed in zones of divergence while folds are more associated with zones of convergence), but they are all ultimately surface expressions of deformation.

Despite the difference in composition, the rheological behavior of ice at the surface of icy satellites in the outer solar system is comparable to that of rocks on the surface of Earth, owing to similar homologous temperatures in their respective environments (Collins et al., 2009). However, the terrestrial plate tectonics model does not fully apply to Titan, or other icy moons. The criteria for modern plate tectonics require that the plates are a part of a global network, that the mobile plates are rigid, and that the plates are eventually recycled at boundaries. There appears to be global deformation on many icy moons, even putative rigid plate motion is argued for Europa, but no strong evidence exists for lithosphere recycling. However, we can nonetheless use our understanding of terrestrial tectonics to study the tectonics of Titan, invoking analogues to ultimately inform our interpretations.

The distribution and morphometry of Titan's mountainous regions share similarities with Earth's orogenic belts, where crustal uplift, erosion, and sediment deposition occur together. As such, terrestrial analog sites for Titan's crustal landforms may include compressional tectonic settings. However, Titan's tectonic diversity suggests that it is also appropriate to draw comparisons to extensional and lateral terrestrial settings (see **Table 4**).

**Table 4**: Terrestrial analogs to Titan Structural Tectonics

| Titan location(s) | Terrestrial Analog Site(s) | Analog Type | Measurements | Notes |
|---|---|---|---|---|
| Titan's Adiri region (centered 144°E, 8°S) | The Basin and Range Province, western USA. | Extension | Angle of fault dip, direction of slip, crustal thinning, stratigraphy, fault petrology, seismic survey | Titan's Adiri region is characterized by parallel, linear mountain belts and represents the best example of Titan's E-W trending mountains, which may have formed by compression. The radar bright mountains are separated by radar dark |
| | | | | organic plains in adjacent valleys. The Basin and Range Province is similarly characterized by parallel and linear mountain ranges. Yet, the topography of the province is the result of extension, where parallel normal faults link at depth to create horst and graben morphology (Eaton, 1982; Dickinson, 2007). |
| Parallel mountain chains located south of Kraken Mare (centered 50°E, 50°N). | The Lost River, Lehmi, and Beaverhead Ranges, Idaho, USA. | Contraction | Angle of fault dip, direction of slip, crustal shortening, stratigraphy, fault petrology, seismic survey | These high-latitude mountains near Titan's mare demonstrate that the E-W linear mountains are not just equatorial features. They also display more pronounced granulation than their equatorial counterparts, characteristic of advanced fluvial erosion. The terrestrial examples are parallel mountain ranges that are part of the northern Rocky Mountains. The Lost River, |
| | | | | Lehmi, and Beaverhead Ranges represent fault blocks uplifted as part of a large-scale thrusting event during the Sevier Orogeny (Fisher and Anastasio, 1994). |
| Titan's Xanadu Annex (centered 62°W, 30°S) | The Northwestern Zagros Mountains, Iran | Contraction | Angle of fault dip, direction of slip, crustal shortening, stratigraphy, fault petrology, seismic survey | The Xanadu Annex is characterized by tightly spaced, remarkably linear mountain belts. The Zagros Mountains formed from convergence between the Arabian and Eurasian plate in the Late Cretaceous-Early Miocene. Zagros is divided by the Kazerun strike slip: the topography NW of the strike-slip differs from the topography SE of it. The northwestern belt is characterized by a narrow zone of deformation with tightly spaced folding. The difference in topography is a result of a difference in basal friction, where the NW Zagros belt lacks a salt layer |
| | | | | decollement (Nilforoushan et al., 2008). |
| SW Xanadu (centered on 140°W, 10°S) | Wallace Creek, California, USA | Lateral | Angle of fault dip, direction of slip, stratigraphy, fault petrology, seismic survey, surface offset | The dog-leg bends (red arrows) in these two large, bright channels in SW Xanadu have been attributed to strike-slip faulting in the region (Burkhard et al.,2022; Matteoni et al., 2020). Wallace Creek crosses the San Andreas fault and is offset at a right-angle bend along the fault trace. The offset follows the right slip of the fault (Sieh and Wallace, 1987). |
| E Xanadu (centered on 97°W, 17°S) | Salta Rift Basin, Argentina | Contraction and Extension | Angle of fault dip, direction of slip, stratigraphy, petrology, seismic survey | Radebaugh et al. (2011) hypothesized that the Xanadu region underwent two tectonic events: an initial stage of crustal thickening that created E-W trending mountains belts, then a subsequent stage of extension that created the linear valleys of Eastern |
| | | | | Xanadu. The Andean foothills of northwestern Argentina suggest a complex structural rift system that developed during the late Jurassic to Eocene times. Subsequent Andean compressive, linked to the subduction of the Farallones plate under the South American plate (Pardo-Casas and Molnar, 1987), resulted in tectonic inversion of the Salta rift system (Monaldi et al., 2008). |

*Field measurements*: Linear features can be identified and measured in the field and can be used to infer kinematic histories. For example, glacial striations indicate direction of ice movement (e.g., Klassen, 1994), while foliation of a metamorphic rock can indicate direction of deformation (e.g., Brown and Solar, 1997). Linear features in remote view can similarly be measured and used to identify regional tectonic influence; basic qualities such as alignment, spacing, and spatial extent of lineaments can provide higher order information on stress regimes and lithologies (e.g., Frank and Head, 1990).

Faulting produces displacement across a surface or plane. The orientation of a plane is characterized by its strike – the compass orientation of a horizontal line on the surface – and by the dip – the angle at which the surface dips from the horizontal, perpendicular to the strike. Like lineaments, these qualities can be measured in the field (i.e., with a compass clinometer). In map view, proper fault categorization requires not just remote sensing data in the form of visual, infrared, or SAR imaging, but requires third-dimensional information provided by topography –whether from laser or radar altimetry, or stereophotogrammetry. Broadly, faults can then be categorized by their dip angle and dynamic histories can be inferred (e.g., Anderson, 1905).

For Titan, a detailed remote sensing survey, with higher resolution SAR imaging and more comprehensive altimetry dataset, would allow for proper tectonic characterization even without direct field measurements. In-situ mineralogy could also be used to analyse Titan minerals and investigate whether deformational signatures are preserved in the mineral grain (whether ice or organic or both). A terrestrial field site that is appropriately similar in map view to a given Titan region can then be selected; terrestrial field measurements can be made that can inform the formation histories of its Titan counterpart.

## 2.4 Dunes and Yardangs

*Titan Dunes*: The large equatorial dune fields of Titan were first identified early in the Cassini mission by the RADAR instrument (Lorenz et al. 2006). The dunes, which are thought to be made of solid hydrocarbon material produced in the atmosphere, have heights up to ~100 m, spacings of ~1-2 km and lengths stretching to 100's of km (Radebaugh et al. 2008; Neish et al., 2010a). Dunes are dark to Cassini Radar, because the particles and dune surface are smooth at Cassini SAR wavelengths of 2 cm. This is similar to X-band imagery of Earth's dunes, which are also dark, or smooth, at 3 cm (**Fig. 12**). Sand supply on Titan varies such that many regions contain dunes of extreme length and close spacing, especially found near Titan's equator (**Fig. 12**), while higher latitude regions have less sand and exhibit dunes that are short and widely spaced and are overprinted upon the underlying substrate (**Fig. 13**).

Using the correlation between Radar (resolving locally the dune fields, but with a partial coverage) and VIMS dark brown units (mapped at the global scale, i.e; Rodriguez et al., 2014; Le Mouélic et al., 2019), it is possible to evaluate the total extent of dune fields on Titan. They are estimated to cover 17% of the surface, and are found entirely within 30 degrees latitude of the equator (Rodriguez et al., 2014). The dunes of Titan are thought to be similar to other dunes in the Solar System in that they are made of solid particulates that saltate through the action of wind and collect into self-organized forms (Lorenz et al. 2006).

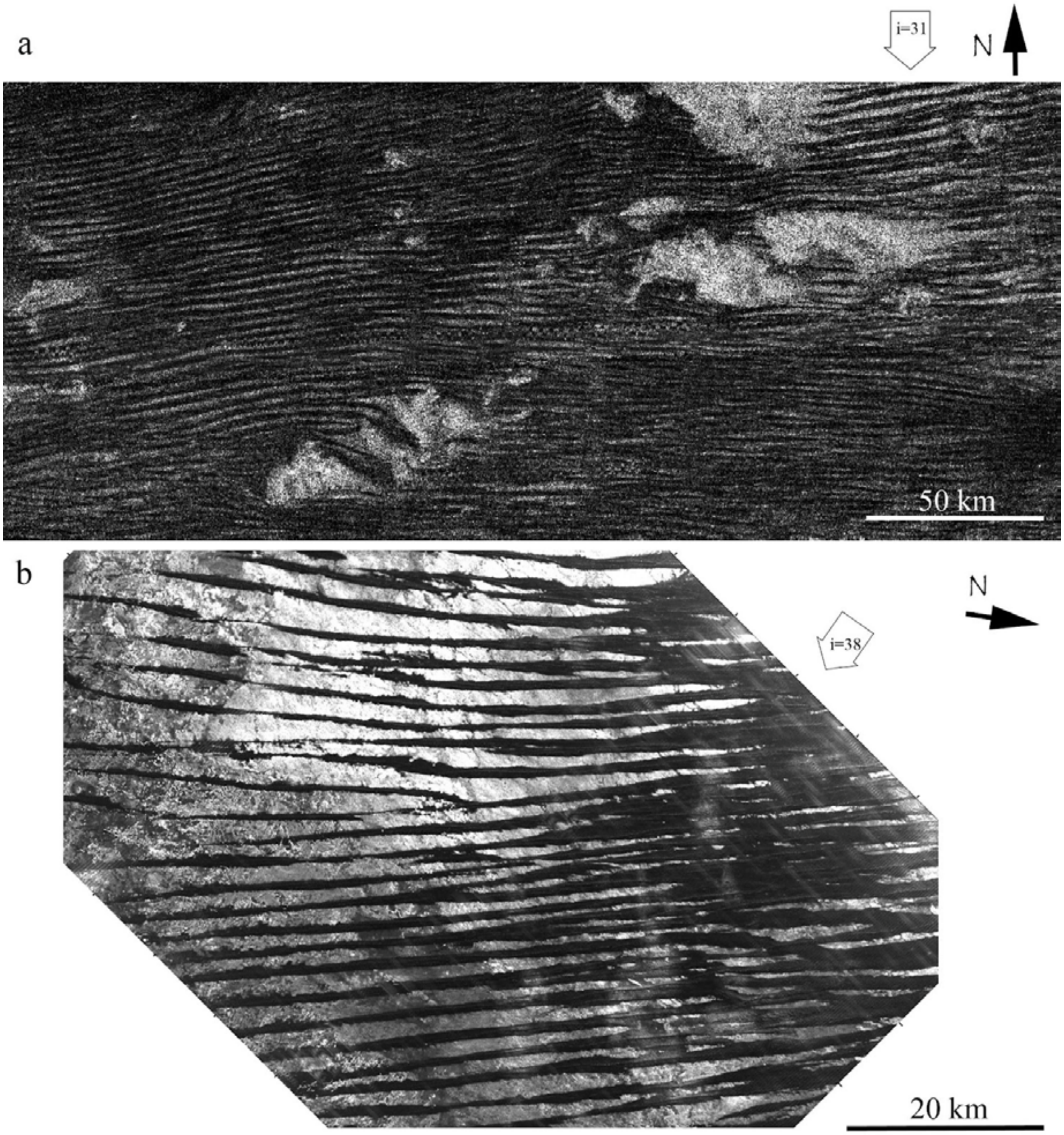

**Figure 12**. Dunes of Titan from Cassini Radar. Dunes are dark, or smooth, at Cassini SAR's 2 cm wavelength. The dunes are dense near Titan's equator and show halting and divergence at obstacles that preclude the W-E transport. b. Dunes of Egypt at X-band, or 3 cm Radar. Silicate dunes on Earth, buff in color, are also dark or smooth to X band Radar and also overlie bright or rough substrate. From Radebaugh et al. (2013).

Given atmospheric pressure similarities and estimates of compositions, Titan’s dune sands are likely sand-sized, 0.6-2.0 mm (Lorenz et al. 2006). However, Titan’s sands are not composed of silicates or carbonates as on Earth, but rather are determined by Cassini remote sensing data to have a component of organics, derived from atmospheric fallout of methane-derived organics (Soderblom et al. 2007). These organic compounds are far too small to directly make sands. Thus, some process must sinter these molecules together in the atmosphere or on the surface. Sedimentary processes may consolidate, harden and solidify the organics into surface deposits that are then eroded by methane rainfall into an appropriate size to be saltated by wind into dunes (Barnes et al. 2015; Radebaugh 2013). Alternatively, materials may be the result of cometary disintegration in Titan’s atmosphere (Bottke, 2024).

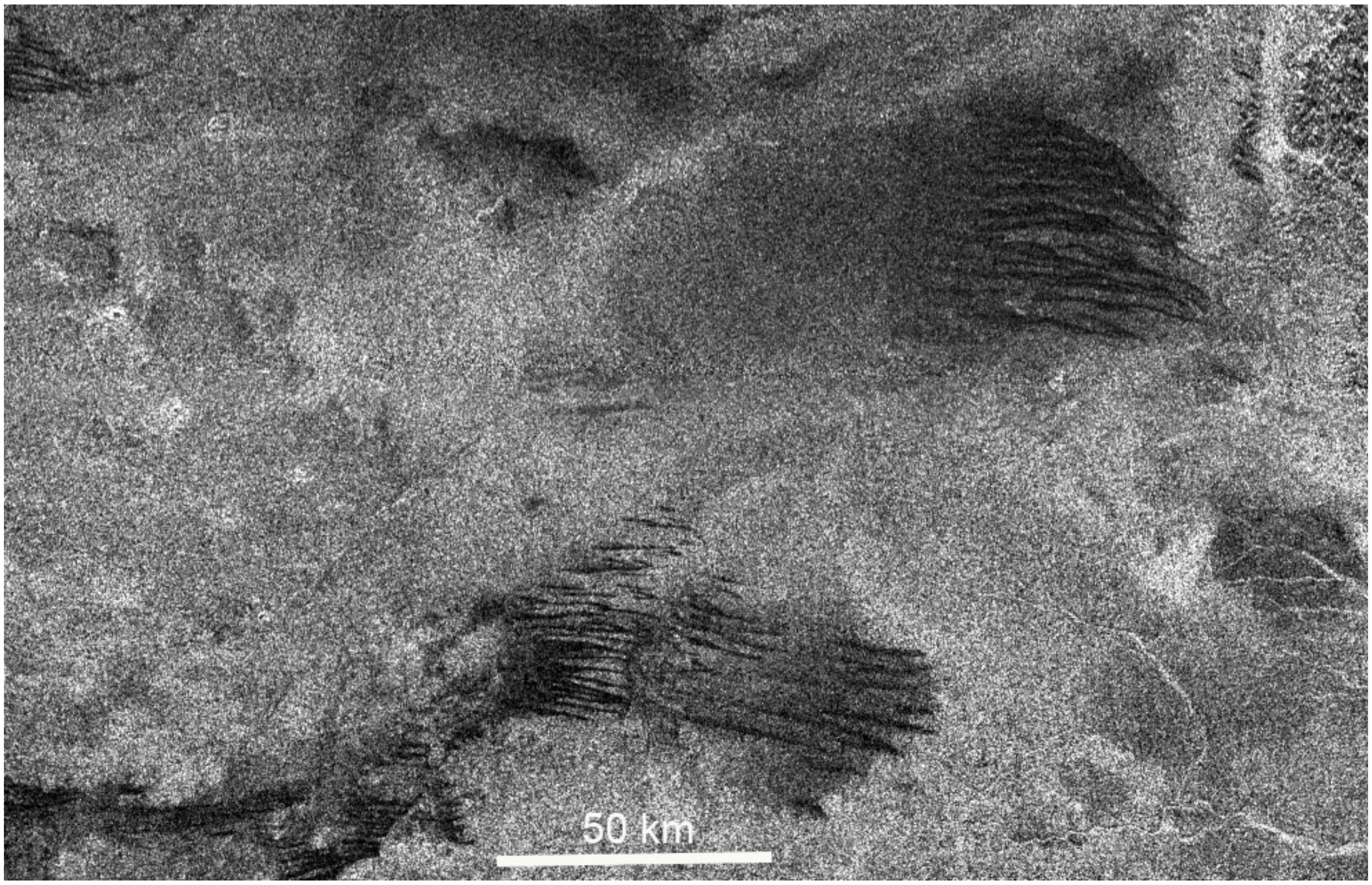

**Figure 13**. Dunes at Titan's higher equatorial latitudes (30 degrees north) are sparser, smaller and more widely spaced than between 10 degrees north and south of the equator. These overlie radar brighter, or rougher, landscapes and fluvial channels.

Titan's dunes are of a unique but widespread type on Earth: large linear or longitudinal (referring to the direction of sand transport along the dune long axis; **Fig. 14;** Lorenz et al. 2006; Radebaugh et al. 2008). On Titan, as on Earth, they are 1-2 km wide, spaced by 2-4 km, up to hundreds of kilometers long, and up to 150 m high (Neish et al. 2010a). This vast collection of dunes represents a huge volume of organic materials, freely available and possibly moving on the surface of Titan today (Lorenz et al. 2008a). The dunes are organized into large sand seas, all found ringing the equator between ±30 degrees latitude and all moderately connected to each other (Radebaugh et al. 2010;

Barnes et al. 2015). Dunes across Titan indicate widespread down-axis transport (**Fig. 14**) and vector sum winds from W-E (Lorenz and Radebaugh 2009; Lucas et al. 2014), which may be accommodated by fast westerlies during seasonal changes (Tokano 2010).

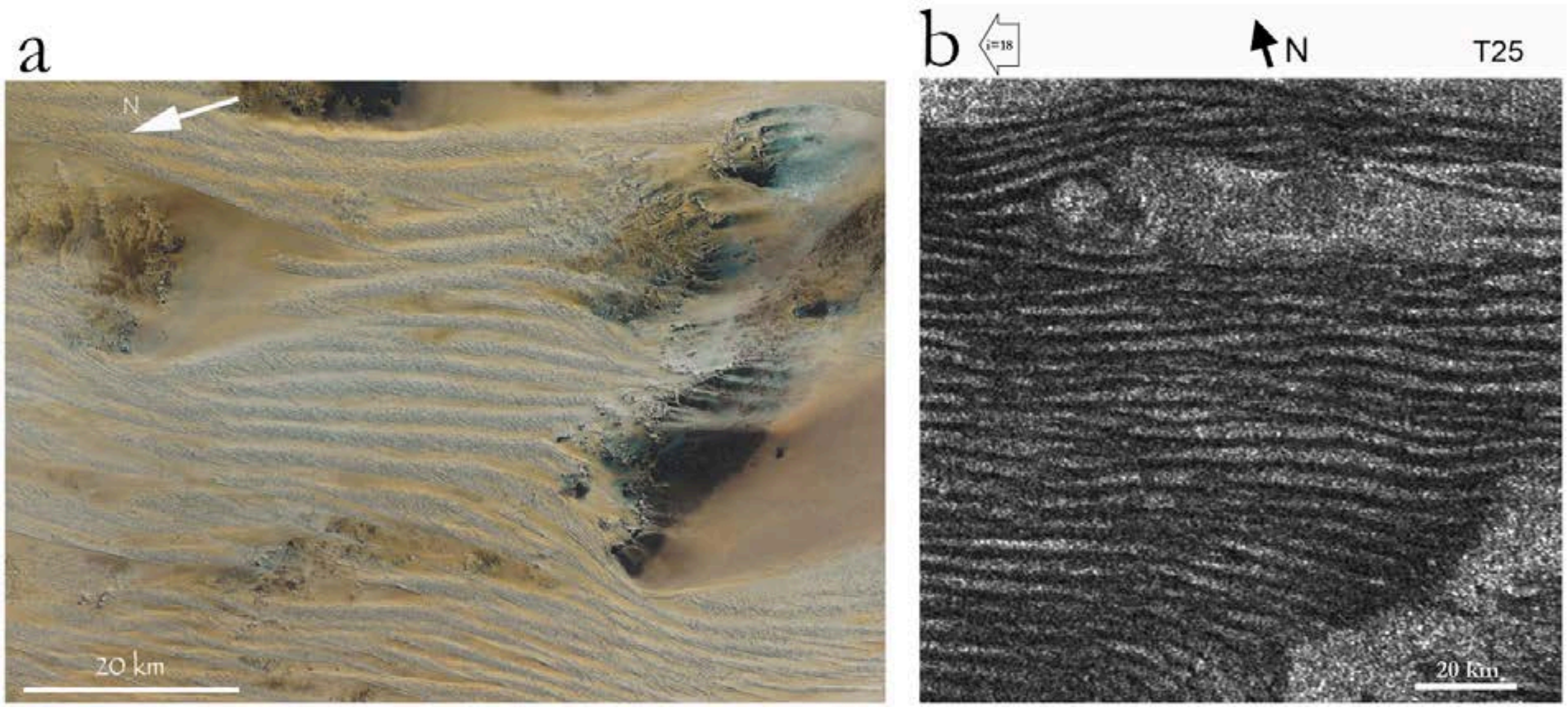

**Figure 14**: a. Visible Landsat image of dunes in Libya. Light colored dunes can be seen overlying a tan substrate and overriding rugged, dark mountains at right. Dust reveals the transport direction from left to right. b.Titan dunes in Cassini Radar, with dark (smooth at 2 cm) dunes overriding a Radar bright (rough) substrate. Transport direction is similar to that of Libya, as dunes stop upwind of bright obstacles. Open arrow in radar image indicates illumination direction and angle.

| **Table 5**: Terrestrial analogs to Titan Dunes |
|---|

| Titan location(s) | Terrestrial Analog Site(s) | Analog Type | Measurements | Notes |
|---|---|---|---|---|
| Belet sand sea densely spaced dunes | Namib Sand Sea | Morphological including size | Dune height, spacing, age, speed | Dunes here closely resemble the dunes of Belet where spacing is close and interdunes may have a sandy component. |
| Belet, Fensal/Aztlan topographic interactions | Namib Sand Sea inselbergs | Morphologic and wind | Orientations, spacing, height | Response of dunes to topography causes deviation of orientations; heights of obstacles may control deviations. |
| Shangri-La and other sand sea sparse dune or interdune dominant locales | Egypt Great Sand Sea, Australia Great Sandy Desert | Morphologic and causal | Interdune characteristics, dune spacing, width, length | Sand supply may drive density of dunes in each locale. |
| Shangri-La, Belet cross hatched or deviated dunes | Namib Sand Sea NE corner trellis dunes and raked dunes, United | Morphologic | Changes over time of dune orientation, cause of shape change | Change in pattern usually indicates wind controlled by topography or change in wind pattern over time. |
| | Arab Emirates SE corner | | | |
| Northern midlatitude yardangs | Argentine Puna mega and meso yardangs, Iran Lut desert megayardangs | Morphologic | Size and scale, location of current wind impact on yardangs | Megayardangs may be a result of wind and fluid interactions over long time periods. |

*Earth analogs:* Parallels to the dunes were soon recognized on the Earth and are among the strongest morphological analogues of any planetary feature. The large, linear/longitudinal dunes of the Namib sand sea are identical in size (which is rare in planetary analogs) and obstacle interaction to those of Titan (**Fig. 14**; Lorenz et al. 2006; Radebaugh et al. 2010). A number of studies have analyzed Cassini/VIMS data over Titan's major equatorial dune fields, including Fensal–Aztlan and Belet, and confirmed that these regions exhibit low albedo and a consistent near-infrared spectral behavior indicative of organic-rich materials (e.g., Brossier et al. 2018; Solomonidou et al. 2018). Their studies also showed that dune and interdune units are spectrally similar at VIMS resolution, possibly due to mixing or partial coverage, complicating compositional distinctions. Notably, the distribution of Titan's infrared units is similar to the transition from mountains to stony and sandy deserts on Earth, suggesting parallels in sediment transport and surface evolution (Brossier et al. 2018). Indeed, these spectral patterns and spatial associations are also reminiscent of terrestrial linear dune systems, such as those in the Namib Desert, where mobile sand units can partially mantle interdune substrates, supporting the use of Earth desert analogs for investigating surface-atmosphere interactions and sedimentary processes on Titan.

Additional terrestrial dune fields that may provide useful analogs include the gypsum dunes of White Sands (USA), the Rub’ al Khali (Arabian Peninsula) (Radebaugh et al. 2010) , the Simpson Desert (Australia), and Southwest Africa, which together span a wide range of wind regimes, sediment cohesion, and dune morphologies(**Table 5**; Telfer et al. 2019). Field studies to date have included examination of the dune sands, overall morphologies, changes over time, relationship to underlying topography and obstacles, and subsurface structure using Ground Penetrating Radar (Chandler et al. 2022). Such studies reveal slow, down-axis movement of sand and long reconstitution timescales that likely translate to the condition of dunes on Titan (Ewing et al. 2015). Dunes are quite different from interdunes, or the between-dune materials, which on Earth are more often clays or evaporites (**Fig. 15**; Barnes et al. 2008; 2011).

Longitudinal dunes often halt or diverge around obstacles to sand transport, which helps reveal the dominant winds and transport direction (**Fig. 15**). This relationship helped establish the dunes of Titan as being of the longitudinal type (Radebaugh et al. 2008), since such behavior was recognized in dunes on Earth (Lancaster 1994).

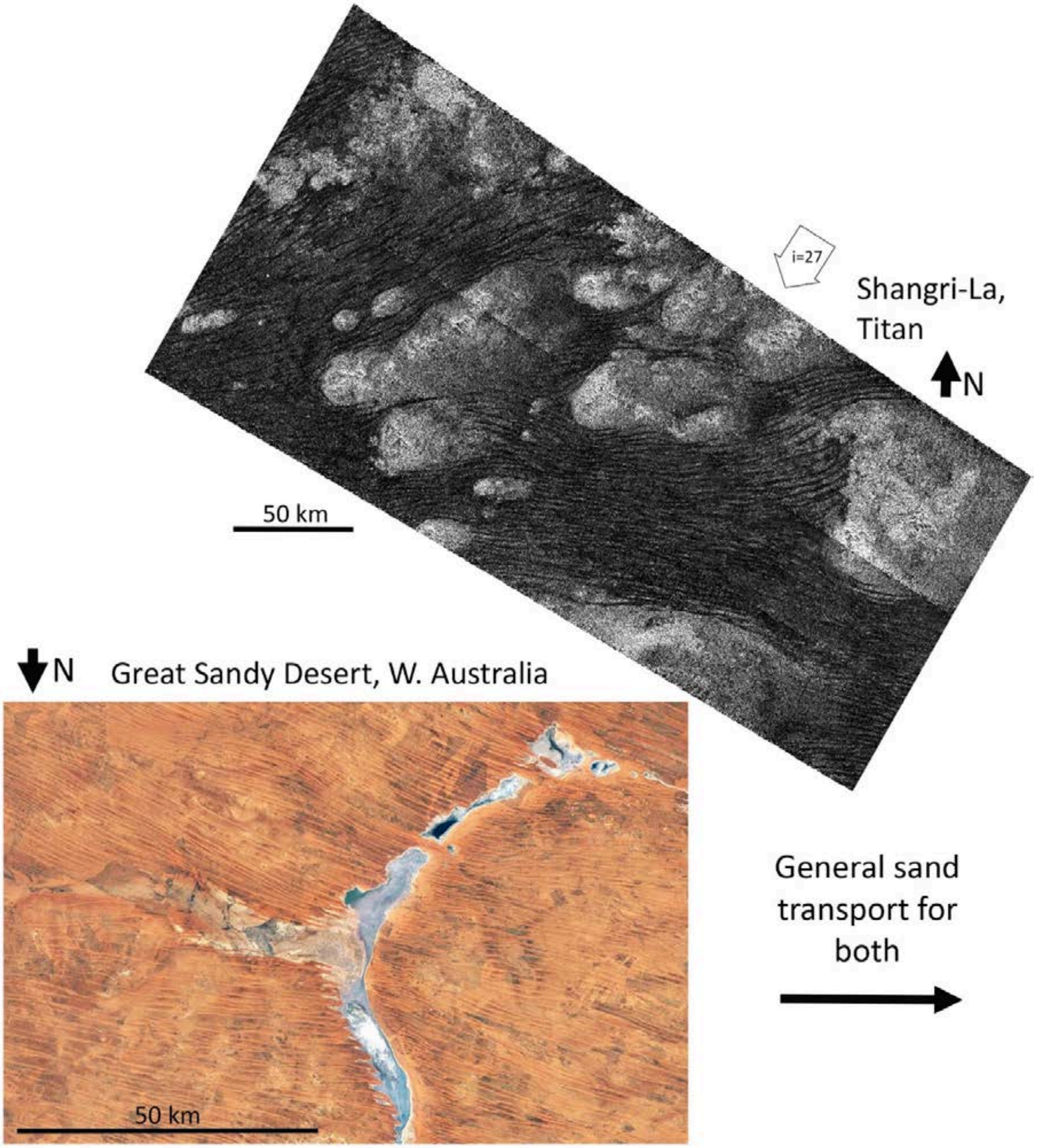

**Figure 15**: Dunes of Titan (top) in Cassini Radar are shown in comparison with visible imagery of dunes in western Australia (bottom). Both demonstrate dune movement from left to right, and halting at obstacles, such as hills or mountains (Titan) or clay pans or ephemeral lakes (Australia). Note Australia's dunes are about half the size of Titan's, while those of Africa and Arabia are similarly sized to Titan's.

*Titan Yardangs*: Large, linear features with Cassini Radar characteristics and shapes not consistent with dunes were observed at the high midlatitudes and were determined to be yardangs, or wind-carved ridges (**Fig. 16**; Paillou et al. 2016). These features on Earth are highly linear (more so than dunes) and parallel, with spacing similar to or smaller than that of dunes and lengths generally shorter than for dunes (Goudie 2007), though there are notable exceptions, such as the 100+ km yardangs of the Lut desert in Iran (**Fig. 16**). The width and spacing appear to arise from interactions between the wind and previously uniform substrate, while the lengths are determined by the amount of sediment available to be eroded by eolian action; in the case of the Lut desert, there was a large basin of lakebed clays exposed to wind action (Goudie 2007). Other materials are similarly soft and easily carved by wind, but they must also exist in hyperarid regions so that fluvial erosion doesn’t rapidly erode and redistribute the materials. On Titan, such features were found to exist on a dome, perhaps as exposed volcanic ash deposits or lakebed clays (Paillou et al. 2016). The presence of yardangs on Titan reveals (i) there is a variety of sediments on Titan, (ii) that deposits may be youthful compared to surrounding materials, and (iii) the action of wind on the surface.

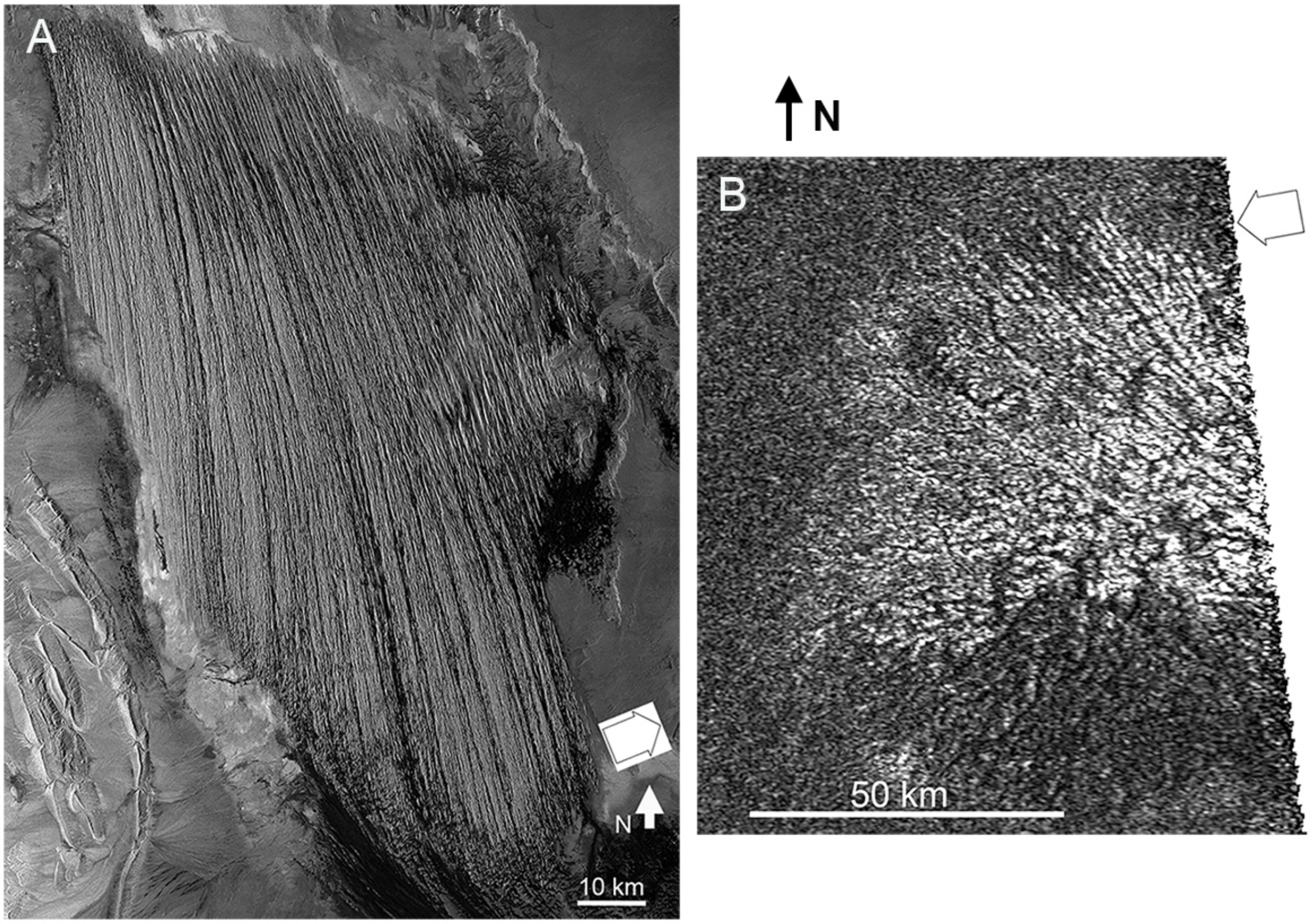

**Figure 16**: a. Yardangs of the Lut desert in Iran as seen by TerraSAR-X at 3 cm, b. Yardangs on Titan as seen by Cassini SAR at 2 cm. Both yardangs are bright, or rough, to SAR. Open arrows indicate direction of SAR illumination.

*Field measurements*: Field measurements of dunes and yardangs can include sampling the material to determine sediment origin and exposure or overturn timescale, Airborne, spaceborne or handheld LIDAR imaging can be performed and in situ morphology measurements, such as dune width, height and interdune spacing can be made. 3D terrain mapping at small scales, e.g. by drone, can help to determine the dune morphology and changes over time, providing information on the formation mechanisms (Solazzo et al. 2018, Guisado-Pintado et al. 2019). A further useful technique is ground-penetrating radar (GPR) which can be used to 'peel back' dune layers and detect compositional or physical layering (Zhao et al. 2018; Chandler et al. 2022).

## 2.5 Labyrinths

*Titan:* Labyrinth terrains on Titan are characterized as highly incised and dissected elevated plateaus composed of organic material (Malaska et al, 2020). They have morphological similarities to karst terrains on Earth (Saloman, 2006; Ford and Williams, 2007) . Various types of labyrinths have been proposed: valleyed, radial, finely dissected, and polygonal (Malaska et al. 2020).

Valleyed labyrinths appear as a mixture of plateaux or remnant ridges interspersed with valleys that often have radar-dark floors or fill suggestive of smoother materials. Some valleys appear closed (at the resolution limit of Cassini radar), while others are interconnected in networks with amphitheater-type heads (Malaska et al., 2020). The valley network patterns are interpreted to be formed from fluvial activity and vary from rectangular to dendritic with undifferentiated plains at the termini. The largest valleys are reminiscent of the largest river-carved canyons and valley networks seen on Earth. Radial labyrinths are a distinct type of valleyed labyrinth found in the mid-latitudes with a valley network pattern integrating away from the center of the feature. Radial labyrinths are found in two clusters in the Afekan and Tseghi regions (Malaska et al., 2019). The circular to elliptical planform spatial extent, dome-like overall topographic shape, and the radial nature of the valleys and fractures are suggestive that doming occurred prior to erosion, possibly caused by injection of cryomagma at depth or a density-driven diapir (Lopes et al. 2019; Schurmeier et al. 2017; 2023). Finely dissected labyrinths are likely valleyed labyrinths that have networks just at or below the resolution limit of SAR.

Polygonal labyrinths have the appearance of a series of cells inset into a plateau. They are morphologically similar to polygonal karstic labyrinth terrain found here on Earth (Williams, 1972; Ford and Williams, 2007). The valley networks are of low order with few connections. Polygonal labyrinths on Titan are defined as labyrinths having more than 50% of the valleys closed at the resolution of Cassini SAR. On Earth, closed valleys are distinctive for karstic dissolution on Earth (Ford & Williams, 2007) and are similarly thought to suggest karstic processes on Titan (Malaska et al., 2011; Cornet et al., 2015; Malaska et al 2020).

The labyrinth terrains are speculated to have formed from a process of uplift followed by dissection via both physical and chemical erosion due to the flow or percolation of methane rain (Malaska et al 2020). As liquid moves across the surface and into the subsurface, it would dissolve soluble organic materials and transport any insoluble icy and organic clastic materials, analogous to how water on Earth can dissolve carbonate (or other soluble materials) plateaux (Malaska et al 2022). Where dissolution dominates, sinkholes and dolines prevail and create polygonal karst (on Earth) or polygonal labyrinths on Titan. Where physical erosion dominates, fluviokarst or fluvial canyons would incise into the uplifted terrains (see Groves (1992) for how physical erosion could occur in a terrestrial carbonate karst system). As the system develops, subsurface drainage and material removal would cause collapse and reveal fluviokarst valley networks (see **Fig. 17(b)** - Sikun labyrinth) that widen to remnant ridges standing above an eroded plain coated with residuum (see **Fig. 17(c)** - Tupile labyrinth).

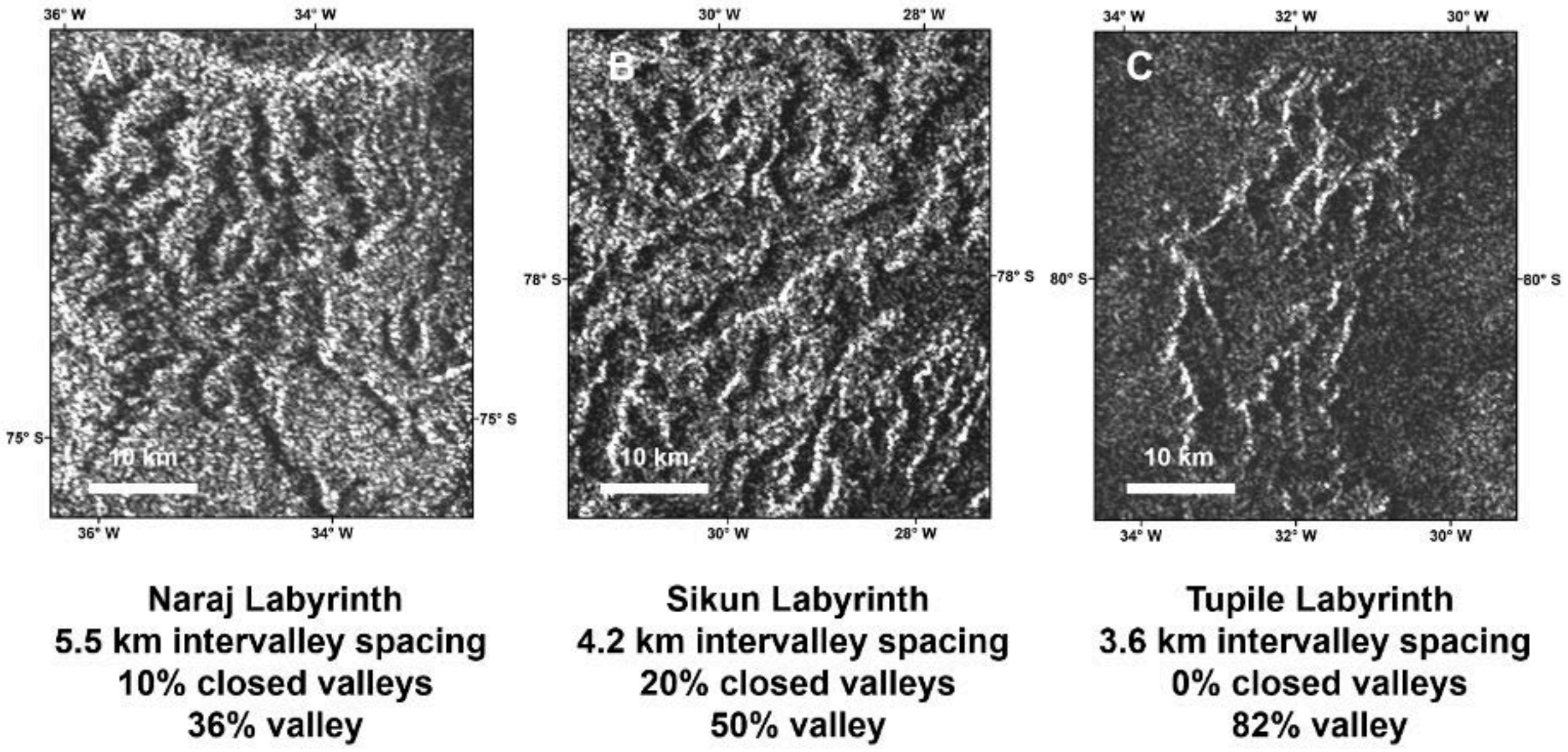

**Figure 17**: Evolution sequence showing a series of Titan labyrinths of nearly constant valley spacing showing the number of closed valleys decreasing while the valley floor spatial extent increases. For Tupile Labyrinth (C), this is nearly the end-stage sequence with the valley floors blending into the surrounding undifferentiated plain and only the remnant plateaux appearing as a series of thin remnant ridges standing proud above the plain. Figure taken from Malaska et al., 2020

The uplift mechanisms of the Titan labyrinths are enigmatic. Comparison between Titan's radial labyrinths and their terrestrial analogs could help resolve that mystery. Radial labyrinths appear compositionally (to the extent that can be determined from remote sensing) similar to their surroundings and are often clustered, dome-shaped and rise to similar maximum heights. This suggests that they may form in fields of rising intrusive materials, so two kinds of intrusions have been proposed: magma bodies and density-driven solid-state flow (Schurmeier et al. 2023). Terrestrial analogs of the radial labyrinths include large intrusive laccoliths and salt diapirs  (Gilbert 1877; Baker 1936; Cook 1954; Corry 1988; Arian 2012; Lopez-Mir et al. 2018; Maghsoudi 2021).

*Earth:* Karst terrain is widespread on the Earth (LeGrand 1973), and many subtypes have been identified (Veress 2020). On Earth, karst formation occurs as materials dissolve and form either vadose (above the liquid table) or phreatic (below the liquid table) fissures and conduits that enable interconnected subsurface transport of soluble and insoluble materials. As dissolution proceeds along existing cracks and faults, material is dissolved leaving a characteristic landscape of sinkholes, dissected plateaux and ridges (Ford & Williams, 2007). Many types of terrestrial materials may be eroded by aqueous solutions into karst, including dolomite, gypsum, halite, and silica (Ford and Williams, 2007; Wray and Sauro 2017). Karstic terrain often includes caves on Earth, and although suspected, have yet to be definitively identified on Titan. However, locations for further investigation have been identified (Malaska et al. 2022). On Titan, it is likely that methane from hydrocarbon rains is the solvating agent at least in the gravity driven vadose zone, while deeper aquifers of higher order hydrocarbons (ethane, propane or 1-butene) may contribute to diffusive phreatic conduit development. Soluble organic hydrocarbon materials such as acetylene, propene, and butane may play the role of the Titan equivalent of gypsum or calcite on Earth, while more insoluble or refractory materials such as tholins may need to be removed as clastic particles. In general, higher order hydrocarbons such as ethane have saturation concentrations as much as 100x the equivalent volume of methane, and would therefore be more efficient at dissolving soluble materials on Titan (see Malaska et al., 2022 and references therein.)

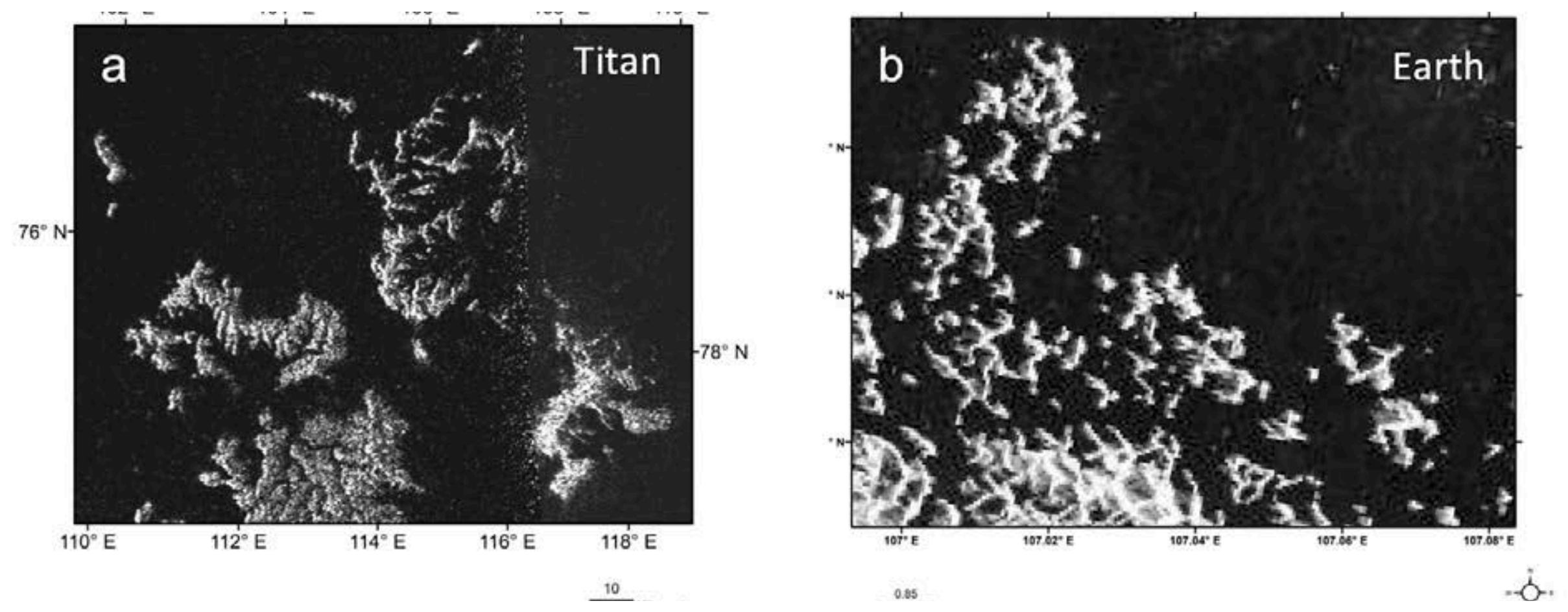

**Figure 18**: Caption: Flooded remnant karstic ridges compared to labyrinths. (a) Cassini SAR (2.2 cm wavelength) image of labyrinth terrains in Ligeia Mare. Image on right at 76°N, 110°E. N is at upper right, polar stereographic projection centered at 200°W, scale as indicated. SAR illumination is from the left with an incidence angle of 38°. Figure reproduced from Malaska et al., 2020 (Fig. 8 in text.) (b) TerraSar-X X-band (2.4 – 4 cm wavelength) terrestrial image of a portion of Ha Long Bay,Vietnam. Image centered at 20.88°N, 107.04°E. North is at top, WGS 1984 projection, scale as indicated. Radar illumination from right with minimum incidence angle of 37.944°. Image is shown at a reduced scale - approximately 56 m per pixel.

It is likely that the plateaux of Titan’s labyrinth terrains are uplifted sedimentary layers that have been dissected and eroded. The presence of closed valleys in many of the labyrinth terrains, at least to the image scale of the Cassini SAR images (roughly 200 m per pixel) is consistent with dissolution processes (Malaska et al., 2020). Many of Titan's organic materials may be soluble in methane or ethane liquids, but some materials, including water ice, may be insoluble. Similar to mixed karstic systems on Earth, or even silica-karst systems, chemical weathering of intragranular connections weakens sedimentary bedrock while insoluble grains are physically eroded and transported out of the system as clastic materials through surface and subsurface channels and conduits (Wray, 1997). Some dissolution-dominated terrestrial analogs for Titan's labyrinth include the flooded remnant karstic ridges of Hai Long Bay in Vietnam (**Fig. 18**) (Khang 1985; Waltham, 2005; Malaska et al., 2020), the uplifted and dissected karstic terrains like Croatia's Biokovo Range (**Fig. 19b**) (Garasic 2021) or large karstic provinces such as those in Gunung Kidul of Java (**Fig. 20**) (Haryono and Day, 2004) .

In the case of a purely clastic erosion system, the dissected canyonlands of the Utah including Grand Staircase, White Canyon (Figure 19b), and Glen Canyon regions would serve as an apt analog, where an immense sequence of sedimentary rocks are fluvially dissected, some displaying evidence of groundwater sapping (Laity and Malin, 1985). An interesting difference between the two worlds is that on Earth some of the grains may have originated from eroding crust, while on Titan some of the organic grains may have originated from skyfall organics. Therefore, on both worlds, sedimentary deposition, induration, uplift, and subsequent erosion could play a major role in the continued recycling of sedimentary deposits. A study of the layers exposed in Titan's labyrinth canyons can reveal the deep history of how Titan's chemical production, sedimentary deposition, and climate cycles have evolved through time, but will require higher resolution imaging and compositional data from a future mission (Malaska et al, 2020, 2022).

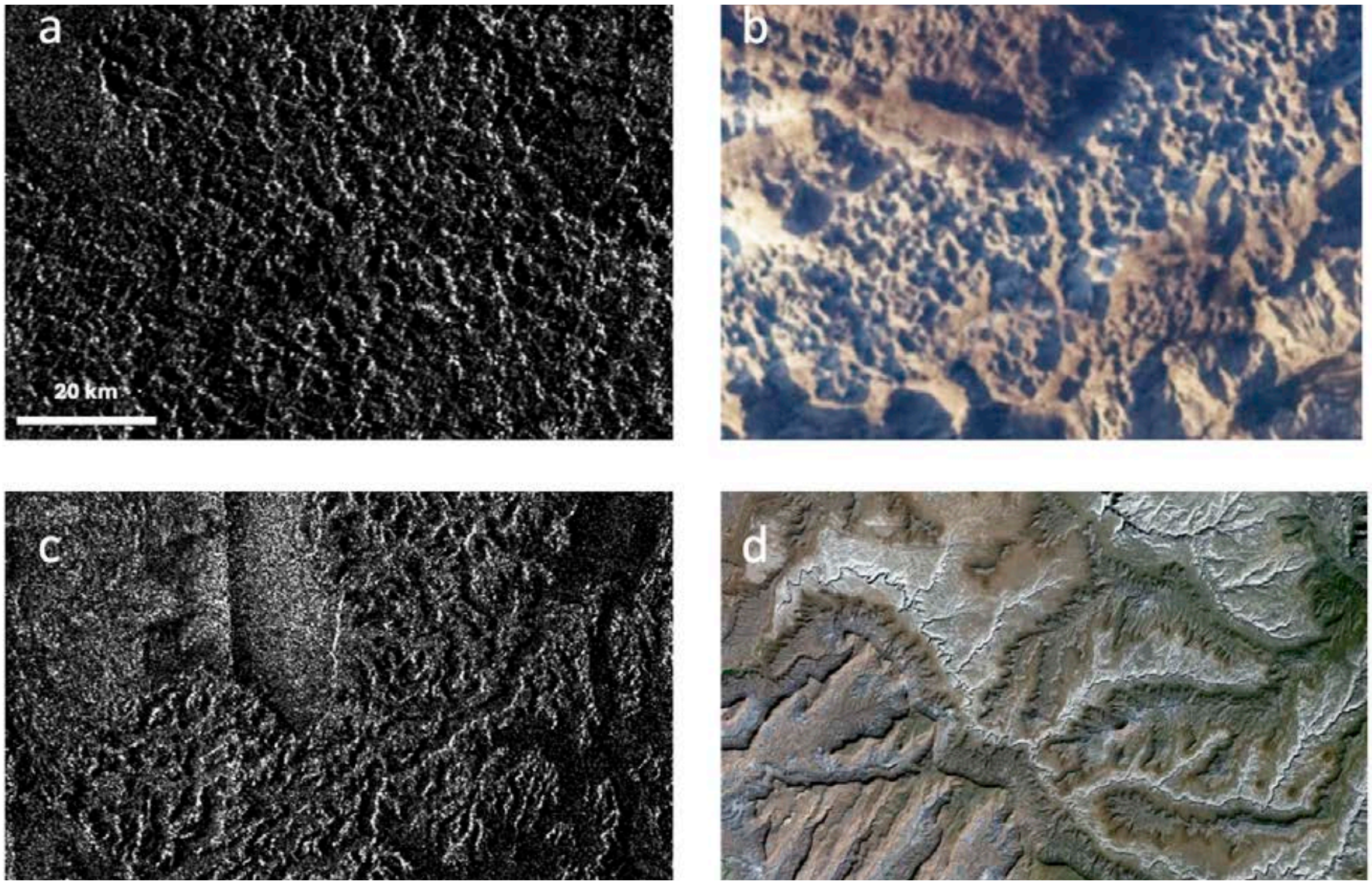

**Figure 19**: (a) Ecaz labyrinth (37°W, 82°S) (b) Croatia's Biokovo Range (NASA Earth Observatory). (c) Sikun Labyrinth (33°W, 77°S) (d) White Canyon, Utah.

Uplift mechanisms on Titan require further study and could provide information about key geophysical processes of the ice crust of Titan. A process similar to laccolith formation has recently been proposed for Titan (Schurmeier et al., 2017; 2023). On Earth, laccoliths exist where conduits of large volumes of magma rise until reaching a prominent rheological contrast (on Earth, a strong rock layer, on Titan this could be the brittle-ductile transition of ice). There, the magma spreads horizontally as sill to a radius determined by the thickness of the overlaying layer, and then bulges upwards and flexes the overlying layers (Corry, 1988; Jackson, 1997). The flexed upper layers experience flexure, tensile stretching, bedding plane slip, and form radial fractures (Jackson and Pollard, 1990). These fractures are more prone to erosion and can form valley networks. We see such fields of very large laccoliths spread across Utah, including the Henry Mountains laccoliths (**Fig. 21a**), Pine Valley Mountain laccolith (**Fig. 21b**), and Navajo Mountain (**Fig. 21c**) (Gilbert 1877; Baker 1936; Cook 1954; Corry 1988).

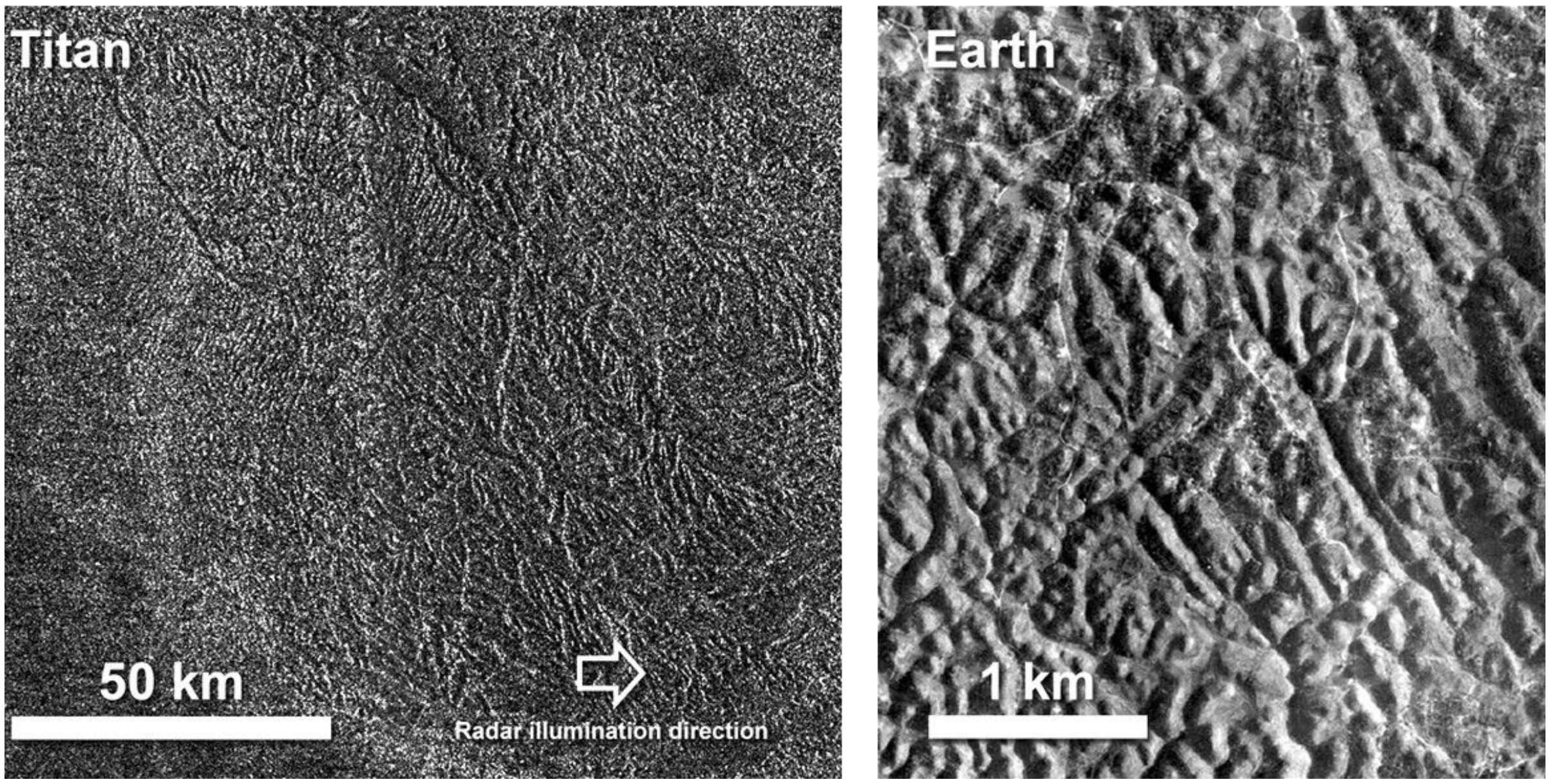

**Figure 20**: (A) Synthetic-aperture radar image of Lankiviel Labyrinth [LS1] by NASA's Cassini spacecraft during its T-120 pass over Titan's southern latitudes on June 7, 2016. Image is centered near 47 degrees south, 153 degrees west and covers an area of 87 by 75 miles (140 by 120 kilometers) with a resolution of about 1,300 feet (400 meters). Radar illuminates the scene from the left at a roughly 36-degree incidence angle. (B) Aerial photograph of the Gunung Kidul polygonal karst region in southern Java. Image: NASA.

In an alternate mechanism, clusters of diapirs topped with radially eroded valley networks can form on Earth through the buoyant upwelling of a relatively low-density layer that has the ability to flow on relatively short timescales. The spacing between diapirs may be related to the thickness of the rising layer (Turcotte and Schubert 2002). The height of the diapir is also controlled by the density contrast with the surrounding materials, and driven by differential loading (e.g., lateral differences in overburden thickness). As these diapirs rise, they also stretch, fracture and weaken the overlying layers, and are more prone to forming radially eroding channel networks. On Earth, examples include fields of salt diapirs in Iran's Dasht-e Kavir (Great Salt Desert) (**Fig. 22b**) and northern Canada's Sverdrup Basin (Arian 2012; Lopez-Mir et al. 2018; Maghsoudi 2021). The studies of all these features can provide insight to the uplift mechanisms that occur on Titan.

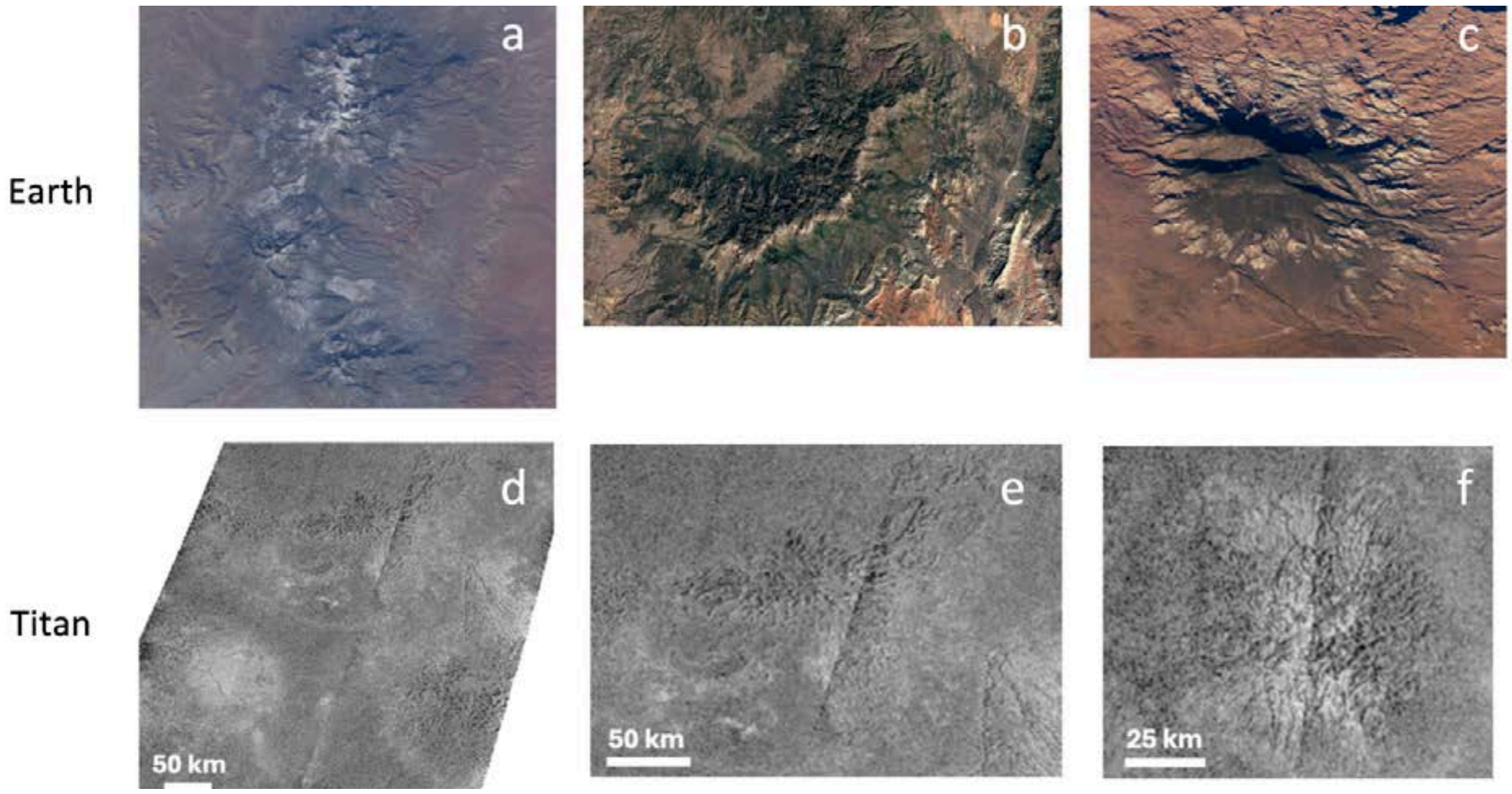

**Figure 21**: Radial labyrinths compared to laccoliths on Earth (a) Henry mountains laccoliths in Utah (~95 km long). (b) Pine valley mountain laccolith (~30 km wide) (Google earth). (c) Navajo Mountain laccolith in Utah (~14 km wide). (d) Richese, “hot cross bun” and other unnamed labyrinths (159.2°E 38.3°N) (e) Richese labyrinth (f) Anbus labyrinth. See Data Availability for sources.

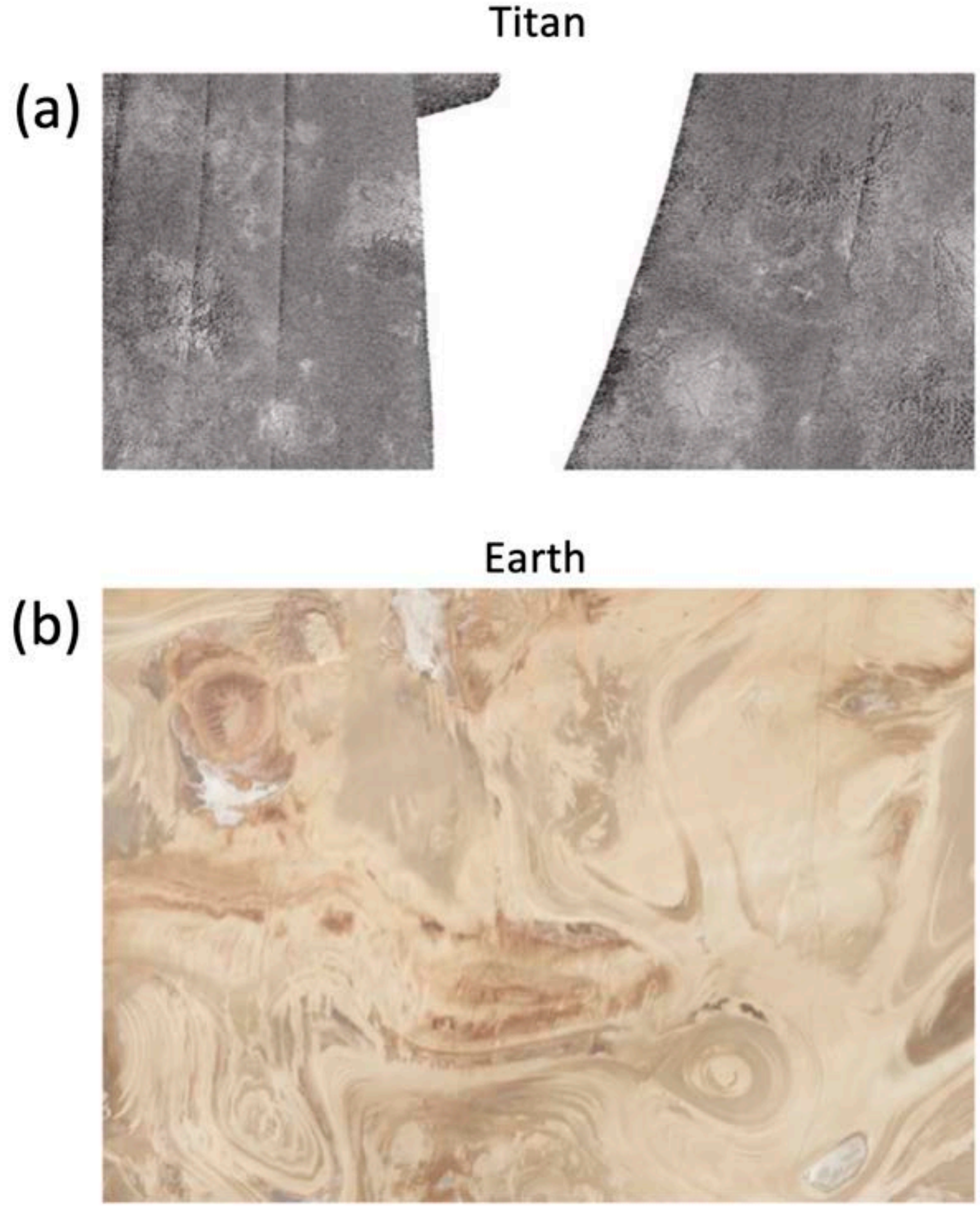

**Figure 22**: Radial labyrinths compared to diapirs on Earth  a) Cluster of radial labyrinth terrains including Anbus, Richese, “hot cross bun” and other unnamed labyrinths. (b) Salt domes in Iran’s Dasht-e Kavir,  or Great Salt Desert (~3-10 km wide). See Data Availability for sources.

*Field measurements*: Comparisons and contrasts between the different types of karstic/fluviokarst erosive valleyed terrains on Earth and Titan can help determine formation mechanisms (**Table 6**). On Earth, many of the karstic terrains are “pure” and result from biological (carbonate reefs), or evaporative (gypsum) deposits that are later uplifted and eroded. On Titan, it is more likely that the sediments had a more complex and mixed composition, thus analogous complex terrestrial systems may provide better comparison. Morphological studies can also try to explain the clear scale disparity between terrestrial karst and Titan karst. Key measurements would include valley widths and scales, intervalley distances, compositional measurements, and overall karstic development styles. A spectrum from pure karstic to pure clastic (erosive) processes exist on Earth, and comparison of examples to Titan terrains would be fruitful. Study of key distinctive morphological features at small scale (such as predicted cave features for Titan, see Malaska et al., 2022) could create expectations and predictions for observations by future in situ missions such as *Dragonfly*.

**Table 6:** Terrestrial analogs to Titan labyrinths

| **Titan location(s)** | **Terrestrial Analog Site(s)** | **Analog Type** | **Measurements** | **Notes/ limitations** |
|---|---|---|---|---|
| Cluster of Richese, Anbus, “hot cross bun” and adjacent unnamed labyrinths Cluster of Richese, Anbus, “hot | Laccoliths in Utah: Henry Mountains laccoliths, Pine Valley Mountain, and Navajo Mountain | Formation process; morphological (Radial labyrinths and dome-shaped “hot cross bun” ) | Spatial dimensions of uplifted area. Uplift vertical distance. Spacings between uplifted regions. | Cryovolcanism is more difficult to initiate than regular silicate volcanism because cryomagmas are negatively buoyant. It may require a pressurized melt source. |
| cross bun" and adjacent unnamed labyrinths | | | | Despite these differences, they are a similar scale as some large laccoliths on Earth. |
| | Density driven diapir fields in Dasht-e Kavir and northern Canada's Melville Island (Sverdrup Basin) | Formation process; morphological (Radial labyrinths and dome-shaped "hot cross bun") | Spatial dimensions of uplifted area. Uplift vertical distance. Spacings between uplifted regions. Composition, densities, and viscosities. | Thick and mobile low-density deposits have yet to be confirmed in Titan's upper crust. The scales of these features are different. Salt domes tend to be smaller, from a few to ten kilometers in length. |
| Sikun labyrinth | Labyrinth Karst, Gunung Kidul kegelkarst, Java, Indonesia White Canyon, Utah; Grand | Morphological (Valleyed labyrinth) | Composition, dimensions and densities of closed valleys. Amount of soluble vs. insoluble materials. Surficial | Since there are some closed valleys in Sikun Labyrinth. Dissolution may play a larger role in these valley formations. |
| | Staircase region | | armoring. Valley spacing and dimensions. | |
| Ecaz labyrinth | Croatia’s Biokovo Range | Morphological (polygonal labyrinth) | Composition; polygon dimensions, feature densities | |
| Lankiviel and Lampadas labyrinths | Karst in southern Java Indonesia and Guangxi Province, China | Morphological (valleyed labyrinth) | Composition; amount of solubles vs. clastics. Valley bottom materials. Valley dimensions, feature densities. | |
| Ligeia labyrinths | Karst in Ha Long Bay | Morphological (near lake/sea labyrinth) | Composition. | |

Using field measurements to characterize known terrestrial regions of laccoliths and salt domes could inform our ability to differentiate between the cryovolcanic laccolith vs. clathrate diapir hypotheses using Titan datasets. Morphologically, the domes produced by either process are similar (e.g., Vera-Arroyo and Bedle). Measurements should include seismic reflection and refraction to interrogate structure at depth (i.e., sill feeding a laccolith versus strata piercements by salt; Jackson and Pollard, 1988), as well as structural analysis of associated folding and faulting (which can be seen both in the seismic data, but also as part of a mapping campaign both on the ground and using remote sensing).

For example, a Titan seismic survey conducted near the radial labyrinth could reveal the crustal structure and distinguish between liquid intrusion at the brittle-ductile transition zone. The laccolith hypothesis necessitates a 1-km thick crust of insulating methane clathrate at the lower heat flow of 4 mW/m2, while a thicker thicker clathrate crust would favor the diapir hypothesis; *Dragonfly* may be able to constrain the thickness of a clathrate crust (Marusiak et al., 2022). Additionally, dedicated gravity measurements from a future Titan orbiter (e.g., Sotin et al., 2017; Barnes et al., 2021; Seltzer et al., 2025) could be used to identify gravity anomalies of varying magnitude that could differ between a cryovolcanic laccolith versus clathrate diapir at depth.

## 2.6 Fluvial features

*Titan Overview*: Titan's fluvial systems have been extensively mapped using Cassini's remote sensing instruments. The highest resolution data were acquired by the Synthetic Aperture Radar (SAR), which reliably resolved valley networks wider than ~700 meters (Miller et al., 2021). Broader-scale fluvial features and/or indirect evidence of fluvial transport of sediment were also detected in data from the Visual and Infrared Mapping Spectrometer (VIMS) (e.g., Barnes et al., 2007; Brossier et al., 2018; Kutsop et al., 2024) and Imaging Science Subsystem (ISS) (Porco et al., 2005). Numerous studies analyzed these datasets individually or in combination, with a more complete review provided in Birch et al. (2025a).

While these datasets have enabled significant inferences about Titan's hydrologic activity (Hayes et al., 2018; Birch et al., 2025a), it is important to first recognize that the resolution of Cassini instruments is insufficient to distinguish the channel bed and banks from the broader valley and surrounding terrain (**Fig. 23**). This means that observations of Titan's rivers are necessarily indirect. Indeed, much of our current understanding of Titan's fluvial system is built on geomorphic and sedimentologic assumptions developed on Earth. Accordingly, the term "river valley" or "valley network" should be used in place of "river channel" when referring to Titan's fluvial features, to reflect the unresolved nature of the active channel (if present) within these broader landforms. This distinction is critical, as channel (not valley) widths are used for hydraulic reconstructions of flow, and the fluvial valley is often significantly wider than the actual channel, a relationship well documented on Earth and presumed to apply to Titan as well.

In practice, the observable component of Titan's fluvial system is limited to the widest, most prominent segments of the drainage network (Miller et al., 2021) – those that can be resolved at the catchment scale. The only exception is the Huygens landing site (**Fig. 23a**), where in situ observations offered the sole opportunity to examine features closer to the channel scale (Tomasko et al., 2005; Perron et al., 2006; Soderblom et al., 2007; Daudon et al., 2020). Even here, however, evidence for an actual river channel bed and banks does not exist. Though rounded clasts imaged at the landing site (**Fig. 23c**) were interpreted as possible indicators of fluvial transport (Tomasko et al., 2005), deriving a magnitude of the flow was not possible. It will not be until *Dragonfly* (Barnes et al., 2021) that we will be afforded direct observations of a channel bed and banks that will enable more definitive, quantitative studies of river dynamics on Titan.

Nevertheless, Cassini and Huygens observations offered strong evidence for widespread fluvial activity. The dendritic valley networks carved into a small hummock in Titan's equatorial dunes (Tomasko et al., 2005; Soderblom et al., 2007; Karkoschka & Schröder, 2016) – one of Titan's driest regions – suggests that fluvial incision and sediment transport is pervasive across a range of latitudes and elevations. Given that Titan is expected to experience 1–2 major methane rain storms per summer (Faulk et al., 2020; Olim et al., 2025), periodic flow and transport of sediment within rivers is likely ongoing. Compared to Earth, these flows may be especially efficient due to the buoyant, low-density nature of Titan's sediment (Burr et al., 2008), such that the beds and geometry of Titan's rivers may appear distinct from Earth's (Birch et al., 2023).

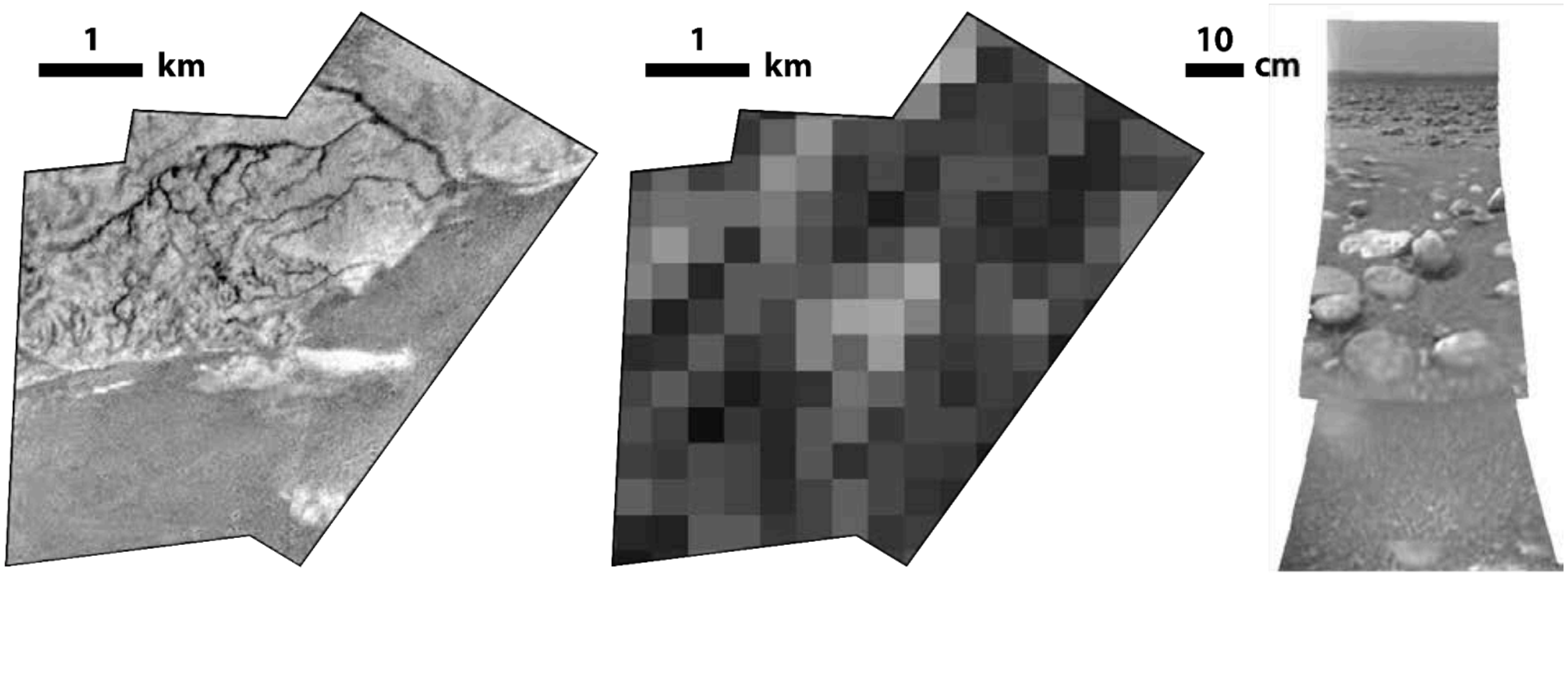

**Figure 23**: Left: Huygens image of Titan during descent, January 14th 2005; Middle: Cassini

SAR image of the same exact region covered by the Huygens image. The bright pixels in the middle correspond to a hummock, with no other detail afforded; Right: Huygens image of Titan's surface, with gravel- and sand-sized sediment both present.

Taken together, these observations point to a planet-scale fluvial system, shaped by intermittent but geologically significant methane rainfall (Perron et al., 2008; Burr et al., 2013). While most specifics remain unknown (Birch et al., 2025a), the Cassini dataset provides a foundation for future investigations, and *Dragonfly* promises to usher in a new era of quantitative fluvial geomorphology on Titan.

Valley networks on Titan are primarily concentrated in the polar regions, particularly near the large northern seas, with a secondary cluster in the mountainous Xanadu terrain and scattered examples elsewhere (Miller et al., 2021). This distribution is consistent with expectations: greater relief and/or precipitation enhances discharge and therefore erosion and sediment transport, thereby increasing the detectability of fluvial features. Titan's valley networks exhibit a range of planform morphologies, including dendritic and rectangular patterns (Burr et al., 2013), as well as features indicative of sapping (Soderblom et al., 2007) and river capture (Burr et al., 2013). These networks are certainly evolving in response to spatial variations in discharge, lithology, tectonic activity, and sea-level fluctuations. As no single terrestrial analog captures this diversity, a suite of comparisons is required; Table 7 presents a list of terrestrial analogs with catchments in tectonically quiescent regions.

*Titan Polar Valley Networks & Fluvial Deposits:* The largest valley networks on Titan, in both width and total length, are located near the poles, primarily within the eastern hemispheres of the north and south polar regions (Burr et al., 2013; Miller et al., 2021). These networks drain toward Titan's major seas—both filled and dry—and are particularly prominent due to the presence of liquid hydrocarbons, which enhance radar contrast and facilitate detection. In contrast, the western halves of the polar regions are dominated by Titan's sharp edged depressions (SEDs), where no fluvial networks were observed (Miller et al., 2021). These lake-rich terrains are instead hypothesized to evolve primarily through some type of dissolution-based process, potentially with subsurface flow of fluids through an underlying soluble layer (Hayes et al., 2018).

The morphology of Titan's polar valleys strongly resembles terrestrial flooded river systems (**Fig. 24**). Many display broad valley mouths and liquid-filled valleys that extend far inland. Some, such as Xanthus Flumen, extend well into the adjacent seas, incised into the seafloor (Birch et al., 2025b). Cassini altimeter observations of the liquid surface elevations of one such network, Vid Flumina (**Fig. 24e**), were equal to sea level (the geoid-corrected liquid surface elevation) of Ligeia Mare (Poggiali et al., 2016), supporting the morphological interpretations that these are flooded, and the backwater zone extends a significant distance upstream. These same altimetry observations suggest that some, at least Vid Flumina, are incised within the surrounding terrains (Poggiali et al., 2016). Despite their scale, and the presence of sediment on Titan, Titan's northern valley networks exhibit no obvious river deltas (Birch et al., 2025b, **Fig. 24**). Only one example has been confidently identified along the shoreline of Ontario

Lacus at Titan's south, where Saraswati Flumen empties into the lake (Wall et al., 2010; Birch et al., 2018, **Fig. 24g**). The absence of deltas at most polar termini stands in contrast to terrestrial expectations and remains unexplained (Birch et al., 2025b).

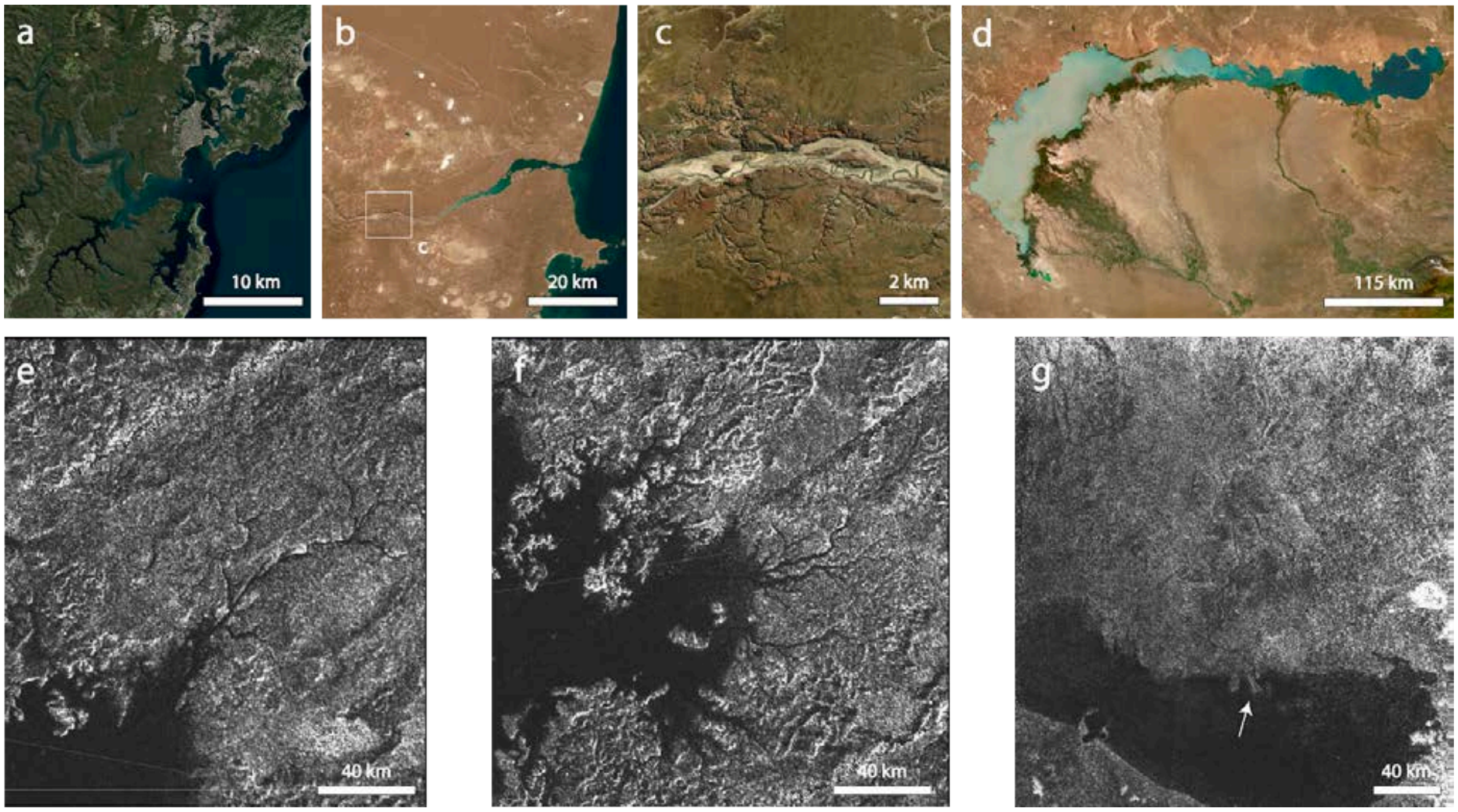

**Figure 24**: (a) Hawkesbury River, Australia; (b) Rio Deseado, Argentina. Both are flooded river valleys. The Rio Deseado has gravel along its bed almost until it reaches the coast (c), a significant length, and potentially analogous to rivers in Xanadu; (d) Lake Balkash, Kazakhstan. Its southern shoreline is potentially analogous to that of Ontario Lacus, as both are broad alluvial plains shaped by avulsing rivers; (e/f) Cassini SAR images of Vid Flumina and an unnamed river network, both of which are flooded and drain into Ligeia Mare; (g) SAR image of Saraswati Flumen and the southwestern margin of Ontario Lacus, interpreted as a broad alluvial plain. Titan's only known delta is marked by a white arrow. Panels a-d are visible image data, freely available from the ESRI World Imagery dataset.

There are several terrestrial analogs to Titan's polar valley networks (**Table 7**). The coastlines of Bar Harbor Maine (Belknap et al. 2002), the Bay of Fundy near Moncton Canada (Greenlaw et al. 2013), and the Hawkesbury River estuary north of Sydney Australia (Nichol et al. 1997, Collis 2013) are all flooded rocky coastlines, with water extending far up the reaches of their coastal river valleys (**Fig. 24a**). None host obvious river deltas. Valley networks like Saraswati and Xanthus Flumina may also be large alluvial (sediment-covered) rivers that are not incised deeply into the surrounding landscape. If so, the Congo River in the Democratic Republic of the Congo provides a compelling analog: it hosts a large river delta (like Saraswati Flumen) (Runge 2022), but is also actively incising the continental shelf offshore (like Saraswati Flumen and Xanthus Flumen) via submarine turbidity currents (Isupova and Dolgopolova 2016). Similarly, the SAR-dark plain that extends 100's of kilometers from the western coastline of Ontario Lacus, over which Saraswati Flumen flows, was interpreted as a broad sedimentary plain (Birch et al., 2025b, **Fig. 24g**). The southern shoreline of Lake Balkhash in Kazakhstan (**Fig. 24d**) (Sala et al., 2020) and the large inland deltas on Lakes Winnipeg, Athabasca, and Slave in Canada, all exhibit extensive low-relief depositional systems along lake margins and are good analogs for Saraswati Flumen and its surroundings (Bennet et al, 1973; Vanderburgh and Smith 1988; Peters et al. 2006; Ferguson et al. 2007). Celadon Flumina, also at Titan's south, clearly meanders (Malaska et al., 2011b) toward the margins of one of the dry paleoseas (Birch et al., 2018). Comparable features on Earth include the Río Colorado in the southern Altiplano, Bolivia, and the Suling Guole River in China, both of which are nearly unvegetated meandering rivers (Li et al., 2020). Numerous other meandering rivers on Earth could similarly serve as viable analogs, provided they too are lightly vegetated, and not dammed/affected by human development (Lapôtre et al., 2019).

Titan's methane-ethane-nitrogen fluids are also unique, in that their densities and viscosities may be highly variable over short distances (Steckloff et al., 2020). Where methane-rich and ethane-rich flows converge – such as at tributary junctions or along lake margins – the fluids will mix. Though Earth lacks such drastic changes in its fluid properties, analogs still exist where rivers of different sediment concentrations interface. A striking example is the Rio Negro–Amazon River confluence near Manaus, Brazil, where waters of differing properties remain distinct over long distances (Filizola et al. 2009; Laraque et al. 2009; Park et al. 2015). While mixing dynamics on Titan remains speculative at the time of this writing, their potential influence on surface hydrology warrants consideration with both field and laboratory analog studies.

*Titan Equatorial Valley Networks & Fluvial Deposits:* Titan's equatorial landscapes are markedly drier, with less abundant fluvial features, though many are still present (Miller et al., 2021). Part of this can be attributed to detectability biases (Miller et al., 2021). Specifically, large sand-bedded alluvial rivers with sandy floodplains would be invisible to Cassini SAR unless they are deeply incised (e.g., Celadon Flumina) and/or have enough fluid within them (e.g., Saraswati Flumen) to provide contrast in SAR images (Birch et al., 2025b). Even extensive, multi-kilometer-wide rivers would remain invisible if that river is not filled up to its bankfull width at the time of observation and/or its materials are similar in roughness to their surroundings (Miller et al., 2021). As a result, wherever deep standing fluids are absent, detection is biased toward gravel-bedded systems and/or incised valleys with rough valley margins (Miller et al., 2021). This is always the case outside the immediate vicinity of the liquid-filled lakes and seas. Similarly, deltas at the margins of now-dry paleoseas (Birch et al., 2018), or fine-grained fluvial deposits interfacing with SAR-dark dune fields or undifferentiated plains, would all be invisible to Cassini.

Despite these detectability biases, several notable low-latitude valley networks were both inferred and observed. Xanadu, in particular, hosts the densest concentration of valley networks on Titan outside the poles (Miller et al., 2021), including exceptionally long, SAR-bright rivers that extend for hundreds of kilometers (Radebaugh et al., 2011). The sharp western margin of Xanadu that interfaces with the Shangri-La dune field (**Fig. 25g**) has been speculated to result from a river limiting further eastward sand transport (Barnes et al., 2015; Marvin et al., 2025). On the eastern ejecta blanket of Menrva Crater, Elvigar Flumina forms a large braided valley network that terminates in an alluvial fan deposit (Birch et al., 2015; Radebaugh et al., 2015), which then further grades into small linear dunes. The bright appearance of SAR-bright channels has been interpreted to result from the presence of coarse gravel on the river's bed (Le Gall et al., 2010) and/or rough surrounding valley walls. If these networks do transport gravel over such long distances, this would also imply a mechanically durable sediment and significant fluid discharges. Unless gravel is consistently supplied locally along these networks, this would suggest minimal attrition of water ice under Titan-like conditions, counter to recent laboratory experiments (Maue et al., 2022).

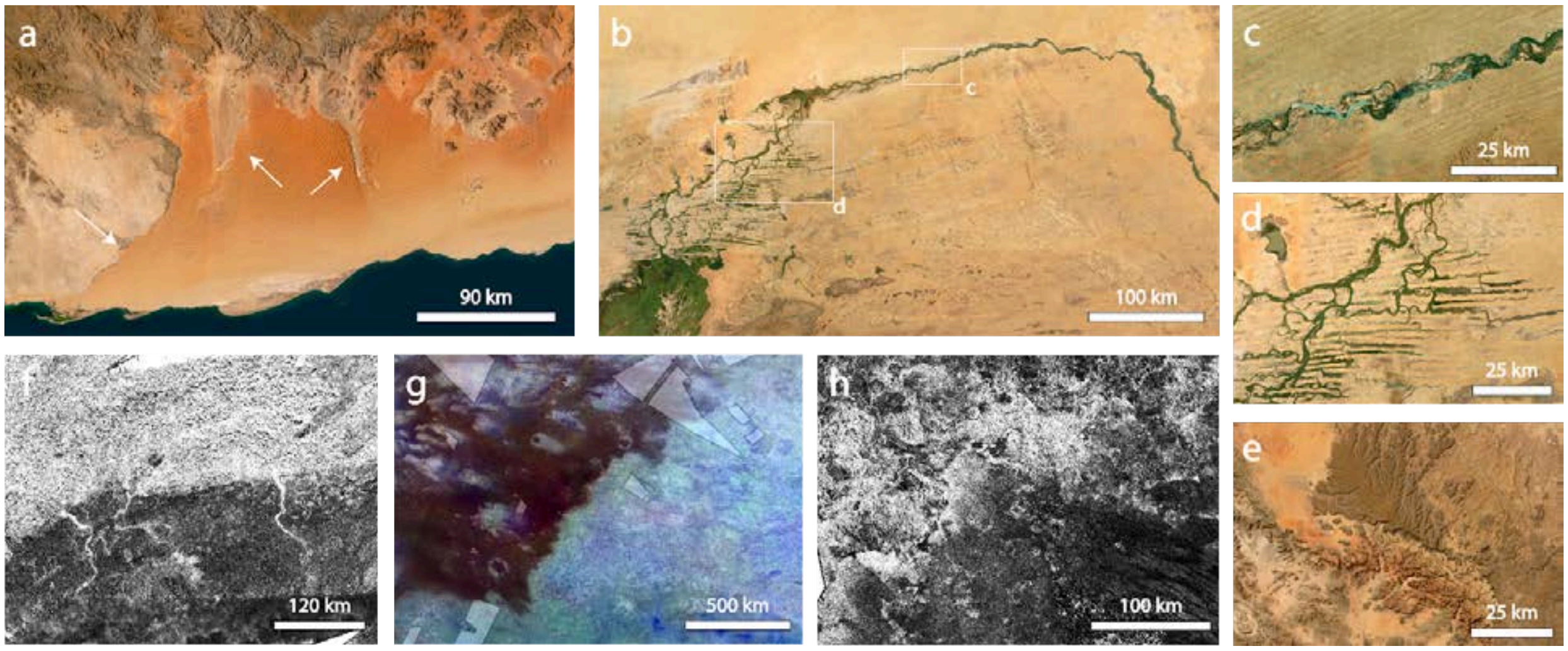

**Figure 25**: (a) Namib sand sea, Namibia. The Kuiseb river (left arrow) blocks the northward (to the left, only in this image) migration of the sand sea, while the Tsondab and Tsauchab rivers (middle and right arrows) puncture the dune fields; (b) The Niger River, Mali, blocks the southward migration of the Sahara sand sea (panels c and d). Complex, poorly-understood fluvial-aeolian interactions occur at boundaries like these. (e) Al-Disah, Saudi Arabia is a highly dissected plateau with active gravel transport out into the Arabian sand sea. Rivers (wadi) have eroded down to the elevation of the surrounding landscape; (f) Cassini SAR images of unnamed rivers flowing out of the southern margin of Xanadu vanish into the surrounding plains, potentially analogous to the Tsondab and Tsauchab rivers; (g) The western margin of Xanadu (here with VIMS data overlain SAR) has a sharp boundary, not unlike the northern Namib and southern Sahara sand seas, both of which are shaped by rivers; (h) Cassini SAR images of an unnamed collection of alluvial fans draining highlands with dunes nearby, not unlike the fluvially-eroded blocks in dune fields (e.g., panel e). Panels a-e are visible image data, freely available from the ESRI World Imagery dataset.

Some rivers also disappear a short distance after they exit highlands into the surrounding plains and dunes (**Fig. 25f**). The cause of this remains unresolved, but likely reflects a competition between fluvial and aeolian sediment transport timescales. A transition from gravel- to sand-sized bed materials may also play a role in reducing radar visibility downstream. Finally, at the Huygens landing site, we observed steeply incised valley networks (Tomasko et al., 2005; Soderblom et al., 2007; Daudon et al., 2020, **Fig. 23a**) in a bright elevated hummocky terrain (<200 m; Soderblom et al., 2007; Daudon et al., 2020). While no clear fluvial features appear in SAR data of the same region (**Fig. 23b**), co-deposited sand and gravel observed in the final Huygens images (Tomasko et al., 2005; Keller et al., 2008, **Fig. 23c**) suggest localized fluvial activity, again hinting at an ongoing interplay between fluvial and aeolian processes at these low latitudes.

Analog fluvial features within arid environments are common across Earth (**Table 7**), with the caveat that Titan's solids and fluids are fundamentally different those on Earth. The Niger River in Mali (Abrate et al. 2013), the Tarim River in China (Chen et al. 2013), and the Kuiseb River in Namibia (Grodek etal. 2020) are all sand-bedded alluvial rivers that limit the expansion of the Sahara, Gobi, and Namib sand seas respectively (**Fig. 25a-e**). The complex fluvial-aeolian interactions in the upper Niger river in Mali are particularly intriguing (**Fig. 25b**) (McIntosh 1983). Analogs for Titan's large SAR-bright, potentially gravel-bedded alluvial rivers, are also common on Earth, and include Rio Deseado in southern Argentina (Desiage et al. 2018), the Fitzroy River Australia (Bostock et al. 2007), and the Smoky River in Alberta, Canada (Fanti and Catuneanu 2009). All maintain coarse gravel beds over long distances and would appear SAR-bright to Cassini. Incised hummocky terrains are also widespread within Earth's deserts. Across Algeria, Saudi Arabia, Chile, and Yemen, for example, dunes advance into fluvially-incised valleys (**Fig. 25**). Meanwhile, in Namibia, alluvial rivers like the Tsauchab and Tsondab extend well into the Namib Sand Sea (Teller et al. 1990), keeping mud flats such as Sossusvlei from getting buried (**Fig. 25a**). All of these serve as equally good analogs for Titan's midlatitude hummocks (Lalich et al., 2022), which are similarly incised down to the level of the surrounding plains and can host alluvial fans (Birch et al., 2015, **Fig. 25**). Intriguingly, such fluvial-aeolian dynamics remain poorly understood on Earth (Liu & Coulthard, 2017) and require further field and numerical studies. To that end, *Dragonfly*'s observations near Selk Crater could clarify how fluvial channels flowing off the crater rim interface with dunes, and even help understand systems like the Tsauchab and Tsondab.

**Table 7**: Selected terrestrial analog sites for river valley types observed or proposed on Titan.

| **Titan example(s)** | **Terrestrial Analog(s)** | **Analog Type** | **Measurement** | **Notes/ Coordinates** |
|---|---|---|---|---|
| Any of the rivers around the northern seas like Vid Flumina, etc. | 1. Bar Harbor Maine, US<br>2. Bay of Fundy, Canada<br>3. Hawkesbury River, Australia | Flooded, rocky river valleys | Channel width; aspect ratio; sediment sizes; channel bed layering | 1. 44°26'N, 68°12'W<br>2. 45°49'N, 64°33'W<br>3. 33°34'S, 151°15'E |
| Xanthus Flumen, Ligeia Mare, among many | 4. Congo River canyon, Democratic Republic of Congo | Incised sub-aqueous channels | Channel width; aspect ratio; sediment | 4. 06°01'S, 12°23'E |
| | 5. Hudson River canyon, US | | sizes; channel bed layering | 5. 39º39'N, 72º28'W |
| Saraswati Flumen delta, Ontario Lacus | 6. Belize lobe of the Mississippi River delta, US (1 of many)<br>7. Rio Grande delta, US | River-dominated deltas | Channel width; aspect ratio; sediment sizes; channel bed layering | 6. 29º11'N, 89º17'W<br>7. 25º56'N, 97º14'W |
| Deltas would be unrecognizable in Cassini SAR, but are speculated to form the eastern shoreline of Ontario Lacus (Birch et al., 2025b). | 8. South coast of Lake Balkhash, Kazakhstan<br>9. River Neales delta, Australia<br>10. Saskatchewan River delta, Canada<br>11. Slave River delta, Canada<br>12. Athabasca River delta, Canada<br>13. Okavango Delta, Botswana<br>14. Inner Niger Delta, Mali | Inland river deltas | Channel width; aspect ratio; sediment sizes; channel bed layering | 8. 46º11'N, 75º32'E<br>9. 28º04'S, 136º54'E<br>10. 53º46'N, 100º47'W<br>11. 61º17'N, 113º36'W<br>12. 58º44'N, 111º26'W<br>13. 19º04'S, 22º34'E<br>14. 14º54'N, 04º08'E |
| Speculated to occur where two rivers of different composition/ | 15. Rio Negro-Amazon confluence, Brazil | Mixing fluids at river confluences | Channel width; aspect ratio; sediment sizes; channel bed layering | 15. 03º08'S, 59º54'W |
| temperature interface | | | | |
| Saraswati Flumen; Celadon Flumina; Speculated dry sandy channels in the undifferentiated plains | 16. Niger River, Mali<br>17. Tarim River, China<br>18. Kuiseb River, Namibia<br>19. Río Colorado, Bolivia<br>20. Suling Guole River, China | Sand-bedded alluvial rivers meandering within broad sedimentary plains | | 16. 16°46’N, 02°26’E<br>17. 40°51’N, 82°08’E<br>18. 23°22’ S, 15°02’E<br>19. 21°13’ S, 67°08’W<br>20. 36°32’ N, 97°15’W |
| SAR-bright rivers emanating from, and flowing within, Xanadu | 21. Smoky River, Canada<br>22. Fitzroy River, Australia<br>23. Rio Deseado, Argentina | Broad alluvial gravel rivers | Channel width; aspect ratio; sediment sizes; channel bed layering | 21. 55°42’N, 117°35’W<br>22. 18°30’S, 125°07’E<br>23. 47°11’S, 67°15’W |
| Huygens channels; Elvigar Flumina; Alluvial fans | 24. Naukluft Mountain Zebra Park – Namib Sand Sea, Namibia<br>25. Quebrada de Arcas, Chile<br>26. Tassili N’Ajjer National Park, Algeria<br>27. Badwater Basin, US | Incised bedrock valleys in steep, arid landscapes, with termini resulting in alluvial fan-dune interactions | Channel width; aspect ratio; sediment sizes; channel bed layering | 24. 24°56’ S, 16°22’E<br>25. 21°42’ S, 69°12’W<br>26. 24°56’ N, 08°50’E<br>27. 36°15’N, 116°15’W<br>28. 27°36’ N, 36°25’E |
| | 28. Al-Disah, Saudi Arabia | | | |

*Field measurements*: Fluvial-aeolian interactions remain understudied on Earth, but have obvious importance to Titan. In particular, quantifying the competing timescales between dune overprinting of rivers and fluvial erosion of dunes would help elucidate the underlying dynamics at river-dune margins. Given that the magnitude and timescale of precipitation events on Titan differs drastically from Earth (Olim et al., 2025), the dynamics may be similar but even more rich on Titan. Further, when doing such studies, development of dimensionless relations (e.g., Birch et al., 2023) are particularly useful for application to Titan, as they permit a generalization of the underlying physics that properly account for differences in the solid and fluid materials, along with gravity.

More analog remote sensing studies would also be of value for Titan science, especially for wavelengths other than the SAR instrument (e.g., Miller et al., 2019) and for portions of the valley network beyond just the largest branch of the river itself, including their deposits, floodplains, and upstream networks. Finally, the arid river analogs in Table 7 would be especially useful for testing new instruments that could be deployed on a future landed Titan mission. Instruments capable of making in situ measurements of the sediment flux(es) from an active Titan river(s), and/or make precise measurements of bed grain size and geometry would allow for answering fundamental outstanding questions in fluvial morphodynamics (e.g., What sets a river channel’s width? − Dunne & Jerolmack, 2020).

## 2.7 Surface Hydrocarbon Lakes and Seas

Following the findings of Voyager 1 about the composition of Titan's atmosphere (Hanel 1981), it was predicted that a long era of photochemistry during Titan's past would result in extensive seas, perhaps even a global ocean of hydrocarbons (Flasar 1983; Lunine et al. 1983). However, RADAR experiments in the 1990s began to accumulate evidence against a global ocean, and instead pointed to more isolated pockets of standing surface liquids (Muhleman et al. 1990; Campbell et al., 2003).

The first confirmation of standing bodies of liquid on Titan's surface came in 2005 when a sea, later named Ontario Lacus, was detected in the far south by ISS (Brown et al., 2008, **Fig. 24g**). A series of follow-up observations targeted the lake with RADAR and VIMS. From VIMS, Ontario Lacus was confirmed to have at least a partial composition of ethane (Brown et al. 2008) and bright annuli around the shoreline thought to represent evaporites (Barnes et al., 2009). RADAR and altimetry observations provided the highest spatial and topographic resolution of the lake and its environs. Altimetry data showed that the lake is quite deep and methane-rich (~80 m; Mastrogiuseppe et al., 2018b), counter to earlier work that speculated the lake to be exceedingly shallow (Cornet et al., 2012). These observations also mean that the ethane layer must be restricted only to the very top-most liquids. Meanwhile, SAR data show that Ontario Lacus' shorelines were dominated by sedimentary structures, including river deltas and potential wave-sculpted beaches (Wall et al., 2010; Birch et al., 2025b, **Fig. 24g**). Some evidence also pointed to ongoing fall of Ontario Lacus' liquid levels (Turtle et al., 2011b), though the evidence remains debated. The rest of the southern hemisphere is largely dry at the present day, with large paleoseas and many dry lake beds dominating the landscape (Birch et al., 2018). This would be consistent with long-term cycling of fluids between Titan's poles, driven by Croll-Milankovitch cycles (Aharonson et al., 2009; Hayes et al., 2018). However, recent global climate modeling casts some doubt on this explanation since the routing of surface runoff and distribution of basins are shown to largely dictate the location and volume of resulting surface liquids (Lora, 2024).

Shortly after the discovery of Ontario Lacus, in 2006 Cassini's RADAR discovered the northern lakes and seas (Stofan et al. 2007, **Fig. 26**). By 2009, the Sun illuminated the north, permitting acquisition of VIMS and ISS data of the north as well. Collectively, these data, acquired systematically over the next 10+ years, showed that the northern liquids were far more extensive than in the south (Aharonson et al. 2009; Lopes et al. 2019).

The large seas are predominantly located on the eastern hemisphere, and have multiple large rivers that drain into them (Stofan et al., 2007; Hayes 2016; Birch et al., 2017). Morphologically, the shorelines of the seas show evidence for sea level rise (Hayes 2016), with fluids now inundating the surrounding landscape and extending far up the surrounding drainage basins (**Fig. 25 & 26**). Evidence for modern wave erosion of these flooded landscapes also exists (Palermo et al., 2024). Rivers that drain into the seas also lack any obvious river deltas (**Fig. 25**), and instead appear to incise significant distances into the sea interiors (Birch et al., 2025b). Circular, 5-km wide pits are also pervasive across the floors and shorelines of the seas (Birch et al., 2025b).

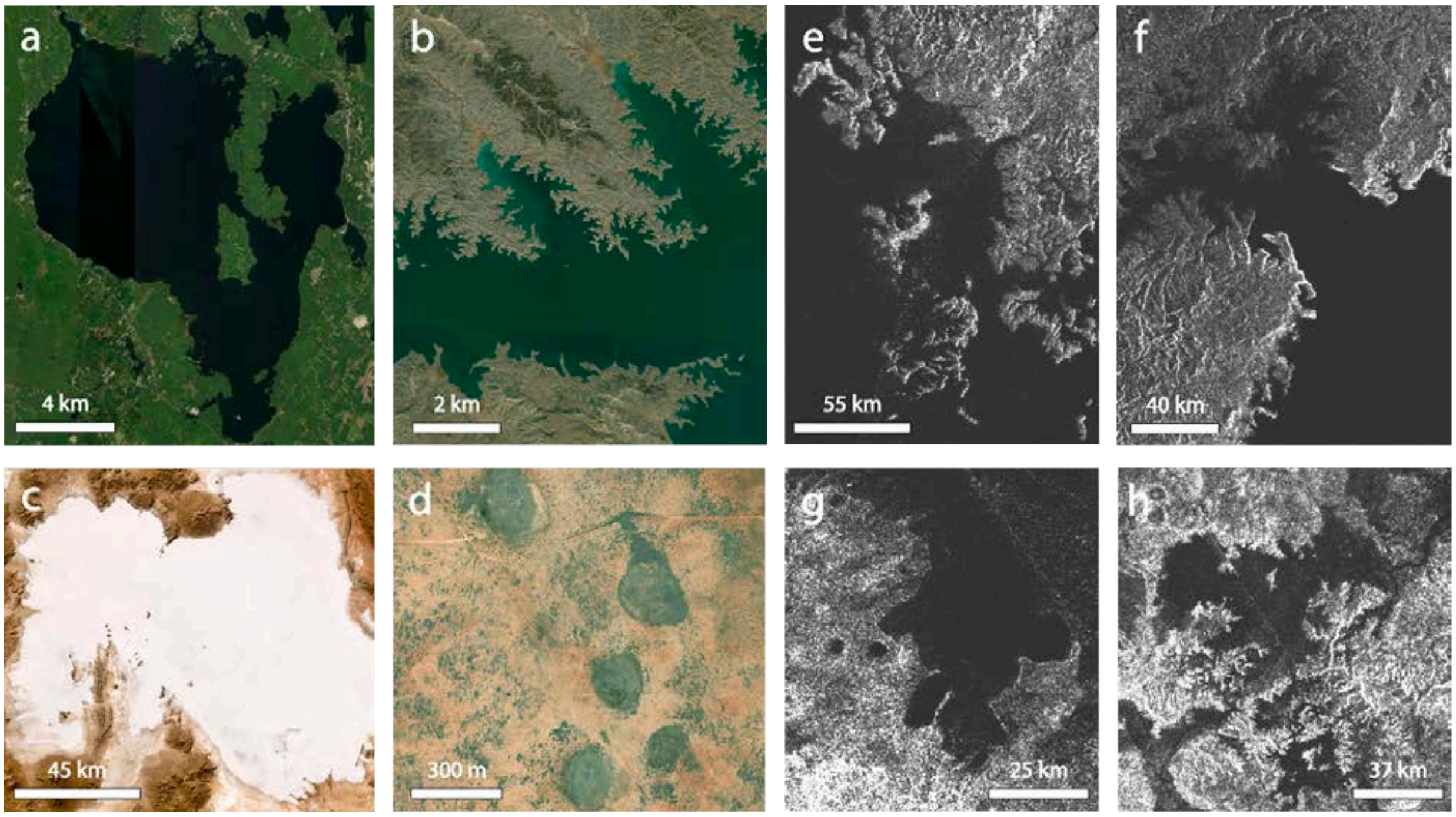

**Figure 26**: (a) Sebago Lake, Maine US hosts characteristically wave-eroded shorelines (Thompson et al. 1995); (b) Fort Peck Lake, Montana US is a flooded reservoir along the Missouri River with characteristically flooded shorelines (Jensen and Varnes 1964); (c) Salar de Uyuni, Bolivia is the world's largest salt flat, serving as a compositional analog for Titan's empty SEDs and/or seas (Risacher and Fritz 1991); (d) Bottomless Lakes State Park, New Mexico US are home to deep karst-formed lakes that are surrounded by cliffs, potentially analogous to Titan's SEDs and/or small circular lakes (panel g) (McLemore 1999); (e) Wave-eroded and flooded shorelines (Palermo et al., 2024) of Ligeia Mare; (f) Wave-eroded and flooded shorelines of Kraken Mare (Palermo et al., 2024), with the large island Mayda Insula in the bottom left; (g) 5-km wide circular pits of Jingpo Lacus; (h) Flooded shorelines of north Kraken Mare. Panels a-d are visible image data, freely available from the ESRI World Imagery dataset.

The smaller lakes, meanwhile, are scattered across the pole, but are dominantly located in the western hemisphere in the “lake district” (Birch et al., 2017). Some of these smaller lakes are referred to as sharp-edged depressions (SEDs) in the literature (**Fig. 27**), and are unlike anything on Earth (Hayes et al., 2017). They characteristically have steep walls, flat floors (Michaelides et al., 2016), and raised rims (Birch et al., 2019) around their perimeters. They appear both filled and empty, but few have rivers draining into them (**Fig. 27**). Analysis of the elevation data of the floors of empty lakes compared to the liquid levels of local filled lakes with the same lake level was also consistent with interlake connectivity through a subsurface methane aquifer (Hayes et al., 2017). A fraction of the SEDs also show evidence for evaporitic material - either on their floors or surrounding them (Barnes et al., 2011; MacKenzie et al. 2014, 2016, **Fig. 27**). Many evaporite deposits are also found in local lows away from any SEDs (MacKenzie et al. 2014, 2016). Altimetry data also show that empty lake floors are compositionally distinct from their surroundings (Michaelides et al., 2016). Together, these coarse compositional constraints further the notion that a dissolution-based process controls their formation and evolution.

Cassini VIMS, altimetry, and bi-static radar observations revealed the composition of the fluids within the lakes and seas at various locations and times (Brown et al. 2008; Barnes et al., 2009; Mastrogiuseppe et al. 2014, 2016, 2018a, 2018b, 2019; Poggiali et al., 2020; Poggiali et al., 2024; among many others). VIMS and the bi-static observations only sensed the upper surface of the liquids, and radar altimetry measurements provided a bulk average of the fluid column’s composition through which the signal traversed. Broadly, all fluids are methane-dominated, with larger methane abundances in the liquid bodies residing in the north polar region than the south polar region. Up to 25 mole% dissolved nitrogen may also be present in Titan's liquids depending on the methane composition and temperature (Malaska et al., 2017; Steckloff et al., 2020).  Cassini radar altimeter observations also directly measured the depths of Ligeia Mare, Ontario Lacus, Punga Mare, Baffin Sinus, Winnipeg Lacus, Moray Sinus, and several smaller lakes to be around 20-200m in depth (Mastrogiuseppe et al. 2014, 2016, 2018a, 2018b, 2019; Le Gall et al. 2016; Poggiali et al., 2020; 2024). Attempts were made to measure the depth of Kraken Mare as well, but the signal never detected the seafloor, suggesting depths >200 m and/or very ethane-rich compositions (Poggiali et al., 2020; Lorenz et al., 2021).

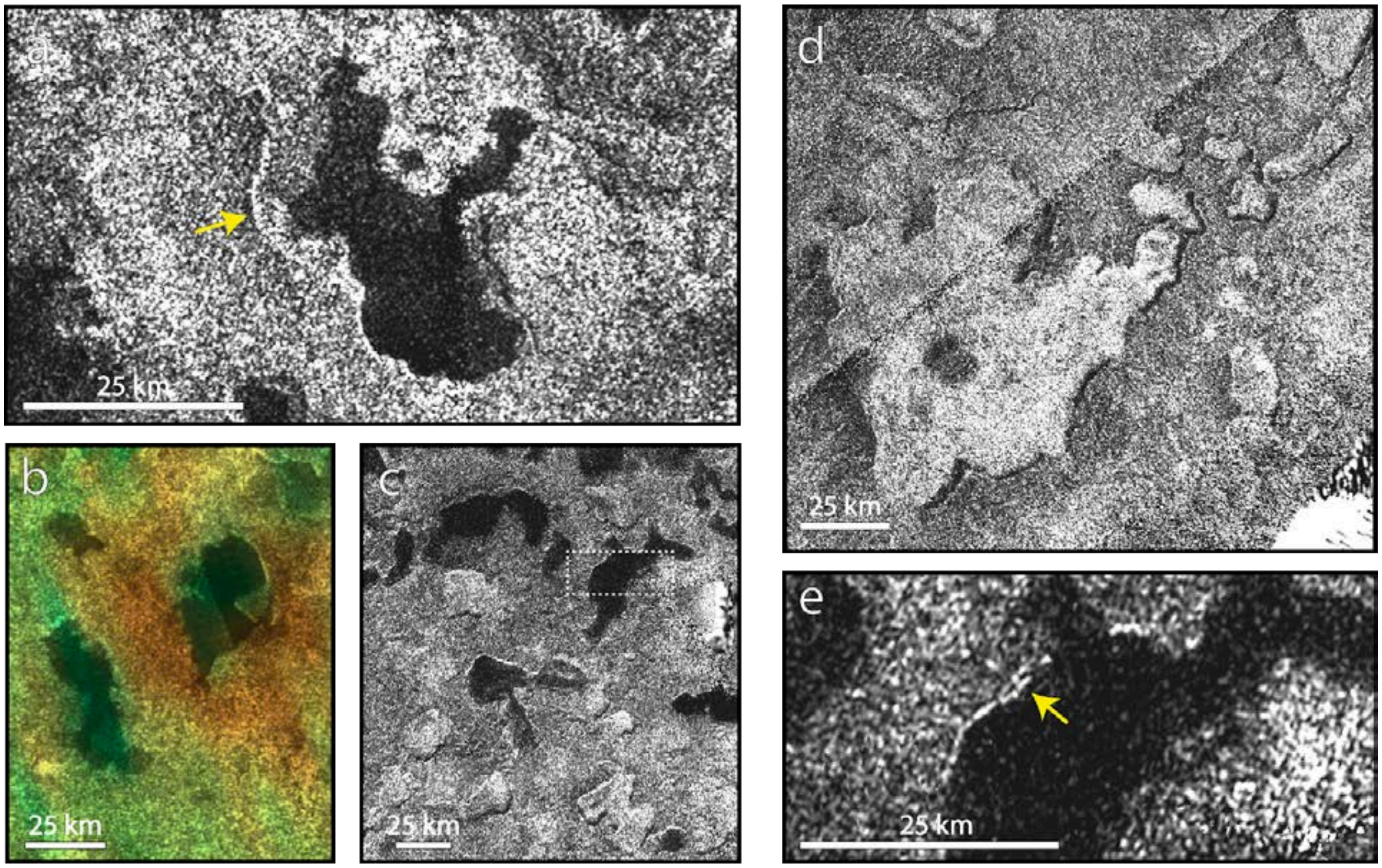

**Figure 27**: Titan’s small lakes (SEDs), figure adapted from Birch et al., (2025a). (a) Viedma Lacus at Titan’s north, surrounded by a bright rampart annulus that is compositionally distinct from its surroundings (Solomonidou et al., 2020a). A smaller scale raised rim is highlighted by the yellow arrow; (b) VIMS image overlaying SAR of a cluster of SEDs at Titan’s north, where the red material is bright at 5-microns, interpreted to be evaporitic in origin (Barnes et al., 2011). (c) A cluster of filled and empty SEDs in Titan’s lake district. The dashed box centered on Winnipeg Lacus highlights panel e. (d) An empty SED at Titan’s south, highlighting their characteristic flat floors. (e) A zoom on Winnipeg Lacus, with its raised rim marked by the yellow arrow.

Finally, transient changes in Titan’s lakes and seas were observed throughout the Cassini mission. Possible shoreline changes may have occurred at Ontario Lacus (Turtle et al., 2011b) and Kraken Mare (Hayes et al., 2011), while some of the periphery of select small SED lakes – at the south (Hayes et al., 2011), and north (MacKenzie et al. 2019) changed, indicating loss of shallow fluids along their margins to evaporation/infiltration. The liquid surfaces of the lakes and seas were also remarkably flat for most of the Cassini mission, down to millimeter scales (Stefan et al., 2010; Wye et al., 2009; Hayes et al., 2013). Transient changes to the sea surface, however, did begin to occur later in the mission as northern summer approached (Barnes et al., 2014; Hofgartner et al., 2014, 2016), and could be the result of waves, suspended solids (Yu et al., 2024), and/or bubbles (Hofgartner et al., 2016; Malaska et al., 2017; Farnsworth et al., 2019). The significance and the exact origin of these transient changes remain a topic of debate.

At the present time, much remains to be understood about Titan's seas and lakes. Significant research regarding Titan's lakes and seas continues today through numerical modeling and laboratory experiments. A select few topics of research are listed here: (1) the physical chemistry and composition of Titan's liquids (Cordier et al. 2009; Cordier et al. 2012; Glein & Shock 2013; Diez-y-Riega et al. 2014; Malaska and Hodyss, 2024; Malaska et al. 2017; Farnsworth et al. 2019; Steckloff et al. 2020; Engle et al. 2021, 2024; Farnsworth et al. 2023) and how it varies spatially, seasonally, and how groundwater-crustal processes play a role (Hayes et al. 2008; Horvath et al. 2016; Hayes et al. 2017; Mousis et al. 2014, also Section 2.8); (2) the evolution of Titan's shorelines and what processes are capable of modifying them (Palermo et al., 2024; Birch et al., 2025b; Black et al. 2012; Tewelde et al 2013), with implications for Titan's climate and sediment cycles; (3) the origin of transient surface features (Hofgartner et al. 2014; Hofgartner et al. 2016; Malaska et al. 2017; Cordier and Liger-Belair, 2018; Cordier and Carrasco, 2019; Farnsworth et al., 2019; Yu et al. 2024), and whether waves, suspended solids, and/or bubbles are responsible for the observed activity within Titan's seas; (4) the origin of the SEDs, a unique feature to Titan whose formation is not known (Birch et al. 2019; Schoenfeld et al. 2023); and (5) the magnitude and variability - both seasonally and spatially - of currents and circulation in the seas (Tokano and Lorenz 2015; Vincent et al. 2018; Heslar et al. 2020).

In addition, there are important questions about the chemistry and pre-biology that could occur in the liquid hydrocarbon environment, including the formation of co-crystals, chemical reactions, and the potential formation of micelles/membranes that could contain and confine biological material (Kawai et al 2013; Corrales et al. 2017; Cable et al 2019; Czaplinski et al., 2019, Cable et al., 2021; Czaplinski et al., 2023; Hirai et al. 2023; Czaplinski et al., 2025). One such proposed structure is an 'azotosome', possibly built from acetonitrile (Stevenson et al. 2015), a molecule that has been confirmed to exist in Titan's atmosphere (Palmer et al. 2017). However, the likelihood of the existence of acetonitrile membranes has been challenged (Sandström and Rahm 2020), leaving much room for further exploration of this parameter space (Mayer and Nixon 2025).

*Earth analogs*: We outline below the Earth lake environments, both natural and anthropogenic, that can act as analogs to lake environments on Titan (see **Table 8)**.

| **Table 8**: Terrestrial analogs to Titan lakes and seas | | | | |
|---|---|---|---|---|
| **Titan location(s)** | **Terrestrial Analog Site(s)** | **Analog Type** | **Measurement** | **Notes/ limitations** |
| Ontario Lacus | Lake Balkhash (Sala et al. 2020) Gulf Coast US; Black Sea; Sea of Azov | Sedimentary coastlines, with deltas, and other wave-sculpted sedimentary features | Surface level fluctuations; wave amplitude; liquid composition; suspended particulates and solutes; pH | Different liquids: water on Earth vs methane/ethane on Titan.; different sediments/particulates and dissolved gases. |
| Ligeia Mare, Kraken Mare, Punga Mare, Jingpo Lacus | Waves: Bar Harbor coast, US (Belknap et al. 2002) Lake Rotoehu, New Zealand (Jolly 1968); Sebago Lake, US (Thompson et al. 1995) Flooded: Fort Peck Lake, | Flooded coastlines with wave-eroded rocky shores and/or shores eroded by uniform erosion. Analogs drawn from Palermo et al. 2024, and | Surface level fluctuations; wave amplitude; liquid composition; suspended particulates and solutes; pH | Different liquids. Earth lakes not stratified like Titan's may be (Malaska et al., 2017; Steckloff et al., 2020) |
| | US (Jensen and Varnes 1964)<br>Lake Murray, US<br>Lake Lanier, US<br><br>Uniform:<br>Prošćankso Jezero, Croatia<br>Kozak Jezero, Croatia (Palermo et al. 2024) | Birch et al. 2025b. | | |
| “Sharp-Edged Depressions” (SED). Large inset steep-sided liquid filled depressions - morphologies suggest growth by coalescence. with elevated (~100-m-high | No real analog, as discussed in Hayes et al., 2017<br><br>Maar craters, tuff rings and tuff cones produced during phreatic or phreatomagmatic eruptions (for the | Formation process, morphology | Morphology; stratigraphy | Formation mechanism still in question: studying earth analogs could help: Different cause for the explosion (by nitrogen dissolved in methane), different styles of karstic lake development |
| ) rims. Evidence of multiple lake stands. Can be empty. | rims and making a hole only)<br><br>Karst lakes such as Crveno Jezero, Croatia (Garasic 2012); Bottomless Lakes State Park, NM (morphologically similar, minus the rims) (McLemore 1999)<br><br>Mantled karst lakes in Florida, USA (e.g. Beck 1986) | | | |
| All lakes | Liquid hydrocarbon deposit (La Brea, CA, Zechmeister and Lijinsky 1953; Pitch Lake, | Liquid hydrocarbon deposit | Temperature; viscosity; dissolved oxygen; composition; isotopic ratios | Earth's pitch lakes are much warmer than Titan's and created by different mechanisms. |
| | Trindad, Schulze-Makuch et al. 2011) | | | |
| All lakes | McMurdo, Antarctica (Kennicutt et al. 2010); Ny-Ålesund, Spitsbergen (Jadwiga Krzyszowska 1989) | Interactions between hydrocarbons (e.g. methane) and water ice | Geochemistry; biology | Earth temperatures are much higher (by 180 K) than Titan's surface, changing reaction and growth rates. |
| Dry Lakebeds | Salt flats like Salar de Uyuni (Risacher and Fritz 1991); Devil's Golf Course (Umurhan et al. 2019); Racetrack Playa, Death Valley, CA (Lorenz et al. 2010); Bonneville | Composition | Evaporite processes | |
| | Salt Flats, UT (Turk et al. 1973); Etosha Pan, Namibia, Africa (Cornet et al. 2012) | | | |

Ontario Lacus appears as a broad lake filling the bottom of a sedimentary basin (**Fig. 24g**). Its shorelines are, therefore, most probably covered in sediment that is transportable by the lake's fluid(s) (Schneck et al. 2022). Its eastern shoreline is most consistent with being wave shaped, while its western shoreline hosts Titan's only river deltas (Wall et al., 2010; Birch et al., 2025b). Lake Balkhash, Kazakhstan (**Fig. 24d**) (Sala et al. 2020), is similar to Ontario Lacus in that it too is slowly drying and hosts a broad sedimentary complex on its southern margin. Modern deltas currently form along this shoreline, with sediment sourced from that broader sedimentary complex just upstream.

The flooded shorelines of Titan's vast northern seas have numerous analogs on Earth, which is also experiencing ongoing sea-level rise (**Fig. 26**). For this reason, Titan's seas also most likely have rocky shorelines, with the sea flooding the surrounding drainage basins (Black et al., 2012; Tewelde et al., 2013). Whether their coasts are eroded or not, and whether they are shaped by uniform erosion or wave erosion is not certain without higher resolution image data. Nevertheless, current data, coupled with landscape evolution models, suggests they are most likely wave eroded (Palermo et al., 2024). Palermo et al. (2024) also compare Titan's northern seas to multiple analogs across Earth. Recently flooded, minimally eroded shoreline analogs include any number of lake reservoirs created by dams (**Fig. 26**). Lakes with wave eroded shorelines include Lake Rotoehu, New Zealand (Jolly 1968), and Sebago Lake, US (**Fig. 26a**) (Thompson et al. 1995). The coastline of Bar Harbor Maine is also morphologically similar to the coasts of Ligeia and Kraken Mare (Birch et al., 2025b).

The mantled karst lakes in Florida (Beck 1986) have planform similarities to Titan's SEDs, although the scale is vastly different between the two worlds. From analyses of the empty SEDs, we know that their floors are flat, with some hosting evaporitic materials (Hayes et al., 2017; Barnes et al. 2011), analogous to salt pans like Salar de Uyuni, Bolivia (**Fig. 26c) (**Risacher and Fritz 1991**)**, Etosha Pan, Namibia (Cornet et al., 2012), and the Racetrack and Bonnie Claire playas in the United States (Lorenz et al. 2010). Solomonidou et al. (2020a) also proposed that some of Titan's elevated SED margins may be shaped by evaporitic hardening and landscape deflation, drawing comparisons to salt pan formation processes on Earth (Lowenstein and Hardie 1985). The presence of indurated ramparts around some of the empty lake basins could reflect chemical and physical mechanisms similar to those observed in terrestrial arid environments (Hurlbert et al. 1976). None of these terrestrial analogs, however, host rims on the scale of those surrounding Titan's lakes, which have heights comparable to the lakes' depth. Similarly, karst lakes such as Crveno Jezero (**Fig. 28**) (Garasic 2012), Croatia, and Bottomless Lakes State Park, US (**Fig. 26d**) (McLemore 1999) match

Titan’s smaller lakes in planform and also do not have rivers draining into them, but they also lack any raised rims.

Brouwer et al. (2025) modeled the formation of Titan's rampart craters/depressions (**Fig. 27**), a sub-type of SED, to assess if they could form as maars (Mitri et al., 2019; Wood and Radebaugh, 2020) or gas-emission craters (Schurmeier et al. 2023), potential terrestrial analogs that have been suggested. They found that a maar-like process could form the raised ramparts. On Titan, this would require that hydrocarbon or liquid nitrogen aquifer fluids within the ice shell to interact with a thermal anomaly (warm upwelling ice or cryomagma) which has yet to be identified. Maar craters, tuff rings and tuff cones produced during phreatic or phreatomagmatic eruptions, and pingos all create raised rims, but all are inconsistent with other known aspects of the lakes (Hayes et al., 2017).

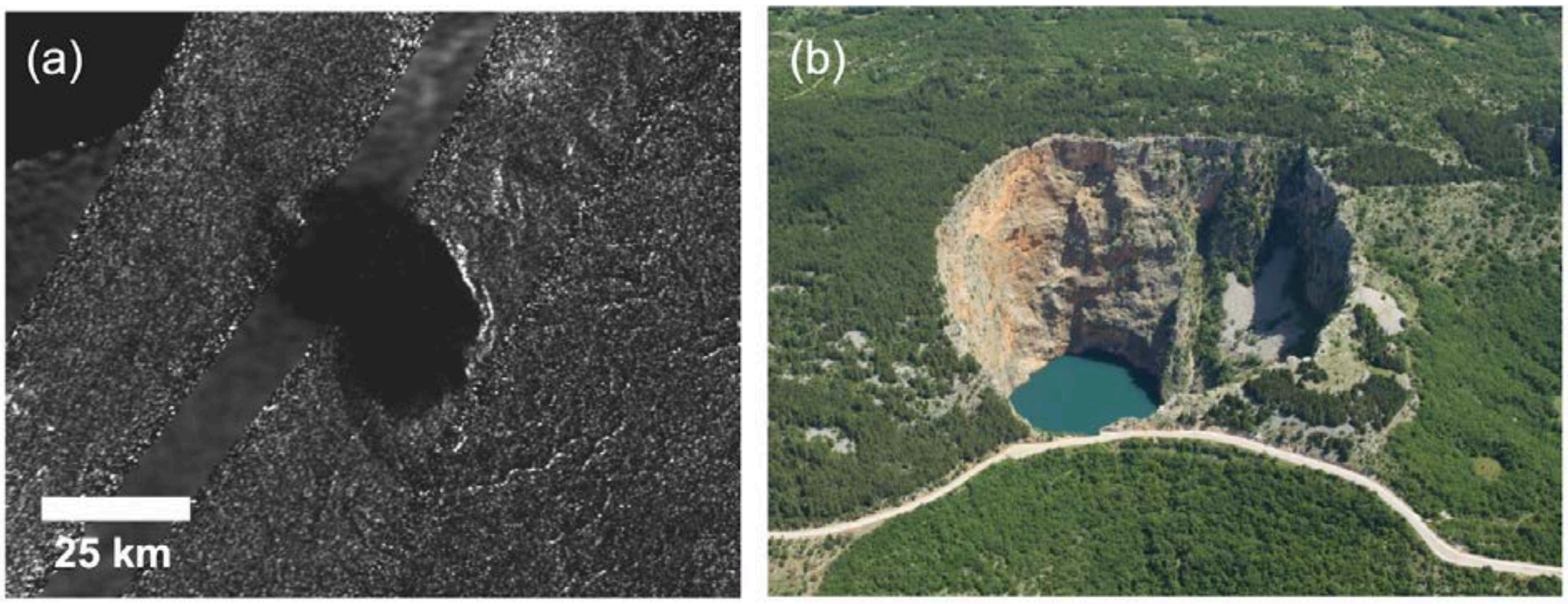

**Fig. 28**: (a) Crveno Lacus, Titan imaged by Cassini RADAR (79.5 S, 185 W) (NASA/JPL). (b) Crveno Jezero, Croatia (43.4541N, 17.1993E). (Google Earth).

Additionally, atmospheric processes resulting from the lakes are interesting to compare on Titan and the Earth. Both worlds have breeze circulations with similar dimensions, though with different energy exchanges and diurnal cycle (Chatain et al. 2024). As on Earth, the evaporation of the lake and local evaporative cooling induced could potentially form clouds and fog above the lakes on Titan (Brown et al. 2009a; Brown et al. 2009b).

Some compositional analogs arguably exist, including liquid surface hydrocarbon pools both natural (e.g. La Brea, Zechmeister and Lijinsky 1953; Pitch Lake Trinidad, Schulze-Makuch et al. 2011) and anthropogenic (see **Fig. 29**). The latter category includes oil spills and leaks onto icy parts of the Earth's surface, including well-studied examples as at McMurdo Station, Antarctica (Kennicutt et al. 2010; Klein et al. 2012), Ny-Ålesund, Spitsbergen (Jadwiga Krzyszowska 1989), and at the US Navy Arctic

Research Lab (NARL) in Alaska (Wagner and Baker 2019) amongst others. There are even examples of intentional oil release onto permafrost and into the ocean (sea ice) to study subsequent distribution and effects on ecosystems (e.g. Martin 1979, Collins et al. 1976) which could provide Titan analogs. Further study of the geochemistry and biology at these hydrocarbon liquid/water ice boundaries may provide important new ideas for future instruments and missions to Titan, as well as lab experiments.

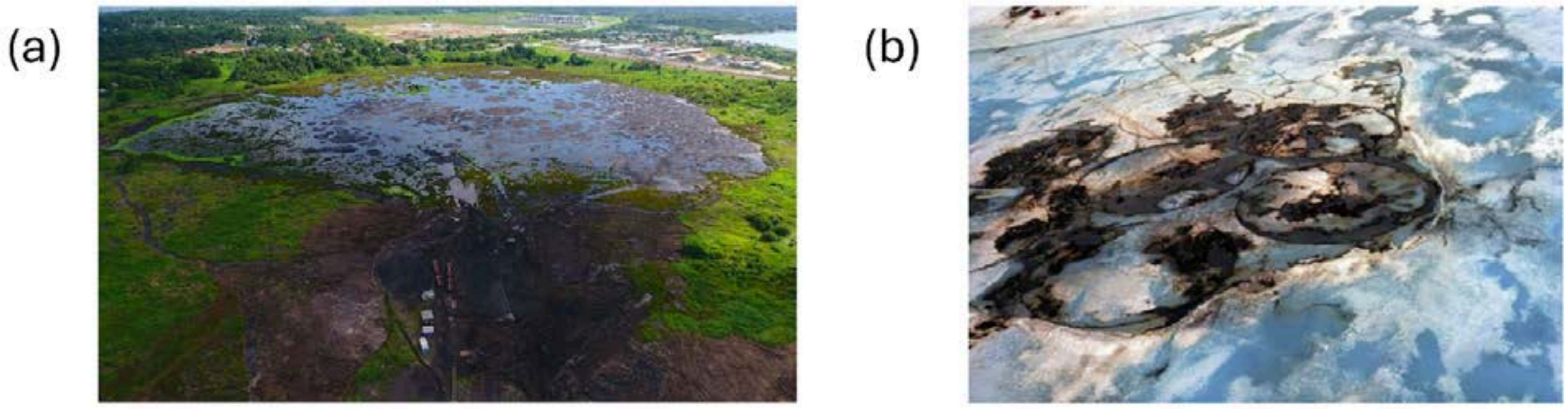

**Figure 29**: (a) Pitch Lake, Trinidad (wikimedia). (b) Oil resurfacing onto sea ice, Balaena Bay, Beaufort Sea, Canada in 1975 after 1974 release (researchgate images).

*Field measurements*: Key measurements needed of Titan lakes/seas include liquid composition (e.g. GC-MS); depth (e.g. sonar); wave and tide amplitude (e.g. GPS); and atmospheric boundary layer properties such as temperature, pressure, wind direction and speed, and even rainfall. Standard meteorological packages include sensors, anemometers and rainfall gauges. An example payload for a Titan lake surface probe is given in Stofan et al. (2013) and for a submarine in Oleson et al. (2015).

Terrestrial in-situ measurements can encompass similar properties, although sampling can also be made allowing liquids to be returned to the lab for more complex analysis, including filtration for biological and non-biological materials (Janasch 1958, Francy et al. 2013), and extraction of dissolved gases (Lelekakis et al 2013). Precise isotopic and compositional measurements can be made. In addition, terrestrial field work has the option to emplace sensors or meters for long durations, either to be recovered later or as 'permanent' installations (e.g. tide gauges - Costanza et al. 2013, flow meters - Staff 1962). A long duration study of a planetary lake lander analog for Titan was performed at Laguna Negra in central Chile. (Pedersen et al., 2014)

Of particular scientific interest would be studies examining the geological and hydrological context and subsurface terrains of terrestrial analogs in order to differentiate formation hypotheses; whether maar or karstic or both.

## 2.8 Alkanofers

Sub-surface reservoirs of hydrocarbons - 'alkanofers', by analogy with 'aquifers' on Earth – have been hypothesized to exist on Titan, primarily to provide sources and sinks for methane photolysis in the atmosphere (Corlies et al. 2017, Hayes et al. 2017, Hayes et al. 2008).

*Shallow volatile hydrocarbon cycling*: the term 'alkanofers' was introduced for Titan's subsurface hydrocarbon liquids to express the potentially similar behaviour to Earth's aquifers. Earth's terrestrial porewater aquifers typically exist either (i) in the shallow subsurface as freshwater to low (below average seawater) saline systems, replenished by rain and its percolation into unconsolidated sediments; (ii) as surface exposures and bedrock via pore spaces or fissures; or (iii) as deep saline systems (Wendland et al. 2008). Globally, there is no unique depth at which systems become deep aquifers which are saline at seawater levels of salinity (3.47% TDS - total dissolved solids), or potentially much higher salinities up to several 10s of percent TDS. As the amount of dissolved solids increases, the density also increases. On Earth, this porewater density is important for both the migration of hydrocarbons generated in the subsurface and their accumulation in reservoirs, due to the density difference between the induced buoyancy of hydrocarbons and the primary water phase filling the subsurface pore-network (Saffer and Tobin 2011).

The need for a methane source on Titan has been recognized since the early post-Voyager years of the 1980s. This was based on photochemical modeling that showed that the current atmospheric inventory of methane will be destroyed in a few 10s of Myr – much less than the age of the solar system – based on current photolysis rates (Yung et al. 1984). This led to ideas such as long-term clathrate release of methane (Lunine & Stevenson 1987; Tobie et al. 2006; Mousis et al. 2014; Davies et al. 2016) although methane was also proposed to be buffered through sub-surface liquid reservoirs of hydrocarbons (Lorenz & Lunine 1996, Lorenz et al. 1997, 1999; Mousis et al. 2016; Kalousova & Sotin 2020).

Since predictions of a global surface ethane ocean – as the product of ~4.6 Gyr of methane photolysis – were not borne out by observation (Muhleman 1990, Muhleman et al. 1995), the need for a hydrocarbon sink became evident. Furthermore, when the hydrocarbon lakes and seas were finally detected (Stofan et al. 2007) and measured, they were much smaller than required, and were primarily methane. This did not match the predicted ethane accumulation, leading to ideas that ethane may be seeping into the crust.

Some evidence for a sub-surface 'ex-hydraulic' (or 'alkanolic', Hayes et al. 2017), connection between the larger lakes via porous organic mantling deposit fill has been put forward based on Cassini radar altimetry measurements of lake levels (Hayes et al. 2008, 2017), although more accurate measurements are required.

On Titan, the lower atmosphere was proposed to be composed of dozens or more different C, H, N carrying components (Wilson & Atreya 2000) overlying the liquid surface accumulations (lakes, rivers) of ethane and methane. From these surficial hydrocarbon accumulations a downward migration has been proposed, due to its higher density than Titan's atmosphere in the vadose (atmospheric gaseous pore-filling of the outer solids) zone - into Titan's crust. This proposed downward migration is in the opposite direction compared to the migrations of hydrocarbons on Earth. This density (specific gravity) controlled process may also provide a means to displace methane from clathrate and provide an answer to the methane supply problem (Choukroun & Sotin 2012).

While the pressure and temperature (PT) conditions would not thermodynamically favour a solidification of ethane and methane, under Titan's conditions both hydrocarbon species may become trapped in a solid water ice clathrate cage structure. This layer must be lower in hydrocarbon concentration than the liquid state alkanofer layer overlying it. If at the surface ethane is percolating downwards within an ethane-methane mixed liquid phase, the mechanism for immiscibility of the two is yet to be proposed and resolved, since on Earth under the full range of PT conditions both are fully miscible. Since both are non-polar their interaction with the polar porous ice crust will be comparable, though overall being negligible. As such, any terrestrial-analog alkanofer effect - where liquid or gaseous hydrocarbons chemically interact with the mineral matrix - would not be found within Titan's porous icy crust and its liquid hydrocarbon fill.

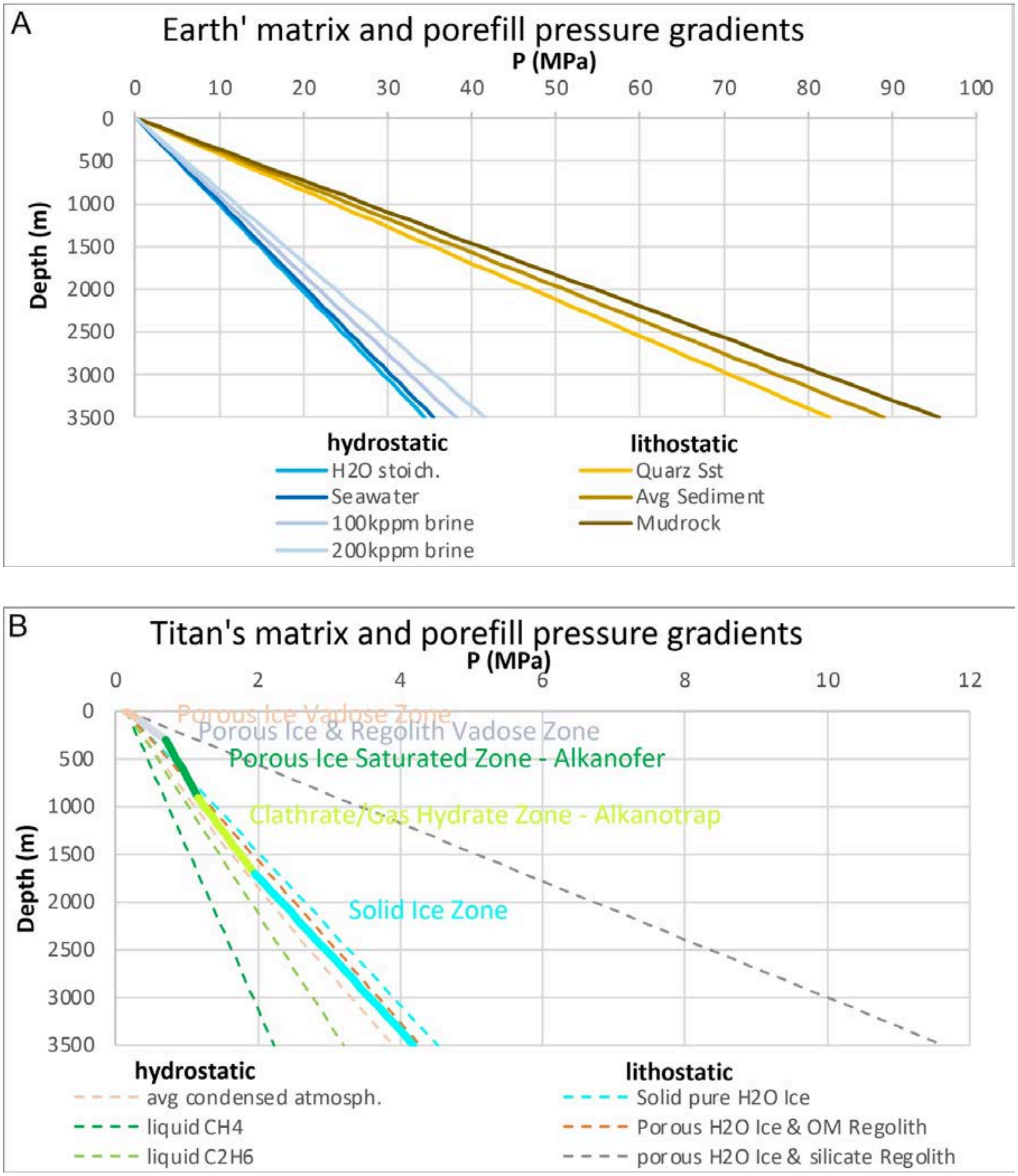

**Figure 30**: (A) Earth hydrostatic (porefill) and lithostatic (matrix) pressure gradients, prior ones for stoichiometric water, seawater and different brines, latter one for three different lithotypes; (B) Titan porefill and matrix pressure gradients (dashed) for different condensed liquid phases and matrix types. The solid multicolour lines is a model

combined pressure gradient for the proposed stacking of upper crustal layers. Models plotted by authors.

**Fig. 30A** presents liquid porous fill (or Earth “hydrostatic”) and solid matrix (on earth lithostatic”) pressure profiles. These are typically used to assess subsurface pore-network connected fluid pressures systems versus the grain-grain load lithostatic profiles as endmembers. The pressures in-between the two endmember gradient curve groups is used to assess subsurface overpressure strata, (under)compaction and pore-filling fluid gradient pressures and help to differentiate fractured from pore-migration pathways. Depending on the fluid type (freshwater, seawater, brines, hydrocarbon liquids, hydrocarbon and inorganic gases) the gradients deviate from the endmember lines and indicate zones containing pressure risks. In comparison, **Fig. 30B** shows the same calculations applied to Titan. These calculations include densities at Titan’s subsurface and surface PT conditions, and take into account Titan’s weaker gravity and denser atmosphere. As expected, the curves look different on Titan in comparison with those for the terrestrial subsurface. A model profile of a stacked layer system as proposed for Titan’s upper crust and combining their different pore-filling fluids and solid matrix phases allows for a quick identification of phase and migration behaviours to be expected in Titan’s subsurface. This would predict a percolation of ethane downwards into Titan’s crust or at least  to a shallow level fracture in absence of silicate regolith.

Earth aquifers can also provide an analog to some processes, especially as far as subsurface hydraulics/alkanolics are concerned and for a better understanding of solubility of light-end (C1-C14) hydrocarbons in an aqueous phase. While beyond the scope of this article, an important topic of further work is to investigate the thermally insulating effect that the solid ice and the expanding clathrate layer would have, allowing for capture of heat from the core to the uppermost crustal layers and enabling a thermal field and gradients to support the liquid interior ocean. This effect has been proposed both for Pluto (Kamata et a. 2019) as well as for Titan (Kalousova & Sotin, 2020).

Titan's crustal layered structure (see Section 1.1) constitutes the major difference with the Earth: comparisons to terrestrial hydrocarbon accumulations are now described.

*Deep core organic transformations*: on Earth, deeply buried biogenically-derived refractory complex organics are chemically transformed to produce more geologically mobile complex hydrocarbon mixtures (e.g. Galimov 1988; Glasby 2006) . Being less dense than the surrounding rock, these migrate upward and interact with an inorganic water-based saline pore-filling phase as well as a siliceous and/or carbonaceous porous matrix. The latter is controlled in its composition by primary sediment deposition and its diagenesis (lithification) alteration during burial induced by aqueous pore fluids. In other words, a non-polar (aliphatic) to partially polar (aromatic) hydrocarbon assemblage of millions of different compounds interacts with polar water molecules, dissolved salt ions and a partially surface charged matrix of primary particles and their thermodynamically controlled alteration and cement precipitates.

On Titan, initial accretion delivered cometary material which partially segregated out to create a hydrous silicate inner core containing cometary organic material (Tobie et al.

2014a). The organics were subjected to heat due to formation and may have chemically "cracked" similar to processes of terrestrial organic processing. Once "cracked", the lighter hydrocarbons and ammonia/nitrogen may have migrated upwards to the deep ocean and overlying crust to then degas to form Titan's methane and nitrogen atmosphere (Glein 2015, Miller et al. 2019).

*Earth analogs:* natural hydrocarbon reservoirs exist on Earth in the form of oil, oil & gas or gas accumulations in a variety of depth settings and accordingly with a large variety of phase mixtures (Zhang et al. 2022).

Three analog subsurface processes (emanations) for hydrocarbons could potentially serve as analogues for Titan's alkanofers. The are listed here in order of increasing depth on Earth, not in their respective ordering in Titan's shallow volatile hydrocarbon cycles and deep core organic transformations:

(1) Seeping hydrocarbon occurrences on Earth (**Fig. 31A**), though upward in nature, and their remnants at the surface, may contribute to an interface between hydrocarbons and regolith. This may also cause downward migration of 'heavy' liquid hydrocarbons from source rocks – instead of the usually dominant upward buoyant migration. The former have been documented since historic times in many places globally, and were indeed the first sources of hydrocarbons for use by humans. Examples are the Caspian petroleum system onshore occurrences of gaseous ('eternal fires') and liquid remains (pitch-springs) in (pre-)Persian times in the eastern Caucasus (Marvin 1884). ‘Pitch’ is usually denser than water and referred to either as ‘bitumen’ (and as remains of thermal alteration called ‘pyro-bitumen’, both by definition soluble in organic solvents) or called asphalt or asphaltene (if non-soluble in organic solvents, **Fig. 31C**) (AlHumaidan et al. 2020; Soziniov et al. 2020; Zuo et al. 2022; Wang et al. 2024). The density is due to the highly complex composition of all these compound assemblages and their molecular size: having at least 30 C atoms per molecule, and up to hundreds of C atoms with associated H, some O and even fewer N and S atoms per molecule depending on the precursor primary organic matter. Phoenicians, the dominant traders in the Mediterranean sea in ancient times, used such pitch to seal their ships, this usage was indeed the dominant one for all naval nations until the mid 18$^{th}$ century. Occurrences are spread globally (Bois et al. 1982; Rogner 1997) and the biggest continental/onshore deposits are the Western Canada Sedimentary Basin (WCSB) tar sands and Jordan oil shale along Jordan river valley (dead sea rift) and quarries on Jordan plateau, and the famous Santa Barbara/Los Angeles Basin (La Brea) tar seeps (**Fig. 31C**). For submarine/offshore occurrences on the ocean floor south of Texas is one of the most ‘leaky’ natural systems due to its complex salt tectonic overprint with several hundred documented seeps (**Figs. 31B**) (Locker and Hine 2019).

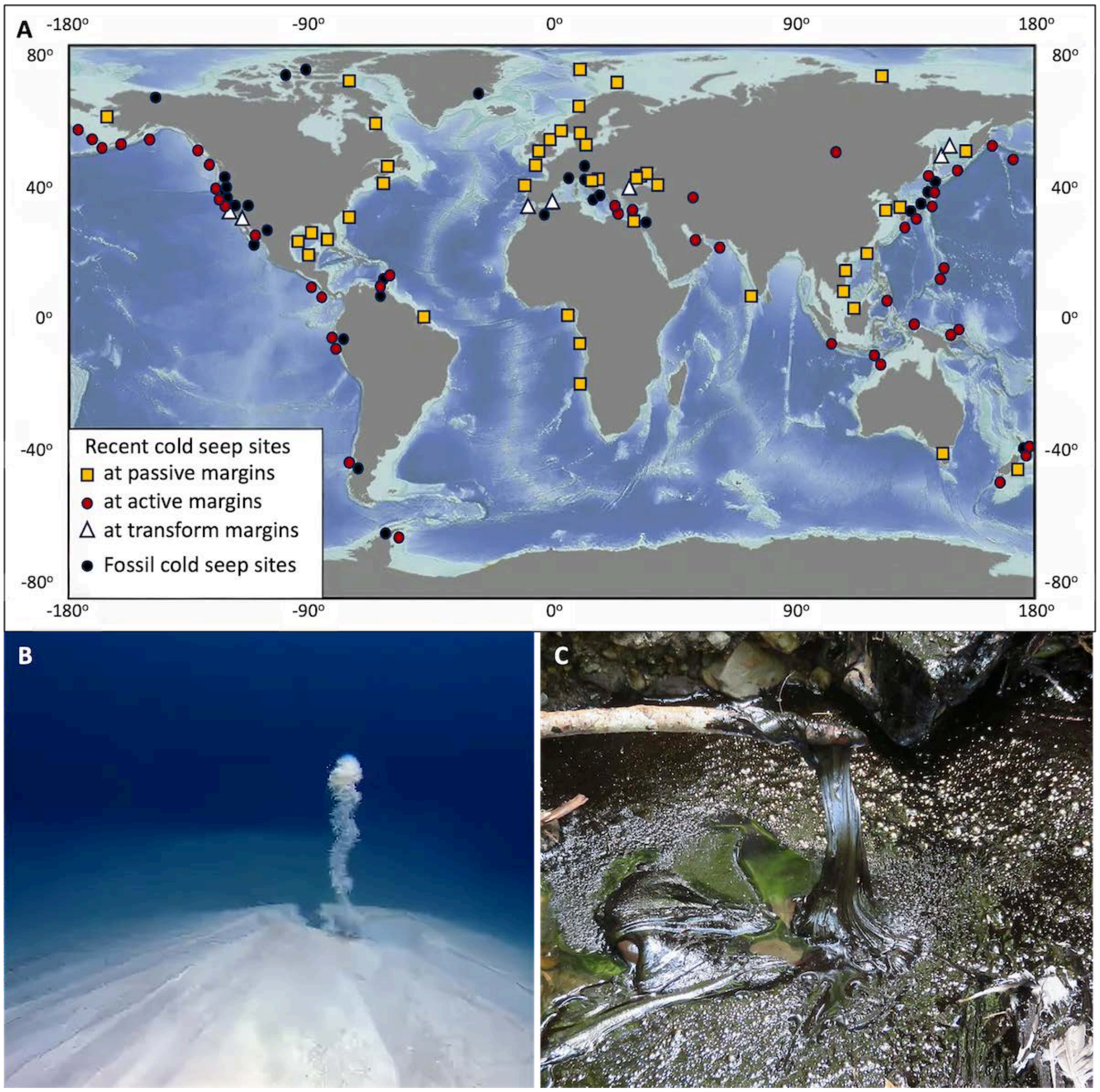

**Fig. 31**: New source map (A) of global occurrences of onshore and offshore cold seeps and their respective plate tectonic relation (NOAA); (B) submarine cold seep (NOAA image Flower Garden Banks, outboard 50 Miles zone south of Port Arthur, Texas. (C) onshore oil seep (USGS image, Tarwater Creek, San Mateo Ct., California). See Data Availability for details.

(2) Shallow bitumen: occurrences would be too viscous in their natural state to seep to the surface, and therefore immobile on human time scales (Mohammadi et al. 2021), and too deep for traditional mining to access. However, these can be mobilized by steam injection ( the rarely-used process of ‘heavy oil steam maturation gravity drainage’, Butler and Stephens 1981; Butler 1994; ), where the heat induced reduces viscosity and allows the hydrocarbon substance to migrate in the pore network. Note that these would not rise but migrate downwards, similar to the proposed liquid ethane downward seeping on Titan, to production well assemblages making use of the heavier-than-water specific gravity of the hydrocarbons (Gates 2007).

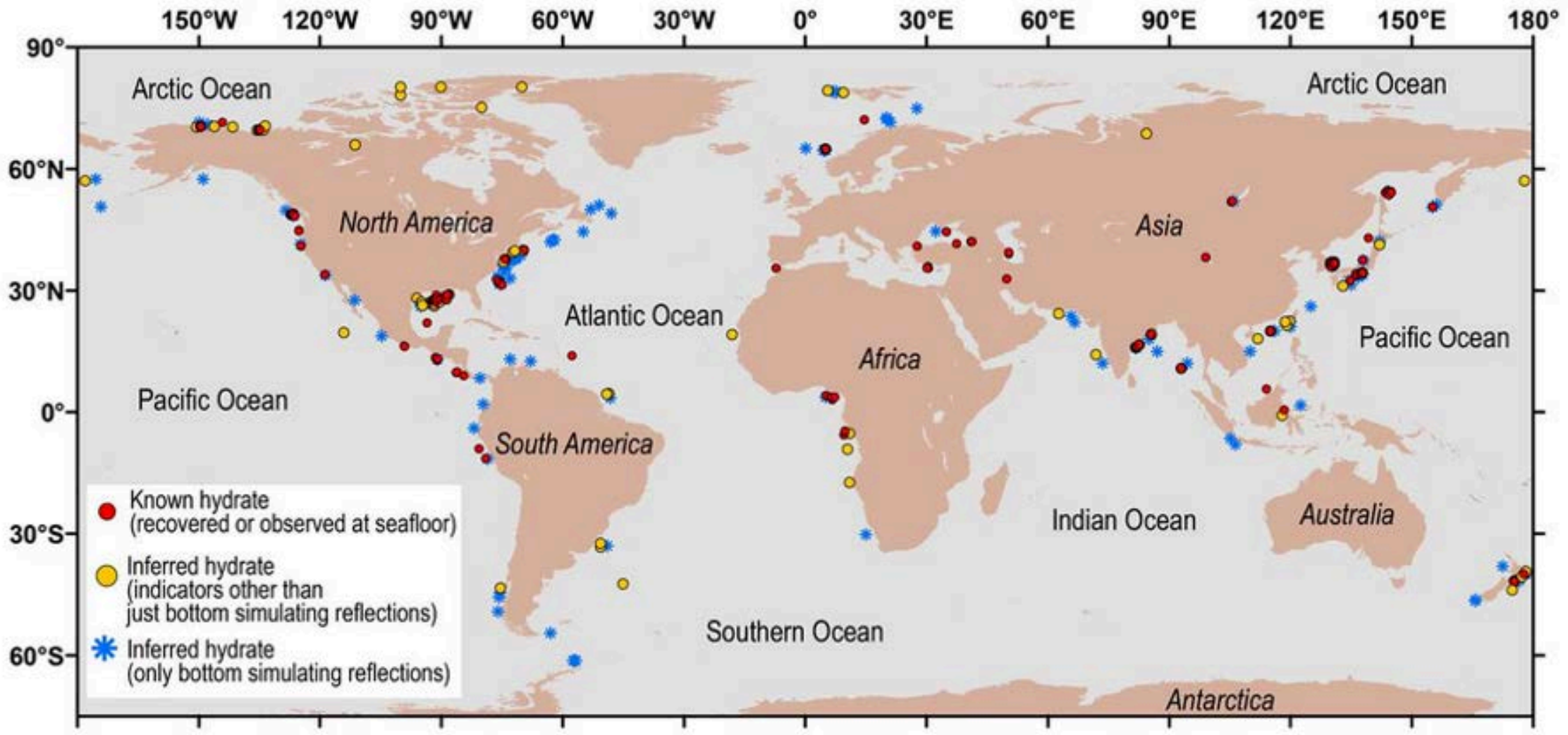

**Figure 32**: Worldwide distribution of observed and inferred gas hydrates in marine and permafrost associated settings that have been the subject of drilling programs. The color coding refers to the primary sediment type in each location and therefore designates the likely type of gas hydrate reservoir at each site. (USGS image: see Data Availability)

(3) Clathrate presence on Earth (**Fig. 32**): typically, methane clathrates are present under permafrost and in a zone along both passive margin shelf and active margin accretionary wedge slopes where the seawater is helping to maintain cool slope sediment temperatures and a favourable pressure regime (Ruppel & Wait 2019), i.e., in the PT thermodynamic stability field of such aggregates. They are at shallow sediment depth, typically 100 to 300 meters in permafrost and in marine environments below 300-600 meters water depth (Hester & Brewer, 2009) (**Fig. 33**). Utilising carbon isotope fingerprints such clathrates were found to contain both or either methane generated microbially (as long as pasteurization temperatures of 75-80$^{o}$C were not exceeded) and methane generated thermally from source rocks. In addition, in onshore settings pseudomorphoses after hydrogen bonded Ikaite calcium carbonates, which can be ascribed to the group of semiclathrates, have been reported: e.g., in the tufa around Mono Lake, California (Whiticar & Suess 1998).

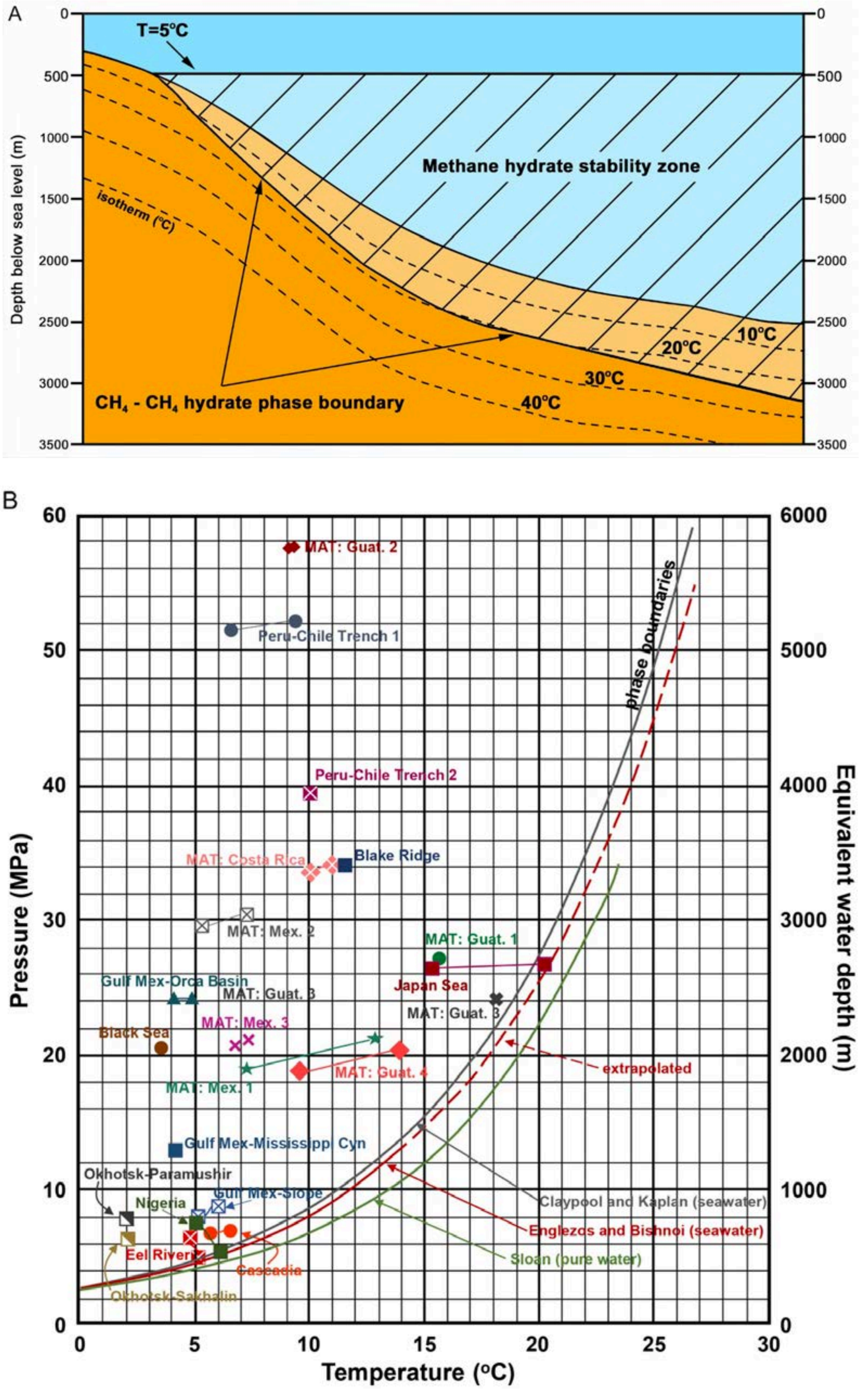

**Figure 33**: Gas Hydrate Stability Zone (GHSZ) and its PT relation in an offshore setting (Booth et al. 1996). (A) schematic cross section of a continental margin slope system exhibiting the GHSZ and free gas – gas hydrate boundary (eq. to BSR) in a depth vs. isotherm context;; (B) Free gas – gas hydrate phase boundary curves vs. hydrate occu sites' pressure and temperature data vs. equivalent water depth. See Data Availability.

Next to these three potential analogue settings for alkanofers on Earth, it is worth following the makeup of a petroleum system on Earth and its potential similarities and known distinct differences on Titan. This could help to identify more analogue settings on Earth or other extra-terrestrial mission settings on planets and moons. These are listed in **Table 9** below, lines 4-10.

Methane and hydrocarbons potentially have been thermally generated from chondritic material stemming from the accretionary phase of the Titan core. Any migration of the hydrocarbons from the core to the surface will have been controlled by various processes: (a) fractionated dissolution leaving the core into the interior ocean waters, (b) partial exsolution from these waters at the transition from liquid (interior ocean) to overlying solid water (crustal ice). In both the interior ocean (in dissolved state) and in the crustal ice (in the partially-exsolved state), a third control process for migration would be (c) diffusion of hydrocarbons. In both states diffusion is a slow but over geological time there is potential for differentiation of hydrocarbons from atmosphere and crust. Finally, liquid hydrocarbons within the porous media (crustal ice and regolith) can migrate by  induced buoyancy due to density difference or by colloidal transport whose effectiveness is controlled by (d) colloidal aggregation into nanodroplets with flocculation across pore throats vs. (e) Brownian motion momentum transfer and deflocculation during in-pore passages.

Unfortunately, many existing published papers report erroneous facts about Earth's hydrocarbon reservoirs, such as stating that the majority would exhibit segregated columns of hydrocarbons (Stainforth, 2009; Stainforth 2013a, b). If the present day PT conditions allow for a subcritical existence of two separate phases, Earth hydrocarbon reservoirs would contain an oil leg on top of a brine and then a free gas phase towards the top of the accumulation. Many reservoir fillings, though, are in a supercritical state, for example all gas is dissolved in the oil phase under reservoir conditions, if the accumulation is to the left of the critical point in a PT diagram, or all liquids are in vapour phase for the combined liquid composition and PT conditions to the right of the critical point (Danesh, 1998). Graded columns with, for example, heavy oil leg, medium oil leg and light oil leg (in terms of specific gravity) are an absolute exception, the reasons being complex (Ross et al., 2010) and beyond the scope of this paper. In the end, each of these phases themselves would hold a mixture of hundreds of thousands of different compounds (Peters et al., 2013), so far different from any extraterrestrial occurrence of – at best – a few individual components.

**Table 9**: Terrestrial analogs to Titan Alkanofers

| Titan location(s) | Terrestrial Analog Site(s) | Analog Type | Measurements | Notes/ limitations |
|---|---|---|---|---|
| Seeps into subsurface | WCSB Tar Sands (Athabasca River Bank outcrops; Birks et al. 2018), Jordan oil shales (Yarmouk river banks and El-Lajjun quarry; Alali et al. 2014) Offshore: South Texas shelf (Hood et al., 2002) | Seeps from the subsurface | Well logging; sample recovery | Terrestrial tar sands are emplaced onto a silicate crust, contrasting Titan's water-ice crust |
| Lacustrine segregation of $CH_4$ and $C_2H_6$ layers | Green River oil shale project (Evacuation creek outcrop "Condo" section; Greenberger et al. 2016) Jordan oil shale project (Sultani mine; Azzam et al., 2016) | Maturation and gravity downward drainage of heavy organic substance (bitumen/asphaltene) | Well logging; sample recovery | Titan hydrocarbons contain lighter liquid phases not sustainable at terrestrial temperatures |
| Clathrate layer in porous icy crust | Eastern India and Myanmar margin (Monteleone et al. 2022), British Columbia and Western US margin (ODP leg 169S: Puget Sound, Saanich Inlet; Whiticar & Elgert 2001; IODP leg 311: North Cascadia Margin; Riedel et al. 2006), Chile accr. wedge | Marine accretionary wedge / shelf slope clathrates | Well logging; sample recovery | Titan crust much colder than terrestrial permafrost |

| **Table 10**: Terrestrial Petroleum Systems (PS) Elements as potential process/stratal analogues to Titan | | | | |
|---|---|---|---|---|
| **PS element** | **Object Type** | **Titan** | **Earth** | **Notes/ limitations** |
| | Hydrocarbon Source (Itata & | · Chondrites<br>· Titan Core | · OM in Source Rocks | Early Titan lifecycle limited or dissolved |
| Strata | Valdivia offshore area; Vargas-Cordero et al., 2017) | · photolyses of higher CNH compounds from Methane | · OM microbially degraded shallow during early burial | into interior ocean water phase |
| | Reservoir | · Porous icy crust<br>· Clathrate layer | · Porous crust | Crust type vs. pore-filling HC interaction difference Permeability contrast more an Earth property |
| | Seal | · Non-porous icy crust at bottom<br>· Liquid phase density and atmosphere as natural self-sealing top | Low permeability sediment strata (e.g. mud rocks, shales) required to withstand losses upward of less dense HC phases | Vadose and saturated layer porous zone physically different Density directional drive (percolation downwards vs. seepage upwards) different |
| Strata and Process | Overburden/ Burial | Burial less important since surface | Burial required for thermal maturation of organic matter | Thermal gradient on Titan potentially |
| | | temperature too low and constant | and therefore HC generation | controlled by clathrate as insulation layer |
| Processes | Trap formation | Omnipresent and growing trap: Clathrate Alkanofer porous icy crust extension unknown, potentially global | Trap types:<br>· Structural (anticlines, faults)<br>· Stratigraphic (facies change, pinch-outs)<br>· Combination traps | Titan : omnipresent to widespread trap layer with no final subsurface – atmospheric loss due to HC in atmosphere<br>Earth: individualised traps depending on local structural style and strata |
| | HC Generation/ Migration/ Accumulation | No publications yet on early history of Titan and its HC generation potential Hypothesis | Geothermal gradient and burial rate controlled, critical moment need of trap-in-place at HC arrival | Earth: migration& accumul. diffusion mostly irrelevant<br>Density difference / buoyancy vs. permeability control |
| | PS preservation | Enigma on Methane resupply Latest papers on thermal gradient, early ocean composition and other outer planetary bodies offer potential for analogues on hydrocarbon genesis and fate | Mainly controlled by:<br>· Tectonics<br>· Climate<br>In effect, compressive settings, orogenesis and unroofing erosion as well as post glacial isostatic uplift are major effects for destruction | · Latest papers on topography and isostacy (e.g., Cadek et al. 2021; Durante et al. 2019) as work to follow up on.<br>· Future work on potential for thermal convection as an induction mechanism for horizontal stress differences could prove useful (assuming crust decoupled from interior ocean). |

*Field measurements*: Since sample recovery and testing of rocks (drillcores) is expensive and time-consuming, apparatus exists for ‘well logging’ which allows for characterization of bulk properties of a bore hole and its penetrated formations, typically in oil and gas exploration (Fertl and Chilingarian 1978). Well-logging measurements include: (i) spring caliper measurements of width; (ii) spontaneous potential; (iii) focused resistivity; (iv) gamma rays; (v) neutron logs, (vi) acoustic measurements, and (vii) image logging. Such measurements can give information on properties such as bulk density, hydrogen atom abundance, lithology, porosity, pore filling fluid (gas/liquid hydrocarbons/saline porewater) contents, resistivity and more. More sophisticated characterization of terrestrial hydrocarbon deposits requires extraction devices and apparatus (GC, MS and various spectroscopy types) often not practical to apply in-situ (Oblad et al. 1987). Therefore sampling for full laboratory analysis of hydrocarbon composition and phase for a minimum of reservoir and stock tank conditions is beneficial (McCain 2017). Nevertheless, future experiments sent to Titan must operate in situ and in closed-loop setups, so terrestrial hydrocarbon deposits represent an excellent field test environment (Arens et al. 2025) as demonstrated for other extraterrestrial lander missions, e.g., on Mars (Eigenbrode et al. 2018).

## 3.9 Interior water ocean

*Titan:* Interior models of Titan have long been consistent with the presence of an interior water ocean (e.g. Sohl et al. 1995; Tobie et al. 2006; Grindrod et al. 2008). Support for this interpretation was strengthened by Cassini's gravity science investigation, which found that Titan was deforming in a periodic way over its orbit in response to Saturn's gravity -- behavior that is most readily explained by the presence of an interior ocean (Iess et al. 2012; cf. Petricca et al., 2025). Independently, the observation of Extreme Low Frequency (ELF) electromagnetic waves by the Huygens probe during its descent through Titan's atmosphere suggests the existence of a resonant cavity between Titan's ionized stratospheric layers and an electrically conductive layer beneath the non-conductive surface. This lower reflecting boundary is best explained by the transition from ice to a salty ocean at a depth of 55 to 80 km below the surface (Béghin et al. 2012; cf. Lorenz & Le Gall 2020). The ocean thickness estimates range over ~50-400 km, with poorly constrained density, composition, and salinity (e.g., Mitri et al. 2014, Vance et al. 2018, Idini & Nimmo 2024, Goossens et al. 2024). The ocean bottom may have been a rocky mantle in the past and more likely a layer of high pressure ice at present time (e.g., Journaux et al. 2020).

More recent analyses of Cassini data, however, challenge these conclusions. Lorenz & Le Gall (2020) demonstrate that the ELF time series correlates strongly with the history of mechanical vibrations (e.g., parachute release), indicating that it is unlikely to represent a robust detection of an ELF wave and thus does not provide firm evidence for a deep conductive layer. Similarly, a recent reanalysis of Titan's gravity field measured by Cassini radiometry, using improved techniques, suggests that tidal dissipation in Titan's interior may be significantly larger than can be readily reconciled with the presence of a subsurface water ocean (Petricca et al., 2025). For the remainder of this section, we nevertheless assume the presence of a subsurface ocean in order to discuss potential analog environments.

Titan's surface is rich in organics and its rocky mantle may have a large fraction of insoluble organic matter (e.g., Sotin et al. 2021). Surface organics can be delivered to the ocean via impact cratering, with complex organic haze particles being the largest component (Neish et al. 2024, Krasnopolsky 2009). Meanwhile, moderate heating of accreted insoluble organic matter in the interior can produce more volatile organic compounds including aromatic molecules that may be extracted into fluid phases (Yabuta et al. 2007) and play an important role in its habitability (Affholder et al. 2024). Little is known about how these materials will be mixed into the bulk ocean or concentrated along the top and/or bottom interfaces, illustrating the importance of ocean dynamics.

With no direct measurements of ocean circulation, predictions rely on theoretical and numerical modeling (e.g., see Soderlund et al. 2020 and Soderlund et al. 2023 for reviews). The dominant mechanism driving ocean dynamics is thought to be convection, presumably thermal with the potential for a compositional contribution being poorly constrained (e.g., Soderlund 2019, Amit et al. 2020, Kvorka and Cadek 2022, Terra-Nova et al. 2023). Tidal forcing may also contribute (Hay et al. 2024), but libration is not expected to be a significant driver (Lemasquerier et al. 2017). Given the large uncertainties in Titan's ocean properties and the challenges of computational limitations for numerical models, there exists a relatively wide range of ocean circulation possibilities.

*Earth analogs:* As illustrated in **Fig. 34**, a wide range of temperature and pressure conditions are possible at the top of the ocean. Titan's ocean in the “warm” and “modest” scenarios have pressures comparable at shallow depths to those found in the deepest ocean waters on Earth in the Mariana Trench (e.g., Taira et al. 2005). The temperatures, however, are not analogous as Titan's oceans are likely ~5-20 K colder. Analogous temperatures may, contrastingly, be found in hypersaline subglacial lakes such as Lake Vida in Antarctica (e.g., Doran et al. 2002), those below the Devon Ice Cap in the Canadian Arctic (Rutishauser et al. 2018; cf. Killingbeck et al. 2024), and that below the northwest Greenland Ice Sheet (Maguire et al. 2021). Here, high water salinity can depress the freezing temperature to reach more extreme temperatures that are comparable to those predicted for Titan's ocean. These conditions provide unique natural laboratories for oceanographic processes and their biological implications (e.g., Horikoshi 1998, Murray et al. 2012).

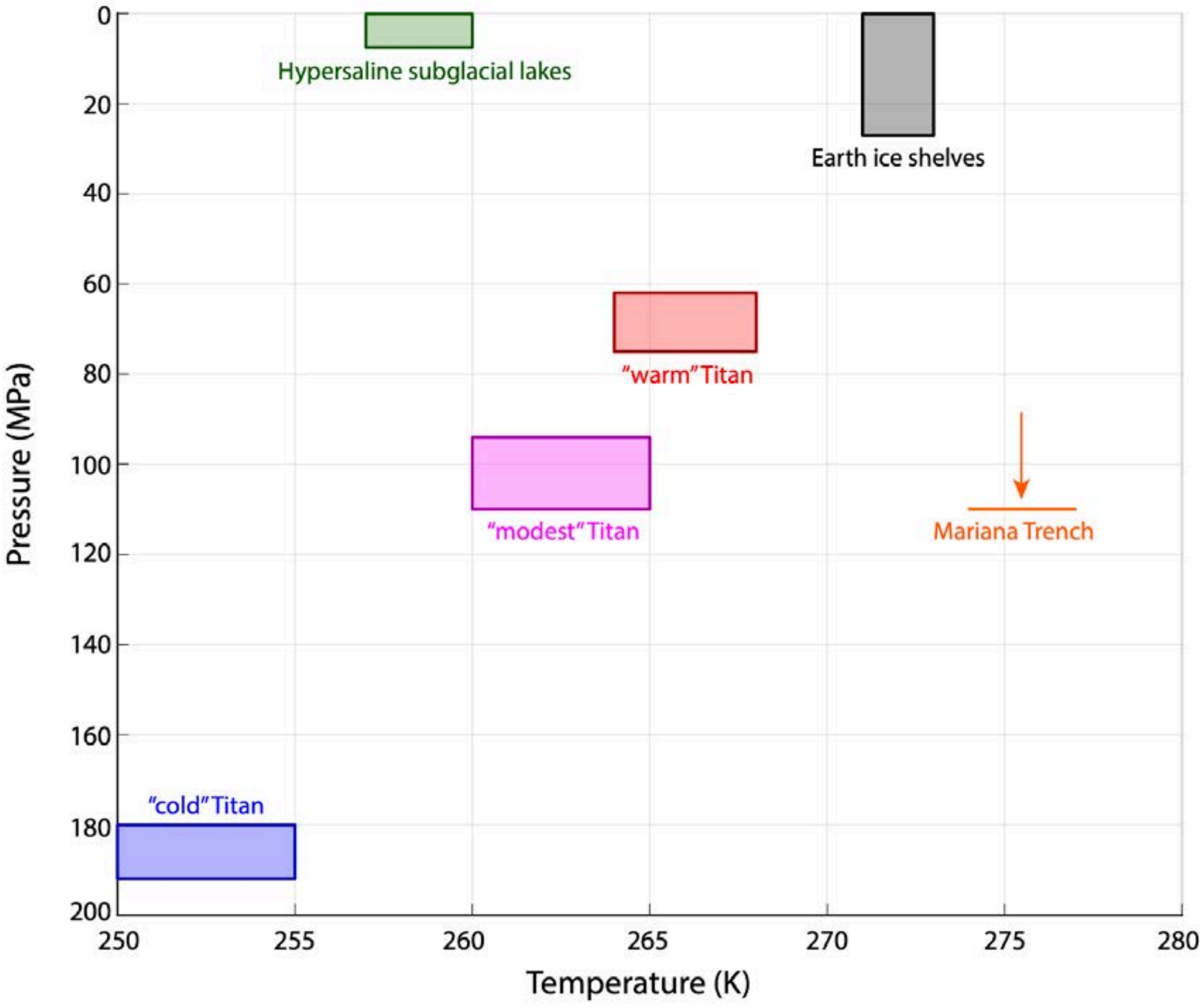

**Figure 34**: Pressure-temperature ranges predicted at the top of Titan's ocean for different internal structure models from Vance et al. (2018) compared against the Mariana Trench (Taira et al. 2005), terrestrial ice shelves (Wolfenbarger et al. 2023), and hypersaline subglacial lakes (Doran et al. 2002, Rutishauser et al. 2018, Maguire et al. 2021). Three possible scenarios for Titan are depicted: "warm" models have outer ice shells less than 60 thick, "modest" models have shells 70-90 km thick, and "cold" models have shells more than 140 km thick; each set spans three different ocean compositions: pure water, 10 wt% $MgSO_4$, and 3 wt% $NH_3$.

Contrastingly, despite the lack of achieving neither analogous pressures nor temperatures, the basal interfaces beneath terrestrial ice shelves will offer insights into ice-ocean coupling processes generally (e.g., Buffo et al. 2018, Wolfenbarger et al. 2023). However, comparisons with Titan systems need to be carefully considered since many of the chemical inputs of terrestrial ice shelves are derived from atmospherically transported continental or terrestrial biotic inputs deposited and later scraped into ocean at the interface; these can form the basis for nitrate-driven ecosystems (see Martínez-Pérez et al., 2020).

On Earth, organics, gas bubbles, and microbes originally existing in liquid water have been found to become incorporated into ice during freezing (e.g., Adams et al., 1998; Collins et al., 2008; Wik et al., 2011; Santibanez et al. 2019). It is therefore likely that Titan may experience a similar incorporation of solutes into its ice shell, as well. Even though Titan has a thick ice shell, Earth locations such as arctic sea ice and ice-covered lakes, are intriguing analogs for Titan's ice-ocean interface (**Fig. 35**). Additionally, the affinity of molecular incorporation into the ice may allow sampling of Titan's ancient ocean via in situ measurements of Titan’s ice crust. Understanding the differential incorporation of microbes and other solutes on Earth may aid in our understanding of such in-situ measurements by a future Titan lander.

(a)

(b)

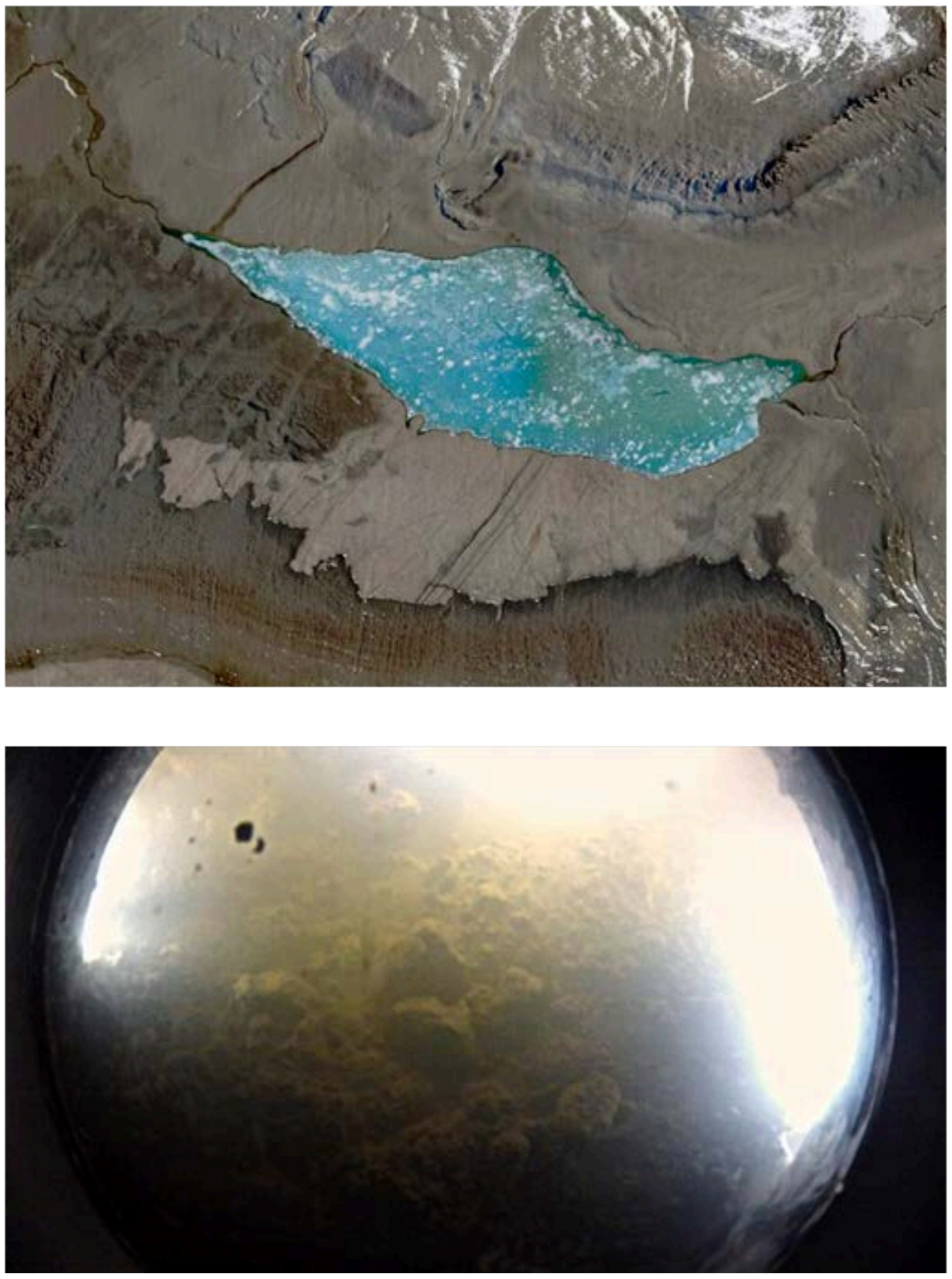

**Figure 35:** (a) Lake Vida, a permanently ice-covered hypersaline lake in Antarctica’s McMurdo Dry Valleys (Sentinel-2 image via ESA). (b) Subglacial Lake Whillans, Antarctica, lies beneath 800 m of ice and was first directly sampled in 2013 (ASUN/NSF/NASA).

Terrestrial microbes can survive in the hypersaline veins and pockets between ice grains following ocean freezing (Price 2007). During freezing, crystalline ice excludes chemical impurities as well as solid grains less than 5 microns in size. This effect can chemically concentrate important ions and nutrients by several orders of magnitude creating favorable microhabitats in the ice (Mader et al. 2006). The physical and chemical processes of oceanic freezing on both Titan and Earth may be very similar. The study of Earth's microhabitats at the ocean-ice interface may be a direct analog of potentially habitable microhabitats on Titan that exist at the deep ice-ocean interface on Titan. The main difference would be in pressure and possibly temperature: with processes on Titan occurring at 100 MPa compared to Earth's 1 MPa, and freezing points potentially depressed due to the anomalous pressure dependence of the ice Ih freezing temperature, and also due to unknown amounts of ammonia in the ocean. Terrestrial deep sea clathrate-ocean interfaces at 50 MPa could be a more direct analog, but the low temperature stability ceiling of methane clathrate seems to generally preclude its formation above about 33 MPa.

If briny fluids rapidly descend through porous channels in the Titan ice, for example as proposed due to foundering after impacts (Carnahan et al. 2022; Kalousová et al. 2024), the supercooled fluid entering the ocean can create inverted chimney structures reminiscent of hydrothermal chimneys (Vance et al., 2019, **Fig. 36**). On Earth, the attendant chemical fluxes in these systems have caused speculation about a possible role supporting or even contributing to an origin of life (Cartwright et al., 2013). For chaos features on Europa, for which the volume of transported brine might approach that created after a large Titan impact, the timescale for the outflow of the brine material. The estimated lifetime of the potential brinicles is on the order 10-$10^4$ years. Whether such structures would be stable against ablation or mechanical degradation from lateral fluid currents is an open question.

The combined extremes of salinity, low temperature, and very high pressure in Titan's subsurface fluids (>> 200 MPa) may limit its biological potential compared to that of the above analogs, each of which hosts microbial populations. Microbial viability under this combination has been sparsely characterized, but the individual extremes themselves are within the ranges life is known to tolerate (Harrison et al. 2013, Merino et al. 2019, Affholder et al. 2025). However, polyextreme conditions often pose significant additional challenges (and occasional benefits) to life which individual extreme analogs may not capture (Cockell et al., 2016), so a purely additive or mean measure of habitability between multiple extremes is often not appropriate. Mindful of that caveat, microbiological habitability models (e.g., Higgins and Cockell 2020; Affholder et al., 2021, Affholder et al. 2025) could be employed to assess the biological potential of Titan's subsurface ocean using analog data as a basis. The required information includes: i) organism-specific biokinetic parameters collected in situ or constrained via laboratory experiments, ii) organism-specific stress responses to the individual or combined extremes, iii) physico-chemical parameters of both analog and target environments; iv) differences in energy and nutrient availability between the laboratory medium/analog and target environments.

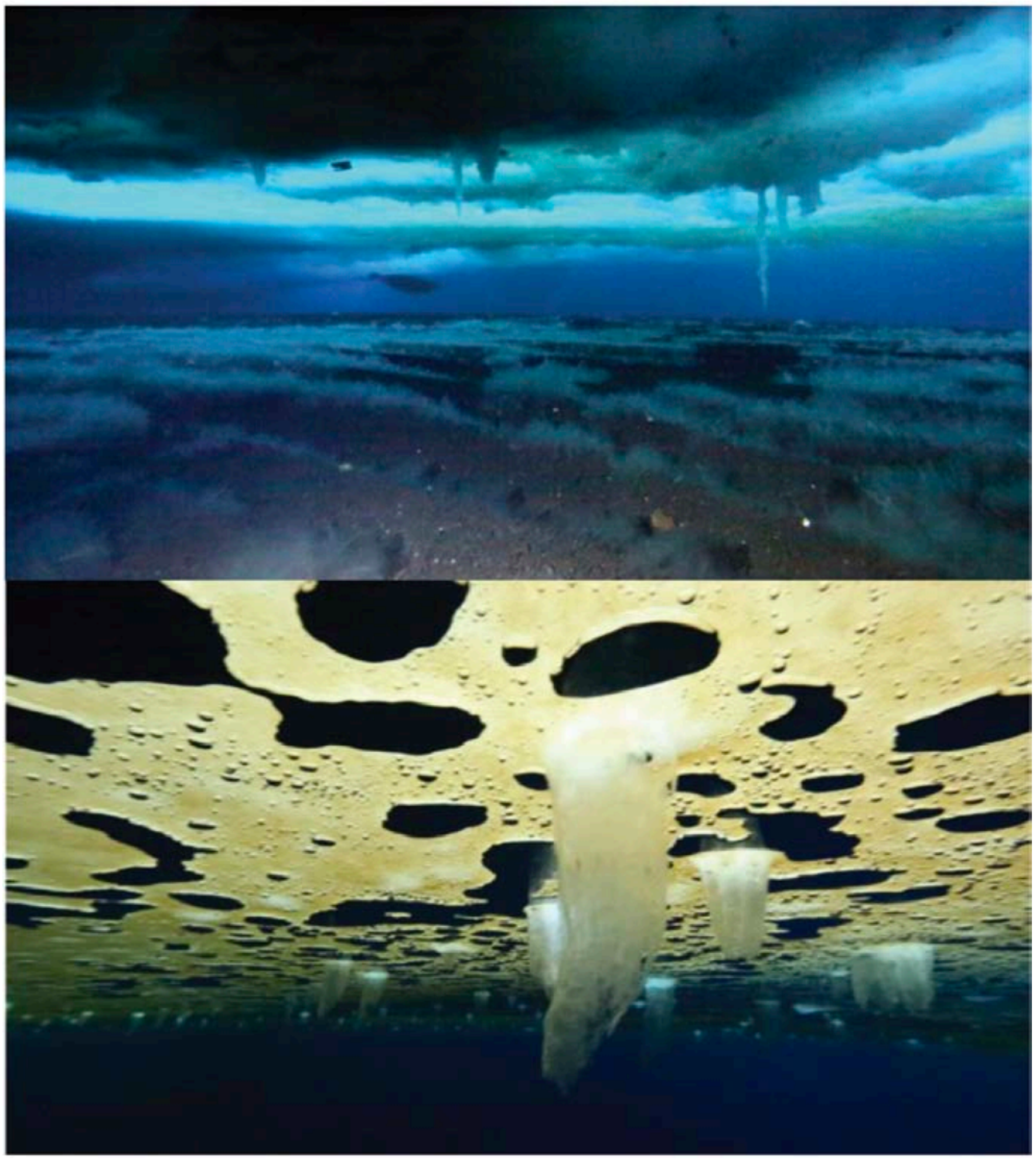

**Figure 36:** Ice brinicles near McMurdo station in Antarctica. From Vance et al. (2019).

Considering the biologically relevant injection of complex organic haze particles into Titan's ocean, possible analog sites may be oil spills and natural oil seeps since haze particles are generally insoluble, viscous, and denser than water. On Earth, this release can produce a variety of viscous hydrocarbons in the deep ocean, from small tar balls to asphalt volcanoes (e.g., Valentine et al. 2010). These environments are well-suited to investigate mixing processes, as well as chemical degradation and habitability (e.g., Kujawinski et al. 2020, Joye 2020).

*Field measurements (**Table 11**)*: Many instrument technologies already exist for in situ ocean measurements, including conductivity, oxygen, pH, and carbon dioxide in sea water as well as depth profiling instruments. Auxiliary sensors may also measure dissolved oxygen, pH, turbidity, chlorophyll a, rhodamine, blue-green algae, ammonia and nitrate (e.g., Mills and Fones 2012). These sensors are mounted on buoys and ships to make global near-surface measurements (e.g., Legler et al. 2015). Additionally, specialized submersibles are used to reach deeper regions of the ocean, sampling the seafood (e.g., Jamieson et al. 2013), and sub-sea-ice regions (e.g., Spears et al. 2016). Finally, drilling/coring can be used to reach sub-glacial lakes (e.g., Tulaczyk et al. 2014), as well to extract data from ice itself (e.g., Thompson 2000; Santibanez et al. 2019).

**Table 11**: Terrestrial analogs to Titan Subsurface Ocean Sites

| **Titan location(s)** | **Terrestrial Analog Site(s)** | **Analog Type** | **Measurement** | **Notes/ limitations** |
|---|---|---|---|---|
| Top of ocean | Mariana Trench | Physical, chemical, and biological oceanography | Water pressure | Different ocean geometry, Mariana Trench is at warmer temperatures; 70 MPa while Titan ocean starts at 200 MPa. Counterintuitively, Earth's Hadal regions are sinks for continental inputs (Liu et al 2018). |
| Top of ocean | Hypersaline subglacial lakes | Physical, chemical, and biological oceanography | Water temperature | Different ocean geometry, lakes are at lower pressures; organic continental biotic materials deposited in ice, inorganic rock inputs from glacier scrapings delivered into subglacial environments. |
| Ice-ocean interfaces | Ice shelves and subglacial lakes | Physical, chemical, and biological oceanography | Phase change | Different pressures and temperatures; possible photic inputs if ice is thin. Materials scraped off anchored land masses. |
| Organic injections into the ocean | Oil spills and seeps | Physical, chemical, and biological oceanography | Dynamical | Complex organics are found in both environments; most organics are composed of terrestrial biotics (steranes, hopanoids, pristanes) that have undergone diagenesis. |

## 3.0 Limitations Of Terrestrial Analogs

In the previous sections we considered in detail the range of overlap between Titan environments and those on Earth, to highlight research topics that would benefit from additional analog field work. Here, we re-assess the limitations of terrestrial sites as Titan analogs – the chemical, physical, and dynamical environments that do not have a good parallel on Earth (**Table 12**).

| **Table 12**: Titan environments not found on Earth | |
|---|---|
| **Environment** | **Analogs not found on Earth** |
| Atmosphere | Chemically reducing atmosphere; atmospheric superrotation; multi-fluid condensation. |
| Craters | Impact craters on Earth are composed of silicate materials, not water ice, and have a different simple-to-complex transition diameter. Larger craters have different morphologies on the two worlds. Impact melt may sink into the crust on Titan due to the density contrast. |
| Surface icy materials (mountains, craters) | Global crustal deformation found on Titan contrasts with plate tectonics on Earth, although some processes (faulting, rifting) appear similar. |
| Surface dunes and solid hydrocarbons | Earth dunes are not organics; likely to have different formative, erosive and transport properties. |
| Labyrinths | Solid hydrocarbon incised labyrinths. |
| Rivers | Rivers with extreme density and viscosity contrasts, especially over short distances. |
| Surface lakes and seas | The SEDs appear to be a uniquely a Titan feature. |
| Alkanofers | Alkanofers at Titan's surface temperature. |
| Interior ocean | Cold ocean, high water pressures, possibly ammonia rich ocean. |

*Atmosphere*: Titan's atmosphere, while possessing some remarkable similarities to Earth's - especially surface pressure and majority composition - exists in a very different temperature regime, with temperatures not exceeding 180 K in the LTE part of the atmosphere. Compositionally, despite being majority-$N_2$ like the terrestrial atmosphere, the Titanian atmosphere is lacking in $O_2$ which leads to a very different chemistry. Whereas Titan's atmosphere is mildly reducing, allowing for complex organics to build up, the Earth's is heavily oxidizing and does not support a long tenure for organics. Some physical processes appear superficially similar, such as the meteorological cycle, although further data is needed to determine how far this analog extends. It already appears clear that the diversity of condensable atmospheric gases on Titan will lead to clouds and raindrops of mixed composition, unlike on Earth. At the global scale, Titan's slow rotation leads to cyclostrophic rather than geostrophic balance.

*Craters:* Large impact features on Earth tend to be older, and hence heavily eroded, while younger simple craters are still well preserved. On Titan, biases may lean the opposite way, with fewer small craters being created due to the weaker impactor material (porous ice vs. silicate rock) breaking up in its moderately denser atmosphere . The difference in target properties and gravity on Titan and Earth will also lead to differences in crater morphologies, and differences in the transition diameters between these morphologies. For example, the largest craters on Earth have a “peak ring” morphology, while the largest craters on icy worlds have a “palimpsest” morphology. Icy worlds also have central pit craters that are rarely seen on terrestrial bodies. Importantly, because Titan's crust is primarily composed of water ice, impact melt may sink into its crust, as liquid water is denser than ice. This is not true for silicate rocks, so we may see less near-surface melt deposits on Titan compared to Earth.

*Tectonic Features*: Unlike Earth - where distinct continental and oceanic plates drift and collide due to underlying mantle convection - Titan's crust is likely to be simpler and made entirely of water ice, and lacking discrete plates. This implies that plate-edge events such as subduction are unlikely to contribute to any large-scale surface geophysics on Titan. However, some features may form in similar ways: uneven contraction or expansion may lead to rifts, folds and mountain chains. Tidal stresses on Titan however may play a much greater role in surface geophysics than on Earth.

*Dunes*: Titan's dunes appear to be predominantly longitudinal while Earth dunes appear also as transverse, star and barchan types in addition to longitudinal. However this may be an observational bias: it may simply be that longitudinal dunes are larger and more prevalent on Titan, and that other types will be revealed by more detailed inspection (e.g. by *Dragonfly*). The formation mechanisms of dunes on Titan and Earth are undoubtedly tightly coupled to the dynamics of the lower atmosphere and planetary boundary layer, and closer investigation of this regime on both worlds will be instructional. Compositionally, it is important to note that Earth dunes are predominantly silicate (quartz) while Titan dunes are refractory hydrocarbons, and will therefore exist different physical (e.g. tribocharging, hardness) and geochemical properties (alteration by rain etc).

*Labyrinths*: Labyrinths on Titan remain poorly known at present, hinted at only by analysis of still-inadequate remote sensing mapping. On Earth, labyrinths are predominantly associated with karstic terrain, but we do not know for sure whether similar sapping processes lead to Titan's labyrinths. Until then, we can only surmise as to the appropriateness of the analogs.

*Rivers*: Titan is overall significantly dryer than Earth, lacking in the 70% global oceans that Earth possesses. This applies at least to today, although Titan may have been wetter in the past, evidenced by the apparently dried-up seabeds at the Huygens landing site (Soderblom et al. 2007). Methane rivers on Titan and water rivers on Earth will both contain dissolved 'impurities' - or minority substances - such as ethane on Titan and carbon dioxide on Earth, leading to different properties of chemical weathering.

*Surface Lakes and Seas*: Titan's seas and lakes appear both familiar and alien. Compositionally, they are much different from Earth's water seas/lakes, and it remains to be seen whether there is any similar dichotomy between 'fresh water' and 'salt water' as on Earth. Titan's seas and lakes of liquid organics are majority methane, but with significant dissolutions of nitrogen, ethane, propane and perhaps other organics, while larger hydrocarbons may solidify to sink, or float. An interesting area of Titan lake science is likely to be stratification of different substances: a phenomenon also seen on Earth with haloclines and thermoclines. On the other hand, given the relatively sluggish lower atmosphere, large and violent surface movements (waves) appear unlikely on Titan, although currents may be important.

*Alkanofers*: On the Earth, crustal hydrocarbon deposits are largely created from processes acting on dead biological matter (plants and animals), rather than an abiological origin. On Titan however, crustal organics are presumed to originate either from atmospheric photochemical processes or be trapped from the time of formation in clathrates. Nevertheless, the degree of complexity reached by Titan organics may be significant, and in this respect the ‘end product’ may not be too different. The abode of these hydrocarbons is different however - rock on Earth, water crust on Titan - and this will lead to significantly different outcomes for subsequent geochemical evolution and alteration. Alkanofers provide an exciting potential habit for life, rich in CHON elements, and therefore an important direction for further research.

*Ocean*: Due to the calculated depth of the Titan deep water ocean within the crust, its minimum pressures may correspond to the highest pressures on Earth, while most of the ocean is probably at even greater pressures, and significantly colder than on Earth. The combination of these two differences pushes Titan’s ocean to the very extremes - or beyond - the realms of habitability explored on Earth. Nevertheless, the existence of life in once-presumed inhabitable environments has pushed the boundaries of our credulity, indicating that we should not quickly dismiss any deep, cold (but still liquid) watery environment as habitable. A larger problem may be the lack of nutrients available in an ocean bounded on both top and bottom by relatively inert water ice (e.g., Affholder et al 2025), but even here our knowledge of Titan’s interior processes is so sparse that it may be best to reserve judgement until more data is available. In the interim, the study of oceanic and lacustrine extremophiles on Earth will be of broad benefit to astrobiology across multiple worlds.

# 4.0 Summary and Conclusions

In this paper we have delved into the potentially similar environments for Titan that exist on Earth, and that may provide fertile grounds for analog research of various kinds. These include locations from the high atmosphere to the deep oceans. Our purpose for exploring these parallels is to point out that significantly greater scope for Titan analog research exists than is currently being exploited, providing many opportunities for fresh research initiatives. **Fig. 37** compares the range of possible analog terrain types across five diverse solar system worlds, showing that Titan and Earth share the greatest number of parallels in geophysical processes at the present day.

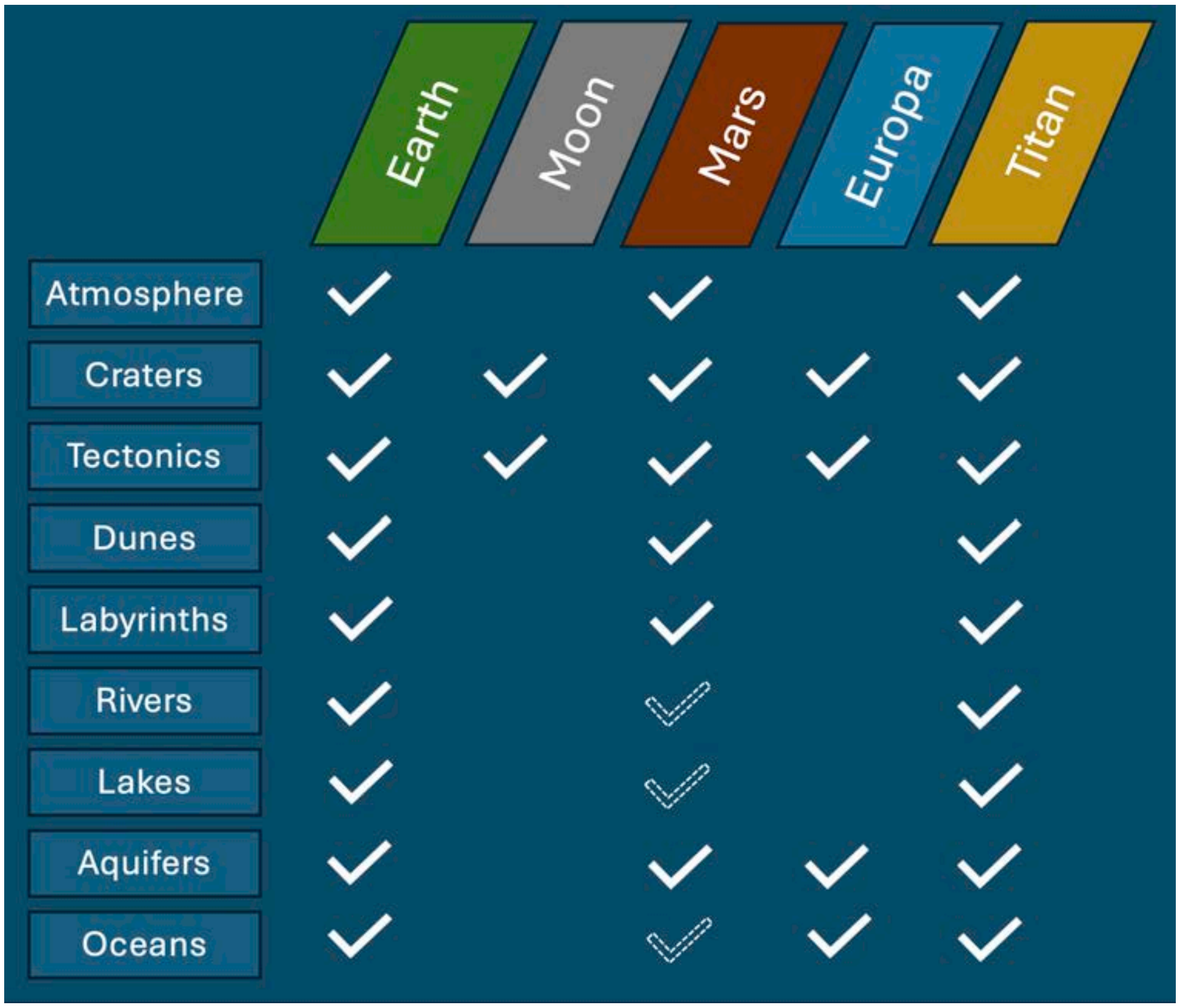

**Figure 37:** Planet-scale comparison between terrain types found on five different solar system bodies. Empty check marks indicate ancient rivers, lakes and ocean on Mars that are no longer filled with liquids at the present day.

We also note that important synergies exist between analog research and other types of indirect investigation of Titan environs (excluding missions and telescopes that collect Titan data) – namely theoretical modeling and laboratory investigations. These three types of investigations may be considered to exist along a continuum, with theoretical investigations being able to access the widest range of possible conditions with the greatest control of parameters, but analog research providing the most richness and complexity, along with the greatest cost.

As an example of such synergy, we can consider research in karstic systems on Titan. Theoretical calculations (Raulin 1987) showed that ice may be dissolved in methane solvent. This was later confirmed by observations (Malaska 2010) while further lab work expanded knowledge of kinetics (Malaska and Hodyss 2014). Field work in terrestrial settings for quartzite karst has revealed further details of how dissolution works, including the non-obvious response that materials that dissolve more slowly are better for karstic development. Therefore, the slow kinetics on Titan makes it more likely for karst to form, but further work is needed in all areas to understand this process.

To date, field analog experiments for Titan have been the least explored of these three possibilities, and our hope is that this paper will elevate the use of Titan analog in appropriate circumstances to complement theoretical and laboratory work. These combined approaches will help us to better understand the data that is returned from past missions such as Voyager and Cassini, helping to answer lingering questions (Nixon et al. 2018, 2025). Multi-disciplinary work will also be invaluable for setting new science goals and interpreting data from future missions (Tobie et al. 2014b; Barnes et al. 2021; Lorenz et al. 2021; Sulaiman et al. 2021; Rodriguez et al. 2022) providing more rapid progress for the entire field.

# Glossary

| | |
|---|---|
| adiabatic | A change of physical properties (volume, temperature etc) that occurs without a change of overall internal energy. |
| aeolian | Arising due to action of wind. |
| alluvial | Pertaining to sediments and residues. |
| clathrate | A crystalline structure where a cage of one material (often water) traps molecules of a more volatile substance (e.g. methane). |
| clastic (rock) | A rock composed of fragmentary pieces (clasts) of other rocks. |
| cyclostropic | An atmospheric regime where pressure forces are balanced by centrifugal forces. |
| diagenesis | Physical and chemical transformations of sedimentary rocks. |
| diapir | An upwelling plume of deformable material into a brittle layer. |
| evaporites | Dissolved solid material left behind after a liquid evaporates. |
| endogenic | Processes occurring due to internal factors, such as mantle convention. |
| exogenic | Processes occurring due to external factors, such as impacts. |
| facies | Distinguishing physical, compositional and formation characteristics of a particular rock type. |
| fluvial | Resulting from the action of rivers. |
| geostrophic | An atmospheric regime where pressure forces are balanced by Coriolis forces. |
| halocline | Gradient of salinity with salt water depth. |
| karst | Terrain sculpted by chemical dissolution. |
| labyrinths | Terrain with recurring complex eroded shapes, often karstic. |
| laccolith | A lens-shaped intrusion of igneous material between rock layers. |
| lacustrine | Relating to lakes. |
| lithification | Transformation of sediments etc into rock. |
| maar | A broad, shallow crater formed when upwelling magma meets groundwater, often water-filled. |
| meridional | Along a line of constant longitude. |
| oblateness | The flattening of a planet’s shape relative to a perfect sphere. |
| obliquity | The tilt between a planet’s rotation axis and the normal axis of its orbital plane. |
| orogeny | The geologic process of mountain formation. |
| phreatic | Part of the crust lying below the water (liquid) table. |
| planform | Two-dimension projected outline of a three-dimensional object. |
| pluvial | Resulting from the action of rain. |
| refractory | Residues left behind after application of heat when volatiles have been removed, such as sooty grains. |
| regolith | The unconsolidated surface covering of a planetary body, such as soil, sand, dust etc. |
| rheological | Pertaining to the fluid deformation of matter. |
| sintering | A process where smaller grains are joined into a large solid mass, usually by application of heat and/or pressure. |
| stratigraphy | The relative positioning and ordering of geologic layers. |
| tectonic | Relating to large-scale crustal movements and processes. |
| thermocline | Temperature gradient with water depth. |
| tribocharging | Electrical charging of sand or dust grains due to collisions, induced by wind. |
| tuff | Rock formed from lithification of volcanic ash. |
| vadose | Part of the crust lying above the water (or liquid) table. |
| volatiles | Chemicals that can transition from a liquid or solid to a gas in a particular environment. |
| yardangs | Wind-eroded ridges. |

## Acknowledgements

This research was supported by the International Space Science Institute (ISSI) in Bern, through ISSI International Team project #539 (The habitability of Titan's subsurface water ocean). C.A.N., R.M.C.L., M.J.M., S.D.V., and A.M.S. were supported in part for their work on this paper by the NASA Astrobiology grant "Habitability of Hydrocarbon Worlds". Portions of the research were carried out at the Jet Propulsion Laboratory, California Institute of Technology, under a contract with the National Aeronautics and Space Administration (80NM0018D0004). C.A.N. was supported in part by a NASA GSFC Strategic Science funding award. K.K.F. was supported in part by the Center for Research and Exploration in Space Science and Technology II cooperative agreement with NASA and the University of Maryland, Baltimore County, under award number 80GSFC24M0006. K.K.F was supported in part by an appointment to the NASA Postdoctoral Program at NASA Goddard Space Flight Center, administered by Universities Space Research Association, and Oak Ridge Associated Universities, both under contract with NASA. K.M.S. was supported by the NASA Network for Life Detection project "Oceans Across Space and Time" (Grant No. 80NSSC18K130). N.A.T was supported by UK Science and Technology Facilities Council grant number ST/Y000676/1.

## Open Research

*Data Availability:*

- Figure 1: images are taken from the USGS photo archive: https://library.usgs.gov/photo/index.html#/ and may be accessed using their image numbers: (a) USGS 768227-3, pap0056a; (b) USGS: P645, F768228, pap0056d; (c) USGS: pap0097d, NASA: S-72-54471; (d) USGS: pap0090e, NASA: AP16-S71-51616. The associated report document is given in Schaber (2005) which can be accessed at: https://pubs.usgs.gov/of/2005/1190/
- Figure 2: image sources (a) https://astrobiology.nasa.gov/news/fossil-preservation-in-one-of-the-driest-place-on-earth/ (b) https://www.flickr.com/photos/nasadesertrats/6127812826/in/photostream (c) https://commons.wikimedia.org/wiki/File:Rovers_Red_and_Black_at_Houghton_Crater_Devon_Island_Canada.jpg (NASA/Matt Deans) (d) https://www.space.com/23228-europe-mars-rover-test-drive-photos.html
- Figure 3, 4: new diagrams. No associated data.
- Figure 5: Source for background image map: https://data.caltech.edu/records/8q9an-yt176. See also Le Mouélic et al. (2019).
- Figure 6: Earth profile from https://data.ceda.ac.uk/badc/cira/data/nhant.lsn (Rees et al. 1988, Rees et al. 1990). Titan profile digitized from Raulin et al. (2012), Fig. 3.

- Figure 7: Image sources (a) https://www.nasa.gov/image-article/titans-arrow-shaped-storm/ (NASA/JPL-Caltech/Space Science Institute) (b) https://spectrumnews1.com/oh/columbus/weather/2021/06/30/ohio-valley-and-mid-atlantic-2012-derecho- (G. Carbin, NWS Storm Prediction Center). (c) https://www.nasa.gov/missions/nasas-cassini-finds-monstrous-ice-cloud-in-titans-south-polar-region/ (NASA/JPL-Caltech/Space Science Institute) (d) https://earth.esa.int/eogateway/success-story/earthcare-s-incredible-images-of-winter-in-the-stratosphere (ESA)
- Figure 8: (a) Landsat image LC08_L2SP_041007_20220826_20220909_02_T1 from https://earthexplorer.usgs.gov/ (b) Sentinel-1 image S1A_IW_GRDH_1SDH_20240819T124333_20240819T124358_055282_06BD5C_27AB from https://search.asf.alaska.edu/ (c) image of Selk crater on Titan cropped from https://astrogeology.usgs.gov/search/map/titan_cassini_sar_hisar_global_mosaic_351m using coordinates: 5°10'N to 8°35'N, 158°30'E to 163°20'E.
- Figure 9: (a) Landsat image LC08_L2SP_108073_20250504_20250512_02_T1 from https://earthexplorer.usgs.gov/ (b) Sentinel-1 image S1A_IW_GRDH_1SDV_20250430T210636_20250430T210701_058991_0750E3_A46D from https://search.asf.alaska.edu/ (c) image of Titan Crater 26 cropped from https://astrogeology.usgs.gov/search/map/titan_cassini_sar_hisar_global_mosaic_351m using coordinates: 9°40'S to 6°50'S, 88°05'W to 84°0'W.
- Figure 10: (a) Landsat image LC08_L2SP_178079_20250412_20250416_02_T1 from https://earthexplorer.usgs.gov/ (b) Sentinel-1 image

S1C_IW_GRDH_1SDV_20250430T173441_20250430T173511_002125_004831_E7FB from https://search.asf.alaska.edu/ (c) image of Ksa crater on Titan cropped from https://astrogeology.usgs.gov/search/map/titan_cassini_sar_hisar_global_mosaic_351m using coordinates: 11°40’N to 15°25’N, 67°55’W to 62°40’W.

- Figure 11: Earth images cropped from Google Earth. b) centered on 39°19’12”N,115°16’30”W; d) centered on 44°23’42”N, 113°26’54”W; f) centered on 28°10’28”N, 53°15’28”E; h) centered on 35°16’16”N, 119°49’30”W. Titan images from the Cassini SAR dataset, available on the PDS: (https://pds-imaging.jpl.nasa.gov/volumes/radar.html). a) centered 144°E, 8°S; c) centered on 50°E, 50°N; e) centered on 62°W, 30°S; g) centered on 140°W, 10°S.
- Figure 12: Previously published in Radebaugh et al. (2013), as Fig. 2.
- Figure 13: Image from https://images.nasa.gov/ image number PIA07009.
- Figure 14: Reproduced from Radebaugh, J., (2010).
- Figure 15: Australia image from Landsat/Google Earth, 22 23’ S, 123 40’ W, Titan image from NASA Image Archive (https://images.nasa.gov/), image number PIA20710.
- Figure 16: From Northrup, D.S., 2018. A Geomorphological Study of Yardangs in China, the Altiplano/Puna of Argentina, and Iran as Analogs for Yardangs on Titan. Brigham Young University.
- Figure 17: Figure reproduced from Malaska et al. (2020) (Figure 17 in text).
- Figure 18:A) Figure reproduced from Malaska et al. (2020) (Figure 8B in text) B) Figure reproduced from Malaska et al., (2020) (Figure 8A in text).

- Figure 19: A) Figure reproduced from Malaska et al., (2020) (Figure 3 in text). B) Image of Biokovo Range, Croatia from NASA Earth Observatory https://www.earthobservatory.nasa.gov/images/36849/biokovo-range-croatia C) Figure reproduced from Malaska et al., (2020) (Fig. 1 in text). D) Image of White Canyon, Utah, from https://science.nasa.gov/resource/white-canyon-utah-simulated-natural-color-view/ Credit: NASA/GSFC/METI/ERSDAC/JAROS and U.S./Japan ASTER Science Team
- Figure 20: Images reproduced from Planetary Photojournal image PIA20708 (https://photojournal.jpl.nasa.gov/catalog/PIA20708). Java photo from Haryono and Day (2004), courtesy of Eko Haryono.
- Figure 21: (a) https://eol.jsc.nasa.gov/SearchPhotos/photo.pl?mission=ISS006&roll=E&frame=17451 (b) (c) https://visibleearth.nasa.gov/images/5978/navajo-mountain-utah/5978t) (d), (e), (f), Cassini RADAR data available at: https://pds-imaging.jpl.nasa.gov/volumes/radar.html
- Figure 22: (a) Schurmeier et al. (2023) (b) https://earthobservatory.nasa.gov/images/80435/irans-great-salt-desert
- Figure 23: Left image from ESA Huygens DISR instrument: https://sci.esa.int/web/cassini-huygens/-/36397-river-channel-on-titan. Right image from NASA image PIA06440: https://science.nasa.gov/resource/view-from-titans-surface/. Middle image: new by authors from Cassini RADAR SAR data available on the PDS (https://pds-imaging.jpl.nasa.gov/volumes/radar.html) and co-registering the Huygens images to the SAR images.

- Figures 24-26: created by authors. Left side: visible image data from the ESRI World Imagery dataset: (https://www.esri.com/en-us/capabilities/imagery-remote-sensing/capabilities/content). Right side: Cassini SAR and VIMS data, available on the PDS (https://pds-imaging.jpl.nasa.gov/volumes/radar.html).
- Figure 27: previously published in Birch et al., (2025a).
- Figure 28: Right side: new image by authors from Cassini RADAR SAR data available on the PDS (https://pds-imaging.jpl.nasa.gov/volumes/radar.html). Left side: Google Earth image centered at 43.4554 N, 17.1993 E.
- Figure 29: (a) image by Grueslayer @Wikipedia, CC BY-SA 4.0 https://commons.wikimedia.org/wiki/File:STAPP_114_La_Brea_Pitch_Lake.jpg (b) https://www.researchgate.net/publication/279427044_An_Overview_of_In-Situ_Burning/figures?lo=1(with permission of uploader, Merv Fingas).
- Figure 30: Models plotting by authors based on fundamental physical data available at NIST: https://webbook.nist.gov/chemistry/
- Figure 31: (A) NOAA image. URL: https://oceanexplorer.noaa.gov/wp-content/uploads/2024/05/what-are-cold-seeps-fact-sheet.pdf (B) NOAA Image. URL: https://oceanexplorer.noaa.gov/wp-content/uploads/2024/05/what-are-cold-seeps-fact-

sheet.pdf (C) USGS image https://d9-wret.s3.us-west-2.amazonaws.com/assets/palladium/production/s3fs-public/styles/full_width/public/thumbnails/image/OilSeepTarwaterCreek.JPG?itok=5c1fNYg4

- Figure 32: USGS image: hydrate_database_browsegraphic.jpg. URL: https://www.sciencebase.gov/catalog/file/get/5eb413a282ce25b5135a9f2a?f=__disk__e9%2F54%2F7b%2Fe9547b14cbfce846136a002378bdee1f4fc2d693&width=580&height=282
- Figure 33: (a) redrawn after USGS Woods Hole Science Center, Open File Report 96-272, fig7_800B0.jpg. (b) redrawn after USGS Woods Hole Science Center, Open File Report 96-272, fig8_700C0.jpg. Sources available at: https://woodshole.er.usgs.gov/operations/ia/index.html
- Figure 34: see citations in figure caption.
- Figure 35: (a) Sentinel-2 image via ESA, available at: https://apps.sentinel-hub.com/eo-browser 77° 23′ 12,44″ S, 161° 55′ 37,99″ W (b) source: https://astrobiology.nasa.gov/news/antarcticas-subglacial-lakes-support-prospects-for-life-on-icy-moons/ (JPL/ASUN/NSF/NASA/Dr A Behar)
- Figure 36: Figure reproduced from Vance et al. (2019).
- Figure 37: new diagram by authors, no associated data.

## Conflict of Interest Disclosure

The authors declare no competing conflicts of interest for this manuscript.

# References

Abrate T, Hubert P, Sighomnou D. A study on hydrological series of the Niger River. Hydrological Sciences Journal. 2013 Feb 1;58(2):271-9.

Achterberg RK, Conrath BJ, Gierasch PJ, et al. Titan's middle-atmospheric temperatures and dynamics observed by the Cassini Composite Infrared Spectrometer. Icarus. 2008 Mar 1;194(1):263-77.

Achterberg RK, Gierasch PJ, Conrath BJ, et al. Temporal variations of Titan’s middle-atmospheric temperatures from 2004 to 2009 observed by Cassini/CIRS. Icarus. 2011 Jan 1;211(1):686-98.

Adams EE, Priscu JC, Fritsen CH, et al. Permanent ice covers of the McMurdo Dry Valley Lakes, Antarctica: bubble formation and metamorphism. Ecosystem dynamics in a polar desert: the McMurdo dry valleys, Antarctica. 1998 Jan 28;72:281-95.

Affholder A, Guyot F, Sauterey B, et al. Bayesian analysis of Enceladus’s plume data to assess methanogenesis. Nature Astronomy. 2021 Aug;5(8):805-14.

Affholder A, Higgins PM, Cockell CS, et al. The Viability of Glycine Fermentation in Titan’s Subsurface Ocean. The Planetary Science Journal. 2025 Apr 7;6(4):86.

Aharonson O, Hayes AG, Lunine JI, et al. An asymmetric distribution of lakes on Titan as a possible consequence of orbital forcing. Nature Geoscience. 2009 Dec;2(12):851-4.

Alali J, Abu Salah, A, Yasin SM, Al Omari W. Oil shale. Jordan Ministry of Energy and Mineral Resources. 2014: 23 p.

AlHumaidan FS, Rana MS, Lababidi HM, Hauser A. Pyrolysis of asphaltenes derived from residual oils and their thermally treated pitch. ACS omega. 2020 Sep 18;5(38):24412-21.

Amit H, Choblet G, Tobie G, et al. Cooling patterns in rotating thin spherical shells—Application to Titan's subsurface ocean. Icarus. 2020 Mar 1;338:113509.

Anderson EM. The dynamics of faulting. Transactions of the Edinburgh Geological Society. 1905;8(3):387-402.

Anderson CM, Samuelson RE, Yung YL, McLain JL. Solid-state photochemistry as a formation mechanism for Titan's stratospheric C4N2 ice clouds. Geophysical Research Letters. 2016 Apr 16;43(7):3088-94.

Anderson CM, Samuelson RE, Nna-Mvondo D. Organic ices in Titan's stratosphere. Space Science Reviews. 2018 Dec;214:1-36.

Arens FL, Uhl J, Schmitt-Kopplin P, Karger C, Mangesldorf K, Sager C, Airo A, Valenzuela B, Zamorano P, Schulze-Makuch D. Exploring organic compound preservation through long-term in situ experiments in the Atacama desert and the relevance for Mars. Nature Scientific Reports. 2025 Aug 15; 15 (29957).

Arian, M. 2012. Clustering of diapiric provinces in the Central Iran Basin. Carbonates Evaporites 27, 9–18. https://doi.org/10.1007/s13146-011-0079-9.

Artemieva, N. and Lunine, J., 2003. Cratering on Titan: impact melt, ejecta, and the fate of surface organics. Icarus, 164(2), pp.471-480.

Atreya, S.K., Donahue, T.M. & Kuhn, W.R., 1978, Evolution of a nitrogen atmosphere on Titan, Science, 201, 611.

Avduevskii VS, Vishnevetskii SL, Golov IA, et al. Measurement of wind velocity on the surface of Venus during the operation of stations Venera 9 and Venera 10. Cosmic Research. 1977 Mar;14(5):710-3.

Azzam MOJ, Al-Ghazawi Z, Al Otoom A. Incorporation of Jordanian oil shale in hot mix asphalt. Journal of Cleaner Production. 2016 Jan 20; 112(4): 2259-2277.

Baker VR, Nummedal D, editors. The Channeled Scabland: a guide to the geomorphology of the Columbia Basin, Washington. NASA; 1978.

Baker, A. A. (1936). Geology of the Monument Valley-Navajo Mountain Region, San Juan County, Utah (No. 865). US Government Printing Office.

Barnes JW, Brown RH, Soderblom L, et al. Spectroscopy, morphometry, and photoclinometry of Titan's dunefields from Cassini/VIMS. Icarus. 2008 May 1;195(1):400-14.

Barnes JW, Soderblom JM, Brown RH, Buratti BJ, Sotin C, Baines KH, Clark RN, Jaumann R, McCord TB, Nelson R, Le Mouélic S. VIMS spectral mapping observations of Titan during the Cassini prime mission. Planetary and Space Science. 2009 Dec 1;57(14-15):1950-62.

Barnes JW, Bow J, Schwartz J, et al. Organic sedimentary deposits in Titan’s dry lakebeds: Probable evaporite. Icarus. 2011 Nov 1;216(1):136-40.Barnes JW, Buratti BJ, Turtle ET, et al. 2013 January 14. Planetary Science, 2, 1.

Barnes JW, Sotin C, Soderblom JM, Brown RH, Hayes AG, Donelan M, Rodriguez S, Mouélic SL, Baines KH, McCord TB. Cassini/VIMS observes rough surfaces on Titan’s Punga Mare in specular reflection. Planetary science. 2014 Aug 21;3(1):3.

Barnes JW, Lorenz RD, Radebaugh J, et al. Production and global transport of Titan's sand particles. Planetary Science. 2015 Dec;4(1):1-9.

Barnes JW, Hayes AG, Soderblom JM, MacKenzie SM, et al. New Frontiers Titan orbiter. Bulletin of the American Astronomical Society. 2021;53(4).

Barnes JW, Turtle EP, Trainer MG, Lorenz RD, MacKenzie SM, Brinckerhoff WB, Cable ML, Ernst CM, Freissinet C, Hand KP, Hayes AG. Science goals and objectives for the Dragonfly titan rotorcraft relocatable lander. The Planetary Science Journal. 2021 Jul 19;2(4):130.

Beck BF. A generalized genetic framework for the development of sinkholes and karst in Florida, USA. Environmental Geology and Water Sciences. 1986 Mar;8(1):5-18.

Béghin C, Randriamboarison O, Hamelin M, et al. Analytic theory of Titan's Schumann resonance: Constraints on ionospheric conductivity and buried water ocean. Icarus. 2012 Apr 1;218(2):1028-42.

Belknap DF, Kelley JT, Gontz AM. Evolution of the glaciated shelf and coastline of the northern Gulf of Maine, USA. Journal of Coastal Research. 2002 Mar 1(36):37-55.

Bennett RM, Card JR, Hornby DM. Hydrology of Lake Athabasca: past, present and future. Hydrological Sciences Journal. 1973 Sep 1;18(3):337-45.

Bercovici D. The generation of plate tectonics from mantle convection. Earth and Planetary Science Letters. 2003 Jan 10;205(3-4):107-21.

Bercovici D, Ricard Y. Plate tectonics, damage and inheritance. Nature. 2014 Apr 24;508(7497):513-6.

Beuthe M. East–west faults due to planetary contraction. Icarus. 2010 Oct 1;209(2):795-817.

Bézard B, Vinatier S, Achterberg RK. Seasonal radiative modeling of Titan's stratospheric temperatures at low latitudes. Icarus. 2018 Mar 1;302:437-50.

Birch SPD, Hayes AG, Howard AD, Moore JM, Radebaugh J. Alluvial fan morphology, distribution and formation on Titan. Icarus. 2016 May 15;270:238-47.

Birch SPD, Hayes AG, Dietrich WE, et al. Geomorphologic mapping of Titan's polar terrains: constraining surface processes and landscape evolution. Icarus. 2017. 282, 214-236.

Birch SPD, Hayes AG, Corlies P, et al. Morphological evidence that Titan's southern hemisphere basins are paleoseas. Icarus. 2018 Aug 1;310:140-8.

Birch SPD, Hayes AG, Poggiali V, et al. Raised rims around Titan's sharp-edged depressions. Geophysical Research Letters. 2019 Jun 16;46(11):5846-54.

Birch SPD, Umurhan OM, Hayes AG, & Malaska MJ (2020). Simulating the evolution of Titan's surface through fluvial and dissolution erosion: II. The details. In Lunar and Planetary Science Conference 51, Abstract 1979. Presented at the Woodlands, TX.

Birch SPD, Parker G, Corlies C, et al. Reconstructing river flow remotely on Earth, Titan, and Mars. PNAS. 2023, 120.

Birch SPD, Hayes AG, Perron JT. Titan's fluvial and lacustrine landscapes. Titan After Cassini-Huygens. 2025a Jan 1:287-324.

Birch SPD, Palermo RV, Schneck UG, Ashton A, Hayes AG, Soderblom JM, Mitchell WH, Perron JT. Detectability of coastal landforms on Titan with the Cassini RADAR. JGR-Planets. 2025b. 130.

Bird MK, Allison M, Asmar SW, et al. The vertical profile of winds on Titan. Nature, 2005 Dec 8, 438, 800-802.

Birks SJ, Moncur MC, Gibson JJ, Yi Y, Fennell JW, Taylor EB. Origin and hydrogeological settings of saline groundwater discharges to the Athabasca River: Geochemical and isotopic characterization of the hyporheic zone. Applied Geochemistry. 2018 Nov; 98: 172-190.

Black BA, Perron JT, Burr DM, Drummond SA. Estimating erosional exhumation on Titan from drainage network morphology. Journal of Geophysical Research: Planets. 2012 Aug;117(E8).

Blasius KR, Cutts JA, Guest JE, Masursky H. Geology of the Valles Marineris: First analysis of imaging from the Viking 1 Orbiter primary mission. Journal of Geophysical Research. 1977 Sep 30;82(28):4067-91.

Bois C, Bouche P, Pelet R. Global geologic history and distribution of hydrocarbon reserves. AAPG Bulletin. 1982 Sep 1;66(9):1248-70.

Booth JA, Rowe, MM, Fischer KM. Offshore Gas Hydrate Sample Database with an Overview and Preliminary Analysis. USGS Woods Hole Science Center Open-File Report 1996 Jun, 96-272, https://pubs.usgs.gov/of/1996/of96-272/index.html

Borg LE, Carlson RW. The Evolving Chronology of Moon Formation. Annual Review of Earth and Planetary Sciences. 2023;51.

Bostock HC, Brooke BP, Ryan DA, Hancock G, Pietsch T, Packett R, Harle K. Holocene and modern sediment storage in the subtropical macrotidal Fitzroy River estuary, Southeast Queensland, Australia. Sedimentary Geology. 2007 Oct 15;201(3-4):321-40.

Bottke WF, Vokrouhlicky D, Nesvorny D, Hayne P. An Exogenic Source for Titan's Dune Particles?. LPI Contributions. 2024 Mar;3040:1550.

Bray VJ, Hagerty JJ, Collins GS. "False peak" creation in the Flynn Creek marine target impact crater. Meteoritics & Planetary Science. 2022 Mar.

Bridges NT, Spagnuolo MG, de Silva SL, et al. Formation of gravel-mantled megaripples on Earth and Mars: Insights from the Argentinean Puna and wind tunnel experiments. Aeolian Research. 2015 Jun 1;17:49-60.

Brossier JF, Rodriguez S, Cornet T, et al. Geological evolution of Titan's equatorial regions: Possible nature and origin of the dune material. Journal of Geophysical Research: Planets. 2018 May;123(5):1089–1112.

Brouwer G. *Craters, Cryovolcanism, and Clathrates: Exploring Interactions Between Titan's Ice Shell, Surface, and Atmosphere* (Doctoral dissertation, University of Hawai'i at Manoa), 2025.

Brouwer, G.E., Schurmeier, L.R. and Fagents, S.A., 2025. Topographic Relaxation of Complex Impact Craters in a Clathrate Crust on Titan. LPI Contributions, 3090, p.2794.

Brown M, Solar GS. Shear-zone systems and melts: feedback relations and self-organization in orogenic belts. Journal of structural geology. 1998 Feb 1;20(2-3):211-27.

Brown ME, Bouchez AH, Griffith CA. Direct detection of variable tropospheric clouds near Titan's south pole. 2002 Dec 19, Nature, 420, 795-797.

Brown RH, Soderblom LA, Soderblom JM, et al. The identification of liquid ethane in Titan's Ontario Lacus. Nature. 2008 Jul;454(7204):607-10.

Brown ME, Schaller EL, Roe HG, et al. Discovery of lake-effect clouds on Titan. Geophysical Research Letters. 2009a Jan 3;36(1);L01103.

Brown ME, Smith AL, Chen C, Ádámkovics M. Discovery of fog at the south pole of Titan. The Astrophysical Journal. 2009b Nov 20;706;L110-L113.

Buchner E, Schmieder M. Multiple fluvial reworking of impact ejecta—A case study from the Ries crater, southern Germany. Meteoritics & Planetary Science. 2009 Jul;44(7):1051-60.

Buffett B, Archer D. Global inventory of methane clathrate: sensitivity to changes in the deep ocean. Earth and Planetary Science Letters. 2004 Nov 15;227(3-4):185-99.

Buffo JJ, Schmidt BE, Huber C. Multiphase reactive transport and platelet ice accretion in the sea ice of McMurdo Sound, Antarctica. Journal of Geophysical Research: Oceans. 2018 Jan;123(1):324-45.

Bull WB. The alluvial-fan environment. Progress in Physical geography. 1977 Jun;1(2):222-70.

Burr DM, Jacobsen RE, Roth DL, Phillips CB, Mitchell KL, Viola D. Fluvial network analysis on Titan: Evidence for subsurface structures and west-to-east wind flow, southwestern Xanadu. Geophysical Research Letters. 2009 Nov;36(22).

Burr DM, Drummond SA, Cartwright R, et al. Morphology of fluvial networks on Titan: Evidence for structural control. Icarus. 2013 Sep 1;226(1):742-59.

Butchart N. The stratosphere: A review of the dynamics and variability. Weather and Climate Dynamics. 2022 Nov 7;3(4):1237-72.

Butler RM. Steam-assisted gravity drainage: concept, development, performance and future. Journal of Canadian Petroleum Technology. 1994 Feb 1;33(02):44-50.

Butler RM, Stephens DJ. The gravity drainage of steam-heated heavy oil to parallel horizontal wells. Journal of Canadian Petroleum Technology. 1981 Apr 1;20(02).

Cable ML, Vu TH, Malaska MJ, et al. A Co-Crystal between Acetylene and Butane: A Potentially Ubiquitous Molecular Mineral on Titan. ACS Earth Space Chem. 2019 Nov 15;3(12);2808–2815.

Cable ML, Runčevski T, Maynard-Casely HE, et al. Titan in a test tube: Organic co-crystals and implications for Titan mineralogy. Accounts of Chemical Research. 2021 Jul 23;54(15):3050-9.

Campbell DB, Black GJ, Carter LM, et al. Radar evidence for liquid surfaces on Titan. Science. 2003 Oct 17;302(5644):431-4.

Carnahan E, Vance SD, Hesse MA, Journaux B, Sotin C. Dynamics of mixed clathrate-ice shells on ocean worlds. Geophysical Research Letters. 2022 Apr 28;49(8):e2021GL097602.

Carr MH, Greeley R. Volcanic features of Hawaii: A basis for comparison with Mars. National Aeronautics and Space Administration; 1980.

Cartwright JH, Escribano B, González DL, Sainz-Díaz CI, Tuval I. Brinicles as a case of inverse chemical gardens. Langmuir. 2013 29(25):7655–7660.

Chandler CK, Radebaugh J, McBride JH, et al. Near-surface structure of a large linear dune and an associated crossing dune of the northern Namib Sand Sea from Ground Penetrating Radar: Implications for the history of large linear dunes on Earth and Titan. Aeolian Research. 2022 Aug 1;57:100813.

Chatain A, Rafkin SCR, Soto A, et al. Air – Sea Interactions on Titan: Effect of Radiative Transfer on the Lake Evaporation and Atmospheric Circulation. The Planetary Science Journal. 2022 Oct;3(10);232.

Chatain A, Rafkin SCR, Soto A, et al. The impact of lake shape and size on lake breezes and air-lake exchanges on Titan. Icarus. 2024;411;115925.

Chen Y, Xu C, Chen Y, Liu Y, Li W. Progress, challenges and prospects of eco-hydrological studies in the Tarim river basin of Xinjiang, China. Environmental management. 2013 Jan;51(1):138-53.

Choukroun M, Sotin C. Is Titan's shape caused by its meteorology and carbon cycle? Geophys. Res. Let. 2012 Feb 29; 39: 1-5. https://doi.org/10.1029/2011GL050747.

Cockell CS, Bush T, Bryce C, et al. Habitability: a review. Astrobiology. 2016 Jan 1;16(1):89-117.

Collins C, Deneke F, Jenkins T, Johnson L, Mcfadden T. Fate and Effects of Crude Oil Spilled on Permafrost Terrain. First Year Progress Report. 1976 Nov 1.

Collins GS, Wünnemann K. How big was the Chesapeake Bay impact? Insight from numerical modeling. Geology. 2005 Dec 1;33(12):925-8.

Collins GC, McKinnon WB, Moore JM, Nimmo F, Pappalardo RT, Prockter LM, Schenk PM. Tectonics of the outer planet satellites. Planetary tectonics. 2009 Dec 17;11(264):229.

Collins RE, Carpenter SD, Deming JW. Spatial heterogeneity and temporal dynamics of particles, bacteria, and pEPS in Arctic winter sea ice. Journal of Marine Systems. 2008 Dec 1;74(3-4):902-17.

Collis P. The Hawkesbury estuary from 1950 to 2050. InEstuaries of Australia in 2050 and beyond 2013 Jul 25 (pp. 247-257). Dordrecht: Springer Netherlands.

Conrad CP, Lithgow-Bertelloni C. How mantle slabs drive plate tectonics. Science. 2002 Oct 4;298(5591):207-9.

Cook, E. F. 1954. Geology of the Pine Valley Mountains, Utah. University of Washington.

Cook-Hallett C, Barnes JW, Kattenhorn SA, Hurford T, Radebaugh J, Stiles B, Beuthe M. Global contraction/expansion and polar lithospheric thinning on Titan from patterns of tectonism. Journal of Geophysical Research: Planets. 2015 Jun;120(6):1220-36.

Cordier D, Mousis O, Lunine JI, et al. An estimate of the chemical composition of Titan's lakes. The Astrophysical Journal. 2009 Dec 4;707(2):L128.

Cordier D, Mousis O, Lunine JI, et al. Titan’s lakes chemical composition: Sources of uncertainties and variability. Planetary and Space Science. 2012 Feb;61(1);99-107.

Cordier D, Liger-Belair G. Bubbles in Titan’s seas: Nucleation, growth, and RADAR signature. The Astrophysical Journal. 2018 May 18;859(1):26.

Cordier D, Carrasco N. The floatability of aerosols and wave damping on Titan’s seas. Nature Geoscience. 2019 May;12(5):315-20.

Cordiner MA, Garcia-Berrios E, Cosentino RG et al. Detection of Dynamical Instability in Titan’s Thermospheric Jet. The Astrophysical Journal Letters. 2020 Nov 20;904(1):L12.

Corfidi SF, Coniglio MC, Cohen AE, Mead CM. A proposed revision to the definition of “derecho”. Bulletin of the American Meteorological Society. 2016 Jun;97(6):935-49.

Corlies, P., Hayes, A. G., Birch, S. P. D., et al. Titan’s Topography and Shape at the End of the Cassini Mission. Geophys. Res. Lett., 2017 Nov 7; 44: 11,754-11,761

Cornet T, Bourgeois O, Le Mouélic S, et al. Geomorphological significance of Ontario Lacus on Titan: Integrated interpretation of Cassini VIMS, ISS and RADAR data and comparison with the Etosha Pan (Namibia). Icarus. 2012 Apr 1;218(2):788-806.

Cornet T, Cordier D, Bahers TL, Bourgeois O, Fleurant C, Mouélic SL, Altobelli N. Dissolution on Titan and on Earth: Toward the age of Titan's karstic landscapes. Journal of Geophysical Research: Planets. 2015 Jun;120(6):1044-74.

Cornet T, Fleurant C, Seignovert B, Cordier D, Bourgeois O, Le Mouélic S. Landscape formation through dissolution proceses on Titan: a 3D landscape evolution model approach. In48th Lunar and Planetary Science Conference 2017.

Corrales LR, Yi TD, Trumbo SK, et al. Acetonitrile cluster solvation in a cryogenic ethane-methane-propane liquid: Implications for Titan lake chemistry. J. Chem. Phys. 2017 Mar 14;146(10);104308.

Corry, C. E. 1988. Laccoliths: mechanics of emplacement and growth (Vol. 220). Geological Society of America.

Costanza A, D'Anna G, D'Alessandro A, Vitale G, Fertitta G, Cosenza P. Mechanical aspects in a tide gauge station design. InOCEANS 2017-Aberdeen 2017 Jun 19 (pp. 1-10). IEEE.

Cottini V, Nixon CA, Jennings DE, et al. Spatial and temporal variations in Titan's surface temperatures from Cassini CIRS observations. Planetary and Space Science. 2012 Jan 1;60(1):62-71.

Coustenis A, Bezard B, Gautier D. Titan's atmosphere from Voyager infrared observations: I. The gas composition of Titan's equatorial region. Icarus. 1989 Jul 1;80(1):54-76.

Coustenis A, Schmitt B, Khanna RK, Trotta F. Plausible condensates in Titan’s stratosphere from Voyager infrared spectra. Planetary and Space Science. 1999 Oct 1;47(10-11):1305-29.

Coustenis A, Achterberg RK, Conrath BJ, et al. The composition of Titan's stratosphere from Cassini/CIRS mid-infrared spectra. Icarus. 2007 Jul 1;189(1):35-62.

Crósta AP, Silber EA, Lopes RM, Johnson BC, Bjonnes E, Malaska MJ, Vance SD, Sotin C, Solomonidou A, Soderblom JM. Modeling the formation of Menrva impact crater on Titan: Implications for habitability. Icarus. 2021 Dec 1;370:114679.

Cui J, Yelle RV, Volk K. Distribution and escape of molecular hydrogen in Titan's thermosphere and exosphere. Journal of Geophysical Research: Planets. 2008 Oct;113(E10).

Cui J, Yelle RV, Vuitton V, et al. Analysis of Titan's neutral upper atmosphere from Cassini Ion Neutral Mass Spectrometer measurements. Icarus. 2009 Apr 1;200(2):581-615.

Currie, K. L. (1971). The geology and petrology of the Mistastin Lake ring structure. Geological Survey of Canada Bulletin, 190, 1–47.

Czaplinski EC, Gilbertson WA, Farnsworth KK, et al. Experimental study of ethylene evaporites under Titan conditions. ACS Earth and Space Chemistry. 2019 Aug 27;3(10):2353-62.

Czaplinski EC, Vu TH, Cable ML, et al. Experimental characterization of the pyridine: acetylene co-crystal and implications for Titan’s surface. ACS Earth and Space Chemistry. 2023 Feb 28;7(3):597-608.

Czaplinski EC, Vu TH, Maynard-Casely H, et al. Formation and Stability of the Propionitrile: Acetylene Co-Crystal Under Titan-Relevant Conditions. ACS earth and space chemistry. 2025 Jan 28;9(2):253-64.

Danesh A. PVT and phase behaviour of petroleum reservoir fluids. Elsevier Science. 1998 May 7; 400p.

Danielson, R.E., Caldwell, J.J. & Larach, D.R., 1973. An inversion in the atmosphere of Titan, Icarus, 20, 437.

Danilov A. Chemistry of the Ionosphere. Springer Science & Business Media; 2012 Dec 6.

Darack E. Weatherscapes: Pinacate Volcanic Field–Geology Curated by the Sky. Weatherwise. 2018 Jul 4;71(4):8-9.

Dartnell LR, Hunter SJ, Lovell KV, et al. Low-temperature ionizing radiation resistance of Deinococcus radiodurans and Antarctic Dry Valley bacteria. Astrobiology. 2010 Sep 1;10(7):717-732.

Dasch P. Apollo Astronaut Training in the Big Bend. Journal of Big Bend Studies. 2016 Jan 1;28.

Davies AG, Sotin C, Choukroun M, Matson DL, Johnson TV. Cryolava flow destabilization of crustal methane clathrate hydrate on Titan. Icarus. 2016 Aug; 274: 23-32.

De Kok RJ, Teanby NA, Maltagliati L, Irwin PG, Vinatier S. HCN ice in Titan's high-altitude southern polar cloud. Nature. 2014 Oct 2;514(7520):65-7.

Dence, M.R., 1972, August. The nature and significance of terrestrial impact structures. In International Geological Congress Proceedings (Vol. 24, pp. 77-89).

Desiage PA, Montero-Serrano JC, St-Onge G, Crespi-Abril AC, Giarratano E, Gil MN, Haller MJ. Quantifying sources and transport pathways of surface sediments in the Gulf of San Jorge, central Patagonia (Argentina). Oceanography. 2018 Dec 1;31(4):92-103.

Dickinson WR. The Basin and Range Province as a composite extensional domain. International Geology Review. 2002 Jan 1;44(1):1-38.

Diez-y-Riega H, Camejo D, Rodriguez AE, Manzanares CE. Unsaturated hydrocarbons in the lakes of Titan: Benzene solubility in liquid ethane and methane at cryogenic temperatures. Planetary and Space Science. 2014 Sept;99(9);28-35.

Dillon WP. Gas Hydrate in the Ocean Environment. Editor(s): Robert A. Meyers, Encyclopedia of Physical Science and Technology (Third Edition), Academic Press, 2002; 473-486

Dhingra RD, Barnes JW, Brown RH, et al. Observational evidence for summer rainfall at Titan's north pole. Geophysical Research Letters. 2019 Feb 16;46(3):1205-12.

Doran PT, Fritsen CH, McKay CP, et al. Formation and character of an ancient 19-m ice cover and underlying trapped brine in an “ice-sealed” east Antarctic lake. Proceedings of the National Academy of Sciences. 2003 Jan 7;100(1):26-31.

Durante D, Hemingway DJ, Racioppa P, et al.. Titan's gravity field and interior structure after Cassini. Icarus. 2019 Jul 1;326:123-132.

Dunne KBJ, and Jerolmack DJ. What sets river width? Science Advances. 2020 Oct 7; 6(41), eabc1505.

Eaton GP. The Basin and Range province: Origin and tectonic significance. Annual Review of Earth and Planetary Sciences, Vol. 10, p. 409. 1982;10:409.

Edberg NJ, Wahlund JE, Ågren K, Morooka MW, Modolo R, Bertucci C, Dougherty MK. Electron density and temperature measurements in the cold plasma environment of Titan: Implications for atmospheric escape. Geophysical Research Letters. 2010 Oct;37(20).

Eigenbrode JL, Summons RE, Steele A, Freissinet C, Maeva Millan RNG, Sutter B, McAdam AC, Franz HB, Glavin DP, Archer PD Jr., Grotzinger JP, Gupta S, Ming DW, Sumner DY, Szopa C, Malespin C, Buch A, Coll P. Organic matter preserved in 3-billion-year-old mudstones at Gale crater, Mars. Science. 2018 Jun 8; 360 (6393): 1096-1101

Engle AE, Hanley J, Dustrud S, Thompson G, Lindberg GE, Grundy WM, Tegler SC. Phase diagram for the methane–ethane system and its implications for Titan's lakes. The Planetary Science Journal. 2021 Jun 21;2(3):118.

Engle AE, Hanley J, Tan SP, Grundy WM, Tegler SC, Lindberg GE, Steckloff JK, Raposa SM, Thieberger CL, Dustrud S, Groven JJ. Ice Formation, Exsolution, and Multiphase Equilibria in the Methane–Ethane–Nitrogen System at Titan Surface Conditions. The Planetary Science Journal. 2024 Oct 9;5(10):224.

Ewing, R.C., Hayes, A.G. and Lucas, A., 2015. Sand dune patterns on Titan controlled by long-term climate cycles. *Nature Geoscience*, *8*(1), pp.15-19.

Fanti F, Catuneanu O. Stratigraphy of the Upper Cretaceous Wapiti Formation, west-central Alberta, Canada. Canadian Journal of Earth Sciences. 2009 Apr 1;46(4):263-86.

Farnsworth KK, Chevrier VF, Steckloff JK, et al. Nitrogen Exsolution and Bubble Formation in Titan's Lakes. Geophysical Research Letters. 2019 Dec 3;46(23);13658-13667.

Farnsworth KK, Soto A, Chevrier VF, et al. Floating Liquid Droplets on the Surface of Cryogenic Liquids: Implications for Titan Rain. ACS Earth Space Chem. 2023 Feb 3;7(2);439-448.

Ferguson GA, Betcher RN, Grasby SE. Hydrogeology of the Winnipeg formation in Manitoba, Canada. Hydrogeology journal. 2007 May;15(3):573-87.

Fertl WH, Chilingarian GV. Formation Evaluation of Tar Sands using Geophysical Well-Logging Techniques. InDevelopments in Petroleum Science 1978 Jan 1 (Vol. 7, pp. 259-276). Elsevier.

Filizola N, Spínola N, Arruda W, Seyler F, Calmant S, Silva J. The Rio Negro and Rio Solimões confluence point–hydrometric observations during the 2006/2007 cycle. River, Coastal and Estuarine Morphodynamics. Santa Fe: Taylor & Francis. 2009 Sep 21:1003-6.

Fisher, D.M. and Anastasio, D.J., 1994. Kinematic analysis of a large-scale leading edge fold, Lost River Range, Idaho. *Journal of Structural Geology*, *16*(3), pp.337-354.

Flasar FM. Oceans on Titan?. Science. 1983 Jul 1;221(4605):55-7.

Flasar FM, Achterberg RK. The structure and dynamics of Titan's middle atmosphere. Philosophical Transactions of the Royal Society A: Mathematical, Physical and Engineering Sciences. 2009 Feb 28;367(1889):649-64.

Ford D, Williams PD. Karst hydrogeology and geomorphology. John Wiley & Sons; 2007 Apr 23.

Forsberg-Taylor NK, Howard AD, Craddock RA. Crater degradation in the Martian highlands: Morphometric analysis of the Sinus Sabaeus region and simulation modeling suggest fluvial processes. Journal of Geophysical Research: Planets. 2004 May;109(E5).

Francy DS, Stelzer EA, Brady AM, Huitger C, Bushon RN, Ip HS, Ware MW, Villegas EN, Gallardo V, Lindquist HA. Comparison of filters for concentrating microbial indicators and pathogens in lake water samples. Applied and Environmental Microbiology. 2013 Feb 15;79(4):1342-52.

Frank SL, Head JW. Ridge belts on Venus: Morphology and origin. Earth, Moon, and Planets. 1990 Jul;50(1):421-70.

Friedson AJ, Yung YL. The thermosphere of Titan. Journal of Geophysical Research: Space Physics. 1984 Jan;89(A1):85-90.

Fudali, R. F., & Pani, E. A. (2013). The Lonar Crater: An impact crater in basalt. Meteoritics & Planetary Science, 48(6), 959–981.

Fulchignoni M, Ferri F, Angrilli F, et al. In situ measurements of the physical characteristics of Titan's environment. Nature. 2005 Dec 8;438(7069):785-91.

Fuller-Rowell T. Physical characteristics and modeling of Earth's thermosphere. Modeling the Ionosphere–Thermosphere System. 2014 Feb 28:13-27.

Fuller-Rowell TJ, Rees D. A three-dimensional time-dependent global model of the thermosphere. Journal of Atmospheric Sciences. 1980 Nov;37(11):2545-67.

Gates ID. Oil phase viscosity behaviour in Expanding-Solvent Steam-Assisted Gravity Drainage. Journal of Petroleum Science and Engineering. 2007; 59: 123-134. doi:10.1016/j.petrol.2007.03.006

Gilbert G.K. 1877. Report on the geology of the Henry Mountains. U.S. Geographical and Geological Survey, Rocky Mountains Region, p 160

Glasby GP. Abiogenic origin of hydrocarbons: An historical overview. Resource Geology. 2006 Mar;56(1):83-96.

Glein CR, Shock EL. A geochemical model of non-ideal solutions in the methane-ethane-propane-nitrogen-acetylene system on Titan. 2013 Geochimica et Cosmochimica Acta. 115;217-240.

Glein CR. Noble gases, nitrogen, and methane from the deep interior to the atmosphere of Titan. Icarus. 2015 Apr 1;250:570-86.

Goldstein AH, Galbally IE. Known and unexplored organic constituents in the earth's atmosphere. Environmental science & technology. 2007 Mar 1;41(5):1514-21.

Goudie, A.S., 2007. Mega-yardangs: A global analysis. Geography Compass, 1(1), pp.65-81.

Galimov EM. Sources and mechanisms of formation of gaseous hydrocarbons in sedimentary rocks. Chemical Geology. 1988 Dec 15;71(1-3):77-95.

Garasic M. Crveno jezero-the biggest sinkhole in Dinaric Karst (Croatia). InAbbasi, A. et Giesen, N., éditeurs: EGU General Assembly Conference Abstracts 2012 Apr (Vol. 14, p. 7132).

Garasic, M. 2021. The Dinaric Karst System in Croatia. In The Dinaric Karst System of Croatia: Speleology and Cave Exploration (pp. 25-45). Springer International Publishing.

Grant JA, Schultz PH. Degradation of selected terrestrial and Martian impact craters. Journal of Geophysical Research: Planets. 1993 Jun 25;98(E6):11025-42.

Grant, J.A., Koeberl, C., Reimold, W.U. and Schultz, P.H., 1997. Gradation of the Roter Kamm impact crater, Namibia. Journal of Geophysical Research: Planets, 102(E7), pp.16327-16338.

Graves SDB, McKay CP, Griffith CA, et al. 2008. Rain and hail can reach the surface of Titan. Planetary and Space Science, 56, 346-357.

Greeley R, editor. Aeolian features of southern California. 1978.

Greeley R, Theilig E, Christensen P. The Mauna Loa sulfur flow as an analog to secondary sulfur flows (?) on Io. Icarus. 1984 Oct 1;60(1):189-99.

Greenberger RN, Ehlmann BL, Jewell PW, Birgenheier LP, Green RO. Detection of organic rich oil shales of the Green River Formation, Utah, with ground-based imaging spectroscopy. WHISPERS 8th Workshop on Hyperspectral Image and Signal Processing: Evolution in Remote Sensing. 2016. https://doi.org/10.1109/WHISPERS.2016.8071807.

Greenlaw ME, Gromack AG, Basquill S, MacKinnon DS, Lynds JA, Taylor RB, Utting DJ, Hackett JR, Grant J, Forbes DL, Savoie F. A physiographic coastline classification of the Scotian Shelf bioregion and environs: the Nova Scotia coastline and the New Brunswick Fundy Shore. Fisheries and Oceans Canada, Science; 2013.

Grieve RA. Astronaut's guide to terrestrial impact craters. Lunar and Planetary Institute; 1988.

Griffith CA, Penteado P, Rannou P, et al. Evidence for a polar ethane cloud on Titan. Science. 2006 Sep 15;313(5793):1620-2.

Grindrod PM, Fortes AD, Nimmo F, Feltham DL, Brodholt JP, Vočadlo L. The long-term stability of a possible aqueous ammonium sulfate ocean inside Titan. Icarus. 2008 Sep 1;197(1):137-51.

Grodek T, Morin E, Helman D, Lensky I, Dahan O, Seely M, Benito G, Enzel Y. Eco-hydrology and geomorphology of the largest floods along the hyperarid Kuiseb River, Namibia. Journal of Hydrology. 2020 Mar 1;582:124450.

Gronoff G, Lilensten J, Desorgher L, Flückiger E. Ionization processes in the atmosphere of Titan - I. Ionization in the whole atmosphere. Astronomy & Astrophysics. 2009;506(2);955-964.

Groves, C.G. Geochemical and Kinetic Evolution of a Karst Flow System: Laurel Creek, West Virginia. Ground Water 1992 March-April, Vol. 30, No. 2, p. 186-191.

Gu G, Adler RF. Spatial patterns of global precipitation change and variability during 1901–2010. Journal of climate. 2015 Jun 1;28(11):4431-53.

Gu H, Cui J, Niu DD, Wellbrock A, Tseng WL, Xu XJ. Monte Carlo calculations of the atmospheric sputtering yields on Titan. Astronomy & Astrophysics. 2019 Mar 1;623:A18.

Guendelman I, Waugh DW, Kaspi Y. Dynamical regimes of polar vortices on terrestrial planets with a seasonal cycle. The planetary science journal. 2022 Apr 26;3(4):94.

Guisado-Pintado E, Jackson DW, Rogers D. 3D mapping efficacy of a drone and terrestrial laser scanner over a temperate beach-dune zone. Geomorphology. 2019 Mar 1;328:157-72.

Gurov EP, Koeberl C, Yamnichenko A. El'gygytgyn impact crater, Russia: Structure, tectonics, and morphology. Meteoritics & Planetary Science. 2007 Mar;42(3):307-19.

Halliday AN. A young Moon-forming giant impact at 70–110 million years accompanied by late-stage mixing, core formation and degassing of the Earth. Philosophical Transactions of the Royal Society A: Mathematical, Physical and Engineering Sciences. 2008 Nov 28;366(1883):4163-81.

Hanel R, Conrath B, Flasar FM, et al. Infrared observations of the Saturnian system from Voyager 1. Science. 1981 Apr 10;212(4491):192-200.

Harms, J.E., Milton, D.J., Ferguson, J., Gilbert, D.J., Harris, W.K. and Goleby, B., 1980. Goat Paddock cryptoexplosion crater, Western Australia. Nature, 286(5774), pp.704-706.

Harrison, K. P., Thermokarst Processes in Titan's Lakes: Comparison with Terrestrial Data, 43th LPSC, 2271-2272, 2012.

Harrison JP, Gheeraert N, Tsigelnitskiy D, Cockell CS. The limits for life under multiple extremes. Trends in microbiology. 2013 Apr 1;21(4):204-12.

Haryono E, Day M. Landform differentiation within the Gunung Kidul Kegelkarst, Java, Indonesia. Journal of Cave and Karst Studies. 2004 Aug 1;66(2):62-9.

Hausrath EM, Treiman AH, Vicenzi E, et al. Short-and long-term olivine weathering in Svalbard: Implications for Mars. Astrobiology. 2008 Dec 1;8(6):1079-1092.

Hay HC, Hewitt I, Katz RF. Tidal forcing in icy-satellite oceans drives mean circulation and ice-shell torques. Journal of Geophysical Research: Planets. 2024 Jun;129(6):e2024JE008408.

Hayes AG, Aharonson O, Callahan P, et al. Hydrocarbon lakes on Titan: Distribution and interaction with a porous regolith. Geophysical Research Letters. 2008 May;35(9).

Hayes AG, Aharonson O, Lunine JI, Kirk RL, Zebker HA, Wye LC, Lorenz RD, Turtle EP, Paillou P, Mitri G, Wall SD. Transient surface liquid in Titan's polar regions from Cassini. Icarus. 2011 Jan 1;211(1):655-71.

Hayes AG, Lorenz RD, Donelan MA, et al. Wind driven capillary-gravity waves on Titan's lakes: Hard to detect or non-existent? Icarus. 2013; 225;403-412.

Hayes AG. The lakes and seas of Titan. Annual Review of Earth and Planetary Sciences. 2016 Jun 29;44:57-83.

Hayes AG, Birch SP, Dietrich WE, et al. Topographic constraints on the evolution and connectivity of Titan's lacustrine basins. Geophysical Research Letters. 2017 Dec 16;44(23):11-745.

Hayes AG, Lorenz RD, Lunine JI. A post-Cassini view of Titan's methane-based hydrologic cycle. Nature Geoscience. 2018 May;11(5):306-13.

Hays LE, Graham HV, Des Marais DJ, et al. Biosignature preservation and detection in Mars analog environments. Astrobiology. 2017 Apr 1;17(4):363-400.

Hedgepeth JE, Neish CD, Turtle EP, et al. Titan's impact crater population after Cassini. Icarus. 2020 Jul 1;344:113664.

Heldmann JL, Pollard W, McKay CP, et al. The high elevation Dry Valleys in Antarctica as analog sites for subsurface ice on Mars. Planetary and Space Science. 2013 Sep 1;85:53-8.

Hergarten S, Kenkmann T. Long-term erosion rates as a function of climate derived from the impact crater inventory. Earth Surface Dynamics. 2019 May 23;7(2):459-73.

Heslar M, Barnes JW, Soderblom JM, et al. Tidal Currents Detected in Kraken Mare Straits from Cassini VIMS Sun Glitter Observations. The Planetary Science Journal. 2020 Aug 14;1(2);35.

Hester KC, Brewer PG. Clathrate Hydrates in Nature. Annu. Rev. Mar. Sci. 2009; 1:303–327. http:// 10.1146/annurev.marine.010908.163824.

Higgins PM, Cockell CS. A bioenergetic model to predict habitability, biomass and biosignatures in astrobiology and extreme conditions. Journal of the Royal Society Interface. 2020 Oct 28;17(171):20200588.

Hipkin VJ, Voytek MA, Meyer MA, et al. Analogue sites for Mars missions: NASA's Mars Science Laboratory and beyond–Overview of an international workshop held at The Woodlands, Texas, on March 5–6, 2011. Icarus. 2013 Jun 1;224(2):261-7.

Hirai E, Sekine Y, Zhang N, et al. Rapid Aggregation and Dissolution of Organic Aerosols in Liquid Methane on Titan. Geophysical Research Letters. 2023;50(12);e2023GL103015.

Hodges KV, Schmitt HH. A new paradigm for advanced planetary field geology developed through analog experiments on Earth. In "Analogs for planetary exploration”, W. Brent Garry (editor) and Jacob E. Bleacher (editor). Special Paper - Geological Society of America (2011) 483: 17-31

Hofgartner JD, Hayes AG, Lunine JI, et al. Transient features in a Titan sea. Nature Geoscience. 2014 June 22;7;493-496.

Hofgartner JD, Hayes AG, Lunine JI, et al. Titan’s “Magic Islands”: Transient features in a hydrocarbon sea. Icarus. 2016 June;271(6);338-349.

Holton JR. An Introduction to Dynamic Meteorology. 2004. Academic Press, San Diego.

Hood KC, Wenger LM, Gross OP, Harrison SC. Hydrocarbon systems analysis of the northern Gulf of Mexico: Delineation of hydrocarbon migration pathways using seeps and seismic imaging. AAPG Studies in Geology. 2002; 48: 25-40.

Horikoshi K. Barophiles: deep-sea microorganisms adapted to an extreme environment. Current opinion in microbiology. 1998 Jun 1;1(3):291-5.

Horowitz NH, Cameron RE, Hubbard JS. Microbiology of the Dry Valleys of Antarctica: Studies in the world's coldest and driest desert have implications for the Mars biological program. Science. 1972 Apr 21;176(4032):242-245.

Hörst SM. Titan's atmosphere and climate. Journal of Geophysical Research: Planets. 2017 Mar;122(3):432-82.

Horvath DG, Andrews-Hanna JC, Newman CE et al. The influence of subsurface flow on lake formation and north polar lake distribution on Titan. Icarus. 2016 Oct;277;103-124.

Houze Jr RA. Types of clouds in earth's atmosphere. InInternational Geophysics 2014 Jan 1 (Vol. 104, pp. 3-23). Academic Press.

Hsu JK, Ip WH. Estimates of the Atmospheric Escape Rates of CH4 from Titan. The Astrophysical Journal. 2019 Jun 6;878(1):3.

Hunten, D.M., 1977, Titan's atmosphere and surface, in Planetary Satellites, J.A. Burns, (Ed.), Univ. of Arizona Press, p. 420.

Hurlbert SH, Berry RW, Lopez M, Pezzani S. Lago Verde and Lago Flaco: Gypsum-bound lakes of the Chilean altiplano 1. Limnology and Oceanography. 1976 Sep;21(5):637-45.

Idini B, Nimmo F. Resonant Stratification in Titan's Global Ocean. Planetary Science Journal. 2024 Jan; 5:15.

Iess L, Rappaport NJ, Jacobson RA et al.. Gravity field, shape, and moment of inertia of Titan. Science. 2010 Mar 12;327(5971):1367-9.

Iess L, Jacobson RA, Ducci M, et al. The tides of Titan. Science. 2012 Jul 27;337(6093):457-459.

Imamura T, Mitchell J, Lebonnois S et al. Superrotation in planetary atmospheres. Space Science Reviews. 2020 Aug;216:1-41.

Isupova MV, Dolgopolova EN. Hydrological–morphological processes in the mouth area of the Congo R. and the effect of submarine canyon on these processes. Water Resources. 2016 Jul;43(4):594-610.

Jackson, M. (1997). Processes of laccolithic emplacement in the southern Henry Mountains, southeastern Utah. In Laccolith Complexes of Southeastern Utah: Time of Emplacement and Tectonic Setting: Workshop Proceedings (Vol. 2158, pp. 51-59). US Geological Survey Bulletin.

Jackson MD, Pollard DD. The laccolith-stock controversy: New results from the southern Henry Mountains, Utah. Geological Society of America Bulletin. 1988 Jan 1;100(1):117-39.

Jackson, M. D., & Pollard, D. D. 1990. Flexure and faulting of sedimentary host rocks during growth of igneous domes, Henry Mountains, Utah. Journal of Structural Geology, 12(2), 185-206.

Jacobson, R.A., 2022. The orbits of the main Saturnian satellites, the Saturnian system gravity field, and the orientation of Saturn's pole. *The Astronomical Journal*, *164*(5), p.199.

Jadwiga Krzyszowska A. Human impact on tundra environment at the Ny-Ålesund Station, Svalbard. Polar Research. 1989 Jan 12;7(2):119-31.

Jamieson AJ, Boorman B, Jones DO. Deep-sea benthic sampling. Methods for the study of marine benthos. 2013 May 14:285-347.

Jannasch HW. Studies on planktonic bacteria by means of a direct membrane filter method. Microbiology. 1958 Jun;18(3):609-20.

Jaumann R, Kirk RL, Lorenz RD, Lopes RM, Stofan E, Turtle EP, Keller HU, Wood CA, Sotin C, Soderblom LA, Tomasko MG. Geology and surface processes on Titan. Titan from Cassini-Huygens. 2010:75-140.

Jennings DE, Flasar FM, Kunde VG, et al. Titan's surface brightness temperatures. The Astrophysical Journal. 2009 Jan 12;691(2):L103.

Jennings DE, Cottini V, Nixon CA, et al. Seasonal changes in Titan's surface temperatures. The Astrophysical Journal Letters. 2011 Jul 21;737(1):L15.

Jennings DE, Cottini V, Nixon CA, et al. Surface temperatures on Titan during northern winter and spring. The Astrophysical Journal Letters. 2016 Jan 4;816(1):L17.

Jennings DE, Tokano T, Cottini V, et al. Titan surface temperatures during the Cassini mission. The Astrophysical Journal Letters. 2019 May 20;877(1):L8.

Jensen FS, Varnes HD. Geology of the Fort Peck area, Garfield, McCone and Valley Counties, Montana. US Government Printing Office; 1964.

Johnson, T.V. and McGetchin, T.R., 1973. Topography on satellite surfaces and the shape of asteroids. Icarus, 18(4), pp.612-620.

Johnson RE. Plasma-induced sputtering of an atmosphere. Space Science Reviews. 1994 Aug;69(3):215-53.

Johnson RE. Sputtering and heating of Titan's upper atmosphere. Philosophical Transactions of the Royal Society A: Mathematical, Physical and Engineering Sciences. 2009 Feb 28;367(1889):753-71.

Johnson RE, Tucker OJ, Volkov AN. Evolution of an early Titan atmosphere. Icarus. 2016 Jun 1;271:202-6.

Jolly VH. The comparative limnology of some New Zealand lakes: 1. Physical and chemical. New Zealand journal of marine and freshwater research. 1968 Jun 1;2(2):214-59.

Jones, W. B., Bacon, M., & Hastings, D. A. (1981). The Lake Bosumtwi impact crater. Geological Society of America Bulletin, 92(6), 342–349.

Journaux B, Kalousová K, Sotin C, et al. Large ocean worlds with high-pressure ices. Space Science Reviews. 2020 Feb;216(1):1-36.

Joye SB. The geology and biogeochemistry of hydrocarbon seeps. Annual Review of Earth and Planetary Sciences. 2020 May 30;48:205-31.

Kalousová K, Sotin C. Melting in high-pressure ice layers of large ocean worlds—implications for volatiles transport. Geophysical Research Letters. 2018 Aug 28;45(16):8096-103.

Kalousová K, Sotin C. The Insulating Effect of Methane Clathrate Crust on Titan's Thermal Evolution. Geoph. Res. Let. 2020 Jun 9; 47(13): 1-9. https://doi.org/10.1029/2020GL087481.

Kalousová K, Wakita S, Sotin C, Neish CD, Soderblom JM, Souček O, Johnson BC. Evolution of impact melt pools on Titan. Journal of Geophysical Research: Planets. 2024 Mar;129(3):e2023JE008107.

Kamata S, Nimmo, F, sekine, Y, Kiyoshi, K, Noguchi, N, Kimura, J, Tani, A. Pluto's ocean is capped by gas hydrate. Nature Geoscience. 2019 May 20; 12(6): 407-410.

Khang, P. The development of karst landscapes in Vietnam. Acta Geologica Polonica. 1985; 35(3-4), 305-324.

Kasting JF. Earth's early atmosphere. Science. 1993 Feb 12;259(5097):920-6.

Kawai J, Jagota S, Kaneko T, et al. Self-assembly of tholins in environments simulating Titan liquidospheres: implications for formation of primitive coacervates on Titan. Astrobiology. 2013;12(4); 282-291.

Kelley DS, Karson JA, Fruh-Green GL, et al. A serpentinite-hosted ecosystem: the Lost City hydrothermal field. Science. 2005 Mar 4;307(5714):1428-1434.

Kennicutt MC, Klein A, Montagna P, Sweet S, Wade T, Palmer T, Sericano J, Denoux G. Temporal and spatial patterns of anthropogenic disturbance at McMurdo Station, Antarctica. Environmental Research Letters. 2010 Sep 16;5(3):034010.

Killingbeck SF, Rutishauser A, Unsworth MJ, Dubnick A, Criscitiello AS, Killingbeck J, Dow CF, Hill T, Booth AD, Main B, Brossier E. Misidentified subglacial lake beneath the Devon Ice Cap, Canadian Arctic: a new interpretation from seismic and electromagnetic data. The Cryosphere. 2024 Aug 20;18(8):3699-722.

Kirk, H.A., Turtle, E.P., Neish, C.D., Stofan, E.R. and Barnes10, J.W., A Global Topographic Map of Titan 4.

Klassen RA. A preliminary interpretation of glacial history derived from glacial striations, central Newfoundland. Current Research, Part D. Geological Survey of Canada, Paper. 1994:117-43.

Koeberl C, Shirey SB. Re–Os isotope systematics as a diagnostic tool for the study of impact craters and distal ejecta. Palaeogeography, Palaeoclimatology, Palaeoecology. 1997 Aug 1;132(1-4):25-46.

Koeberl C. Mineralogical and geochemical aspects of impact craters. Mineralogical Magazine. 2002 Oct;66(5):745-68.

Klein AG, Sweet ST, Wade TL, Sericano JL, Kennicutt MC. Spatial patterns of total petroleum hydrocarbons in the terrestrial environment at McMurdo Station, Antarctica. Antarctic Science. 2012 Oct;24(5):450-66.

Korycansky, D.G. and Zahnle, K.J., 2005. Modeling crater populations on Venus and Titan. Planetary and Space Science, 53(7), pp.695-710.

Korycansky DG, Zahnle KJ. Titan impacts and escape. Icarus. 2011 Jan 1;211(1):707-21.

Koskinen TT, Yelle RV, Snowden DS, et al. The mesosphere and lower thermosphere of Titan revealed by Cassini/UVIS stellar occultations. Icarus. 2011 Dec 1;216(2):507-34.

Krasnopolsky VA. A photochemical model of Titan's atmosphere and ionosphere. Icarus. 2009 May 1;201(1):226-56.

Kräuchi A, Philipona R, Romanens G, Hurst DF, Hall EG, Jordan AF. Controlled weather balloon ascents and descents for atmospheric research and climate monitoring. Atmospheric measurement techniques. 2016 Mar 7;9(3):929-38.

Kujawinski EB, Reddy CM, Rodgers RP, et al. The first decade of scientific insights from the Deepwater Horizon oil release. Nature Reviews Earth & Environment. 2020 May;1(5):237-50.

Kuminov, A.A., Yushkov, V.A., Gvozdev, Y.N. et al. Meteorological Rocket Sounding for Atmospheric Research and Geophysical Monitoring. Russ. Meteorol. Hydrol. 46, 571–578 (2021). https://doi.org/10.3103/S1068373921090028.

Kvorka J, Čadek O. A numerical model of convective heat transfer in Titan’s subsurface ocean. Icarus. 2022 Apr 1;376:114853.

Laity, J.E., Malin, M.C. Sapping processes and the development of theater-headed valley networks in the Colorado Plateau. Geological Society of America Bulletin, 1985 February, V96, p. 203-217.

Lammer H, Bauer SJ. Nonthermal atmospheric escape from Mars and Titan. Journal of Geophysical Research: Space Physics. 1991 Feb 1;96(A2):1819-25.

Lammer H, Bauer SJ. Atmospheric mass loss from Titan by sputtering. Planetary and Space Science. 1993 Sep 1;41(9):657-63.

Lammer H, Stumptner W, Bauer SJ. Dynamic escape of H from Titan as consequence of sputtering induced heating. Planetary and space science. 1998 Oct 1;46(9-10):1207-13.

Lancaster, N. 1994. Geomorphology of Desert Environments, Springer.

Langhans M, Lunine JI, Mitri G. Titan's Xanadu region: Geomorphology and formation scenario. Icarus. 2013 Apr 1;223(2):796-803.

Lapôtre MG, Ielpi A, Lamb MP, Williams RM, Knoll AH. Model for the formation of single-thread rivers in barren landscapes and implications for pre-Silurian and Martian fluvial deposits. Journal of Geophysical Research: Earth Surface. 2019 Dec;124(12):2757-77.

Laraque A, Guyot JL, Filizola N. Mixing processes in the Amazon River at the confluences of the Negro and Solimões Rivers, Encontro das Águas, Manaus, Brazil. Hydrological Processes: An International Journal. 2009 Oct 30;23(22):3131-40.

Lebreton JP, Matson DL. The Huygens probe: science, payload and mission overview. Space Science Reviews. 2002 Jul;104(1-4):59-100.

Lee P, Osinski GR. The Haughton-Mars Project: Overview of science investigations at the Haughton impact structure and surrounding terrains, and relevance to planetary studies. Meteoritics & Planetary Science Archives. 2005 Jan 1;40(12):1755-8.

Lees, T. and O’Donohue, D., 2024. Formation of the Lawn Hill Impact Structure. Australian Journal of Earth Sciences, 71(3), pp.307-318.

Le Gall A, Malaska M, Lorenz R, et al. Composition, seasonal change, and bathymetry of Ligeia Mare, Titan, derived from its microwave thermal emission. Journal of Geophysical Research: Planets. 2016 Jan 20;121(2);233-251.

LeGrand HE. Hydrological and Ecological Problems of Karst Regions: Hydrological actions on limestone regions cause distinctive ecological problems. Science. 1973 Mar 2;179(4076):859-64.

Legler DM, Freeland HJ, Lumpkin R, Ball G, McPhaden MJ, North S, Crowley R, Goni GJ, Send U, Merrifield MA. The current status of the real-time in situ Global Ocean Observing System for operational oceanography. Journal of Operational Oceanography. 2015 Aug 18;8(sup2):s189-200.

Leitner MA, Lunine JI. Modeling early Titan’s ocean composition. Icarus, 2019, 333, 61-70.

Lelekakis N, Martin D, Guo W, Wijaya J. Comparison of dissolved gas-in-oil analysis methods using a dissolved gas-in-oil standard. IEEE Electrical Insulation Magazine. 2011 Sep 22;27(5):29-35.

Lellouch E, Hunten DM, Kockarts G, Coustenis A. Titan's thermosphere profile. Icarus. 1990 Feb 1;83(2):308-24.

Lellouch E, Gurwell MA, Moreno R, et al. An intense thermospheric jet on Titan. Nature Astronomy. 2019 Jul;3(7):614-9.

Le Mouélic, S.; P. Paillou, M. A. Janssen, et al., Mapping and interpretation of Sinlap crater on Titan using Cassini VIMS and RADAR data, J. Geophys. Res., 113, E04003, 2008

Le Mouélic, S., P. Rannou, S. Rodriguez, et al., Dissipation of Titan's north polar cloud at northern spring equinox., Planet. Space Science, 60, pp. 86-92, doi 10.1016/j.pss.2011.04.006, 2012.

Le Mouélic, S., Rodriguez, S., Robidel, et al., Mapping polar atmospheric features with VIMS: from the dissipation of the northern cloud to the onset of the southern polar vortex, Icarus, 311, pp 371–383, 2018

Le Mouélic, S., Cornet, T., Rodriguez, et al., The Cassini VIMS archive of Titan: from browse products to global infrared color maps, Icarus, 319, 121-132, 2019

Le Pichon X. Sea-floor spreading and continental drift. Journal of geophysical research. 1968 Jun 15;73(12):3661-97.

Li J, Grenfell MC, Wei H, Tooth S, Ngiem S. Chute cutoff-driven abandonment and sedimentation of meander bends along a fine-grained, non-vegetated, ephemeral river on the Bolivian Altiplano. Geomorphology. 2020 Feb 1;350:106917.

Liu B, Coulthard TJ. Modelling the interaction of aeolian and fluvial processes with a combined cellular model of sand dunes and river systems. Computers & Geosciences. 2017 Sep 1;106:1-9.

Liu ZY, Radebaugh J, Harris RA, Christiansen EH, Neish CD, Kirk RL, Lorenz RD, Cassini RADAR Team. The tectonics of Titan: Global structural mapping from Cassini RADAR. Icarus. 2016a May 15;270:14-29.

Liu ZY, Radebaugh J, Harris RA, Christiansen EH, Rupper S. Role of fluids in the tectonic evolution of Titan. icarus. 2016b May 15;270:2-13.

Liu R, Wang L, Wei Y, et al. The hadal biosphere: Recent insights and new directions, Deep Sea Research Part II: Topical Studies in Oceanography, 2018 155: Pages 11-18.

Locker SD, Hine AC. An overview of the geologic origins of hydrocarbons and production trends in the Gulf of Mexico. Scenarios and Responses to Future Deep Oil Spills: Fighting the Next War. 2019 Jul 5:60-74.

Lofgren GE, Horz F, Eppler D. Geologic field training of the Apollo astronauts and implications for future manned exploration.

Lopes RM, Kamp LW, Smythe WD, et al. Lava lakes on Io: Observations of Io's volcanic activity from Galileo NIMS during the 2001 flybys. Icarus. 2004 May 1;169(1):140-74.

Lopes RM, Mitchell KL, Wall SD, et al. The lakes and seas of Titan. EOS, Transactions American Geophysical Union. 2007 Dec 18;88(51):569-70.

Lopes RM, Stofan ER, Peckyno R, Radebaugh J, Mitchell KL, Mitri G, Wood CA, Kirk RL, Wall SD, Lunine JI, Hayes A. Distribution and interplay of geologic processes on Titan from Cassini radar data. Icarus. 2010 Feb 1;205(2):540-58.

Lopes, R.M.C., M. J. Malaska, A. Solomonidou, A. Le Gall, M. A. Janssen, C.D. Neish, E.P. Turtle, S.P.D. Birch, A. G. Hayes, J. Radebaugh, A. Coustenis, B. Stiles, R. Kirk, K.L. Mitchell, K. Lawrence and the Cassini RADAR Team (2016). Nature, Distribution, and Origin of Titan’s Undifferentiated Plains. Icarus, pp. 162-182.

Lopes RM, Gregg TK, Harris A, Radebaugh J, Byrne P, Kerber L, Mouginis-Mark P. Extraterrestrial lava lakes. Journal of Volcanology and Geothermal Research. 2018 Oct 15;366:74-95.

Lopes RM, Wall SD, Elachi C, et al. Titan as revealed by the Cassini radar. Space Science Reviews. 2019 Jun; 215(4):1-50.

Lopes RM, Malaska MJ, Schoenfeld AM, et al. A global geomorphologic map of Saturn's moon Titan. Nature astronomy. 2020 Mar;4(3):228-233.

Lopez-Mir, B., Schneider, S., & Hülse, P. 2018. Fault activity and diapirism in the Mississippian to Late Cretaceous Sverdrup Basin: New insights into the tectonic evolution of the Canadian Arctic. Journal of Geodynamics, 118, 55-65.

Lora JM. Moisture transport and the methane cycle of Titan's lower atmosphere. Icarus. 2024 November, 422; 116241.

Lorenz RD,Le Gall A. Schumann resonance on Titan: a critical Re-assessment. Icarus.2020; 351:113942

Lorenz RD, Lunine JI. Erosion on Titan: Past and present. Icarus. 1996 Jul 1;122(1):79-91.

Lorenz RD, McKay CP, Lunine JI. Photochemically-Driven Collapse of Titan's Atmosphere. Science. 1997; 275: 642-644

Lorenz RD, McKay CP, Lunine JI. Analytic Stability of Titan's Climate : Sensitivity to Volatile Inventory. Planetary and Space Science. 1999; 47: 1503-1515.

Lorenz RD, Wall S, Radebaugh J, et al. The sand seas of Titan: Cassini RADAR observations of longitudinal dunes. Science. 2006 May 5;312(5774):724-727.

Lorenz RD, Wood CA, Lunine JI, et al. Titan's young surface: Initial impact crater survey by Cassini RADAR and model comparison. Geophysical Research Letters. 2007 Apr;34(7).

Lorenz RD, Lopes RM, Paganelli F, et al. Fluvial channels on Titan: initial Cassini RADAR observations. Planetary and Space Science. 2008a Jun 1;56(8):1132-44.

Lorenz RD, Mitchell KL, Kirk RL, et al. Titan's inventory of organic surface materials. Geophysical Research Letters. 2008b Jan;35(2).

Lorenz, R.D. and Radebaugh, J., 2009. Global pattern of Titan's dunes: Radar survey from the Cassini prime mission. Geophysical Research Letters, 36(3).

Lorenz RD, Jackson B, Hayes A. Racetrack and Bonnie Claire: southwestern US playa lakes as analogs for Ontario Lacus, Titan. Planetary and Space Science. 2010 Mar 1;58(4):724-31.

Lorenz RD, Gleeson D, Prieto-Ballesteros O, et al. Analog environments for a Europa lander mission. Advances in Space Research. 2011 Aug 16;48(4):689-696.

Lorenz RD, Stiles BW, Aharonson O, Lucas A, Hayes AG, Kirk RL, Zebker HA, Turtle EP, Neish CD, Stofan ER, Barnes JW. A global topographic map of Titan. Icarus. 2013 Jul 1;225(1):367-77.

Lorenz RD. The Challenging Depths of Titan's Seas. Journal of Geophysical Research: Planets. 2021 Apr;126(4);e2020JE006786.

Lorenz RD, et al. Selection and characteristics of the Dragonfly landing site near Selk crater, Titan. Planetary Science Journal. 2021;2(24):13.

Lowenstein TK, Hardie LA. Criteria for the recognition of salt-pan evaporites. Sedimentology. 1985 Oct;32(5):627-44.

Lucas, A., Rodriguez, S., Narteau, C., Charnay, B., du Pont, S.C., Tokano, T., Garcia, A., Thiriet, M., Hayes, A.G., Lorenz, R.D. and Aharonson, O., 2014. Growth mechanisms and dune orientation on Titan. Geophysical Research Letters, 41(17), pp.6093-6100.

Lunine JI, Stevenson DJ, Yung YL. Ethane ocean on Titan. Science. 1983 Dec 16;222(4629):1229-30.

Lunine JI, Atreya SK. The methane cycle on Titan. Nature Geoscience. 2008 Mar;1(3):159-64.

Lunine JI, Stevenson DJ. Clathrate and Ammonia Hydrates at High Pressure: Application to the Origin of Methane on Titan. Icarus. 1987 Apr; 70(1): 61-77.

MacKenzie, S. M., Barnes, J. W., Sotin, C., Soderblom, J. M., Le Mouélic, S., Rodriguez, S., ... & McCord, T. B. (2014). Evidence of Titan’s climate history from evaporite distribution. Icarus, 243, 191-207.

MacKenzie, S. M., & Barnes, J. W. (2016). Compositional similarities and distinctions between Titan’s evaporitic terrains. The Astrophysical Journal, 821(1), 17.

MacKenzie SM, Barnes JW, Hofgartner JD, Birch SP, Hedman MM, Lucas A, Rodriguez S, Turtle EP, Sotin C. The case for seasonal surface changes at Titan’s lake district. Nature Astronomy. 2019 Jun;3(6):506-10.

Mader HM, Pettitt ME, Wadham JL, Wolff EW, Parkes RJ. Subsurface ice as a microbial habitat. Geology. 2006 Mar 1;34(3):169-72.

Maghsoudi, M. 2021. Other Desert Landforms. In: Desert Landscapes and Landforms of Iran. Geography of the Physical Environment. Springer, Cham.

Maguire R, Schmerr N, Pettit E, et al. Geophysical constraints on the properties of a subglacial lake in northwest Greenland. The Cryosphere. 2021 Jul 16;15(7):3279-91.

Malaska M, Radebaugh J, Mitchell K, Lopes R, Wall S, Lorenz R. Surface dissolution model for Titan karst. InFirst International Planetary Caves Workshop: Implications for Astrobiology, Climate, Detection, and Exploration 2011a Oct (Vol. 1640, p. 15).

Malaska, M., Radebaugh, J., Le Gall, A., et al. 2011b, in Lunar and Planetary Inst. Technical Report, Vol. 42, Lunar and Planetary Science Conference, 1562

Malaska MJ, Hodyss R. Dissolution of benzene, naphthalene, and biphenyl in a simulated Titan lake. Icarus. 2014 Nov 1;242:74-81.

Malaska, M. J., Lopes, R. M., Hayes, A. G., Radebaugh, J., Lorenz, R. D., & Turtle, E. P. (2016). Material transport map of Titan: The fate of dunes. Icarus, 270, 183-196.

Malaska M. J, Hodyss R, Lunine JI, et al. Laboratory measurements of nitrogen dissolution in Titan lake fluids. Icarus. 2017 Jun 1;289:94-105.

Malaska, M.J., Schuremeier, L., Lopes, R.M.C., Schoenfeld, A.M., Solomonidou, A., Le Gall, A., Radebaugh, J. (2019) EPSC Abstracts, Vol. 13, Abstract EPSC-DPS2019-1082-1.

Malaska, M.J., Jani Radebaugh, Rosaly M.C. Lopes, et al. Labyrinth Terrain on Titan. Icarus, 344 (2020) 113764.

Malaska, M. J., Schoenfeld, A., Wynne, J. J., Mitchell, K. L., White, O., Howard, A., ... & Umurhan, O. (2022). Potential caves: Inventory of subsurface access points on the surface of Titan. Journal of Geophysical Research: Planets, 127(11), e2022JE007512.

Mangold N, Adeli S, Conway S, Ansan V, Langlais B. A chronology of early Mars climatic evolution from impact crater degradation. Journal of Geophysical Research: Planets. 2012 Apr;117(E4).

Marchant DR, Head III JW. Antarctic dry valleys: Microclimate zonation, variable geomorphic processes, and implications for assessing climate change on Mars. Icarus. 2007 Dec 1;192(1):187-222.

Martinez I, Agrinier P, Schärer U, Javoy M. A SEM-ATEM and stable isotope study of carbonates from the Haughton impact crater, Canada. Earth and Planetary Science Letters. 1994 Feb 1;121(3-4):559-74.

Martínez-Pérez C, Greening C,Bay SK. Phylogenetically and functionally diverse microorganisms reside under the Ross Ice Shelf. Nature Communications.2022 13: Article 117

Martin S. A field study of brine drainage and oil entrainment in first-year sea ice. Journal of Glaciology. 1979 Jan;22(88):473-502.

Marvin C. The Region of the Eternal Fire: an Account of a Journey to the Petroleum Region of the Caspian in 1883. WH Allen and Company; 1884.

Masaitis VL. Review of the Barringer crater studies and views on the crater's origin. Solar System Research. 2006 Dec;40(6):500-12.

Mastrogiuseppe M, Poggiali V, Hayes A, et al. The bathymetry of a Titan sea. Geophysical Research Letters. 2014 Mar 16;41(5):1432-7.

Mastrogiuseppe M, Hayes AG, Poggiali V, et al. Radar Sounding Using the Cassini Altimeter: Waveform Modeling and Monte Carlo Approach for Data Inversion of Observations of Titan's Seas. IEEE Transactions on Geoscience and Remote Sensing. 2016;54(10);5646-5656.

Mastrogiuseppe M, Hayes AG, Poggiali V, et al. Bathymetry and composition of Titan's Ontario Lacus derived from Monte Carlo-based waveform inversion of Cassini RADAR altimetry data. Icarus. 2018b Jan 15;300;203-209.

Mastrogiuseppe M, Poggiali V, Hayes AG, et al. Cassini radar observation of Punga Mare and environs: Bathymetry and composition. Earth and Planetary Science Letters. 2018a Aug 15;496;89-95, doi: 10.1016/j.epsl.2018.05.033

Mastrogiuseppe, M., Hayes, A., Poggiali, V., et al. 2018b, Icarus, 300, 203, doi: 10.1016/j.icarus.2017.09.009

Mastrogiuseppe M, Poggiali V, Hayes AG, et al. Deep and methane-rich lakes on Titan. Nature Astronomy. 2019 Jun;3(6):535-42.

Masursky H, Batson RM, McCauley JF, Soderblom LA, Wildey RL, Carr MH, Milton DJ, Wilhelms DE, Smith BA, Kirby TB, Robinson JC. Mariner 9 television reconnaissance of Mars and its satellites: Preliminary results. Science. 1972 Jan 21;175(4019):294-305.

Matson DL, Spilker LJ, Lebreton JP. The Cassini/Huygens mission to the Saturnian system. Space Science Reviews. 2002 Jul;104(1-4):1-58.

Matteoni P, Mitri G, Poggiali V, Mastrogiuseppe M. Geomorphological analysis of the southwestern margin of Xanadu, Titan: Insights on tectonics. Journal of Geophysical Research: Planets. 2020 Dec;125(12):e2020JE006407.

Mayer C, Nixon CA. A proposed mechanism for the formation of protocell-like structures on Titan. International Journal of Astrobiology. 2025 Jan;24:e7.

McCain WD Jr. The Properties of Petroleum Fluids. 3rd edition. 2017; PennWell Books 592 p.

McCauley JF, Carr MH, Cutts JA, Hartmann WK, Masursky H, Milton DJ, Sharp RP, Wilhelms DE. Preliminary Mariner 9 report on the geology of Mars. Icarus. 1972 Oct 1;17(2):289-327.

McCormick MP, Steele HM, Hamill P, Chu WP, Swissler TJ. Polar stratospheric cloud sightings by SAM II. Journal of Atmospheric Sciences. 1982 Jun;39(6):1387-97.

McIntosh RJ. Floodplain geomorphology and human occupation of the upper inland delta of the Niger. Geographical Journal. 1983 Jul 1:182-201.

McKay CP, Pollack JB, Courtin R. The greenhouse and antigreenhouse effects on Titan. 1991 Sep 6, Science, 253, 1118-1121.

McKenzie, D.P. and Parker, R.L., 1967. The North Pacific: an example of tectonics on a sphere. *Nature*, *216*(5122), pp.1276-1280.

McLemore VT. Bottomless Lakes State Park. New Mexico Geol. 1999;21:51-5.

Melosh, H.J., 1989. Impact cratering: A geologic process. New York: Oxford University Press; Oxford: Clarendon Press.

Merino N, Aronson HS, Bojanova DP, et al. Living at the extremes: extremophiles and the limits of life in a planetary context. Frontiers in microbiology. 2019 Apr 15;10:780.

Michaelides RJ, Hayes AG, Mastrogiuseppe M, Zebker HA, Farr TG, Malaska MJ, Poggiali V, Mullen JP. Constraining the physical properties of Titan's empty lake basins using nadir and off-nadir Cassini RADAR backscatter. Icarus. 2016 May 15;270:57-66.

Miller KE, Glein CR, Waite JH. Contributions from accreted organics to Titan's atmosphere: new insights from cometary and chondritic data. The Astrophysical Journal. 2019 Jan 22;871(1):59.

Miller JW, Birch SP, Hayes AG, et al. Fluvial Features on Titan and Earth: Lessons from Planform Images in Low-resolution SAR. The Planetary Science Journal. 2021 Jul 30;2(4):142.

Mills G, Fones G. A review of in situ methods and sensors for monitoring the marine environment. Sensor Review. 2012 Jan 20;32(1):17-28.

Mitchell DM, Scott RK, Seviour WJ, et al. Polar vortices in planetary atmospheres. Reviews of Geophysics. 2021 Dec;59(4):e2020RG000723.

Mitchell JL, Ádámkovics M, Caballero R, et al. Locally enhanced precipitation organized by planetary-scale waves on Titan. 2011 Aug 14, Nature Geosciences, 4, 589-592.

Mitchell JL, Lora JM. The climate of Titan. Annual Review of Earth and Planetary Science. 2016 June, 44, 353-380.

Mitri G, Showman AP, Lunine JI, Lorenz RD. Hydrocarbon lakes on Titan. Icarus. 2007;186(2);385-394.

Mitri G, Bland MT, Showman AP, Radebaugh J, Stiles B, Lopes RM, Lunine JI, Pappalardo RT. Mountains on Titan: Modeling and observations. Journal of Geophysical Research: Planets. 2010 Oct;115(E10).

Mitri G, Meriggiola R, Hayes A, et al. Shape, topography, gravity anomalies and tidal deformation of Titan. Icarus. 2014 Jul 1;236:169-77.

Mitri G, Lunine JI, Mastrogiuseppe M, Poggiali V. Possible explosion crater origin of small lake basins with raised rims on Titan. Nature Geoscience. 2019 Oct;12(10);791-796.

Mohammadi M, Raad SMJ, Zirrahi M, Hassanzadeh, H. Subsurface Migration of Methane From Oil Sands Thermal Recovery Operations. Water Resources Research. 2021 Apr 15; 57-5, 1-16. https://doi.org/10.1029/2020WR028745

Monaldi CR, Salfity JA, Kley J. Preserved extensional structures in an inverted Cretaceous rift basin, northwestern Argentina: Outcrop examples and implications for fault reactivation. Tectonics. 2008 Feb;27(1).

Monteleone V, Marin-Moreno H, Bayrakci G, Best A, Shaon F, Hossain MM, Al Karim A, Kurshed Alam MD. Seismic characterization and modelling of the gas hydrate system in the northern Bay of Bengal, offshore Bangladesh. Marine and Petroleum Geology. 2022 Jul; 141: 105690.

Morgan, W.J., 1972. Plate motions and deep mantle convection.

Morgan, J.V., Gulick, S.P., Bralower, T., Chenot, E., Christeson, G., Claeys, P., Cockell, C., Collins, G.S., Coolen, M.J., Ferrière, L. and Gebhardt, C., 2016. The formation of peak rings in large impact craters. Science, 354(6314), pp.878-882.

Mousis O, Choukroun M, Lunine JI, et al. Equilibrium composition between liquid and clathrate reservoirs on Titan. Icarus. 2014 Sep 1;239(9);39-45.

Mousis O, Lunine JI, Hayes AG, Hofgartner JD. The fate of ethane in Titan's hydrocarbon lakes and seas. Icarus. 2016 May 15; 270: 37-40.

Muhleman DO, Grossman AW, Butler BJ, et al.. Radar reflectivity of Titan. Science. 1990 May 25;248(4958):975-80.

Muhleman DO, Grossman AW, Butler BJ. Radar Investigation of Mars, Mercury and Titan. Ann. Rev. Earth and Planet. Sci. 1995; 23: 337-374.

Müller-Wodarg IC, Yelle RV, Mendillo M, et al. The thermosphere of Titan simulated by a global three-dimensional time-dependent model. Journal of Geophysical Research: Space Physics. 2000 Sep 1;105(A9):20833-56.

Müller-Wodarg IC, Yelle RV, Cui J, et al. Horizontal structures and dynamics of Titan's thermosphere. Journal of Geophysical Research: Planets. 2008 Oct;113(E10).

Murray AE, Kenig F, Fritsen CH, et al. Microbial life at− 13 C in the brine of an ice-sealed Antarctic lake. Proceedings of the National Academy of Sciences. 2012 Dec 11;109(50):20626-31.

Navarro-González R, Rainey FA, Molina P, et al. Mars-like soils in the Atacama Desert, Chile, and the dry limit of microbial life. Science. 2003 Nov 7;302(5647):1018-1021.

Nichol SL, Zaitlin BA, Thom BG. The upper Hawkesbury River, New South Wales, Australia: a Holocene example of an estuarine bayhead delta. Sedimentology. 1997 Apr;44(2):263-86.

Neish CD, Lorenz RD, Kirk RL, Wye LC. Radarclinometry of the sand seas of Africa's Namibia and Saturn's moon Titan. Icarus. 2010a Jul 1;208(1):385-94.

Neish, C.D., Somogyi, A. and Smith, M.A., 2010b. Titan's primordial soup: formation of amino acids via low-temperature hydrolysis of tholins. Astrobiology, 10(3), pp.337-347.

Neish CD, Lorenz RD. Titan's global crater population: A new assessment. Planetary and Space Science. 2012 Jan 1;60(1):26-33.

Neish CD, Kirk RL, Lorenz RD, et al. Crater topography on Titan: Implications for landscape evolution. Icarus. 2013 Mar 1;223(1):82-90.

Neish CD, Lorenz RD. Elevation distribution of Titan's craters suggests extensive wetlands. Icarus. 2014 Jan 15;228:27-34.

Neish CD, Barnes JW, Sotin C, et al. Spectral properties of Titan's impact craters imply chemical weathering of its surface. Geophysical Research Letters. 2015 May 28;42(10):3746-54.

Neish CD, Molaro JL, Lora JM, et al. Fluvial erosion as a mechanism for crater modification on Titan. Icarus. 2016 May 15;270:114-29.

Neish, C.D., Lorenz, R.D., Turtle, E.P., Barnes, J.W., Trainer, M.G., Stiles, B., Kirk, R., Hibbitts, C.A. and Malaska, M.J., 2018. Strategies for detecting biological molecules on Titan. Astrobiology, 18(5), pp.571-585.

Neish, C.D., Malaska, M.J., Sotin, C., Lopes, R.M.C., Nixon, C.A., Affholder, A., Chatain, A., Cockell, C., Farnsworth, K.K., Higgins, P.M., Miller, K.E., and Soderlund, K.M.

Organic input to Titan’s subsurface ocean through impact cratering. Astrobiology 24(2), 2024.

Niemann HB, Atreya SK, Bauer SJ et al. The abundances of constituents of Titan's atmosphere from the GCMS instrument on the Huygens probe. Nature. 2005 Dec 8;438(7069):779-84.

Niemann HB, Atreya SK, Demick JE et al. Composition of Titan's lower atmosphere and simple surface volatiles as measured by the Cassini-Huygens probe gas chromatograph mass spectrometer experiment. Journal of Geophysical Research: Planets. 2010 Dec;115(E12).

Nimmo F, Pappalardo RT. Ocean worlds in the outer solar system. Journal of Geophysical Research: Planets. 2016 Aug;121(8):1378-99.

Nixon CA. The composition and chemistry of Titan’s atmosphere. ACS Earth and Space Chemistry. 2024 Feb 29;8(3):406-56.

Nixon CA, Lorenz RD, Achterberg RK, Buch A, Coll P, Clark RN, Courtin R, Hayes A, Iess L, Johnson RE, Lopes RM. Titan's cold case files-Outstanding questions after Cassini-Huygens. Planetary and Space Science. 2018 Jun 1;155:50-72.

Nixon CA, Carrasco N, Sotin C. Open questions and future directions in Titan science. In Titan After Cassini-Huygens 2025 Jan 1 (pp. 473-515). Elsevier.

Northrup, DS. A Geomorphological Study of Yardangs in China, the Altiplano/Puna of Argentina, and Iran as Analogs for Yardangs on Titan. Brigham Young University ProQuest Dissertations & Theses, 2018. 28108150.

Oblad AG, Bunger JW, Hanson FV, Miller JD, Ritzma HR, Seader JD. Tar sand research and development at the University of Utah. Annual review of energy. 1987 Jan 1;12(1):283-356.

O'Keefe JD, Ahrens TJ. Complex craters: Relationship of stratigraphy and rings to impact conditions. Journal of Geophysical Research: Planets. 1999 Nov 25;104(E11):27091-104.

Oleson SR, Lorenz R, Paul M. Titan submarine: exploring the depths of Kraken Mare. InAIAA Space 2015 Conference and Exposition 2015 (p. 4445).

Osinski, G.R., Lee, P., Spray, J.G., Parnell, J., Lim, D.S., Bunch, T.E., Cockell, C.S. and Glass, B., 2005. Geological overview and cratering model for the Haughton impact structure, Devon Island, Canadian High Arctic. Meteoritics & Planetary Science, 40(12), pp.1759-1776.

Osinski GR, Tornabene LL, Grieve RA. Impact ejecta emplacement on terrestrial planets. Earth and Planetary Science Letters. 2011 Oct 15;310(3-4):167-81.

Osinski GR, Bunch TE, Flemming RL, Buitenhuis E, Wittke JH. Impact melt-and projectile-bearing ejecta at Barringer Crater, Arizona. Earth and Planetary Science Letters. 2015 Dec 15;432:283-92.

Osinski GR, Grieve RA, Ferrière L, et al. Impact Earth: A review of the terrestrial impact record. Earth-Science Reviews. 2022 Jul 21:104112.

Paillou P, Seignovert B, Radebaugh J, Wall S. Radar scattering of linear dunes and mega-yardangs: Application to Titan. Icarus. 2016 May 15;270:211-21.

Palermo RV, Ashton AD, Soderblom JM, Birch SP, Hayes AG, Perron JT. Signatures of wave erosion in Titan's coasts. Science Advances. 2024 Jun 19;10(25):eadn4192.

Palmer MY, Cordiner MA, Nixon CA, et al. ALMA detection and astrobiological potential of vinyl cyanide on Titan. Science Advances. 2017 Jul 28;3(7):e1700022.

Paniello RC, Day JM, Moynier F. Zinc isotopic evidence for the origin of the Moon. Nature. 2012 Oct 18;490(7420):376-9.

Pantazidis A, Baziotis I, Solomonidou A, et al. Santorini volcano as a potential Martian analogue: The Balos Cove Basalts. Icarus. 2019 Feb; 325:128–140.

Pardo-Casas F, Molnar P. Relative motion of the Nazca (Farallon) and South American plates since Late Cretaceous time. Tectonics. 1987 Jun;6(3):233-48.

Park E, Latrubesse EM. Surface water types and sediment distribution patterns at the confluence of mega rivers: The Solimões-Amazon and Negro Rivers junction. Water Resources Research. 2015 Aug;51(8):6197-213.

Paton M, Rieber R, Cruz S, Gildner M, Abma C, Abma K, Aghli S, Ambrose E, Archanian A, Bagshaw EA, Baroco C. 2023 EELS field tests at Athabasca Glacier as an icy moon analogue environment. In2024 IEEE Aerospace Conference 2024 Mar 2 (pp. 1-18). IEEE.

Pedersen L, Smith T, Lee SY, Cabrol N. Planetary LakeLander—A robotic sentinel to monitor remote lakes. Journal of Field Robotics. 2014 Sep;32(6):860-79.

Perkins, R.P., Trainer, M.G., Neish, C.D. and Osinski, G.R., 2023. Characterization of Impact Melt at the Haughton Impact Structure: Applications for the Dragonfly Mission to Titan. LPI Contributions, 2806, p.2575.

Peter T. Microphysics and heterogeneous chemistry of polar stratospheric clouds. Annual Review of Physical Chemistry. 1997 Oct;48(1):785-822.

Peters DL, Prowse TD, Pietroniro A, Leconte R. Flood hydrology of the peace-Athabasca Delta, northern Canada. Hydrological Processes: An International Journal. 2006 Dec 15;20(19):4073-96.

Peters KE, Walters, CC, Moldowan JM. 2013. The biomarker guide - Volume 2: Biomarkers and Isotopes in Petroleum Systems and Earth History. Cambridge University press, 1155 p.

Petricca, F., Vance, S. D., Parisi, M., Buccino, D., Gascioli, G., Castillo-Rogez, J., Downey, B. G., Nimmo, F., Tobie, G, Journaux, B., Magnanini, A., Jones, U., Panning, M., Bagheri, A., Genova, A., Lunine, J. I., 2025. Titan’s strong tidal dissipation precludes a subsurface ocean. Nature 648, 556-561.

Poag CW, Powars DS, Poppe LJ, Mixon RB. Meteoroid mayhem in Ole Virginny: Source of the North American tektite strewn field. Geology. 1994 Aug 1;22(8):691-4.

Pilkington M, Grieve RA. The geophysical signature of terrestrial impact craters. Reviews of Geophysics. 1992 May;30(2):161-81.

Poelchau MH, Kenkmann T, Kring DA. Rim uplift and crater shape in Meteor Crater: Effects of target heterogeneities and trajectory obliquity. Journal of Geophysical Research: Planets. 2009 Jan;114(E1).

Poggiali V, Mastrogiuseppe M, Hayes AG, et al. Liquid-filled canyons on Titan. Geophysical Research Letters. 2016 Aug 16;43(15):7887-94.

Poggiali, V.,, Hayes, A. G.,, Mastrogiuseppe, M.,, et al. 2020,The Bathymetry of Moray Sinus at Titan's Kraken Mare. Journal of Geophysical Research: Planets, . 2020 Nov 12;125, (12);e2020JE006558, doi: 10.1029/2020JE006558.

Poggiali V, Brighi G, Hayes AG, Nicholson PD, MacKenzie S, Lalich DE, Bonnefoy LE, Oudrhiri K, Lorenz RD, Soderblom JM, Tortora P. Surface properties of the seas of Titan as revealed by Cassini mission bistatic radar experiments. Nature Communications. 2024 Jul 16;15(1):5454.

Porcelli D, Ballentine CJ. Models for distribution of terrestrial noble gases and evolution of the atmosphere. Reviews in Mineralogy and Geochemistry. 2002 Jan 1;47(1):411-80.

Porco CC, Baker E, Barbara J, et al. Imaging of Titan from the Cassini spacecraft. Nature. 2005 Mar 10;434(7030):159-68.

Preston LJ, Dartnell LR. Planetary habitability: lessons learned from terrestrial analogues. International Journal of Astrobiology. 2014 Jan;13(1):81-98.

Price PB. Microbial life in glacial ice and implications for a cold origin of life. FEMS Microbiology Ecology. 2007 Feb 1;59(2):217-31.

Radebaugh J, Lorenz RD, Kirk RL, Lunine JI, Stofan ER, Lopes RM, Wall SD, Cassini Radar Team. Mountains on Titan observed by Cassini RADAR. Icarus. 2007 Dec 1;192(1):77-91.

Radebaugh J, Lorenz RD, Lunine JI, Wall SD, Boubin G, Reffet E, Kirk RL, Lopes RM, Stofan ER, Soderblom L, Allison M. Dunes on Titan observed by Cassini RADAR. Icarus. 2008 Apr 1;194(2):690-703.

Radebaugh J, Lorenz R, Farr T, et al. Linear dunes on Titan and earth: Initial remote sensing comparisons. Geomorphology. 2010 Sep 1;121(1-2):122-132.

Radebaugh J, Lorenz RD, Wall SD, Kirk RL, Wood CA, Lunine JI, Stofan ER, Lopes RM, Valora P, Farr TG, Hayes A. Regional geomorphology and history of Titan's Xanadu province. Icarus. 2011 Jan 1;211(1):672-85.

Radebaugh J. Dunes on Saturn's moon Titan as revealed by the Cassini Mission. Aeolian Research. 2013 Dec 1;11:23-41.

Radebaugh J, Lopes RM, Howell RR, et al. Eruptive behavior of the Marum/Mbwelesu lava lake, Vanuatu and comparisons with lava lakes on Earth and Io. Journal of Volcanology and Geothermal Research. 2016 Aug 15;322:105-18.

Radebaugh, J., Ventra, D., Lorenz, R.D., Farr, T., Kirk, R., Hayes, A., Malaska, M.J., Birch, S., Liu, Z.Y.C., Lunine, J. and Barnes, J., 2018. Alluvial and fluvial fans on Saturn's moon Titan reveal processes, materials and regional geology.

Raulin F. Organic chemistry in the oceans of Titan. Advances in Space Research. 1987 Jan 1;7(5):71-81.

Raulin F, Brassé C, Poch O, Coll P. Prebiotic-like chemistry on Titan. Chemical Society Reviews. 2012;41(16):5380-93.

Read PL, Lebonnois S. Superrotation on Venus, on Titan, and elsewhere. Annual Review of Earth and Planetary Sciences. 2018 May 30;46:175-202.

Rees, D. (ed.), CIRA 1986, Part I Thermospheric Models,Advances in Space Research (COSPAR), Volume 8, Issues 5-6, pages 1-471, 1988

Rees, D., J.J. Barnett, K. Labitzke (ed.), CIRA 1986, Part II Middle Atmosphere Models, Advances in Space Research (COSPAR), Volume 10, Issue 12, pages 1-519, 1990.

Riedel M, Willoughby EC, Chen MA, He T, Novosel I, Schwalenberg K, Hyndman RD, Spence GD, Chapman NR, Edwards RN. Gas hydrate on the northern Cascadia margin: regional geophysics and structural framework. Proceedings of the Integrated Ocean Drilling Program Volume 311. 2006 Oct 28; DOI: 10.2204/iodp.proc.311.109.2006.

Risacher F, Fritz B. Quaternary geochemical evolution of the salars of Uyuni and Coipasa, Central Altiplano, Bolivia. Chemical Geology. 1991 Jun 25;90(3-4):211-31.

Rodriguez S, Le Mouélic S, Rannou P, et al. Titan's cloud seasonal activity from winter to spring with Cassini/VIMS. Icarus. 2011 Nov 1;216(1):89-110.

Rodriguez, S.; Garcia, A.; Lucas, et al., Global mapping and characterization of Titan's dune fields with Cassini: Correlation between RADAR and VIMS observations, Icarus, Volume 230, p. 168-179, 2014.

Rodriguez S, et al. Science goals and new mission concepts for future exploration of Titan's atmosphere, geology, and habitability: Titan POlar scout/orbitEr and in situ lake lander and DrONe explorer (POSEIDON). Experimental Astronomy. 2022.

Rogner HH. An assessment of world hydrocarbon resources. Annual review of energy and the environment. 1997 Nov;22(1):217-62.

Rosenfeld D, Woodley WL, Krauss TW, Makitov V. Aircraft microphysical documentation from cloud base to anvils of hailstorm feeder clouds in Argentina. Journal of applied meteorology and climatology. 2006 Sep;45(9):1261-81.

Roble RG. Energetics of the mesosphere and thermosphere. The upper mesosphere and lower thermosphere: a review of experiment and theory. 1995 Jan 1;87:1-21.

Ross AS, Farrimond P, Erdmann M, Larter SR 2010. Geochemical compositional gradients in a mixed oil reservoir indicative of ongoing biodegradation. Organic Geochemistry. 41(3): 307-320.

Ruff SW, Farmer JD. Silica deposits on Mars with features resembling hot spring biosignatures at El Tatio in Chile. Nature Communications. 2016 Nov 17;7(1):1-0.

Runge J. The Congo River, Central Africa. Large Rivers: Geomorphology and Management, Second Edition. 2022 Jun 13:433-56.

Ruppel CD, Waite, WF. Timescales and Processes of Methane Hydrate Formation and Breakdown, With Application to Geologic Systems. JGR Solid Earth. 2020; 125, 1-43. https://doi.org/10.1029/2018JB016459.

Rusch DW, Thomas GE, McClintock W, Merkel AW, Bailey SM, Russell III JM, Randall CE, Jeppesen C, Callan M. The cloud imaging and particle size experiment on the aeronomy of ice in the mesosphere mission: Cloud morphology for the northern 2007 season. Journal of atmospheric and solar-terrestrial physics. 2009 Mar 1;71(3-4):356-64.

Rutishauser A, Blankenship DD, Sharp M, et al. Discovery of a hypersaline subglacial lake complex beneath Devon Ice Cap, Canadian Arctic. Science advances. 2018 Apr 11;4(4):eaar4353.

Saffer DM, Tobin HJ. Hydrogeology and mechanics of subduction zone forearcs: Fluid flow and pore pressure. Annual Review of Earth and Planetary Sciences. 2011 May 30;39(1):157-86.

Sagan C. Sulphur flows on Io. Nature. 1979 Aug 30;280:750-3.

Sagan C, Thompson WR. Production and condensation of organic gases in the atmosphere of Titan. Icarus. 1984 Aug 1;59(2):133-61.

Sala R, Deom JM, Aladin NV, Plotnikov IS, Nurtazin S. Geological history and present conditions of Lake Balkhash. InLarge Asian Lakes in a Changing World: Natural State and Human Impact 2020 May 12 (pp. 143-175). Cham: Springer International Publishing.

Salameh, E., Khoury, H., Reimold, W.U. and Schneider, W., 2008. The first large meteorite impact structure discovered in the Middle East: Jebel Waqf as Suwwan, Jordan. Meteoritics & Planetary Science, 43(10), pp.1681-1690.

Salem S, Moslem WM, Radi A. Expansion of Titan atmosphere. Physics of Plasmas. 2017 May 1;24(5).

Salomon, J.-N., Précis de Karstologie, 2nd ed., Presses Universitaires de Bordeaux. 2006

Sandström H, Rahm M. Can polarity-inverted membranes self-assemble on Titan?. Science Advances. 2020 Jan 24;6(4):eaax0272.

Santibáñez PA, Michaud AB, Vick-Majors TJ, et al. Differential incorporation of bacteria, organic matter, and inorganic ions into lake ice during ice formation. Journal of Geophysical Research: Biogeosciences. 2019 Mar;124(3):585-600.

Sauro F, Payler SJ, Massironi M, et al. Training astronauts for scientific exploration on planetary surfaces: The ESA PANGAEA programme. Acta Astronautica. 2022 Dec 26.

Schaufelberger A, Wurz P, Lammer H, Kulikov YN. Is hydrodynamic escape from Titan possible?. Planetary and space science. 2012 Feb 1;61(1):79-84.

Scheidegger AE. Morphotectonics. Berlin: Springer; 2004 May.

Schenk, P.M., 2002. Thickness constraints on the icy shells of the Galilean satellites from a comparison of crater shapes. Nature, 417(6887), pp.419-421.

Schlichting HE, Mukhopadhyay S. Atmosphere impact losses. Space Science Reviews. 2018 Feb;214(1):34.

Schmidt B. Exploring Europan Analogs of the McMurdo Sound and Ross Ice Shelf. 42nd COSPAR Scientific Assembly. 2018 Jul;42:B5-3.

Schmidt BE, Lawrence JD, Meister MR, Dicheck DJ, Hurwitz BC, Spears A, Mullen AD, Washam PM, Bryson FE, Quartini E, Ramey CD. Europa in our backyard: under ice robotic exploration of Antarctic analogs. LPI Contrib.. 2020 Mar;2326:1065.

Schneck UG, Ashton AD, Birch S, Perron JT. Sediment entrainment by waves and tides on Titan. In AGU Fall Meeting Abstracts 2022 Dec (Vol. 2022, pp. P46B-08).

Schoenfeld A, Lopes R, Malaska M, et al. Geomorphological map of the south Belet region of Titan. Icarus. 2021 Jul;366:114516.

Schoenfeld A, Solomonidou A, Malaska M, et al. Geomorphological map of the Soi crater region on Titan. Journal of Geophysical Research: Planets. 2023 Apr;128(4):e2022JE007499.

Schaber, G. G. The U.S. Geological Survey, Branch of Astrogeology—A Chronology of Activities from Conception through the End of Project Apollo (1960-1973), U.S. Geological Survey Open-File Report 2005-1190, 2005.

Schoeberl MR, Hartmann DL. The dynamics of the stratospheric polar vortex and its relation to springtime ozone depletions. Science. 1991 Jan 4;251(4989):46-52.

Schulze-Makuch D, Haque S, de Sousa Antonio MR, Ali D, Hosein R, Song YC, Yang J, Zaikova E, Beckles DM, Guinan E, Lehto HJ. Microbial life in a liquid asphalt desert. Astrobiology. 2011 Apr 1;11(3):241-58.

Schurmeier LR, Dombard AJ. Crater relaxation on Titan aided by low thermal conductivity sand infill. Icarus. 2018 May 1;305:314-23.

Schurmeier LR, Dombard AJ, Malaska MJ, Fagents SA, Radebaugh J, Lalich DE. An intrusive cryomagmatic origin for northern radial labyrinth terrains on Titan and implications for the presence of crustal clathrates. Icarus. 2023 Nov 1;404:115664.

Schurmeier, L.R., Brouwer, G.E., Kay, J.P., Fagents, S.A., Marusiak, A.G. and Vance, S.D., 2024. Rapid Impact Crater Relaxation Caused by an Insulating Methane Clathrate Crust on Titan. The Planetary Science Journal, 5(9), p.211.

Schurmeier, L. and Fagents, S., 2025 "Titan's Lithospheric Strength Envelope and Brittle-Ductile Transition: The Importance of Crustal Pore Fluids, Organics, and Clathrates" Journal of Geophysical Research: Planets, 130(9), p.e2025JE009118.

Schurmeier, L.R., Brouwer, G.E., Wakita, S., Soderblom, J.M., Neish, C.D. and Johnson, B.C., 2025. Titan's Methane Clathrate Crust Constrained by Crater Depths Using Combined Impact and Relaxation Modeling. LPI Contributions, 3090, p.2580.

Scott SG, Bui TP, Chan KR, Bowen SW. The meteorological measurement system on the NASA ER-2 aircraft. Journal of Atmospheric and Oceanic Technology. 1990 Aug;7(4):525-40.

Seignovert B, Le Mouélic S, Brown RH, Joseph E, Karkoschka E, Pasek V, Sotin C, Turtle EP. Titan’s global map combining VIMS and ISS mosaics. In50th Annual Lunar and Planetary Science Conference 2019 Sep 15 (No. 2132, p. 1423).

Seltzer, C., Martel, S. J., Perron, J. T., 2025. Topographic stress as a mechanical weathering mechanism on Titan. J. Geophys. Res. Planets 130, e2024JE008873.

Šepić J, Rabinovich AB. Meteotsunami in the Great Lakes and on the Atlantic coast of the United States generated by the “derecho” of June 29–30, 2012. InMeteorological Tsunamis: The US East Coast and Other Coastal Regions 2014 Nov 16 (pp. 75-107). Cham: Springer International Publishing.

Shah, J., Neish, C.D. and Trozzo, S., 2025. An analogue study of impact craters on Titan: Implications for Titan's surface age. Icarus, 433, p.116536.

Sharma, P. and Byrne S., Comparison of Titan's north polar lakes with terrestrial analogs, Geophys. Res. Lett., 38, doi:10.1029/2011GL049577, 2011.

Shematovich VI, Johnson RE, Michael M, Luhmann JG. Nitrogen loss from Titan. Journal of Geophysical Research: Planets. 2003 Aug;108(E8).

Shoemaker, Interpretation of lunar craters. Physics and Astronomy of the Moon. 1962:283-359.

Shoemaker EM, Macdonald FA, Shoemaker CS. Geology of five small Australian impact craters. Australian Journal of Earth Sciences. 2005 Sep 1;52(4-5):529-44.

Shuman NS, Hunton DE, Viggiano AA. Ambient and modified atmospheric ion chemistry: From top to bottom. Chemical reviews. 2015 May 27;115(10):4542-70.

Sieh K, Wallace RE. The San Andreas fault at Wallace Creek, San Luis Obispo County, California.

Sinclair AT. The orbits of tethys, dione, rhea, titan and iapetus. Monthly Notices of the Royal Astronomical Society. 1977 Oct 1;180(3):447-59.

Skjöld M, Beauchemin DE, Magnússon BM, Kroell H, Valencia JJ. Lost in Space: A Framework Analysis on the Space Sector in Iceland. New Space. 2024 Dec 1;12(4):233-44.

Snowden D, Higgins A. A Monte Carlo model of energy deposition, ionization, and sputtering due to thermal ion precipitation into Titan's upper atmosphere. Icarus. 2021 Jan 15;354:113929.

Soderblom LA, Tomasko MG, Archinal BA, Becker TL, Bushroe MW, Cook DA, Doose LR, Galuszka DM, Hare TM, Howington-Kraus E, Karkoschka E. Topography and geomorphology of the Huygens landing site on Titan. Planetary and Space Science. 2007 Nov 1;55(13):2015-24.

Soderblom JM, Brown RH, Soderblom LA, et al. Geology of the Selk crater region on Titan from Cassini VIMS observations. Icarus. 2010 Aug 1;208(2):905-12.

Soderlund KM. Ocean dynamics of outer solar system satellites. Geophysical Research Letters. 2019 Aug 16;46(15):8700-10.

Soderlund KM, Kalousová K, Buffo JJ, Glein CR, Goodman JC, Mitri G, Patterson GW, Postberg F, Rovira-Navarro M, Rückriemen T, Saur J. Ice-ocean exchange processes in the Jovian and Saturnian satellites. Space Science Reviews. 2020 Aug;216:1-57.

Soderlund KM, Rovia-Navarro M, Le Bars M, Schmidt BE, Gerkema T. The physical oceanography of ice-covered moons, Annual Reviews of Marine Sciences 16, (2023).

Sohl F, Sears WD, Lorenz RD. Tidal dissipation on Titan. Icarus. 1995 Jun 1;115(2):278-94.

Solazzo D, Sankey JB, Sankey TT, Munson SM. Mapping and measuring aeolian sand dunes with photogrammetry and LiDAR from unmanned aerial vehicles (UAV) and multispectral satellite imagery on the Paria Plateau, AZ, USA. Geomorphology. 2018 Oct 15;319:174-85.

Solomonidou A, Bampasidis G, Hirtzig M, et al. Morphotectonic features on Titan and their possible origin. Planetary and Space Science. 2013 Mar 1;77:104-17.

Solomonidou A, Coustenis A, Lopes R, et al. The spectral nature of Titan's major geomorphological units: Constraints on surface composition. Journal of Geophysical Research: Planets. 2018 Feb;123(2):489–507.

Solomonidou A, Le Gall A, Malaska M, et al. Spectral and emissivity analysis of the raised ramparts around Titan's northern lakes. Icarus. 2020a Nov;344:113338.

Solomonidou A, Neish C, Coustenis A, et al. The chemical composition of impact craters on Titan I. Implications for exogenic processing. Astronomy & Astrophysics. 2020b Sep;641:A16.

Solomonidou A, Malaska MJ, Lopes RMC,  et al. (2024). Detailed chemical composition analysis of the Soi crater on Titan. Icarus, 421, 116215.

Sotin C, Kalousová K, Tobie G. Titan's interior structure and dynamics after the Cassini-Huygens mission. Annual Review of Earth and Planetary Sciences. 2021 May 30;49:579-607.

Sozinov SA, Sotnikova LV, Popova AN, Hitsova LM, Lyrschikov SY, Efimova OS, Malysheva VY, Rusakov DM. Composition and structure of hexane-insoluble asphaltenes from coal pitch. Coke and Chemistry. 2020 Jan;63(1):26-34.

Spears A, West M, Meister M, Buffo J, Walker C, Collins TR, Howard A, Schmidt B. Under ice in antarctica: The icefin unmanned underwater vehicle development and deployment. IEEE Robotics & Automation Magazine. 2016 Nov 7;23(4):30-41.

Spray JG, Butler HR, Thompson LM. Tectonic influences on the morphometry of the Sudbury impact structure: Implications for terrestrial cratering and modeling. Meteoritics & Planetary Science. 2004 Feb;39(2):287-301.

Staff AW. Water Meters: Selection, Installation, Testing, and Maintenance. American Water Works Association; 1962.

Stainforth JG. Practical kinetic modeling of petroleum generation and expulsion. Marine and Petroleum Geology. 2009 Apr; 26(4): 552-572.

Stainforth JG. Highly efficient secondary migration of petroleum in colloidal dispersions. 26th International Meeting on Organic Geochemistry Abstracts. 2013 Sep 15; 1: 384-385

Stainforth JG. Compositional grading in petroleum fields: Unification of trap-filling and gravitational segregation models. 26th International Meeting on Organic Geochemistry Abstracts. 2013 Sep 15; 1: 111-112

Steckloff J, Soderblom J, Farnsworth K, et al. Stratification Dynamics of Titan's Lakes via Methane Evaporation. The Planetary Science Journal. 2020;1(2);26.

Steele A, Fries MD, Amundsen HE, et al. Comprehensive imaging and Raman spectroscopy of carbonate globules from Martian meteorite ALH 84001 and a terrestrial analogue from Svalbard. Meteoritics & Planetary Science. 2007 Sep;42(9):1549-1566.

Steele HM, Hamill P, McCormick MP, Swissler TJ. The formation of polar stratospheric clouds. Journal of Atmospheric Sciences. 1983 Aug;40(8):2055-68.

Stephan K, Jaumann R, Brown RH, Soderblom JM, Soderblom LA, Barnes JW, Sotin C, Griffith CA, Kirk RL, Baines KH, Buratti BJ. Specular reflection on Titan: liquids in Kraken Mare. Geophysical Research Letters. 2010 Apr;37(7).

Stevenson J, Lunine J, Clancy P. Membrane alternatives in worlds without oxygen: Creation of an azotosome. Science advances. 2015 Feb 27;1(1):e1400067

Stiles BW, Hensley S, Gim Y, Bates DM, Kirk RL, Hayes A, Radebaugh J, Lorenz RD, Mitchell KL, Callahan PS, Zebker H. Determining Titan surface topography from Cassini SAR data. Icarus. 2009 Aug 1;202(2):584-98.

Stofan ER, Elachi C, Lunine JI, et al. The lakes of Titan. Nature. 2007 Jan;445(7123):61-4, doi:10.1038/nature05438.

Stofan E, Lorenz R, Lunine J, Bierhaus EB, Clark B, Mahaffy PR, Ravine M. Time-the titan mare explorer. In2013 IEEE aerospace conference 2013 Mar 2 (pp. 1-10). IEEE.

Strobel DF. The ionospheres of the major planets. Reviews of Geophysics. 1979 Nov;17(8):1913-22.

Strobel DF. Titan's hydrodynamically escaping atmosphere. Icarus. 2008 Feb 1;193(2):588-94.

Strobel DF. Titan's hydrodynamically escaping atmosphere: escape rates and the structure of the exobase region. Icarus. 2009 Aug 1;202(2):632-41.

Strobel DF. Hydrogen and methane in Titan’s atmosphere: Chemistry, diffusion, escape, and the Hunten limiting flux principle. Canadian Journal of Physics. 2012 Aug;90(8):795-805.

Sulaiman A, et al. Enceladus and Titan: emerging worlds of the Solar System. Experimental Astronomy.

Sultana RE, Le Gall A, Tokano T, et al. Taking Titan’s Boreal Pole Temperature: Evidence for Evaporative Cooling in Ligeia Mare. The Astrophysical Journal. 2024 Feb 1;961(2);191.

Tabazadeh A, Turco RP, Drdla K, Jacobson MZ, Toon OB. A study of type I polar stratospheric cloud formation. Geophysical Research Letters. 1994 Jul 15;21(15):1619-22.

Tackley, P.J., 2000. Mantle convection and plate tectonics: Toward an integrated physical and chemical theory. *Science*, *288*(5473), pp.2002-2007.

Taira K, Yanagimoto D, Kitagawa S. Deep CTD casts in the challenger deep, Mariana Trench. Journal of Oceanography. 2005 Jun;61:447-54.

Teanby NA, de Kok R, Irwin PG, et al. Titan's winter polar vortex structure revealed by chemical tracers. Journal of Geophysical Research: Planets. 2008 Dec;113(E12).

Telfer MW, Radebaugh J, Cornford B, Lewis C. Long-wavelength sinuosity of linear dunes on Earth and Titan and the effect of underlying topography. Journal of Geophysical Research: Planets. 2019 Sep;124(9):2369-81.

Teller JT, Rutter N, Lancaster N. Sedimentology and paleohydrology of Late Quaternary lake deposits in the northern Namib Sand Sea, Namibia. Quaternary Science Reviews. 1990 Jan 1;9(4):343-64.

Terra-Nova F, Amit H, Choblet G, et al. The influence of heterogeneous seafloor heat flux on the cooling patterns of Ganymede’s and Titan’s subsurface oceans. Icarus. 2023 Jan 1;389:115232.

Tewelde Y, Perron JT, Ford P, et al. Estimates of fluvial erosion on Titan from sinuosity of lake shorelines. Journal of Geophysical Research: Planets. 2013;118(10);2198-2212.

Thompson AF. The atmospheric ocean: Eddies and jets in the Antarctic Circumpolar Current. Philosophical Transactions of the Royal Society A: Mathematical, Physical and Engineering Sciences. 2008 Dec 28;366(1885):4529-41.

Thompson LG. Ice core evidence for climate change in the Tropics: implications for our future. Quaternary Science Reviews. 2000 Jan 1;19(1-5):19-35.

Thompson, W.B., Johnston, R., Hildreth, C., and Retelle, M.J.,1995, Glacial Geology of the Portland-Sebago Lake area, in Guidebook to Field trips in Southern Maine and adjacent New Hampshire, New England Intercollegiate Geological Conference 87th Annual Meeting, Brunswick, Maine. p. 51-70.

Tobie G, Lunine JI, Sotin C. Episodic outgassing as the origin of atmospheric methane on Titan. Nature. 2006 Mar 2;440(7080):61-4.

Tobie G, Lunine JI, Monteux J, Mousis O, Nimmo F. The origin and evolution of Titan. Titan: interior, surface, atmosphere, and space environment. 2014a Mar:29-62.

Tobie G, et al. Science goals and mission concept for the future exploration of Titan and Enceladus. Planetary and Space Science. 2014b;104:59–77.

Tokano T, Neubauer FM, Laube M, McKay CP. Three-dimensional modeling of the tropospheric methane cycle on Titan. Icarus. 2001 Sep 1;153(1):130-47.

Tokano T. Relevance of fast westerlies at equinox for the eastward elongation of Titan's dunes. Aeolian Research. 2010 Nov 1;2(2-3):113-27.

Tokano T. Precipitation climatology on Titan. Science. 2011 March 18, 331, 1393-1394.

Tokano T, Lorenz RD. Wind-driven circulation in Titan's seas. J. Geophys. Res. Planets. 2015 Jan;120(1);20–33.

Tomasko MG, Archinal B, Becker T, et al. Rain, winds and haze during the Huygens probe's descent to Titan's surface. Nature. 2005 Dec; 438(7069):765-78.

Tornabene, L.L., Moersch, J.E., Osinski, G.R., Lee, P. and Wright, S.P., 2005. Spaceborne visible and thermal infrared lithologic mapping of impact-exposed subsurface lithologies at the Haughton impact structure, Devon Island, Canadian High Arctic: Applications to Mars. Meteoritics & Planetary Science, 40(12), pp.1835-1858.

Torr DG, Torr MR. Chemistry of the thermosphere and ionosphere. Journal of Atmospheric and Terrestrial Physics. 1979 Jul 1;41(7-8):797-839.

Tritscher I, Pitts MC, Poole LR, Alexander SP, Cairo F, Chipperfield MP, Grooß JU, Höpfner M, Lambert A, Luo B, Molleker S. Polar stratospheric clouds: Satellite observations, processes, and role in ozone depletion. Reviews of geophysics. 2021 Jun;59(2):e2020RG000702.

Tucker OJ, Johnson RE. Thermally driven atmospheric escape: Monte Carlo simulations for Titan's atmosphere. Planetary and Space Science. 2009 Dec 1;57(14-15):1889-94.

Tulaczyk S, Mikucki JA, Siegfried MR, Priscu JC, Barcheck CG, Beem LH, Behar A, Burnett J, Christner BC, Fisher AT, Fricker HA. WISSARD at Subglacial Lake Whillans, West Antarctica: scientific operations and initial observations. Annals of Glaciology. 2014 Jan;55(65):51-8.

Turcotte, D.L., Schubert, G., 2002. Geodynamics, 2nd ed. Cambridge University Press, New York.

Turk LJ, Davis SN, Bingham CP. Hydrogeology of lacustrine sediments, Bonneville Salt Flats, Utah. Economic Geology. 1973 Jan 1;68(1):65-78.

Turtle EP, Perry JE, Hayes AG, et al. Rapid and extensive surface changes near Titan's equator: Evidence of April showers. science. 2011a Mar 18;331(6023):1414-7.

Turtle EP, Perry JE, Hayes AG, McEwen AS. Shoreline retreat at Titan's Ontario Lacus and Arrakis Planitia from Cassini imaging science subsystem observations. Icarus. 2011b Apr 1;212(2):957-9.

Turtle EP, Perry JE, Barbara JM, et al. Titan's meteorology over the Cassini mission: Evidence for extensive subsurface methane reservoirs. Geophysical Research Letters. 2018 Jun 16;45(11):5320-8.

Turtle, E.P. and Lorenz, R.D., 2024, March. Dragonfly: In Situ Aerial Exploration to Understand Titan's Prebiotic Chemistry and Habitability. In 2024 IEEE Aerospace Conference (pp. 1-5). IEEE.

Udelhofen PM, Hartmann DL. Influence of tropical cloud systems on the relative humidity in the upper troposphere. Journal of Geophysical Research: Atmospheres. 1995 Apr 20;100(D4):7423-40.

Ulrich M, Hauber E, Herzschuh U, et al. Polygon pattern geomorphometry on Svalbard (Norway) and western Utopia Planitia (Mars) using high-resolution stereo remote-sensing data. Geomorphology. 2011 Nov 15;134(3-4):197-216.

Umurhan OM, Allan MB, Edwards LJ, Tardy A, Welsh TM, Wong U. High resolution digital elevation models of The Devil's Golf Course: a possible terrestrial analog of Europa's surface. InAGU Fall Meeting Abstracts 2019 Dec (Vol. 2019, pp. P53C-3468).

Umurhan, O. M., Birch, S. P. D., Hayes, A. G., & Malaska, M. J. (2020). Simulating the evolution of Titan’s surface through fluvial and dissolution erosion: I. The big picture. In Presented at the Lunar and Planetary Science Conference 51, The Woodlands, TX, Abstract 1552.

Urrutia-Fucugauchi J, Morgan J, Stˆffler D, Claeys P. The Chicxulub scientific drilling project (CSDP). Meteoritics & Planetary Science Archives. 2004 Jan 1;39(6):787-90.

Valentine DL, Reddy CM, Farwell C, et al. Asphalt volcanoes as a potential source of methane to late Pleistocene coastal waters. Nature Geoscience. 2010 May;3(5):345-8.

Vance SD, Panning MP, Stähler S, et al. Geophysical investigations of habitability in ice-covered ocean worlds. Journal of Geophysical Research: Planets. 2018 Jan;123(1):180-205.

Vance SD, Barge LM, Cardoso SS, Cartwright JH. Self-assembling ice membranes on Europa: brinicle properties, field examples, and possible energetic systems in icy ocean worlds. Astrobiology. 2019 May 1;19(5):685-95.

Vanderburgh S, Smith DG. Slave River delta: geomorphology, sedimentology, and Holocene reconstruction. Canadian Journal of Earth Sciences. 1988 Dec 1;25(12):1990-2004.

Vargas-Cordero I, Umberta T, Villar-Munoz L. Gas hydrate and free gas concentrations in two sites inside the Chilean margin (Itata and Valdivia offshores). Energies. 2017 Dec 16; 10: 2154-2165.

Vaughan RG, Schindler K, Stevens J, et al. Walk in the footsteps of the Apollo astronauts: A field guide to northern Arizona astronaut training sites. Geologic Excursions in Southwestern North America. 2019 Sep 4;55:307.

Vennemann TW, Morlok A, von Engelhardt W, Kyser K. Stable isotope composition of impact glasses from the Nördlinger Ries impact crater, Germany. Geochimica et Cosmochimica Acta. 2001 Apr 15;65(8):1325-36.

Vera-Arroyo A, Bedle H. Contrasting faulting styles of salt domes and volcanoes: Can unsupervised learning techniques differentiate fault styles?. Interpretation. 2023 Feb 1;11(1):C1-3.

Veress M. Karst types and their karstification. Journal of Earth Science. 2020 Jun;31(3):621-34.

Vinatier S, Schmitt B, Bezard B, et al. Study of Titan's fall southern stratospheric polar cloud composition with Cassini/CIRS: Detection of benzene ice. Icarus. 2018 Aug 1;310:89-104.

Vincent D, Karatekin Ö, Lambrechts J, et al. A numerical study of tides in Titan's northern seas, Kraken and Ligeia Maria. Icarus. 2018 Aug;310;105–126.

Volkov AN, Johnson RE, Tucker OJ, Erwin JT. Thermally driven atmospheric escape: Transition from hydrodynamic to Jeans escape. The Astrophysical Journal Letters. 2011 Feb 16;729(2):L24.

Vuitton V, Yelle RV, McEwan MJ. Ion chemistry and N-containing molecules in Titan's upper atmosphere. Icarus. 2007 Nov 15;191(2):722-42.

Vuitton V, Lavvas P, Yelle RV, et al. Negative ion chemistry in Titan's upper atmosphere. Planetary and Space Science. 2009 Nov 1;57(13):1558-72.

Vuitton V, Dutuit O, Smith MA, et al. Chemistry of Titan’s atmosphere. In: Titan Interior, Surface, Atmosphere, and Space Environment (edited by Müller-Wodarg I, Griffith CA, Lellouch E, et al.), 2014, Cambridge University Press, Cambridge

Vuitton V, Yelle R, Klippenstein S J, Hörst S M, Lavvas P. Simulating the density of organic species in the atmosphere of Titan with a coupled ion-neutral photochemical model. Icarus. 2019 May 15;324;120-197.

Wagner AM, Barker AJ. Distribution of polycyclic aromatic hydrocarbons (PAHs) from legacy spills at an Alaskan Arctic site underlain by permafrost. Cold Regions Science and Technology. 2019 Feb 1;158:154-65.

Wakita S, Johnson BC, Soderblom JM, Shah J, Neish CD. Methane-saturated layers limit the observability of impact craters on Titan. The Planetary Science Journal. 2022 Feb 28;3(2):50.

Wakita, S., Johnson, B.C., Soderblom, J.M. and Neish, C.D., 2024. Revised Age Estimate for Titan's Surface Assuming a Methane Clathrate Layer. LPI Contributions, 3040, p.1146.

Wall S, Hayes A, Bristow C, Lorenz R, Stofan E, Lunine J, Le Gall A, Janssen M, Lopes R, Wye L, Soderblom L. Active shoreline of Ontario Lacus, Titan: A morphological study of the lake and its surroundings. Geophysical Research Letters. 2010 Mar;37(5).

Wallace JM, Hobbs PV. Atmospheric science: an introductory survey. Elsevier; 2006 Mar 24.

Waltham T. Karst and Caves of Ha Long Bay. Speleogenesis and Evolution of Karst Aquifers, 2005, 3 (2), p.2-9. https://digitalcommons.usf.edu/kip_articles/4799.

Waite Jr JH, Niemann H, Yelle RV, et al. Ion neutral mass spectrometer results from the first flyby of Titan. Science. 2005 May 13;308(5724):982-6.

Wang Z, Zhou Y, Liu Y. The escape mechanisms of the proto-atmosphere on terrestrial planets:“boil-off” escape, hydrodynamic escape and impact erosion. Acta Geochimica. 2022 Aug;41(4):592-606.

Wang M, Li Y, Wang H, Tao J, Li M, Shi Y, Zhou X. The neglected role of asphaltene in the synthesis of mesophase pitch. Molecules. 2024 Mar 27;29(7):1500.

Waugh DW. Fluid Dynamics of Polar Vortices on Earth, Mars, and Titan. Annual Review of Fluid Mechanics. 2023 Jan 19;55.

Werynski A, Neish CD, Le Gall A, et al. Compositional variations of Titan's impact craters indicates active surface erosion. Icarus. 2019 Mar 15;321:508-21.

Wendland F, Blum A, Coetsiers M, Gorova R, Griffioen J, Grima J, Hinsby K, Kunkel R, Marandi A, Melo T, Panagopoulos A. European aquifer typology: a practical framework for an overview of major groundwater composition at European scale. Environmental Geology. 2008 Jul;55(1):77-85.

Whiticar MJ, Elvert ME. Organic geochemistry of Saanich Inlet, BC, during the Holocene as revealed by Ocean Drilling Program Leg 169S. Marine Geology. 2001 Mar 15; 174: 249-271.

Whiticar MJ, Suess E. The Cold Carbonate Connection Between Mono Lake, Californiaand the Bransfield Strait, Antarctica. Aquatic Geochemistry. 1998 Sep; 4, 429–454. https://doi.org/10.1023/A:1009696617671

Williams, P.W., Morphometric analysis of polygonal karst in New Guinea. Geological Society of America Bulletin. 1972, v. 83, p. 761-796.

Williams DA, Radebaugh J, Lopes RM, Stofan E. Geomorphologic mapping of the Menrva region of Titan using Cassini RADAR data. Icarus. 2011 Apr 1;212(2):744-50.

Wik M, Crill PM, Bastviken D, et al. Bubbles trapped in arctic lake ice: Potential implications for methane emissions. Journal of Geophysical Research: Biogeosciences. 2011 Sep;116(G3).

Wilson EH, Atreya SK. Sensitivity studies of methane photolysis and its impact on hydrocarbon chemistry in the atmosphere of Titan. JGR. 2000, Aug 25; 105-E8, 20263-20273

Witek PP, Czechowski L. Dynamical modelling of river deltas on Titan and Earth. Planetary and Space Science. 2015;105;65-79.

Wolfenbarger NS, Buffo JJ, Soderlund KM, Blankenship DD. Ice shell structure and composition of ocean worlds: Insights from accreted ice on Earth. Astrobiology. 2022 Aug 1;22(8):937-61.

Wood CA, Lorenz R, Kirk R, et al. Impact craters on Titan. Icarus. 2010 Mar 1;206(1):334-44.

Wood CA, Radebaugh J. Morphologic evidence for volcanic craters near Titan's north polar region. Journal of Geophysical Research: Planets. 2020 Aug;125(8):e2019JE006036.

Wray, R.A.L. "Quartzite dissolution: karst or pseudokarst?" Cave and Karst Science 24 (2), 1997, 81-86.

Wray RA, Sauro F. An updated global review of solutional weathering processes and forms in quartz sandstones and quartzites. Earth-Science Reviews. 2017 Aug 1;171:520-57.

Wye LC, Zebker HA, Lorenz RD. Smoothness of Titan's Ontario Lacus: Constraints from Cassini RADAR specular reflection data. Geophysical Research Letters. 2009 Aug;36(16).

Yabuta H, Williams LB, Cody GD, et al. The insoluble carbonaceous material of CM chondrites: A possible source of discrete organic compounds under hydrothermal conditions. Meteoritics & Planetary Science. 2007 Jan;42(1):37-48.

Yelle RV, Cui J, Müller-Wodarg IC. Methane escape from Titan's atmosphere. Journal of Geophysical Research: Planets. 2008 Oct;113(E10).

Yu X, Yu Y, Garver J, et al. The Fate of Simple Organics on Titan's Surface: A Theoretical Perspective. Geophysical Research Letters. 2024 Jan 4;51(1); e2023GL106156.

Yung YL, Allen M, Pinto JP. Photochemistry of the atmosphere of Titan: Comparison between model and observations. Astrophysical Journal Supplement Series. 1984;55(3):465-506.

Zebker HA, Stiles B, Hensley S, Lorenz R, Kirk RL, Lunine J. Size and shape of Saturn's moon Titan. Science. 2009 May 15;324(5929):921-3.

Zechmeister L, Lijinsky W. Some neutral constituents of a natural tar originating from the LaBrea Pits. Archives of Biochemistry and Biophysics. 1953 Dec 1;47(2):391-5.

Zhang Z, Wang H, Zhu G, Sun C, Ge X, Chi L, Li J. Phase fractionation and oil mixing as contributors to complex petroleum phase in deep strata: A case study from LG7 block in the Tarim Basin, China. Marine and Petroleum Geology. 2022 Jun 1;140:105660.

Zhao W, Forte E, Fontolan G, Pipan M. Advanced GPR imaging of sedimentary features: Integrated attribute analysis applied to sand dunes. Geophysical Journal International. 2018 Apr;213(1):147-56.

Zuo P, Qu S, Shen W. Molecular growth from coal-based asphaltenes to spinnable pitch. Materials Chemistry and Physics. 2022 Jan 15;276:125427.